\documentclass[fleqn,usenatbib]{mnras}

\usepackage{newtxtext,newtxmath}

\usepackage[T1]{fontenc}
\usepackage{ae,aecompl}

\usepackage{graphicx}	
\usepackage{amsmath}	
\usepackage{amssymb}	
\usepackage{color}
\usepackage[normalem]{ulem}
\usepackage{rotating}
\usepackage{hyperref}
\usepackage{graphicx}
\usepackage{caption}
\usepackage{lscape}
\usepackage{float}
\usepackage{ulem}
\usepackage[caption = false]{subfig}
\usepackage{footnote}
\usepackage{tablefootnote}
\makesavenoteenv{tabular}
\makesavenoteenv{table}
\usepackage{tabularx}
\usepackage{adjustbox}
\usepackage{verbatim}
\usepackage{longtable}
\usepackage{appendix}

\def\aj{AJ}
\def\aaps{A\&AS}
\def\aap{A\&A}
\def\apj{ApJ}
\def\apjs{ApJS}

\def\kms{km\,s$^{-1}$}

\title[1720-MHz OH masers]{MAGMO: Polarimetry of 1720-MHz OH Masers towards Southern Star Forming Regions}

\author[C. S. Ogbodo et al.] 
{C. S. Ogbodo,$^{1,2,9}$\thanks{E-mail: chikaedu.ogbodo@mq.edu.au}
J. A. Green$^{2}$, J. R. Dawson$^{1}$, S. L. Breen$^{3}$, S. A. Mao$^{4}$,
\newauthor{N. M. McClure-Griffiths$^{5}$, T. Robishaw$^{6}$, and L. Harvey-Smith$^{7,8}$} \\
\\
$^{1}$Department of Physics and Astronomy and MQ Research Centre in Astronomy, Astrophysics and Astrophotonics,\\ Macquarie University, NSW 2109, Australia\\
$^{2}$Australia Telescope National Facility, CSIRO Astronomy and Space Science, PO Box 76, Epping, NSW 1710, Australia\\
$^{3}$Sydney Institute for Astronomy (SIfA), School of Physics, University of Sydney, NSW 2006, Australia\\
$^{4}$Max Planck Institute for Radio Astronomy, Auf dem Hugel 69, D-53121, Bonn, Germany\\
$^{5}$Research School of Astronomy and Astrophysics, Australian National University, Canberra ACT 2611, Australia\\
$^{6}$National Research Council, Herzberg Institute of Astrophysics, Dominion Radio Astrophysical
Observatory, PO Box 248,\\ Penticton, BC V2A 6J9, Canada\\
$^{7}$Deans Unit, Faculty of Science, UNSW Sydney, NSW 2052, Australia.\\
$^{8}$School of Computing, Engineering and Mathematics, Western Sydney University, Locked Bay 1797, Penrith NSW 2751, Australia\\
$^{9}$Department of Physics/Geology/Geophysics, Alex Ekwueme Federal University Ndufu-Alike, Ikwo, P.M.B. 1010,\\ Abakaliki, Ebonyi State, Nigeria.\\
}
\date{Accepted XXX. Received YYY; in original form ZZZ}

\pubyear{2018}

\begin{document}
\label{firstpage}
\pagerange{\pageref{firstpage}--\pageref{lastpage}}
\maketitle

\begin{abstract}
From targeted observations of ground-state OH masers towards 702 Multibeam (MMB) survey 6.7-GHz methanol masers, between Galactic longitudes 186$^{\circ}$ through the Galactic centre to 20$^{\circ}$, made as part of the `MAGMO' project, we present the physical and polarisation properties of the 1720-MHz OH maser transition, including the identification of Zeeman pairs. We present  10 new and 23 previously catalogued 1720-MHz OH maser sources detected towards star formation regions. In addition, we also detected 16 1720-MHz OH masers associated with supernova remnants and two sites of diffuse OH emission. Towards the 33 star formation masers, we identify 44 Zeeman pairs, implying magnetic field strengths ranging from $-$11.4 to $+$13.2 mG, and a median magnetic field strength of $|B_{LOS}|$ $\sim$ 6 mG. With limited statistics, we present the in-situ magnetic field orientation of the masers and the Galactic magnetic field distribution revealed by the 1720-MHz transition. We also examine the association statistics of 1720-MHz OH SFR masers with other ground-state OH masers, excited-state OH masers, class I and class II methanol masers and water masers, and compare maser positions with mid-infrared images of the parent star forming regions. Of the 33 1720-MHz star formation masers, ten are offset from their central exciting sources, and appear to be associated with outflow activity. 

\end{abstract}





\section{Introduction}
High-mass star forming regions are familiar hosts to maser emission from the paramagnetic molecule hydroxyl (OH), which is a reliable probe of the small-scale magnetic field in the early phases of star formation. Zeeman splitting measurements of maser emission lines can be used to determine magnetic field strengths and line-of-sight direction in star forming regions. The 1720-MHz ($^{2}\Pi_{3/2} J = 3/2, F = 2 - 1$) OH maser transition is one of the ground-state satellite lines, and is primarily found in star forming sites \citep{Caswell99} and supernova remnants \citep{Green97} but has also been seen towards a small number of post-AGB stars and PNe \citep[e.g.][]{Sevenster2001,qiao16a}. Detections in star forming regions commonly accompany ground and excited-state OH main lines \citep[e.g.][]{Caswell2004}. The 1720-MHz transition requires a much stricter set of conditions compared to the ground-state main lines (e.g. column densities $N$(OH)$/\Delta V>$ 10$^{10}$ cm$^{-3}$ s; \citealt{Cragg2002}).

The four ground-state levels of OH are split in the presence of magnetic fields into magnetic hyperfine substates -- a phenomenon possible because of the magnetic moment of diatomic molecules like OH \citep[e.g][]{Gray1995}. Transitions between these substates are denoted by $\sigma$$^{+}$ which is due to lower$\rightarrow$upper substate transitions and $\sigma$$^{-}$ which is due to upper$\rightarrow$lower substate transitions. 
According to the Institute of Electrical and Electronics Engineers (IEEE) convention, these  $\sigma^{\pm}$ components correspond to right-hand circular polarisation (RHCP) and left-hand circular polarisation (LHCP). OH maser emission is frequently observed to have Zeeman pairs, with the RHCP and LHCP components separated in frequency with consequently different local standard of rest (LSR) velocities. The LSR velocity separation between circular polarization components of OH masers associated with star forming regions are typically larger than their spectral linewidth leading to clearly resolved Zeeman splitting. In this case, magnetic field strength is directly proportional to the velocity separation and is evaluated by applying the Zeeman splitting (Lande) factor corresponding to the transition. The line of sight magnetic field orientation is given by which of the RHCP or LHCP components is shifted to higher and lower frequencies.

Magnetic fields are pervasive across all scales in the Galaxy, but restricting their study to star forming regions allows us to investigate magnetic fields on smaller scales than recent larger scale studies in the diffuse interstellar medium primarily using Faraday rotation measures \citep{Han2006,brown2007,Mao2010,Vaneck2011}. It is to this end that the MAGMO project (Mapping the Galactic Magnetic field through OH masers) was designed, with initial results from the pilot study \citep{Green2012pilot} focusing on the main line transitions of sources located near the Carina-Sagittarius spiral arm tangent ($280^{\circ} < l < 295^{\circ}$). \citet{Green2012pilot} further describes the details and the motivation behind the MAGMO project.

As the first in a series of results, this work undertakes a polarimetric and magnetic field investigation using the 1720-MHz OH maser transition. In this paper, we focus only on the 1720-MHz line, and only include detections in the other three ground-state OH lines when 1720-MHz maser emission is found. Full analysis and results of the main-lines will be the subject of a future publication. 

In Section \ref{section2}, we summarise the observations and data processing; while Section \ref{section3} presents results from the reduced data; and finally, Section \ref{section4} discusses our results and analysis.

\section{Observations, data reduction and analysis}
\label{section2}
Observations were made with a 6-km array configuration (6B) of the Australia Telescope Compact Array (ATCA) over the period 2010 May to 2012 June with project code C2291 as part of the MAGMO project \citep{Green2012pilot}, and are available in the raw ATCA data format online via the Australia Telescope Online Archive\footnote{https://atoa.atnf.csiro.au}. Observations utilised the CFB 1M-0.5k mode of the Compact Array Broadband Backend, \citep[CABB;][]{wilson11} with `zoom' bands configured to cover the four ground-state transitions of hydroxyl at 1612.2310, 1665.4018, 1667.3590 and 1720.5300 MHz \citep{Lovas1979}. The correlator was configured with 4096 channels across 4 $\times$ 2 MHz giving a channel width of 0.5\,kHz and a velocity channel width of 0.091, 0.088, 0.088 and 0.085\,km\,s$^{-1}$ for the transitions as stated in the order above. 
Based on the evidence that 6.7-GHz methanol masers solely trace sites of high-mass star formation \citep[e.g.][]{Minier2003, Breen2013}, targeted observations were made of 6.7-GHz methanol masers, detected in the MMB survey \citep{Caswell2010,Green2010,Caswell2011,Green2012MMB, Breen2015} in the Galactic longitude range 186$^{\circ}$ through the Galactic Centre to 20$^{\circ}$. For each of the MMB targets, five observations of six minutes were typically made across a time span of 10 to 12 hours to obtain full synthesis and an average integration time of 30 minutes. Sources were clustered in groups of 5 or fewer such that they could be bracketed with observations of an appropriate phase calibrator (within 10$^{\circ}$ of each source). PKS B1934$-$638 was observed for both absolute flux calibration and bandpass calibration (with the exception of 189.032+0.809 and 189.776+0.346 for which PKS B0823$-$500 was used). As an integral part of the observations, the ATCA's dipole linear receivers provided the full polarisation products, XX, YY, XY and YX, to enable the generation of the Stokes parameters $I$, $Q$, $U$ and $V$. Multiple observations of the phase calibrators yielded good parallactic angle coverage which enabled  calibration of instrumental polarisation leakages. In some cases the phase calibrator had significant intrinsic polarisation and so PKS B1934$-$638 which has no detected linear polarisation and circular polarisation of the order 0.01\% at these frequencies \footnote{https://www.narrabri.atnf.csiro.au/calibrators}, was used instead. The median value of the leakage solutions with the phase calibrators was 0.05\% for Stokes Q and 0.01\% for Stokes U. PKS B1934$-$638 solutions are typically good to 0.01\% for both Q and U. See Table \ref{Table1} for the list of calibrators used.  
The Australia Telescope National Facility's data processing software package, {\em MIRIAD}, was used for the data reduction. For every targeted field, cleaned Stokes I image cubes were generated, and inspected spatially and spectrally for emission. The size of the synthesized beam at full width at half-maximum (FWHM) in these images was typically $\sim$7 arcsec in right-ascension and between $\sim7$--16 arcsec in declination, depending on source declination. Typical per-channel rms noise levels ($\sigma$) were $\sim$40 mJy. Where emission was present above the $5\sigma$ detection limit of 0.2\,Jy, source positions were obtained from a two-dimensional parabolic fit in the image plane containing the brightest spectral feature. We assumed the rms positional error in each coordinate was $\sim$0.4 arcsec based on \citet[][]{Caswell98}.   

Stokes $I$, $Q$, $U$ and $V$ spectra were then extracted (with an appropriate phase shift to the fitted position) from the $uv$ data. Spectra were scaled by primary beam correction factors for sources offset from the pointing centre. The final set of linear (lin), and circular (RHCP and LHCP) spectra were obtained from the Stokes $I$, $Q$, $U$ and $V$ products using $lin = \sqrt{Q^2+U^2}$, $RHCP = \frac{I+V}{2}$ and $LHCP = \frac{I-V}{2}$. 

The RHCP and LHCP profiles were inspected for Zeeman pairs, which are identified on the basis of their spectral profile shapes, and we have assumed that we are in the `far' splitting regime where the split components are clearly separated in frequency. We note that this approach is common in magnetic field studies with masers \citep[e.g.][]{Davies1974,Fish03b,Caswell2004,caswell2013,caswell2014}, and and does not require the differential of the Stokes I to determine Zeeman pairs.
We then performed Gaussian fits to the spectra to obtain centroid velocities for the identified components. We qualify our certainty of the Zeeman pair association in spectral and spatial domains with letters: class ``A" refers to robustly identified Zeeman pairs; i.e. those with comparable profile shapes and FWHM (identical to within 1$\sigma$), and for which there is minimal ambiguity in the association of corresponding RHCP and LHCP components; and class ``B" refers to cases where multiple components are blended spectrally, and for which the identification of corresponding RHCP and LHCP pairs is ambiguous. These classifications refer purely to the quality of the Zeeman pair identification; we separately define the statistical significance of fields later based on the velocity separation and uncertainties. The line-of-sight magnetic field strengths are derived from the velocity differences of the Zeeman-pairs, using the Lande splitting factor of the 1720-MHz OH line \citep[0.113 \kms mG$^{-1}$,][]{Palmer1967}. Uncertainties associated with the magnetic field strengths are derived in quadrature using half of the spectral velocity channel width (0.085 \kms) and the fitted centroid error of the associated component. 

As a result of the quadrature summation of $Q$ and $U$, the linearly polarised intensity ends up as a positive quantity, leading to a polarisation bias that has to be corrected for. For non-detections ($lin < 3\sigma$), a robust upper limit may be estimated by $lin_{UL} = lin + 2\sigma$ \citep{Vaillancourt2006}. Here we assumed that the uncertainties on Stokes $U$ and $Q$ are identical, given by the spectral rms noise, $\sigma$, and that this value also describes the uncertainty on $lin$. 
For cases with detections above the 4$\sigma$ level, we corrected for polarisation bias using $lin_{corr} = \sqrt{lin^{2} - \sigma^{2}}$ \citep{Simmons1985}. Based on these corrections, the fractional linear polarisation ($P_l$) is then either given by $P_l = \frac{lin_{UL}}{I}$ or $P_l = \frac{lin_{corr}}{I}$. Assuming that the polarisation fraction noise is Gaussian and that Stokes $Q$ and $U$ are uncorrelated, we may adequately quantify the associated uncertainties on the polarisation fraction using the Geometric uncertainty estimator \citep[Equation 27 in][]{Montier2015} for $\frac{lin}{\sigma}$ between 0.5--4, and the Conventional estimator for sources with higher signal-to-noise \citep[Equation 28 in][]{Montier2015}. On the other hand, fractional circular polarisation ($P_c =\frac{V}{I}$) is not affected by these issues, and uncertainties are treated in the normal way.

Source distances were obtained via several methods. For the star formation 1720~MHz OH masers, direct parallax measurements of nearby masers were used where available \citep[four sources;][]{Reid2014,Wu2014,Krishnan2015} and the remaining distances were taken from \citet{Green2011} and \citet{Green2017}, who used HI self-absorption-resolved kinematic distance estimates and the Bayesian method outlined in \citet{Reid2016}, respectively. For completeness and consistency, we also computed distances to our supernova remnant masers and diffuse OH detections using the \citet{Reid2016} method, though we note that these distances are not used in any further scientific analysis.


\begin{table}
\fontsize{8.5}{7}\selectfont
\centering
\caption{List of the flux/bandpass, phase and polarisation calibrators associated with each source.}
\label{Table1}
\setlength{\tabcolsep}{5pt}
\renewcommand{\arraystretch}{1.5}
\begin{tabular}{ lllr  }

\hline
\hline

\multicolumn{1}{l}{Source name} & 
\multicolumn{1}{l}{Flux \&} & 
\multicolumn{1}{l}{Phase} &  
\multicolumn{1}{c}{polarisation} \\ 

\multicolumn{1}{l}{} & 
\multicolumn{1}{l}{Bandpass} & 
\multicolumn{1}{l}{} &  
\multicolumn{1}{c}{} \\

\hline

189.032+0.809 & B0823$-$500 & 0550+032 & 0550+032 \\
189.776+0.346 & B0823$-$500 & 0550+032 & 0550+032 \\
306.322$-$0.334 & B1934$-$638 & 1352$-$63 & 1352$-$63 \\
323.459$-$0.079 & B1934$-$638 & 1352$-$63 & B1934$-$638 \\
328.165+0.586 & B1934$-$638 & 1613$-$586 & 1613$-$586 \\
328.808+0.633 & B1934$-$638 & 1613$-$586 & 1613$-$586 \\
329.339+0.148 & B1934$-$638 & 1613$-$586 & 1613$-$586 \\
330.953$-$0.180 & B1934$-$638 & 1613$-$586 & 1613$-$586 \\
336.941$-$0.156 & B1934$-$638 & 1613$-$586 & 1613$-$586 \\
337.612$-$0.060 & B1934$-$638 & 1613$-$586 & 1613$-$586 \\
339.622$-$0.120 & B1934$-$638 & 1740$-$517 & 1740$-$517 \\
339.884$-$1.259 & B1934$-$638 & 1740$-$517 & 1740$-$517 \\
340.785$-$0.096 & B1934$-$638 & 1740$-$517 & 1740$-$517 \\
344.582$-$0.023 & B1934$-$638 & 1622$-$297 & B1934$-$638 \\
345.003$-$0.224 & B1934$-$638 & 1740$-$517 & 1740$-$517 \\
345.117+1.592 & B1934$-$638 & 1740$-$517 & 1740$-$517 \\
345.495+1.462 & B1934$-$638 & 1622$-$297 & B1934$-$638 \\
345.497+1.461 & B1934$-$638 & 1622$-$297 & B1934$-$638 \\
348.727$-$1.039 & B1934$-$638 & 1622$-$297 & B1934$-$638 \\
350.112+0.095 & B1934$-$638 & 1740$-$517 & 1740$-$517 \\
350.686$-$0.491 & B1934$-$638 & 1710$-$269 & 1710$-$269 \\
351.158+0.699 & B1934$-$638 & 1740$-$517 & 1740$-$517 \\
351.419+0.646 & B1934$-$638 & 1622$-$297 & B1934$-$638 \\
351.774$-$0.537 & B1934$-$638 & 1710$-$269 & 1710$-$269 \\
353.410$-$0.360 & B1934$-$638 & 1740$-$517 & 1740$-$517 \\
357.557$-$0.321 & B1934$-$638 & 1830$-$210 & 1830$-$210 \\
359.970$-$0.456 & B1934$-$638 & 1622$-$297 & B1934$-$638 \\
000.376+0.040 & B1934$-$638 & 1830$-$210 & 1830$-$210 \\
000.665$-$0.036 & B1934$-$638 & 1830$-$210 & 1830$-$210 \\
008.669$-$0.356 & B1934$-$638 & 1830$-$210 & 1830$-$210 \\
010.474+0.027 & B1934$-$638 & 1710$-$269 & 1710$-$269 \\
011.034+0.062 & B1934$-$638 & 1710$-$269 & 1710$-$269 \\
017.638+0.158 & B1934$-$638 & 1908$-$201 & B1934$-$638 \\

\hline
\hline

269.141$-$1.214 & B1934$-$638 & 0823$-$500 & 0823$-$500 \\
337.801$-$0.053 & B1934$-$638 & 1613$-$586 & 1613$-$586 \\
349.729+0.166 & B1934$-$638 & 1830$-$210 & 1830$-$210 \\
349.731+0.173 & B1934$-$638 & 1830$-$210 & 1830$-$210 \\
349.734+0.172 & B1934$-$638 & 1830$-$210 & 1830$-$210 \\
358.936$-$0.485 & B1934$-$638 & 1830$-$210 & 1830$-$210 \\
358.983$-$0.652 & B1934$-$638 & 1830$-$210 & 1830$-$210 \\
359.940$-$0.067 & B1934$-$638 & 1740$-$517  & 1740$-$517 \\
001.010$-$0.225 & B1934$-$638 & 1830$-$210 & 1830$-$210 \\
006.584$-$0.052 & B1934$-$638 & 1710$-$269 & 1710$-$269 \\
006.687$-$0.296 & B1934$-$638 & 1710$-$269 & 1710$-$269 \\
006.696$-$0.284 & B1934$-$638 & 1710$-$269 & 1710$-$269 \\
006.698$-$0.281 & B1934$-$638 & 1710$-$269 & 1710$-$269 \\
006.700$-$0.280 & B1934$-$638 & 1710$-$269 & 1710$-$269 \\
006.703$-$0.283 & B1934$-$638 & 1710$-$269 & 1710$-$269 \\
006.706$-$0.281 & B1934$-$638 & 1710$-$269 & 1710$-$269 \\

\hline
\hline

012.803-0.202 & B1934$-$638 & 1830$-$210 & 1830$-$210 \\
019.610$-$0.234 & B1934$-$638 & 1908$-$201 & 1908$-$201 \\

\hline
\end{tabular}
\end{table}

\section{Results}\label{section3}
\subsection{Detections, Non-Detections and Associations}

1720-MHz OH emission was detected at a total of 51 positions. These detections include maser lines in supernova remnants (SNR), star formation regions (SFR) and also diffuse OH. Considering that 6.7-GHz methanol masers mark regions of high-mass star formation, the angular offset between the targeted 6.7-GHz sources and the masers detected in this study was adopted as the primary criterion to verify association with star formation sites. 1720-MHz OH and 6.7-GHz methanol maser association is further discussed in Section~\ref{1720-MMB}. Secondarily, we searched the SIMBAD Astronomical Database, and available compact source catalogues, primarily, the 870 $\micron$ ATLASGAL survey \citep{Schuller2009,Csengeri2014,Contreras2013,Urquhart2015} for any identifiable association within the broad phase of high-mass star formation, which includes ultracompact H{\sc ii} (UCH{\sc ii}) regions, and compact dust clumps. We also searched the literature for previous detections and published associations.
As a result of this process, we identified 33 SFR masers, of which 10 are new detections, and 16 SNR masers, of which two are new detections. We also made two serendipitous detections of diffuse OH, 012.803$-$0.202 and 019.610$-$0.234, seen in contrast against bright, compact continuum emission. These are discussed further in their individual source notes in Section~\ref{sourcenotes}.

Table~\ref{Table2} lists 11 1720-MHz OH masers reported by \citet{Caswell2004} but not detected within the 5$\sigma$ detection limit (0.2\,Jy) of the current observations. Three of these 11 undetected sources had peak flux densities $\leq$ 0.2\,Jy in the \citet{Caswell2004} observations, and of the three sources that were observed at multiple epochs, two fell below the detection limit on one of the observation epochs. Table \ref{Table3} lists detected sources with their observed properties: source names and coordinates, the angular offset between MAGMO 1720-MHz OH and the MMB's 6.7-GHz CH$_{3}$OH masers \citep{Caswell2010,Green2010,Caswell2011,Green2012MMB}, peak flux densities, peak velocities, velocity range, distances, date of observation and references if previously detected. Polarisation properties and inferred magnetic properties are considered later. 

Figures~\ref{Figure1}, \ref{Figure2} and \ref{Figure3} show the Stokes I, linear polarisation, RHCP and LHCP spectra for the star formation masers, diffuse OH sources and SNR masers, respectively.

\subsection{Flux density distribution}

A histogram of peak flux densities (defined as the brightest channel in each Stokes $\it{I}$ spectrum) is shown in Figure \ref{Figure4}. The observed range of peak flux densities is similar for both SFR and SNR masers, with SFR masers falling in the range 0.5$-$101.3\,Jy and SNR masers between 0.2$-$102.3\,Jy. The medians of the two distributions are 1.3\,Jy for the SFR masers and 4.2\,Jy for the SNR masers.

\subsection{Velocity distribution}\label{3.1.2}
The SFR masers fall within a peak velocity range of $-$106.9 to 64.7 km\,s$^{-1}$, and the SNR masers within a range of $-$44.9 to 66.3~km\,s$^{-1}$. The 1720-MHz masers with the widest velocity spreads in the star formation and SNR categories are 340.785$-$0.096 and 006.706$-$0.281, with velocity ranges of 7.5 km\,s$^{-1}$ and 6.0 km\,s$^{-1}$, respectively. In general we find the star formation masers to have slightly smaller velocity ranges than the SNR masers (mean and median values of 2.7 and 2.3 for the star formation and 3.0 and 3.0~km\,s$^{-1}$ for the SNR masers). Additionally, we report $\sim$0.4 km\,s$^{-1}$ as the median value of the FWHM for both RHCP and LHCP profiles of the SFR masers -- consistent with typical OH SFR maser linewidths \citep[e.g.][]{Fish03b,Caswell2004,caswell2013,caswell2014}. With a median value of $\sim$0.9 km\,s$^{-1}$, the FWHM linewidth of the SNR masers is generally broader than the SFR masers.

\subsection{Magnetic field measurements of present and past studies}

We tested for consistency between the magnetic field strengths derived in this work and those reported for the 26 common sources observed (with the Parkes telescope) by \citet{Caswell2004}. Figure \ref{Figure5} shows a scatter plot of the absolute values of our magnetic field strength measurements and those measured by \citet{Caswell2004}. In cases where multiple components are present, we compare matching Zeeman pairs. On the figure, the unbroken red line corresponds to a one-to-one line between these measurements, while the dashed lines represent the $\pm$3$\sigma$ offsets from the red line, defined from the uncertainties on the magnetic field measurements of this study. Most of the sources fall within $3\sigma$ of the one-to-one line, showing that the general agreement between our values and previous values is excellent, and that there is no suggestion of any systematic error, even though the measurements were obtained at separate times and with different instruments. The extreme outlying source(s) above (000.665$-$0.036) and below (345.495+1.462 and 017.638+0.158) the broken lines  appear to have undergone an increase and decrease in their magnetic field strengths respectively.


\begin{table}
\fontsize{8.5}{7}\selectfont
\centering
\caption{List of previously known 1720-MHz SFR masers and their flux densities (or detection limits) from \citet{Caswell2004}, which were not detected in the current study. The superscripts refer to the epochs at which the sources were observed: $^{1}$1997, $^{2}$2001, $^{3}$April 2002, $^{4}$September 2002, $^{5}$2003.} 

\label{Table2}
\setlength{\tabcolsep}{5pt}
\renewcommand{\arraystretch}{1.5}
\begin{tabular}{lcc} %
\hline
Source name & Velocity & Flux density\\
&  (\kms) & (Jy)\\
\hline

290.375+1.666 & $-$20.5 & 0.6$^{2}$ \\
300.969+1.147 & $-$42.2 & 0.35$^{2}$, 0.35$^{3}$, 0.35$^{4}$ \\
310.146+0.760 & $-$55.0 & 1.20$^{3}$ \\
336.822+0.028 & $-$76.0 & 0.14$^{4}$ \\
336.994$-$0.027 & $-$126.0 & 0.3$^{2}$ \\
338.075+0.012 & $-$53.0 & 0.35$^{3}$, 0.16$^{4}$ \\
338.925+0.557 & $-$61.0 & 0.80$^{1}$, <0.15$^{3}$ \\
348.550$-$0.979 & $-$19.7 & 2.6$^{3}$ \\
005.885$-$0.392 & $-$13.9 & 0.17$^{4}$ \\
012.680$-$0.183 & 63.0 & 0.20$^{2}$ \\
012.908$-$0.260 & 32.5 & 1.20$^{2}$ \\

\hline
\end{tabular}
\end{table}


\begin{table*}
\fontsize{9}{7}\selectfont
\centering
\caption{Properties of detected 1720-MHz OH emission broken into the three categories (separated by a double ruled line) of SFR masers (top), SNR masers (middle) and diffuse OH (bottom). Column 1 is the source name in Galactic coordinates. Positions are given in equatorial coordinates in columns 2 \& 3; column 4 is the angular offset between MAGMO 1720-MHz OH and the MMB's 6.7-GHz CH$_{3}$OH  masers; columns 5 \& 6 are the Stokes I peak flux densities obtained from the spectral channel and peak velocities measured at the maximum pixel. Column 7 gives the minimum and maximum velocity of each source. Column 8 are the distances, where superscripts are references to published works ($^{a}$\citealt{Green2011}, $^{b}$\citealt{Green2017}, $^{c}$\citealt{Reid2014}, $^{d}$\citealt{Krishnan2015}, $^{e}$\citealt{Wu2014}). Entries marked with ``a" are kinematic distances derived using associated methanol maser velocities, with the distance ambiguity broken via the HI self-absorption method; ``b" refers to distances derived using the \citet{Reid2016} tool; entries with no superscripts are distances derived in this work, also using the \citet{Reid2016} method; ``c, d \& e" are directly measured maser parallax distances. In order, the eight and last columns lists the date of observation; and references to previous detections (1: \citealt{Macleod97}, 6: \citealt{Gaume87}, 2: \citealt{Caswell2004}, 3 \citealt{Caswell99}, 4 \citealt{Caswell2001}, 5: \citealt{caswell83}, 7: \citealt{Frail96}, 8: \citealt{zadeh95}, 9: \citealt{zadeh96}, 10: \citealt{zadeh99}, 11: \citealt{Claussen97}, 12: \citealt{Qiao18}, 13: \citealt{Beuther_Thor}) or ``N" if newly detected.
}
\label{Table3}
\setlength{\tabcolsep}{5pt}
\renewcommand{\arraystretch}{1.5}
\begin{tabular}{ lllclrrlll  }

\hline
\hline

\multicolumn{1}{l}{Source name} & 
\multicolumn{1}{l}{RA} & 
\multicolumn{1}{l}{DEC} & 
\multicolumn{1}{l}{CH$_{3}$OH} &  
\multicolumn{1}{c}{Peak} & 
\multicolumn{1}{l}{Peak} &
\multicolumn{1}{l}{Velocity} &
\multicolumn{1}{l}{Dist.} &
\multicolumn{1}{c}{Date} & 
\multicolumn{1}{l}{Ref.} \\ 

\multicolumn{1}{l}{} & 
\multicolumn{1}{l}{(J2000)}                       & 
\multicolumn{1}{l}{(J2000)} & 
\multicolumn{1}{c}{offset} & 
\multicolumn{1}{c}{flux} & 
\multicolumn{1}{l}{velocity} & 
\multicolumn{1}{l}{range} &
\multicolumn{1}{l}{} &
\multicolumn{1}{c}{of Obs.} & 
\multicolumn{1}{l}{} \\

\multicolumn{1}{l}{{}} & 
\multicolumn{1}{l}{$h:m:s$}                       & 
\multicolumn{1}{l}{$^\circ$ : $'$ : $''$} & 
\multicolumn{1}{c}{($''$)} & 
\multicolumn{1}{c}{$(Jy)$} & 
\multicolumn{1}{l}{(kms$^{-1}$)} & 
\multicolumn{1}{l}{(kms$^{-1}$)} &
\multicolumn{1}{l}{(kpc)} &
\multicolumn{1}{c}{} & 
\multicolumn{1}{l}{} \\

\hline

189.032+0.809 & 06:08:46.73 & +21:31:43.97 & 90 & 1.34 $\pm$ 0.04 & 3.5 & 3.0,4.0 &  $^{a}$2.0$^{+1.0}_{-1.0}$ & 01/2012 &  N \\
189.776+0.346 & 06:08:35.20 & +20:39:14.16 & 8 & 1.47 $\pm$ 0.04 & 12.4 & 10.5,14.0  &  $^{a}$2.0$^{+1.0}_{-1.0}$ & 01/2012 & N \\

306.322$-$0.334 & 13:21:23.03 & $-$63:00:29.54 & <1 & 2.04 $\pm$ 0.04 & $-$23.4 & $-$20.5,$-$24.5  &  $^{a}$1.5$^{+0.6}_{-0.5}$ & 09/2011 &  1,2 \\

323.459$-$0.079 & 15:29:19.35 & $-$56:31:21.99 & <1 & 1.78 $\pm$ 0.04 & $-$68.3 & $-$68.0,$-$69.0  &  $^{a}$3.8$^{+0.4}_{-0.4}$ &  01/2012 & 2 \\

328.165+0.586 & 15:52:42.58 & $-$53:09:52.46 & 3 & 0.45 $\pm$ 0.05 & $-$87.9
 & $-$85.5,$-$89.0 &  $^{b}$5.5$^{+0.6}_{-0.6}$ & 01/2012 &  N \\

328.808+0.633 & 15:55:48.52 & $-$52:43:06.21 & <1 & 29.19 $\pm$ 0.05 & $-$43.2 & $-$42.5,$-$43.7 &  $^{a}$2.6$^{+0.4}_{-0.4}$ & 12/2011 &  1,2,3 \\

329.339+0.148 & 16:00:33.12 & $-$52:44:39.77 & <1 & 1.32 $\pm$ 0.04 & $-$106.9 & $-$106.3,$-$108.0  & $^{b}$7.3$^{+1.1}_{-1.1}$ & 12/2011 &  2,4 \\

330.953$-$0.180 & 16:09:52.01 & $-$51:54:52.04 & 7 & 0.75 $\pm$ 0.06 & $-$86.2 & $-$85.0,$-$86.8  &  $^{a}$4.7$^{+0.3}_{-0.3}$ & 01/2012 &  2 \\

336.941$-$0.156 & 16:35:55.18 & $-$47:38:45.75 & <1 & 1.71 $\pm$ 0.06 & $-$69.4 & $-$68.0,$-$70.0  &  $^{b}$11.1$^{+0.4}_{-0.4}$ & 01/2012 &  2 \\

337.612$-$0.060 & 16:38:09.53 & $-$47:05:00.56 & <1 & 0.83 $\pm$ 0.05 & $-$40.2 &  $-$39.5,$-$41.5 & $^{b}$12.5$^{+0.5}_{-0.5}$ & 01/2012 &  2 \\

339.622$-$0.120 & 16:46:05.95 & $-$45:36:43.68 & <1 & 6.75 $\pm$ 0.04 & $-$34.0 & $-$32.0,$-$37.3  & $^{b}$13.1$^{+0.5}_{-0.5}$ & 01/2011 &  1,2,3 \\

339.884$-$1.259 & 16:52:04.66 & $-$46:08:34.13 & <1 & 31.85 $\pm$ 0.04 & $-$37.8 & $-$33.6,$-$39.0  &  $^{d}$2.1$^{+0.4}_{-0.3}$ & 09/2011 &  2,4 \\

340.785$-$0.096 & 16:50:14.81 & $-$44:42:26.40 & <1 & 1.47 $\pm$ 0.04 & $-$105.6 & $-$99.5,$-$107.0  &  $^{a}$10.6$^{+0.2}_{-0.2}$ & 01/2011 &  2,5 \\

344.582$-$0.023 & 17:02:57.73 & $-$41:41:52.98 & <1 & 3.24 $\pm$ 0.04 & $-$3.9 & $-$3.5,$-$4.6  &  $^{b}$17.8$^{+1.1}_{-1.1}$ & 01/2012 &  N \\

345.003$-$0.224 & 17:05:11.21 & $-$41:29:07.04 & <1 & 37.00 $\pm$ 0.06 & $-$29.3
 & $-$28.5,$-$29.8  &  $^{b}$2.7$^{+0.5}_{-0.5}$ & 12/2010 & 1,2  \\

345.117+1.592 & 16:57:57.96 & $-$40:16:50.24 & 818 & 8.13 $\pm$ 0.06 & $-$16.6
 & $-$15.5,$-$17.5  &  1.4$^{+0.1}_{-0.1}$ & 12/2010 & 2  \\

345.495+1.462 & 16:59:43.71 & $-$40:03:53.85 & 20 & 0.69 $\pm$ 0.04 & $-$24.0 & $-$26.5,$-$23.0  &  $^{a}$1.5$^{+0.6}_{-0.8}$ & 01/2012 &  2 \\

345.497+1.461 & 16:59:44.07 & $-$40:03:50.71 & 20 & 1.57 $\pm$ 0.04 & $-$21.3 & $-$23.5,$-$17.5  & $^{a}$1.5$^{+0.6}_{-0.8}$ & 01/2012 &  2 \\

348.727$-$1.039 & 17:20:06.88 & $-$38:57:12.43 & 5 & 1.11 $\pm$ 0.04 & $-$13.4 & $-$12.5,$-$14.7  &  $^{c}$3.4$^{+0.3}_{-0.3}$ & 01/2012 &  N \\

350.112+0.095 & 17:19:25.36 & $-$37:10:06.67 & 49 & 0.66 $\pm$ 0.07 & $-$66.1
 & $-$63.5,$-$67.0  &  10.5$^{+0.4}_{-0.4}$ & 12/2010 &  N \\

350.686$-$0.491 & 17:23:28.65 & $-$37:01:48.73 & <1 & 2.95 $\pm$ 0.04 & $-$15.1 & $-$13.0,$-$16.7  & $^{a}$2.1$^{+0.8}_{-1.0}$ & 06/2012 &  2 \\

351.158+0.699 & 17:19:56.30 & $-$35:57:53.10 & 15 & 8.78 $\pm$ 0.04 & $-$8.6
 & $-$7.5,$-$10.5  &  1.4$^{+0.1}_{-0.1}$ & 12/2010 &  N \\

351.419+0.646 & 17:20:53.33 & $-$35:47:02.08 & 1 & 101.34 $\pm$ 0.08 & $-$9.8 & $-$9.2,$-$11.0  & $^{e}$1.3$^{+0.6}_{-0.5}$ & 01/2012 &  2,6 \\

351.774$-$0.537 & 17:26:42.49 & $-$36:09:19.83 & 2 & 6.28 $\pm$ 0.05 & 4.3 & $-$2.0,5.0  & $^{b}$1.3$^{+0.1}_{-0.1}$ &  06/2012 & 1,2,3 \\

353.410$-$0.360 & 17:30:26.17 & $-$34:41:44.73 & <1 & 13.63 $\pm$ 0.04 & $-$19.4 & $-$18.7,$-$21.0  & $^{b}$3.7$^{+0.8}_{-0.8}$ & 01/2011 &  2,5 \\

357.557$-$0.321 & 17:40:57.11 & $-$31:10:59.67 & 1 & 0.84 $\pm$ 0.05 & $-$0.2 & 0.3,$-$1.0  & $^{b}$17.5$^{+3.1}_{-3.1}$ &  01/2011 & N \\

359.970$-$0.456 & 17:47:20.09 & $-$29:11:56.84 & 3 & 0.46 $\pm$ 0.06 & 15.7 & 14.0,17.5  & $^{b}$5.5$^{+4.5}_{-4.5}$ &  01/2011 & N \\

000.376+0.040 & 17:46:21.42 & $-$28:35:39.39 & <1 & 1.01 $\pm$ 0.07 & 41.2 & 40.5,41.8  & $^{b}$12.9$^{+3.7}_{-3.7}$ & 01/2011 &  N \\

000.665$-$0.036 & 17:47:20.01 & $-$28:23:13.19 & <1 & 1.53 $\pm$ 0.07 & 61.4 & 60.0,64.0  & $^{a}$7.9$^{+0.8}_{-0.7}$ & 01/2011 &  1,2 \\

008.669$-$0.356 & 18:06:19.02 & $-$21:37:33.49 & 1 & 2.25 $\pm$ 0.07 & 39.7 & 39.0,40.5  & $^{a}$4.4$^{+0.4}_{-0.4}$ & 12/2010 &  2 \\

010.474+0.027 & 18:08:38.32 & $-$19:51:45.28  & 5 & 1.57 $\pm$ 0.06 & 64.7 & 64.2,65.4  &  $^{c}$8.5$^{+0.6}_{-0.5}$ & 01/2012 &  2,3 \\

011.034+0.062 & 18:09:39.83 & $-$19:21:21.71  & 1 & 1.63 $\pm$ 0.04 & 23.1 & 21.3,24.0  &  $^{a}$2.4$^{+0.6}_{-0.8}$ &  01/2012 & 2,3 \\ 

017.638+0.158 & 18:22:26.09 & $-$13:30:08.99  & 4 & 0.48 $\pm$ 0.04 & 28.6 & 27.0,29.3  &  $^{a}$2.0$^{+0.5}_{-0.6}$ &  06/2012 & 2,13 \\

\hline
\hline

269.141$-$1.214 & 09:03:07.53 & $-$48:30:56.15 & 309 & 4.35 $\pm$ 0.04 & 10.6 & 8.5,13.0  &  2.4$^{+0.9}_{-0.9}$ &  09/2011 & N \\

337.801$-$0.053 & 16:38:52.14 & $-$46:56:17.06 & 345 & 2.02 $\pm$ 0.05 & $-$44.9 & $-$43.5,$-$46.8  & 12.3$^{+0.5}_{-0.5}$ &  01/2012 &  2 \\

349.729+0.166 & 17:18:01.44 & $-$37:26:24.94 & 599 & 1.58 $\pm$ 0.04 & 16.8 & 16.0,17.5 & 20.9$^{+3.0}_{-3.0}$ &  05/2010 & 7  \\

349.731+0.173 & 17:18:00.00 & $-$37:26:09.08 & 588 & 0.85 $\pm$ 0.04 & 16.4 & 15.0,18.0  & 20.8$^{+3.0}_{-3.0}$ &  05/2010 & 7  \\

349.734+0.172 & 17:18:00.88 & $-$37:26:01.80 & 578 & 2.61 $\pm$ 0.04 & 15.3 & 14.6,16.0  & 20.4$^{+2.9}_{-2.9}$ &  05/2010 & 7  \\

358.936$-$0.485 & 17:44:58.15 & $-$30:05:49.17 & 970 & 3.62 $\pm$ 0.07 & $-$6.8 & $-$5.5,$-$7.7  & 2.7$^{+0.3}_{-0.3}$ &  01/2011 & 8,12 \\

358.983$-$0.652 & 17:45:44.65 & $-$30:08:39.24 & 594 & 0.84 $\pm$ 0.05 & $-$0.2 & $-$5.5,$-$7.5  & 2.7$^{+0.2}_{-0.2}$ &  01/2011 & N \\

359.940$-$0.067 & 17:45:44.33 & $-$29:01:19.97  & 1007 & 9.54 $\pm$ 0.04 & 66.3 & 64.5,68.0  & 10.9$^{+0.2}_{-0.2}$ &  05/2010 & 9  \\

\hline
\end{tabular}
\end{table*}


\addtocounter{table}{-1}
\begin{table*}
\fontsize{9}{7}\selectfont
\centering
\caption{\it{$-$continued}}
\setlength{\tabcolsep}{5pt}
\renewcommand{\arraystretch}{1.5}
\begin{tabular}{ lllclrrlll  }

\hline
\hline

\multicolumn{1}{l}{Source name} & 
\multicolumn{1}{l}{RA} & 
\multicolumn{1}{l}{DEC} & 
\multicolumn{1}{l}{CH$_{3}$OH} &  
\multicolumn{1}{c}{Peak} & 
\multicolumn{1}{l}{Peak} &
\multicolumn{1}{l}{Velocity} &
\multicolumn{1}{l}{Dist.} &
\multicolumn{1}{c}{Date} & 
\multicolumn{1}{l}{Ref.} \\ 

\multicolumn{1}{l}{} & 
\multicolumn{1}{l}{(J2000)}                       & 
\multicolumn{1}{l}{(J2000)} & 
\multicolumn{1}{c}{offset} & 
\multicolumn{1}{c}{flux} & 
\multicolumn{1}{l}{velocity} & 
\multicolumn{1}{l}{range} &
\multicolumn{1}{l}{} &
\multicolumn{1}{c}{of Obs.} & 
\multicolumn{1}{l}{} \\

\multicolumn{1}{l}{{}} & 
\multicolumn{1}{l}{$h:m:s$}                       & 
\multicolumn{1}{l}{$^\circ$ : $'$ : $''$} & 
\multicolumn{1}{c}{($''$)} & 
\multicolumn{1}{c}{$(Jy)$} & 
\multicolumn{1}{l}{(kms$^{-1}$)} & 
\multicolumn{1}{l}{(kms$^{-1}$)} &
\multicolumn{1}{l}{(kpc)} &
\multicolumn{1}{c}{} & 
\multicolumn{1}{l}{} \\

\hline

001.010$-$0.225 & 17:48:52.69 & $-$28:11:19.12 & 45 & 0.71 $\pm$ 0.06 & $-$1.3 & $-$0.5,$-$2.5  & 3.3$^{+0.1}_{-0.1}$ &  01/2011 & 10 \\

006.584$-$0.052 & 18:00:43.86 & $-$23:17:27.46 & 258 & 5.79 $\pm$ 0.06 & 6.1 & 4.0,8.0  & 2.9$^{+0.2}_{-0.2}$ &  01/2012 & 11  \\

006.687$-$0.296 & 18:01:52.65 & $-$23:19:24.27 & 415 & 102.34 $\pm$ 0.06 & 11.3 & 9.0,12.5  & 2.9$^{+0.2}_{-0.2}$ &  01/2012 & 11  \\

006.696$-$0.284 & 18:01:51.07 & $-$23:18:34.38 & 371 & 15.50 $\pm$ 0.06 & 10.4 & 9.0,11.5  &  2.9$^{+0.2}_{-0.2}$  &  01/2012 & 11  \\

006.698$-$0.281 & 18:01:50.57 & $-$23:18:18.59 & 357 & 6.36 $\pm$ 0.06 & 15.7 & 14.0,17.0 & 2.9$^{+0.2}_{-0.2}$ &  01/2012 &  11 \\

006.700$-$0.280 & 18:01:50.67 & $-$23:18:13.25 & 352 & 14.66 $\pm$ 0.06 & 13.4 & 12.0,15.5 & 2.9$^{+0.2}_{-0.2}$ &  01/2012 & 11  \\

006.703$-$0.283 & 18:01:51.66 & $-$23:18:06.92 & 342 & 12.05 $\pm$ 0.06 & 9.4 & 8.0,11.0  &  2.9$^{+0.2}_{-0.2}$  &  01/2012 & 11  \\

006.706$-$0.281 & 18:01:51.58 & $-$23:17:56.25 & 332 & 35.79 $\pm$ 0.05 & 11.8 & 8.5,14.5 & 2.9$^{+0.2}_{-0.2}$ &  01/2012 &  11 \\

\hline
\hline

012.803-0.202 & 18:14:13.96 & $-$17:55:55.27 & 434 & 0.64 $\pm$ 0.05 & 32.5 & 30.0,34.0 & 3.0$^{+0.3}_{-0.3}$ &  01/2011 & N \\

019.610$-$0.234 & 18:27:38.02 & $-$11:56:31.99 & 6 & 0.56 $\pm$ 0.05 & 43.9 & 40.5,47.0 &  3.3$^{+0.2}_{-0.2}$ &  06/2012 & 13 \\

\hline
\end{tabular}
\end{table*}

\begin{table*}
\fontsize{9}{6.2}\selectfont
\centering
\caption{Peak flux densities and velocities of Zeeman pairs with derived magnetic field strengths and directions of SFR masers (top), SNR masers (middle) and diffuse OH sources (bottom).
Columns 2 \& 3 are the Gaussian fitted peak flux density and velocity of RHCP components (where present), while columns 4 \& 5 are the fitted LHCP counterparts. Uncertainties in columns 2, 3, 4 \& 5 are the Gaussian fit errors. Column 6 is the inferred magnetic field strength: positive field strengths indicate a field oriented away from us, negative field strengths indicate a field oriented towards us. Italicized items in column 6 are field strengths detected at less than the 3$\sigma$ level, while those in parenthesis are the measurements of \citet{Caswell2004}. Column 7 is the reliability of the Zeeman pair association, with ``A" indicating a reliable association, and ``B" an ambiguous association (see also in-text description). }
\label{Table4}
\renewcommand{\arraystretch}{1.5}
\begin{tabular}{ llllllcc  }

\hline
\hline

\multicolumn{1}{l}{Source name} & 
\multicolumn{2}{c}{RHCP} &
\multicolumn{2}{c}{LHCP} &
\multicolumn{1}{c}{Magnetic field} &
\multicolumn{1}{c}{Reliability} \\

\multicolumn{1}{c}{} & 
\multicolumn{1}{c}{$S_\mathrm{peak}$ (Jy)} & 
\multicolumn{1}{c}{$V_\mathrm{peak}$ (km\,s$^{-1}$)} & 
\multicolumn{1}{c}{$S_\mathrm{peak}$ (Jy)} & 
\multicolumn{1}{c}{$V_\mathrm{peak}$ (km\,s$^{-1}$)} & 
\multicolumn{1}{c}{(mG)} &
\multicolumn{1}{c}{} \\

\hline

189.032$+$0.809	&	1.06	$\pm$	0.03	&	3.56	$\pm$	0.01	&	0.45	$\pm$	0.04	&	3.50	$\pm$	0.01	&	\it{$+$0.5	$\pm$	0.4}	&	A	\\

189.776$+$0.346	&	0.89	$\pm$	0.05	&	11.73	$\pm$	0.03	&	0.79	$\pm$	0.01	&	12.58	$\pm$	0.02	&	$-$7.5	$\pm$	0.4	&	B	\\
	&	0.77	$\pm$	0.02	&	12.35	$\pm$	0.03	&	0.77	$\pm$	0.01	&	13.52	$\pm$	0.02	&	$-$10.4	$\pm$	0.4	&	B	\\
	
306.322$-$0.334	&	0.34	$\pm$	0.02	&	$-$23.89	$\pm$	0.03	&	$-$	&	$-$	&	$-$	&	$-$	\\
         	&	0.61	$\pm$	0.03	&	$-$22.66	$\pm$	0.01	&	1.85	$\pm$	0.01	&	$-$23.44	$\pm$	0.02	&	$+$6.9	$\pm$	0.4 ($+$6.0)   &	B	\\
         	&	0.98	$\pm$	0.03	&	$-$21.56	$\pm$	0.01	&	1.19	$\pm$	0.01	&	$-$22.28	$\pm$	0.01	&	$+$6.4	$\pm$	0.4	&	B	\\
         	
323.459$-$0.079	&	1.21	$\pm$	0.02	&	$-$68.33	$\pm$	0.01	&	0.62	$\pm$	0.02	&	$-$68.38	$\pm$	0.01	&	\it{$+$0.4	$\pm$	0.4} ($+$0.5)	&	A	\\

328.165+0.586	&	0.31	$\pm$	0.02	&	$-$86.43	$\pm$	0.02	&	0.34	$\pm$	0.02	&	$-$87.92	$\pm$	0.02	&	$+$13.2	$\pm$	0.4 	&	A	\\

328.808$+$0.633	&	17.97	$\pm$	0.03	&	$-$43.115	$\pm$	0.01	&	17.07	$\pm$	0.04	&	$-$43.275	$\pm$	0.01	&	$+$1.4	$\pm$	0.4	($+$1.5) &	A	\\

329.339$+$0.148	&	$-$	&	$-$	&	0.79	$\pm$	0.03	&	$-$106.95	$\pm$	0.01	&	$-$	&	$-$	\\
         	&	0.71	$\pm$	0.02	&	$-$106.79	$\pm$	0.01	&	1.03	$\pm$	0.03	&	$-$107.34	$\pm$	0.01	&	$+$4.9	$\pm$	0.4 ($+$4.0)	&	A	\\
         	
330.953$-$0.180	&	0.38	$\pm$	0.01	&	$-$85.935	$\pm$	0.02	&	0.46	$\pm$	0.01	&	$-$86.12	$\pm$	0.02	&	$+$1.6	$\pm$	0.4 ($+$2.0)	&	A	\\

336.941$-$0.156	&	1.14	$\pm$	0.03	&	$-$68.43	$\pm$	0.01	&	1.76	$\pm$	0.03	&	$-$69.35	$\pm$	0.01	&	$+$8.1	$\pm$	0.4 ($+$6.0)	&	A	\\

337.612$-$0.060	&	0.53	$\pm$	0.02	&	$-$40.91	$\pm$	0.02	&	0.84	$\pm$	0.02	&	$-$40.21	$\pm$	0.01	&	$-$6.2	$\pm$	0.4 ($-$5.7)	&	A	\\

339.622$-$0.120	&	0.18	$\pm$	0.01	&	$-$36.98	$\pm$	0.03	&	0.15	$\pm$	0.01	&	$-$36.67	$\pm$	0.03	&	$-$2.7	$\pm$	0.5	&	A	\\
	&	4.93	$\pm$	0.01	&	$-$33.9	$\pm$	0.01	&	2.48	$\pm$	0.01	&	$-$34.02	$\pm$	0.01	&	\it{$+$1.1	$\pm$	0.4}	&	A	\\
	&	$-$	&	$-$	&	0.26	$\pm$	0.02	&	$-$32.29	$\pm$	0.02	&	$-$	&	$-$	\\
	
339.884$-$1.259	&	29.97	$\pm$	0.02	&	$-$37.85	$\pm$	0.01	&	14.62	$\pm$	0.30	&	$-$36.93	$\pm$	0.01	&	$-$8.1	$\pm$	0.4	&	A	\\
	&	$-$	&	$-$	&	6.93	$\pm$	0.40	&	$-$37.55	$\pm$	0.04	&	$-$	&	$-$	\\
	&	2.2	$\pm$	0.01	&	$-$34.99	$\pm$	0.01	&	4.34	$\pm$	0.80	&	$-$34.16	$\pm$	0.01	&	$-$7.3	$\pm$	0.4	($-$6.0) &	A	\\
	
340.785$-$0.096	&	0.68	$\pm$	0.04	&	$-$106.52	$\pm$	0.01	&	1.41	$\pm$	0.08	&	$-$105.57	$\pm$	0.01	&	$-$8.4 $\pm$ 0.4 ($-$8.0)	&	A	\\
	&	0.28	$\pm$	0.04	&	$-$105.81	$\pm$	0.01	&	0.35	$\pm$	0.05	&	$-$105.06	$\pm$	0.03	&	$-$6.6 $\pm$ 0.4	&	A	\\
	&	0.20	$\pm$	0.03	&	$-$105.03	$\pm$	0.03    &	0.21	$\pm$	0.06	&	$-$104.35	$\pm$	0.03	&	$-$6.0 $\pm$ 0.4 ($-$5.0)	&	A	\\
	&	0.55	$\pm$	0.02	&	$-$99.99	$\pm$	0.01	&	0.55	$\pm$	0.02	&	$-$100.61	$\pm$	0.01	&	$+$5.5	$\pm$	0.4 ($+$5.0)	&	A	\\

344.582$-$0.023	&	3.19	$\pm$	0.02	&	$-$3.85	$\pm$	0.01	&	3.02	$\pm$	0.02	&	$-$4.26	$\pm$	0.01	&	$+$3.6	$\pm$	0.4	&	A	\\

345.003$-$0.224	&	1.54	$\pm$	0.02	&	$-$28.91	$\pm$	0.01	&	37.79	$\pm$	0.03	&	$-$29.27	$\pm$	0.01	&	$+$3.2	$\pm$	0.4	($+$3.5) &	A	\\

345.117$+$1.592	&	5.03	$\pm$	0.04	&	$-$16.65	$\pm$	0.01	&	3.60	$\pm$ 0.04	&	$-$16.44	$\pm$	0.01	&	$-$1.9	$\pm$	0.4 ($-$1.5)	&	A	\\

345.495$+$1.462	&	0.29	$\pm$	0.01	&	$-$24.96	$\pm$	0.02	&	0.63	$\pm$	0.01	&	$-$24.07	$\pm$	0.02	&	$-$7.9	$\pm$	0.4 ($-$12.0)	&	A	\\

345.497$+$1.461	&	0.71	$\pm$	0.01	&	$-$18.99	$\pm$	0.01	&	1.34	$\pm$	0.01	&	$-$18.16	$\pm$	0.01	&	$-$7.3	$\pm$	0.4	&	A	\\
	&	0.54	$\pm$	0.01	&	$-$22.27	$\pm$	0.02	&	0.94	$\pm$	0.01	&	$-$21.13	$\pm$	0.01	&	$-$10.1	$\pm$	0.4 ($-$10.0)	&	A	\\
	
348.727$-$1.039	&	0.98	$\pm$	0.02	&	$-$13.41	$\pm$	0.01	&	0.68	$\pm$	0.02	&	$-$14.01	$\pm$	0.01	&	$+$5.3	$\pm$	0.4	&	A	\\

350.112$+$0.095	&	0.60	$\pm$	0.02	&	$-$65.97	$\pm$	0.02	&	1.30	$\pm$ 0.02	&	$-$64.68	$\pm$	0.04	&	$-$11.4	$\pm$	0.5	&	A	\\

350.686$-$0.491	&	3.01	$\pm$	0.01	&	$-$15.143	$\pm$	0.01	&	1.09	$\pm$	0.02	&	$-$14.686	$\pm$	0.01	&	$-$4.0	$\pm$	0.4 ($-$4.0)	&	A	\\
	&	1.99	$\pm$	0.01	&	$-$14.104	$\pm$	0.01	&	2.06	$\pm$	0.02	&	$-$13.632	$\pm$	0.01	&	$-$4.2	$\pm$	0.4 ($-$4.9)	&	A	\\
	
351.158$+$0.699	&	6.33	$\pm$	0.04	&	$-$8.48	$\pm$	0.01	&	5.21	$\pm$ 0.03	&	$-$9.07	$\pm$	0.01	&	$+$5.2	$\pm$	0.4	&	A	\\	
	
351.419$+$0.646	&	76.98	$\pm$	0.04	&	$-$10.56	$\pm$	0.01	&	102.19	$\pm$	0.04	&	$-$9.86	$\pm$	0.01	&	$-$6.2	$\pm$	0.4 ($-$6.4)	&	A	\\

351.774$-$0.537	&	0.91	$\pm$	0.02	&	$-$1.515	$\pm$	0.01	&	0.86	$\pm$	0.01	&	$-$0.8149	$\pm$	0.01	&	$-$6.2	$\pm$	0.4 ($-$6.0)	&	A	\\
	&	5.85	$\pm$	0.02	&	4.3438	$\pm$	0.01	&	3.32	$\pm$	0.02	&	4.0436	$\pm$	0.01	&	$+$2.7	$\pm$	0.4 ($+$3.0)	&	A	\\

353.410$-$0.360	&	4.16	$\pm$	0.02	&	$-$19.67	$\pm$	0.01	&	11.82	$\pm$	0.02	&	$-$19.36	$\pm$	0.01	&	$-$2.7	$\pm$	0.4	($-$2.3) &	A	\\

357.557$-$0.321	&	0.86	$\pm$	0.02	&	$-$0.25	$\pm$	0.01	&	0.28	$\pm$	0.02	&	$-$0.52	$\pm$	0.01	&	$+$2.4	$\pm$	0.4	&	A	\\

359.970$-$0.457	&	0.3	$\pm$	0.02	&	16.06	$\pm$	0.04	&	0.18	$\pm$	0.02	&	15.18	$\pm$	0.08	&	$+$7.8	$\pm$	0.7	&	B	\\

\hline
\end{tabular}
\end{table*}
	

\addtocounter{table}{-1}	
\begin{table*}
\fontsize{9}{6.2}\selectfont
\centering
\caption{\it{$-$continued}}
\renewcommand{\arraystretch}{1.5}
\begin{tabular}{ llllllcc  }

\hline
\hline

\multicolumn{1}{l}{Source name} & 
\multicolumn{2}{c}{RHCP} &
\multicolumn{2}{c}{LHCP} &
\multicolumn{1}{c}{Magnetic field} &
\multicolumn{1}{c}{Reliability} \\

\multicolumn{1}{c}{} & 
\multicolumn{1}{c}{$S_\mathrm{peak}$ (Jy)} & 
\multicolumn{1}{c}{$V_\mathrm{peak}$ (km\,s$^{-1}$)} & 
\multicolumn{1}{c}{$S_\mathrm{peak}$ (Jy)} & 
\multicolumn{1}{c}{$V_\mathrm{peak}$ (km\,s$^{-1}$)} & 
\multicolumn{1}{c}{(mG)} &
\multicolumn{1}{c}{} \\

\hline

000.376$+$0.040	&	0.43	$\pm$	0.03	&	41.29	$\pm$	0.01	&	0.67	$\pm$	0.03	&	41.12	$\pm$	0.01	&	$+$1.5	$\pm$	0.4	&	A	\\

000.665$-$0.036	&	1.07	$\pm$	0.06	&	61.32	$\pm$	0.01	&	0.23	$\pm$	0.08	&	60.57	$\pm$	0.08	&	$+$6.6	$\pm$	0.4 ($+$3.0)	&	A	\\
	&	-	&	-	&	0.39	$\pm$	0.06	&	61.39	$\pm$	0.04	&	-	&	-	\\
	&	0.38	$\pm$ 0.06	&	63.33	$\pm$	0.04	&	1.06	$\pm$	0.08	&	62.33	$\pm$	0.01	&	$+$8.8	$\pm$	0.4 ($+$9.0)	&	A	\\

008.669$-$0.356	&	1.79	$\pm$	0.06	&	39.31	$\pm$	0.01	&	2.07	$\pm$	0.07	&	39.69	$\pm$	0.01	&	$-$3.4	$\pm$	0.4 ($-$3.0)	&	A	\\

010.474$+$0.027	&	0.79	$\pm$	0.02	&	64.66	$\pm$	0.01	&	0.82	$\pm$	0.02	&	64.62	$\pm$	0.01	&	\it{$+$0.4	$\pm$	0.4} ($-$0.0)	&	A	\\

011.034$+$0.062	&	0.63	$\pm$	0.02	&	21.77	$\pm$	0.02	&	$-$	&	$-$	&	$-$	&	$-$	\\
&		1.34	$\pm$	0.02	&	22.32	$\pm$	0.01	&	1.67	$\pm$	0.02	&	23.17	$\pm$	0.01	&	$-$7.5	$\pm$	0.4 ($-$7.0)	&	B \\

017.638+0.158	&	0.12	$\pm$	0.01	&	27.57	$\pm$	0.05	&	0.47	$\pm$	0.01	&	28.48	$\pm$	0.01	&	$-$8.1	$\pm$	0.5	 ($-$14.0) &	A	\\

\hline
\hline
																							
269.141$-$1.213	&	2.35	$\pm$	0.01	&	9.23	$\pm$	0.01	&	2.52	$\pm$	0.01	&	10.44	$\pm$	0.01	&	$-$10.7	$\pm$	0.4	&	A	\\
	&	2.01	$\pm$	0.01	&	10.68	$\pm$	0.01	&	1.42	$\pm$	0.01	&	11.97	$\pm$	0.01	&	$-$11.4	$\pm$	0.4	&	A	\\
	
337.801$-$0.053	&	1.04	$\pm$	0.01	&	$-$45.09	$\pm$	0.01	&	1.12	$\pm$	0.01	&	$-$45.16	$\pm$	0.01	&	\it{$+$0.6	$\pm$	0.4}	&	A	\\

349.729$+$0.166	&	0.74	$\pm$	0.04	&	16.8	$\pm$	0.01	&	0.74	$\pm$	0.04	&	16.83	$\pm$	0.01	&	\it{$-$0.3	$\pm$	0.4}	&	A	\\

349.731$+$1.731	&	0.41	$\pm$	0.02	&	15.63	$\pm$	0.04	&	0.37	$\pm$	0.02	&	15.57	$\pm$	0.07	&	\it{$+$0.5	$\pm$	0.6}	&	A	\\
	&	0.42	$\pm$	0.02	&	16.48	$\pm$	0.06	&	0.33	$\pm$	0.02	&	16.58	$\pm$	0.09	&	\it{$-$0.9	$\pm$	0.8}	&	A	\\
	
349.734$+$0.172	&	1.25	$\pm$	0.03	&	15.3	$\pm$	0.01	&	1.21	$\pm$	0.04	&	15.29	$\pm$	0.01	&	\it{$+$0.1	$\pm$	0.4}	&	A	\\

358.936$-$0.485	&	1.95	$\pm$	0.07	&	$-$6.78	$\pm$	0.01	&	1.83	$\pm$ 0.07	&	$-$6.78	$\pm$	0.02	&	\it{0.0	$\pm$	0.4}	&	A	\\

358.983$-$0.652	&	2.09	$\pm$ 0.04	&	$-$6.30	$\pm$	0.01	&	2.13	$\pm$ 0.04	&	$-$6.30	$\pm$	0.01	&	\it{0.0	$\pm$	0.4}	&	A	\\

359.940$-$0.067	&	4.74	$\pm$	0.07	&	66.56	$\pm$	0.01	&	5.18	$\pm$	0.06	&	66.17	$\pm$	0.01	&	$+$3.5	$\pm$	0.4	&	A	\\

001.010$-$0.225	&	0.39	$\pm$	0.02	&	$-$1.35	$\pm$	0.02	&	0.33	$\pm$	0.01	&	$-$1.41	$\pm$	0.02	&	\it{$+$0.5	$\pm$	0.4}	&	A	\\

006.584$-$0.052	&	0.52	$\pm$	0.02	&	4.8	$\pm$	0.02	&	0.49	$\pm$	0.02	&	4.82	$\pm$	0.03	&	\it{$-$0.2	$\pm$	0.4}	&	A	\\
	&	2.96	$\pm$	0.02	&	6.19	$\pm$	0.01	&	2.89	$\pm$	0.02	&	6.16	$\pm$	0.01	&	\it{$+$0.3	$\pm$	0.4}	&	A	\\
	
006.686$-$0.296	&	28.83	$\pm$	0.03	&	10.02	$\pm$	0.01	&	23.23	$\pm$	0.03	&	9.95	$\pm$	0.01	&	\it{$+$0.6	$\pm$	0.4}	&	A	\\
	&	52.37	$\pm$	0.03	&	11.37	$\pm$	0.01	&	48.91	$\pm$	0.03	&	11.31	$\pm$	0.01	&	\it{$+$0.5	$\pm$	0.4}	&	A	\\
	
006.696$-$0.284	&	7.339	$\pm$	0.03	&	10.51	$\pm$	0.01	&	7.68	$\pm$	0.03	&	10.44	$\pm$	0.01	&	\it{$+$0.6	$\pm$	0.4}	&	A	\\

006.698$-$0.281	&	3.16	$\pm$	0.03	&	15.64	$\pm$	0.01	&	3.18	$\pm$	0.03	&	15.63	$\pm$	0.01	&	\it{$+$0.1	$\pm$	0.4}	&	A	\\

006.700$-$0.280	&	7.49	$\pm$	0.03	&	13.35	$\pm$	0.01	&	7.43	$\pm$	0.03	&	13.31	$\pm$	0.01	&	\it{$+$0.4	$\pm$	0.4}	&	A	\\

006.703$-$0.283	&	6.25	$\pm$	0.02	&	9.42	$\pm$	0.01	&	5.94	$\pm$	0.02	&	9.37	$\pm$	0.01	&	\it{$+$0.4	$\pm$	0.4}	&	A	\\

006.706$-$0.281	&	18.93	$\pm$	0.08	&	11.9	$\pm$	0.01	&	19.51	$\pm$	0.08	&	11.83	$\pm$	0.01	&	\it{$+$0.6	$\pm$	0.4}	&	A	\\

\hline
\hline
																							
012.803$-$0.202	&	0.21	$\pm$	0.02	&	32.13	$\pm$	0.07	&	0.23	$\pm$	0.02	&	32.13	$\pm$	0.07	&		\it{$+$0.1 $\pm$	0.7}	&	A	\\

019.610$-$0.234	&	0.22	$\pm$	0.01	&	43.67	$\pm$	0.06	&	0.21	$\pm$	0.01	&	43.66	$\pm$	0.01	&	\it{$+$0.1	$\pm$	0.5}	&	A	\\

\hline
\end{tabular}
\end{table*}


\begin{table*}
\fontsize{9}{7}\selectfont
\centering
\caption{1720-MHz flux densities of all Stokes parameters and corresponding estimated percentage fractional polarisations, for SFR masers (top), SNR masers (middle) and diffuse OH sources (bottom). Components are identified in Stokes I by fitting multiple Gaussians. Column 2 is the Gaussian fitted peak velocity of the Stokes $I$ component; columns 3, 4, 5 \& 6 are the flux densities of Stokes $I$, $Q$, $U$ and $V$ at the velocity channel of the fitted Stokes $I$ peak; columns 7 \& 8 are the fractional linear ($P_l$) and circular ($P_c$) polarisations respectively. The quoted flux density uncertainties are $1\sigma_\mathrm{rms}$. Uncertainties on $P_c$ are propagated through assuming uncorrelated errors on the input parameters; uncertainties on $P_l$ are derived as described in Section \ref{section2}. Italicized items are less than 5$\sigma$ and flagged as unreliable.}
\label{Table5}
\renewcommand{\arraystretch}{1.5}
\begin{tabular}{ llllllllc  }

\hline
\hline

\multicolumn{1}{l}{Source name} &
\multicolumn{1}{c}{Velocity} &
\multicolumn{1}{c}{I} & 
\multicolumn{1}{c}{Q} & 
\multicolumn{1}{c}{U} & 
\multicolumn{1}{c}{V} & 
\multicolumn{1}{c}{$P_l$} &
\multicolumn{1}{c}{$P_c$} \\ 

\multicolumn{1}{c}{} &
\multicolumn{1}{c}{km\,s$^{-1}$} &
\multicolumn{1}{c}{(Jy)} & 
\multicolumn{1}{c}{(Jy)} & 
\multicolumn{1}{c}{(Jy)} & 
\multicolumn{1}{c}{(Jy)} & 
\multicolumn{1}{c}{(per cent)} &
\multicolumn{1}{c}{(per cent)} \\

\hline

189.032$+$0.809	&	3.55  $\pm$  0.01	&	1.34  $\pm$  0.04	&	0.45  $\pm$  0.04	&	0.42  $\pm$  0.04	&	0.41  $\pm$  0.04	&	46  $\pm$  3	&	31  $\pm$  3	\\

189.776$+$0.346	&	12.46  $\pm$  0.01	&	1.46  $\pm$  0.04	&	0.08  $\pm$  0.04	&	0.01  $\pm$  0.04	&	0.03  $\pm$  0.04	&	\it{11  $\pm$  3}	&	\it{2  $\pm$  3}	\\

306.322$-$0.334	&	$-$23.42  $\pm$  0.01	&	2.04  $\pm$  0.04	&	0.05  $\pm$  0.04	&	0.09  $\pm$  0.04	&	1.64  $\pm$  0.04	&	\it{9  $\pm$  2}	&	80  $\pm$  3	\\
         	    &	$-$22.65  $\pm$  0.01	&	0.90  $\pm$  0.04	&	0.09  $\pm$  0.04	&	0.02  $\pm$  0.04	&	0.33  $\pm$  0.04	&	\it{19  $\pm$  4}	&	37  $\pm$  5	\\
         	    &	$-$22.47  $\pm$  0.01	&	0.96  $\pm$  0.04	&	0.13  $\pm$  0.04	&	0.05  $\pm$  0.04	&	0.63  $\pm$  0.04	&	\it{14  $\pm$  4}	&	66  $\pm$  5	\\
         	    &	$-$21.57  $\pm$  0.01	&	0.99  $\pm$  0.04	&	0.02  $\pm$  0.04	&	0.05  $\pm$  0.04	&	1.01  $\pm$  0.04	&	\it{14  $\pm$  4}	&	102  $\pm$  6	\\
         	    
323.459$-$0.079	&	$-$68.35  $\pm$  0.01	&	1.85  $\pm$  0.04	&	0.05  $\pm$  0.04	&	0.15  $\pm$  0.04	&	0.61  $\pm$  0.05	&	\it{8  $\pm$  2}	&	33  $\pm$  3	\\

328.165$+$0.586	&	$-$87.92  $\pm$  0.02	&	0.45  $\pm$  0.05	&	0.02  $\pm$  0.05	&	0.08  $\pm$  0.06	&	0.40  $\pm$  0.06	&	\it{43 $\pm$ 12}	&	89  $\pm$  17	\\

                &	$-$86.47  $\pm$  0.03	&	0.43  $\pm$  0.05	&	0.03  $\pm$  0.05	& 0.02 $\pm$  0.06	&	0.27  $\pm$  0.06	&	\it{34 $\pm$ 13}	&	63  $\pm$ 16 	\\

328.808$+$0.633	&	$-$43.18  $\pm$  0.01	&	30.43  $\pm$  0.05	&	0.01  $\pm$  0.05	&	0.19  $\pm$  0.05	&	7.17  $\pm$  0.05	&	\it{0.6  $\pm$  0.2}	&	24.0  $\pm$  0.2	\\
329.339$+$0.148 &	$-$106.86  $\pm$  0.01	&	1.32  $\pm$  0.04	&	0.01  $\pm$  0.04	&	0.1  $\pm$  0.04	&	0.30  $\pm$  0.04	&	\it{14  $\pm$  3}	&	23  $\pm$  3	\\
         	    &	$-$107.33  $\pm$  0.01	&	1.07  $\pm$  0.04	&	0.01  $\pm$  0.04	&	0.01  $\pm$  0.04	&	1.11  $\pm$  0.04	&	\it{9  $\pm$  4}	&	104  $\pm$  5	\\
         	    
330.953$-$0.180	&	$-$86.03  $\pm$  0.01	&	0.81  $\pm$  0.06	&	0.04  $\pm$  0.06	&	0.05  $\pm$  0.06	&	0.15  $\pm$  0.05	&	\it{23  $\pm$  7}	&	\it{19  $\pm$  6}	\\

336.941$-$0.156	&	$-$69.37  $\pm$  0.01	&	1.71  $\pm$  0.06	&	0.01  $\pm$  0.06	&	0.05  $\pm$  0.06	&	1.73  $\pm$  0.07	&	\it{10  $\pm$  4}	&	101  $\pm$  5	\\
         	    &	$-$68.39  $\pm$  0.01	&	1.16  $\pm$  0.06	&	0.11  $\pm$  0.06	&	0.01  $\pm$  0.06	&	1.13  $\pm$  0.07	&	\it{20  $\pm$  5}	&	97  $\pm$  8	\\
         	    
337.612$-$0.060	&	$-$40.84  $\pm$  0.05	&	0.66  $\pm$  0.05	&	0.03  $\pm$  0.05	&	0.05  $\pm$  0.05	&	0.41  $\pm$  0.05	&	\it{24  $\pm$  8}	&	62  $\pm$  9	\\
         	    &	$-$40.15  $\pm$  0.02	&	0.85  $\pm$  0.05	&	0.01  $\pm$  0.05	&	0.08  $\pm$  0.05	&	0.80  $\pm$  0.05	&	\it{21  $\pm$  6}	&	94  $\pm$  8	\\
         	    
339.622$-$0.120	&	$-$36.85  $\pm$  0.01	&	0.25  $\pm$  0.04	&	0.08  $\pm$  0.04	&	0.1  $\pm$  0.04	&	0.12  $\pm$  0.04	&	\it{49  $\pm$  16}	&	\it{48  $\pm$  18}	\\
         	    &	$-$33.96  $\pm$  0.01	&	6.75  $\pm$  0.04	&	0.01  $\pm$  0.04	&	0.17  $\pm$  0.04	&	3.00  $\pm$  0.04	&	\it{2  $\pm$  1}	&	44  $\pm$  1	\\
         	    &	$-$32.32  $\pm$  0.01	&	0.28  $\pm$  0.04	&	0.09  $\pm$  0.04	&	0.09  $\pm$  0.04	&	0.28  $\pm$  0.04	&	\it{43  $\pm$  14}	&	100  $\pm$  20	\\

\hline
\end{tabular}
\end{table*}


\addtocounter{table}{-1}
\begin{table*}
\fontsize{9}{7}\selectfont
\centering
\caption{\it{$-$continued}}
\label{Table4}
\renewcommand{\arraystretch}{1.5}
\begin{tabular}{ llllllllc  }

\hline
\hline

\multicolumn{1}{l}{Source name} &
\multicolumn{1}{c}{Velocity} &
\multicolumn{1}{c}{I} & 
\multicolumn{1}{c}{Q} & 
\multicolumn{1}{c}{U} & 
\multicolumn{1}{c}{V} & 
\multicolumn{1}{c}{$P_l$} &
\multicolumn{1}{c}{$P_c$} \\ 

\multicolumn{1}{c}{} &
\multicolumn{1}{c}{km\,s$^{-1}$} &
\multicolumn{1}{c}{(Jy)} & 
\multicolumn{1}{c}{(Jy)} & 
\multicolumn{1}{c}{(Jy)} & 
\multicolumn{1}{c}{(Jy)} & 
\multicolumn{1}{c}{(per cent)} &
\multicolumn{1}{c}{(per cent)} \\

\hline

339.884$-$1.259	&	$-$37.82  $\pm$  0.03	&	32.54  $\pm$  0.04	&	0.31  $\pm$  0.04	&	3.45  $\pm$  0.04	&	27.9  $\pm$  0.04	&	11.0  $\pm$  0.1	&	86.0  $\pm$  0.2	\\
         	    &	$-$36.94  $\pm$  0.06	&	15.44  $\pm$  0.04	&	0.21  $\pm$  0.04	&	0.90  $\pm$  0.04	&	15.25  $\pm$  0.04	&	6.0  $\pm$  0.3	    &	99.0  $\pm$  0.4	\\
         	    &	$-$34.99  $\pm$  0.09	&	2.10  $\pm$  0.04	&	0.02  $\pm$  0.04	&	0.14  $\pm$  0.04	&	2.14  $\pm$  0.04	&	\it{6  $\pm$  2}	&	102  $\pm$  3	\\
         	    &	$-$34.16  $\pm$  0.05	&	4.11  $\pm$  0.04	&	0.24  $\pm$  0.04	&	0.19  $\pm$  0.04	&	4.32  $\pm$  0.04	&	7  $\pm$  1	        &	105  $\pm$  1	\\
         	    
340.785$-$0.096	&	$-$106.53  $\pm$  0.01	&	0.71  $\pm$  0.04	&	0.08  $\pm$  0.05	&	0.07  $\pm$  0.05	&	0.68  $\pm$  0.04	&	\it{29  $\pm$  7}	&	96  $\pm$  8	\\
         	    &	$-$105.58  $\pm$  0.01	&	1.47  $\pm$  0.04	&	0.10  $\pm$  0.05	&	0.02  $\pm$  0.05	&	1.39  $\pm$  0.04	&	\it{14  $\pm$  3}	&	95  $\pm$  4	\\
         	    &	$-$105.03  $\pm$  0.01	&	0.57  $\pm$  0.04	&	0.03  $\pm$  0.05	&	0.05  $\pm$  0.05	&	0.24  $\pm$  0.04	&	\it{28  $\pm$  9}	&	42  $\pm$  8	\\
         	    &	$-$100.61  $\pm$  0.01	&	0.57  $\pm$  0.04	&	0.15  $\pm$  0.05	&	0.01  $\pm$  0.05	&	0.59  $\pm$  0.04	&	\it{25  $\pm$  9}	&	104  $\pm$  10	\\
         	    &	$-$99.99  $\pm$  0.02	&	0.54  $\pm$  0.04	&	0.07  $\pm$  0.05	&	0.07  $\pm$  0.05	&	0.59  $\pm$  0.04	&	\it{37  $\pm$  9}	&	109  $\pm$  11	\\

344.582$-$0.023	&	$-$4.26  $\pm$  0.01	&	3.06  $\pm$  0.04	&	0.20  $\pm$  0.04	&	0.11  $\pm$  0.04	&	2.96  $\pm$  0.03	&	7  $\pm$  1	&	97  $\pm$  2	\\
         	    &	$-$3.85  $\pm$  0.01	&	3.33  $\pm$  0.04	&	0.46  $\pm$  0.04	&	0.15  $\pm$  0.04	&	3.07  $\pm$  0.03	&	14  $\pm$  1	&	92  $\pm$  1	\\
         	    
345.003$-$0.224	&	$-$29.26  $\pm$  0.01	&	37.00  $\pm$  0.06	&	2.61  $\pm$  0.06	&	1.16  $\pm$  0.06	&	36.63  $\pm$  0.06	&	8.0 $\pm$ 1	&	99  $\pm$  1	\\         	    
         	    
345.117$+$1.592	&	$-$16.56  $\pm$  0.01	&	8.13  $\pm$  0.06	&	0.12  $\pm$  0.06	&	0.12  $\pm$  0.06	&	1.99  $\pm$  0.06	&	\it{4 $\pm$ 1}	&	24  $\pm$  1	\\         	    
         	   
345.495$+$1.462	&	$-$24.1  $\pm$  0.01	&	0.69  $\pm$  0.04	&	0.01  $\pm$  0.04	&	0.05  $\pm$  0.04	&	0.56  $\pm$  0.04	&	\it{19  $\pm$  6}	&	81  $\pm$  7	\\

345.497$+$1.461	&	$-$18.47  $\pm$  0.01	&	1.23  $\pm$  0.04	&	0.13  $\pm$  0.04	&	0.08  $\pm$  0.04	&	0.42  $\pm$  0.04	&	\it{12  $\pm$  3}	&	34  $\pm$  3	\\
         	    &	$-$21.35  $\pm$  0.01	&	1.60  $\pm$  0.04	&	0.09  $\pm$  0.04	&	0.07  $\pm$  0.04	&	1.11  $\pm$  0.04	&	12  $\pm$ 2 	&	69  $\pm$  3	\\
         	    
348.727$-$1.039	&	$-$14.11  $\pm$  0.02	&	0.75  $\pm$  0.04	&	0.01  $\pm$  0.04	&	0.02  $\pm$  0.04	&	0.64  $\pm$  0.04	&	\it{14  $\pm$  5}	&	85  $\pm$  7	\\
         	    &	$-$13.46  $\pm$  0.02	&	1.17  $\pm$  0.04	&	0.11  $\pm$  0.04	&	0.05  $\pm$  0.04	&	0.79  $\pm$  0.04	&	\it{10  $\pm$  3}	&	68  $\pm$  4	\\
         	    
350.112$+$0.095	&	$-$66.06  $\pm$  0.02	&	0.66  $\pm$  0.07	&	0.12  $\pm$  0.07	&	0.01  $\pm$  0.07	&	0.63  $\pm$  0.07	&	\it{39  $\pm$  11}	&	95  $\pm$  15	\\         	    
         	    
350.686$-$0.491	&	$-$15.14  $\pm$  0.01	&	2.89  $\pm$  0.05	&	0.59  $\pm$  0.05	&	0.11  $\pm$  0.05	&	2.86  $\pm$  0.06	&	21  $\pm$  2	&	99  $\pm$  3	\\
         	    &	$-$14.67  $\pm$  0.01	&	1.21  $\pm$  0.05	&	0.15  $\pm$  0.05	&	0.15  $\pm$  0.05	&	1.08  $\pm$  0.06	&	\it{17  $\pm$  4}	&	89  $\pm$  6	\\
         	    &	$-$14.11  $\pm$  0.01	&	2.02  $\pm$  0.05	&	0.41  $\pm$  0.05	&	0.13  $\pm$  0.05	&	1.82  $\pm$  0.06	&	21  $\pm$  2	&	90  $\pm$  4	\\
         	    &	$-$13.63  $\pm$  0.01	&	2.15  $\pm$  0.05	&	0.15  $\pm$  0.05	&	0.29  $\pm$  0.05	&	2.04  $\pm$  0.06	&	15  $\pm$  2	&	95  $\pm$  4	\\
         	    
351.158$+$0.699	&	$-$8.62  $\pm$  0.01	&	8.78  $\pm$  0.04	&	0.25  $\pm$  0.04	&	0.57  $\pm$  0.04	&	4.87  $\pm$  0.04	&	7.0  $\pm$  0.3	&	55  $\pm$  1	\\         	    
         	    
351.419$+$0.646	&	$-$9.85  $\pm$  0.01	&	101.89  $\pm$  0.08	&	11.61  $\pm$  0.08	&	1.89  $\pm$  0.08	&	99.3  $\pm$  0.08	&	12.0  $\pm$  0.1	&	97.0  $\pm$  0.1	\\
         	    &	$-$10.56  $\pm$  0.01	&	75.77  $\pm$  0.08	&	2.28  $\pm$  0.08	&	1.04  $\pm$  0.08	&	75.38  $\pm$  0.08	&	2.5  $\pm$  0.1	&	99.0  $\pm$  0.1	\\
         	    
351.774$-$0.537	&	$-$1.52  $\pm$  0.01	&	0.84  $\pm$  0.05	&	0.03  $\pm$  0.05	&	0.01  $\pm$  0.05	&	0.75  $\pm$  0.06	&	\it{16  $\pm$  6}	&	89  $\pm$  9	\\
         	    &	$-$0.81  $\pm$  0.01	&	0.80  $\pm$  0.05	&	0.01  $\pm$  0.05	&	0.02  $\pm$  0.05	&	0.83  $\pm$  0.06	&	\it{15  $\pm$  6}	&	104  $\pm$  10	\\
         	    &	3.97  $\pm$  0.01	    &	3.45  $\pm$  0.05	&	0.10  $\pm$  0.05	&	0.10  $\pm$  0.05	&	3.15  $\pm$  0.06	&	7  $\pm$  1	&	91  $\pm$  2	\\
         	    &	4.32  $\pm$  0.01	    &	6.78  $\pm$  0.05	&	0.28  $\pm$  0.05	&	1.01  $\pm$  0.05	&	4.13  $\pm$  0.06	&	15  $\pm$  1	&	61  $\pm$  1	\\
         	    
353.410$-$0.360	&	$-$19.36  $\pm$  0.01	&	13.85  $\pm$  0.04	&	1.34  $\pm$  0.04	&	0.16  $\pm$  0.04	&	10.1  $\pm$  0.04	&	10.0  $\pm$  0.3	&	73.0  $\pm$  0.4	\\

357.557$-$0.321	&	$-$0.29  $\pm$  0.01	&	0.84  $\pm$  0.05	&	0.14  $\pm$  0.05	&	0.14  $\pm$  0.05	&	0.86  $\pm$  0.05	&	\it{23  $\pm$  6}	&	102  $\pm$  9	\\

359.970$-$0.457	&	15.76  $\pm$  0.04	    &	0.48  $\pm$  0.06	&	0.11  $\pm$  0.06	&	0.07  $\pm$  0.06	&	0.25  $\pm$  0.06	&	\it{52  $\pm$  13}	&	\it{52  $\pm$  14}	\\
000.376$+$0.040	&	41.24  $\pm$  0.01	    &	1.01  $\pm$  0.07	&	0.1  $\pm$  0.07	&	0.21  $\pm$  0.07	&	0.25  $\pm$  0.07	&	\it{22  $\pm$  7}	&	\it{25  $\pm$  7}	\\

000.665$-$0.036	&	60.71  $\pm$  0.08	    &	0.52  $\pm$  0.07	&	0.04  $\pm$  0.07	&	0.06  $\pm$  0.07	&	0.12  $\pm$  0.07	&	\it{41 $\pm$ 13}	&	\it{23  $\pm$  14}	\\
                &	61.36  $\pm$  0.01	    &	1.53  $\pm$  0.07	&	0.12  $\pm$  0.07	&	0.05  $\pm$  0.07	&	0.80  $\pm$  0.07	&	\it{18 $\pm$ 5}	&   52  $\pm$  5	\\

                &	62.33  $\pm$  0.04	    &	1.03  $\pm$  0.07	&	0.11  $\pm$  0.07	&	0.06  $\pm$  0.07	&	1.00  $\pm$  0.07	&	\it{26 $\pm$ 7}	&	97  $\pm$  9	\\


008.669$-$0.356	&	39.32  $\pm$  0.02	    &	1.89  $\pm$  0.07	&	035  $\pm$  0.07	&	0.26  $\pm$  0.07	&	1.68  $\pm$  0.07	&	23  $\pm$  4	&	89  $\pm$  5	\\
	&	39.69  $\pm$  0.01	    &	2.26  $\pm$  0.07	&	0.03  $\pm$  0.07	&	0.04  $\pm$  0.07	&	1.85  $\pm$  0.07	&	\it{8  $\pm$  3}	&	82  $\pm$  4	\\
	
010.474$+$0.027	&	64.64  $\pm$  0.01	    &	1.62  $\pm$  0.06	&	0.03  $\pm$  0.05	&	0.04  $\pm$  0.06	&	0.04  $\pm$  0.05	&	\it{10  $\pm$  3}	&	\it{2  $\pm$  3}	\\

011.034$+$0.062	&	21.78  $\pm$  0.02	    &	0.71  $\pm$  0.05	&	0.04  $\pm$  0.05	&	0.05  $\pm$  0.05	&	0.65  $\pm$  0.05	&	\it{23  $\pm$  7}	&	92  $\pm$  10	\\
         	    &	22.32  $\pm$  0.01	    &	1.41  $\pm$  0.05	&	0.07  $\pm$  0.05	&	0.10  $\pm$  0.05	&	1.23  $\pm$  0.05	&	\it{16  $\pm$  4}	&	87  $\pm$  5	\\
         	    &	23.17  $\pm$  0.01	    &	1.64  $\pm$  0.05	&	0.03  $\pm$  0.05	&	0.01  $\pm$  0.05	&	1.58  $\pm$  0.05	&	\it{8  $\pm$  3}	&	96  $\pm$  4	\\
         	    
017.638+0.158	&	28.42  $\pm$  0.02	    &	0.48  $\pm$  0.04	&	0.02  $\pm$  0.04	&	0.02  $\pm$  0.04	&	0.48  $\pm$  0.04	&	\it{23  $\pm$  8}	&	100  $\pm$  12	\\

\hline
\hline

269.141$-$1.213	&	9.23  $\pm$  0.01	&	2.37  $\pm$  0.04	&	0.07  $\pm$  0.04	&	0.04  $\pm$  0.04	&	2.37  $\pm$  0.04	&	\it{7  $\pm$  2}	&	100  $\pm$  3	\\
         	    &	10.51  $\pm$  0.01	&	4.32  $\pm$  0.04	&	0.03  $\pm$  0.04	&	0.01  $\pm$  0.04	&	0.51  $\pm$  0.04	&	\it{3  $\pm$  1}	&	12  $\pm$  1	\\
         	    &	11.86  $\pm$  0.01	&	1.67  $\pm$  0.04	&	0.02  $\pm$  0.04	&	0.02  $\pm$  0.04	&	1.24 $\pm$  0.04	 &	\it{6  $\pm$  2}	&	74  $\pm$  3	\\
         	    
337.801$-$0.053	&	$-$45.13  $\pm$  0.01	&	2.02  $\pm$  0.05	&	0.16  $\pm$  0.05	&	0.03  $\pm$  0.05	&	0.08  $\pm$  0.05	&	\it{8  $\pm$  2}	&	\it{4  $\pm$  2}	\\

349.729$+$0.166	&	16.82  $\pm$  0.01	&	1.59  $\pm$  0.04	&	0.05  $\pm$  0.04	&	0.01  $\pm$  0.04	&	0.03  $\pm$  0.03	&	\it{8  $\pm$  3}	&	\it{2  $\pm$  2}	\\

349.731$+$1.731	&	15.6  $\pm$  0.01	&	0.85  $\pm$  0.04	&	0.05  $\pm$  0.04	&	0.18  $\pm$  0.04	&	0.04  $\pm$  0.03	&	\it{21  $\pm$  5}	&	\it{5  $\pm$  4}	\\
         	    &	16.52  $\pm$  0.01	&	0.83  $\pm$  0.04	&	0.05  $\pm$  0.04	&	0.01  $\pm$  0.04	&	0.09  $\pm$  0.03	&	\it{16  $\pm$  5}	&	\it{11  $\pm$  4}	\\         	    

349.734$+$0.172	&	15.29  $\pm$  0.01	&	2.61  $\pm$  0.04	&	0.06  $\pm$  0.04	&	0.01  $\pm$  0.04	&	0.04  $\pm$  0.03	&	\it{5  $\pm$  2}	&	\it{2  $\pm$  1}	\\

358.936$-$0.485 	&	$-$6.80  $\pm$  0.01	&	3.62  $\pm$  0.07	&	0.22  $\pm$  0.07	&	0.31  $\pm$  0.07	&	0.58  $\pm$  0.07	&	10 $\pm$ 2	&	16  $\pm$  2	\\

358.983$-$0.652	&	$-$6.29  $\pm$  0.01	&	4.14  $\pm$  0.07	&	0.06  $\pm$  0.07	&	0.02  $\pm$  0.07	&	0.09  $\pm$  0.07	&	\it{5 $\pm$ 2}	&	\it{2  $\pm$  2}	\\

359.940$-$0.067	&	66.35  $\pm$  0.01	&	9.54  $\pm$  0.04	&	0.41  $\pm$  0.04	&	0.13  $\pm$  0.04	&	1.45  $\pm$  0.03	&	4.5  $\pm$  0.4	&	15.0  $\pm$  0.3	\\

001.010$-$0.225	&	$-$1.38  $\pm$  0.01	&	0.72  $\pm$  0.06	&	0.01  $\pm$  0.05	&	0.01  $\pm$  0.05	&	0.07  $\pm$  0.05	&	\it{16  $\pm$  7}	&	\it{10  $\pm$  7}	\\

006.584$-$0.052	&	4.81  $\pm$  0.01	&	1.06  $\pm$  0.06	&	0.12  $\pm$  0.06	&	0.01  $\pm$  0.06	&	0.06  $\pm$  0.06	&	\it{23  $\pm$  6}	&	\it{6  $\pm$  6}	\\
         	    &	6.17  $\pm$  0.01	&	5.91  $\pm$  0.06	&	0.03  $\pm$  0.06	&	0.16  $\pm$  0.06	&	0.17  $\pm$  0.06	&	5  $\pm$  1	&	\it{3  $\pm$  1}	\\
         	    
006.686$-$0.296	&	9.9  $\pm$  0.01	&	53.13  $\pm$  0.06	&	4.29  $\pm$  0.06	&	0.29  $\pm$  0.06	&	4.81  $\pm$  0.06	&	8.0  $\pm$  0.1	&	9.0  $\pm$  0.1	\\
         	    &	11.4  $\pm$  0.01	&	103.32  $\pm$  0.06	&	6.22  $\pm$  0.06	&	3.01  $\pm$  0.06	&	6.67  $\pm$  0.06	&	7.0  $\pm$  0.1	&	6.5  $\pm$  0.1	\\
         	    
\hline
\end{tabular}
\end{table*}

\clearpage
\newpage

\addtocounter{table}{-1}

\begin{table*}
\fontsize{9}{7}\selectfont
\centering
\caption{\it{$-$continued}}

\renewcommand{\arraystretch}{1.5}
\begin{tabular}{ llllllllc  }

\hline
\hline

\multicolumn{1}{l}{Source name} &
\multicolumn{1}{c}{Velocity} &
\multicolumn{1}{c}{I} & 
\multicolumn{1}{c}{Q} & 
\multicolumn{1}{c}{U} & 
\multicolumn{1}{c}{V} & 
\multicolumn{1}{c}{P$_l$} &
\multicolumn{1}{c}{P$_c$} \\ 

\multicolumn{1}{c}{} &
\multicolumn{1}{c}{km\,s$^{-1}$} &
\multicolumn{1}{c}{(Jy)} & 
\multicolumn{1}{c}{(Jy)} & 
\multicolumn{1}{c}{(Jy)} & 
\multicolumn{1}{c}{(Jy)} & 
\multicolumn{1}{c}{(per cent)} &
\multicolumn{1}{c}{(per cent)} \\

\hline

006.696$-$0.284	&	10.04  $\pm$  0.01	&	15.07  $\pm$  0.06	&	0.10  $\pm$  0.06	&	0.02  $\pm$  0.06	&	0.38  $\pm$  0.06	&	\it{1  $\pm$  1}	&	\it{2  $\pm$  1}	\\

006.698$-$0.281	&	15.63  $\pm$  0.01	&	6.49  $\pm$  0.06	&	0.22  $\pm$  0.06	&	0.3  $\pm$  0.06	&	0.04  $\pm$  0.06	&	6  $\pm$  1	&	\it{1  $\pm$  1}	\\

006.700$-$0.280	&	13.38  $\pm$  0.01	&	15.15  $\pm$  0.06	&	0.17  $\pm$  0.06	&	0.24  $\pm$  0.06	&	0.05  $\pm$  0.06	&	\it{1.9  $\pm$  0.4}	&	\it{0.3  $\pm$  0.3}	\\

006.703$-$0.283	&	9.4  $\pm$  0.01	&	12.3  $\pm$  0.06	&	0.08  $\pm$  0.06	&	0.19  $\pm$  0.06	&	0.27  $\pm$  0.06	&	\it{1.6  $\pm$  0.5}	&	\it{2.0  $\pm$  0.5}	\\

006.706$-$0.281	&	11.86  $\pm$  0.01	&	38.63  $\pm$  0.06	&	0.09  $\pm$  0.06	&	0.04  $\pm$  0.06	&	0.63  $\pm$  0.06	&	\it{0.6  $\pm$ 0.2}  	&	2.0  $\pm$  0.1	\\

\hline
\hline

012.803$-$0.202	&	32.13  $\pm$  0.01	&	0.44  $\pm$  0.05	&	0.01  $\pm$  0.05	&	0.15  $\pm$  0.05	&	0.13  $\pm$  0.05	&	\it{32  $\pm$  11}	&	\it{30  $\pm$  12}	\\

019.610$-$0.234	&	43.67  $\pm$  0.01	&	0.43  $\pm$  0.05	&	0.01  $\pm$  0.05	&	0.08  $\pm$  0.05	&	0.03  $\pm$  0.05	&	\it{42  $\pm$  12}	&	\it{7  $\pm$  12}	\\

\hline
\end{tabular}
\end{table*}



\begin{figure*}
\subfloat{\includegraphics[width = 3.5in]{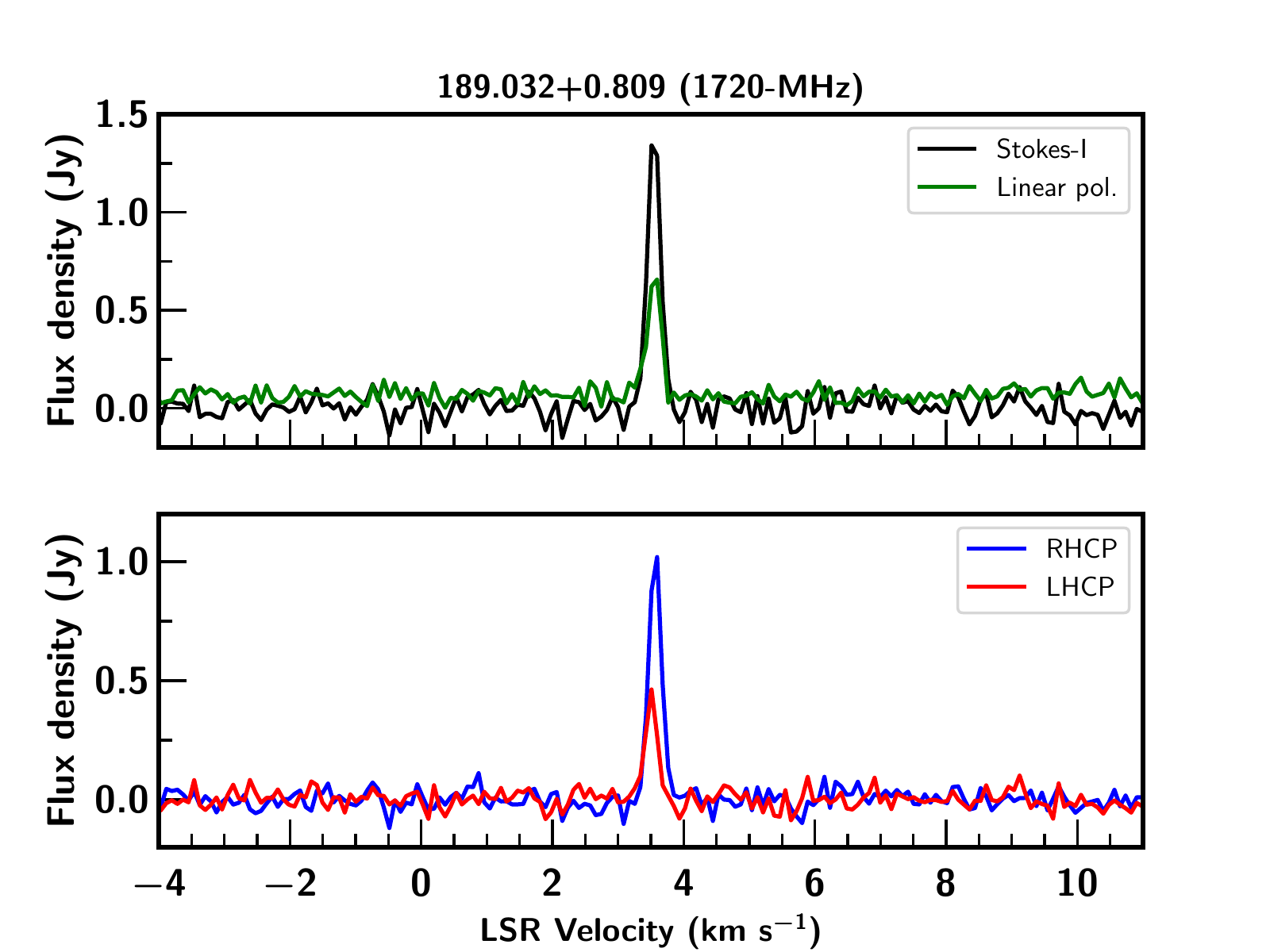}} 
\subfloat{\includegraphics[width = 3.5in]{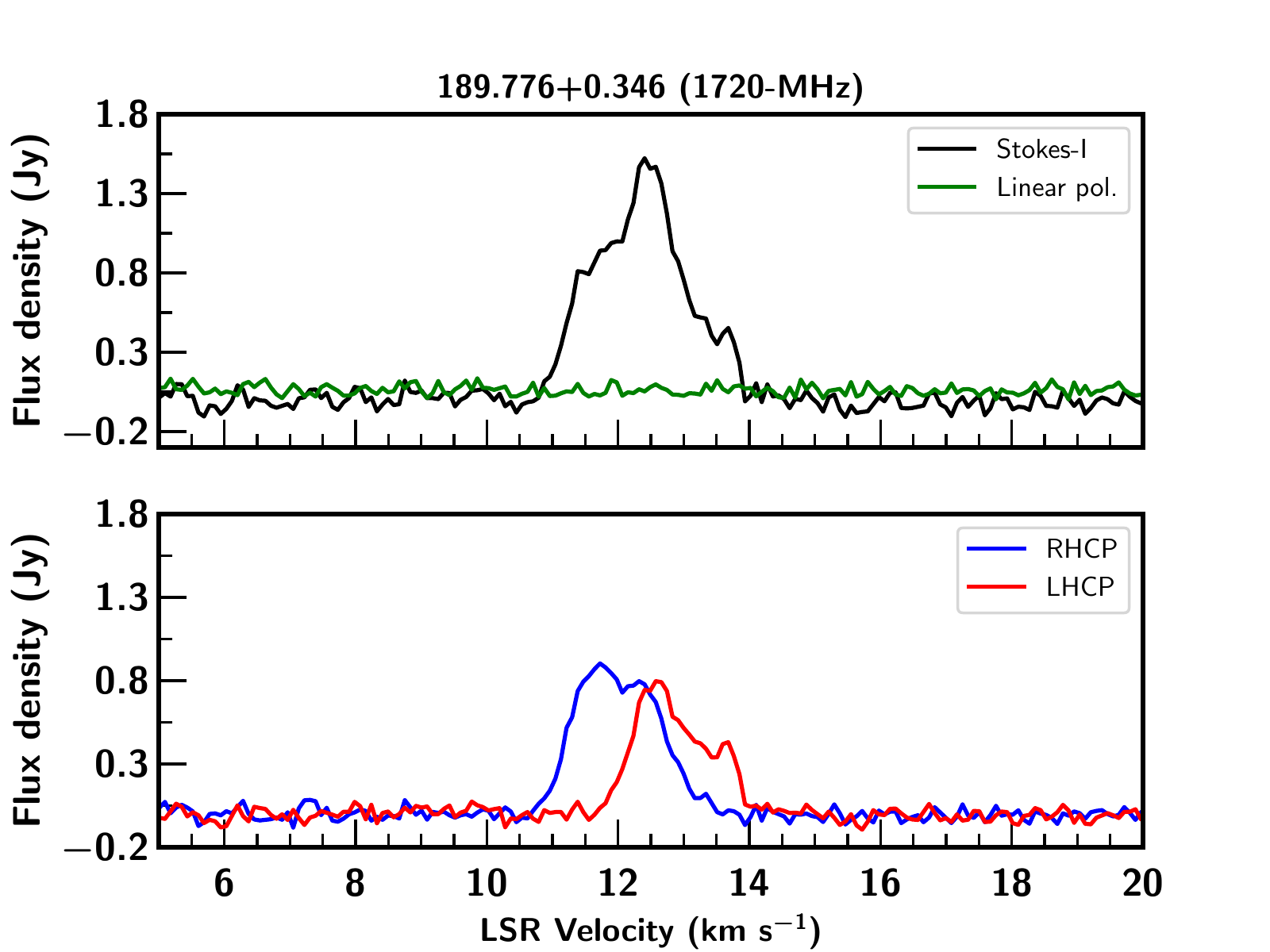}}\\
\subfloat{\includegraphics[width = 3.5in]{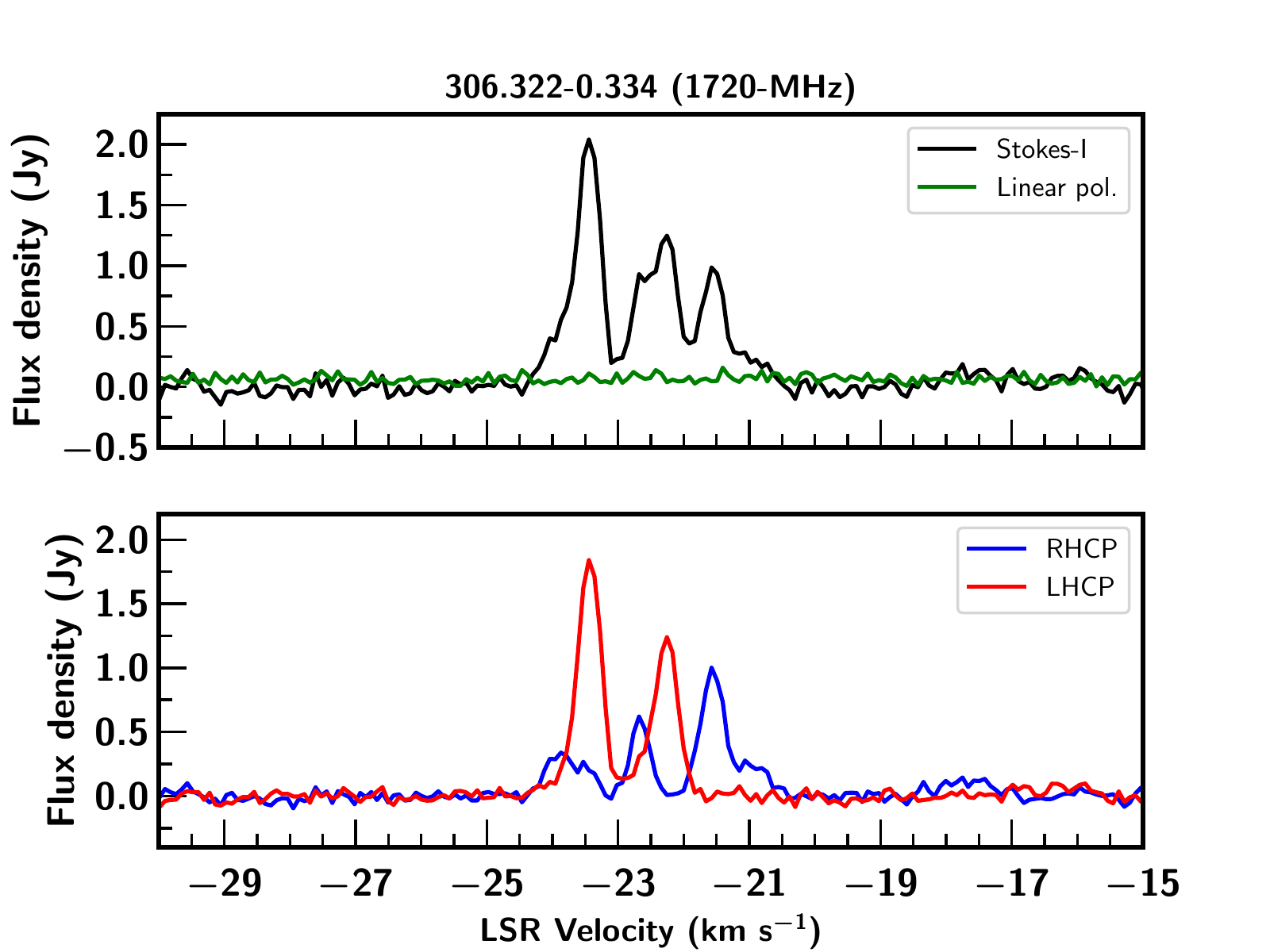}}
\subfloat{\includegraphics[width = 3.5in]{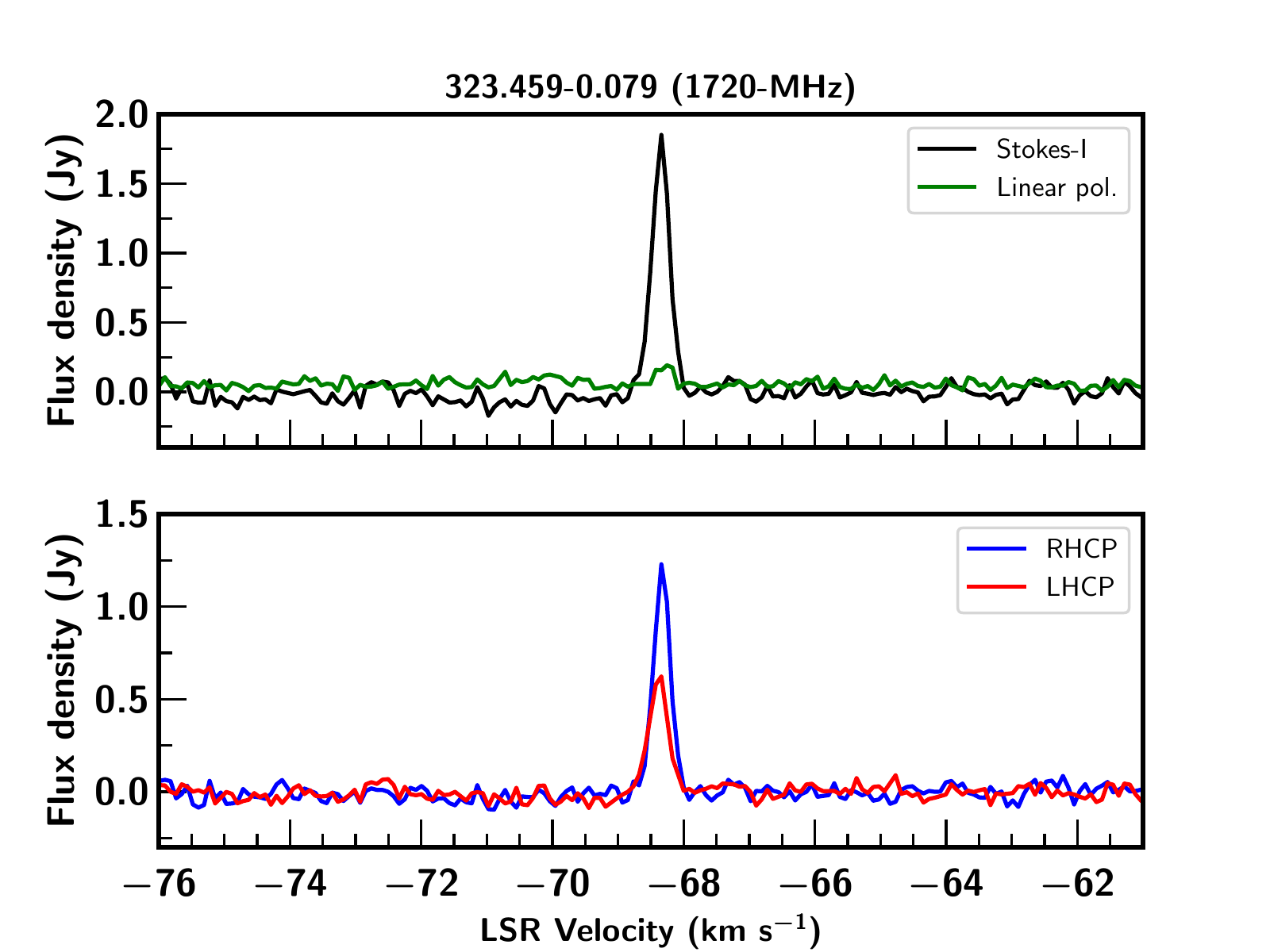}}\\
\caption{Spectra of 1720-MHz OH masers associated with star forming regions}
\label{Figure1}
\end{figure*}

\clearpage
\newpage


\addtocounter{figure}{-1}
\begin{figure*}
\subfloat{\includegraphics[width = 3.5in]{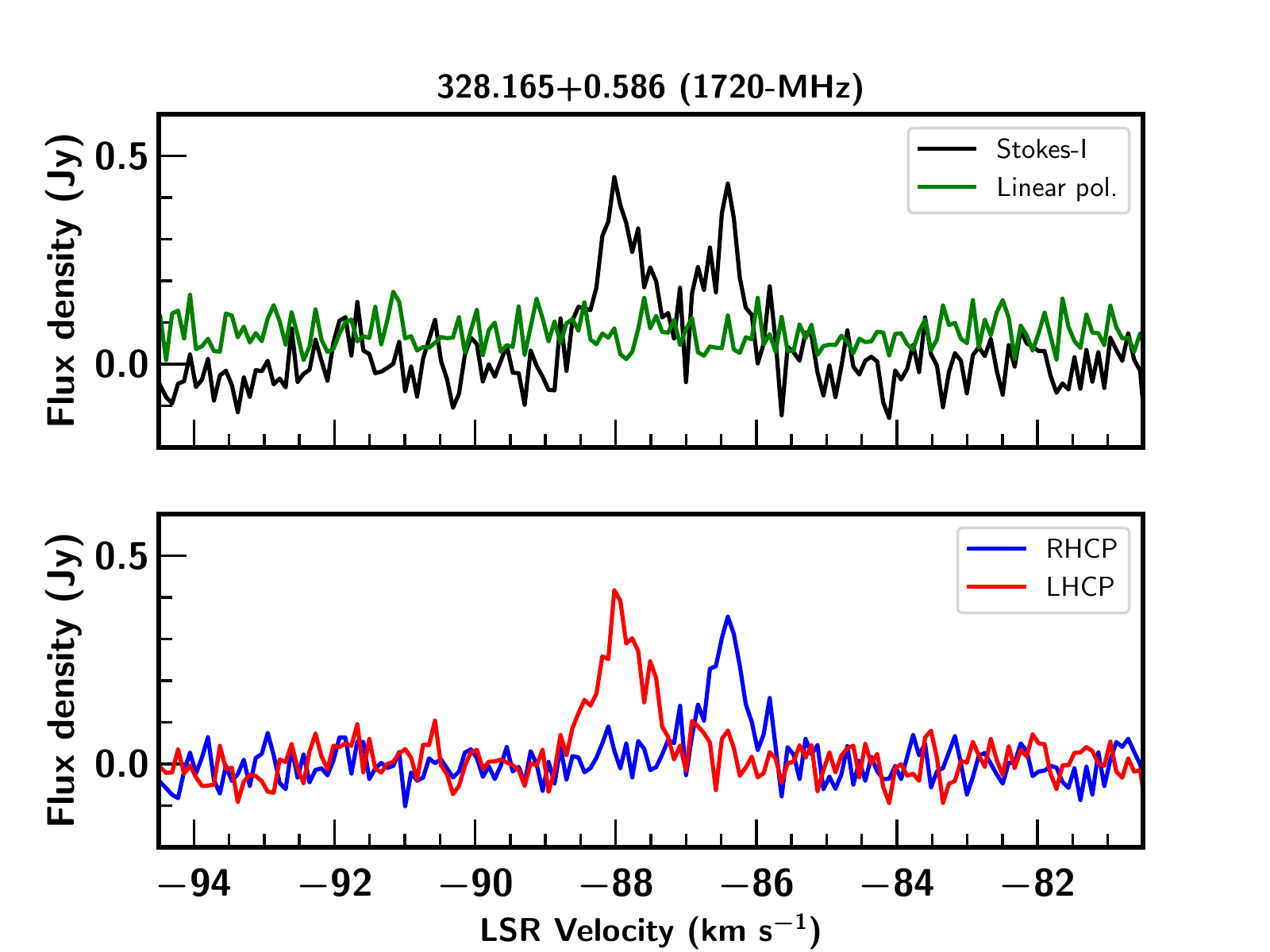}}
\subfloat{\includegraphics[width = 3.5in]{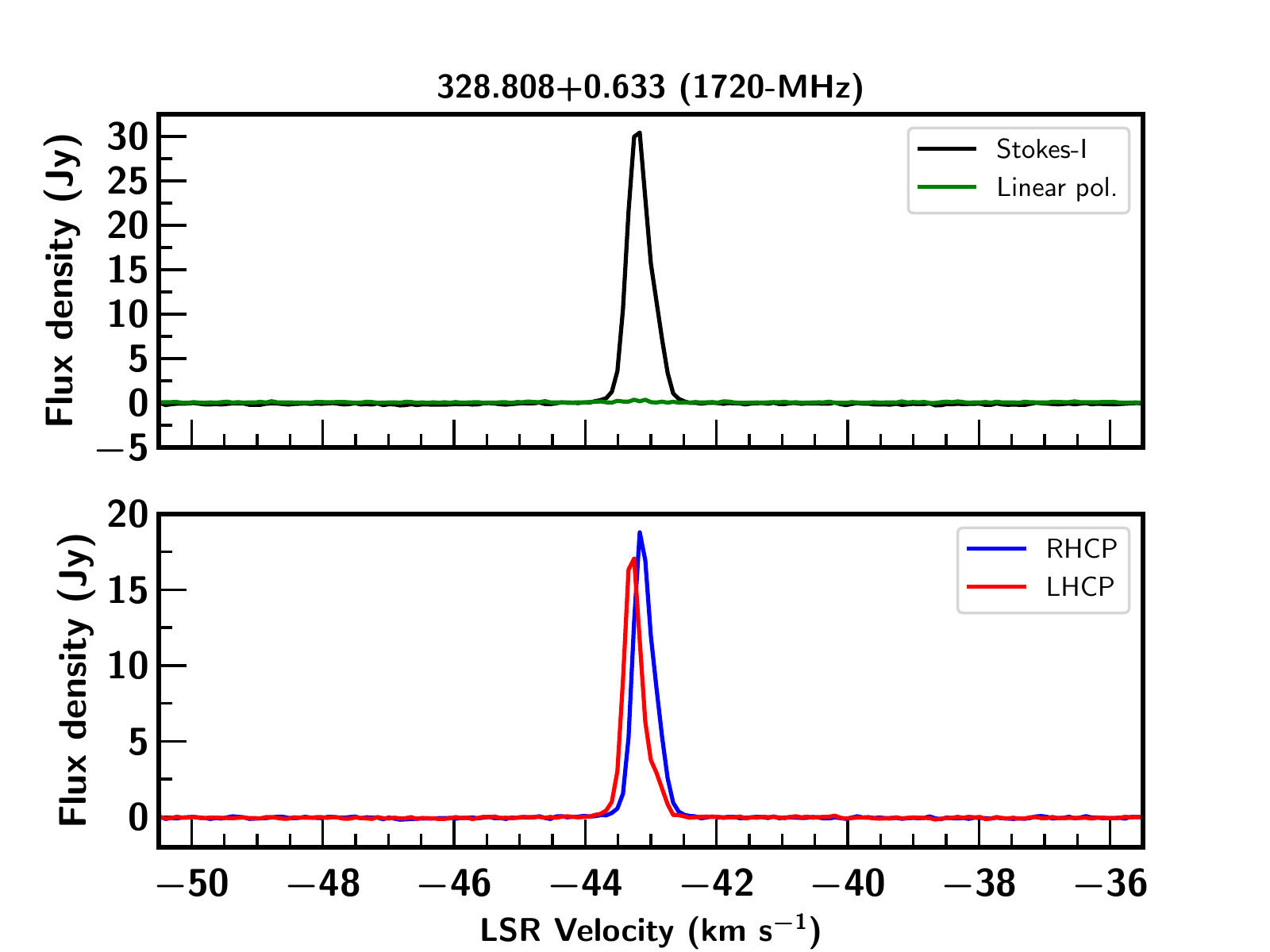}}\\
\subfloat{\includegraphics[width = 3.5in]{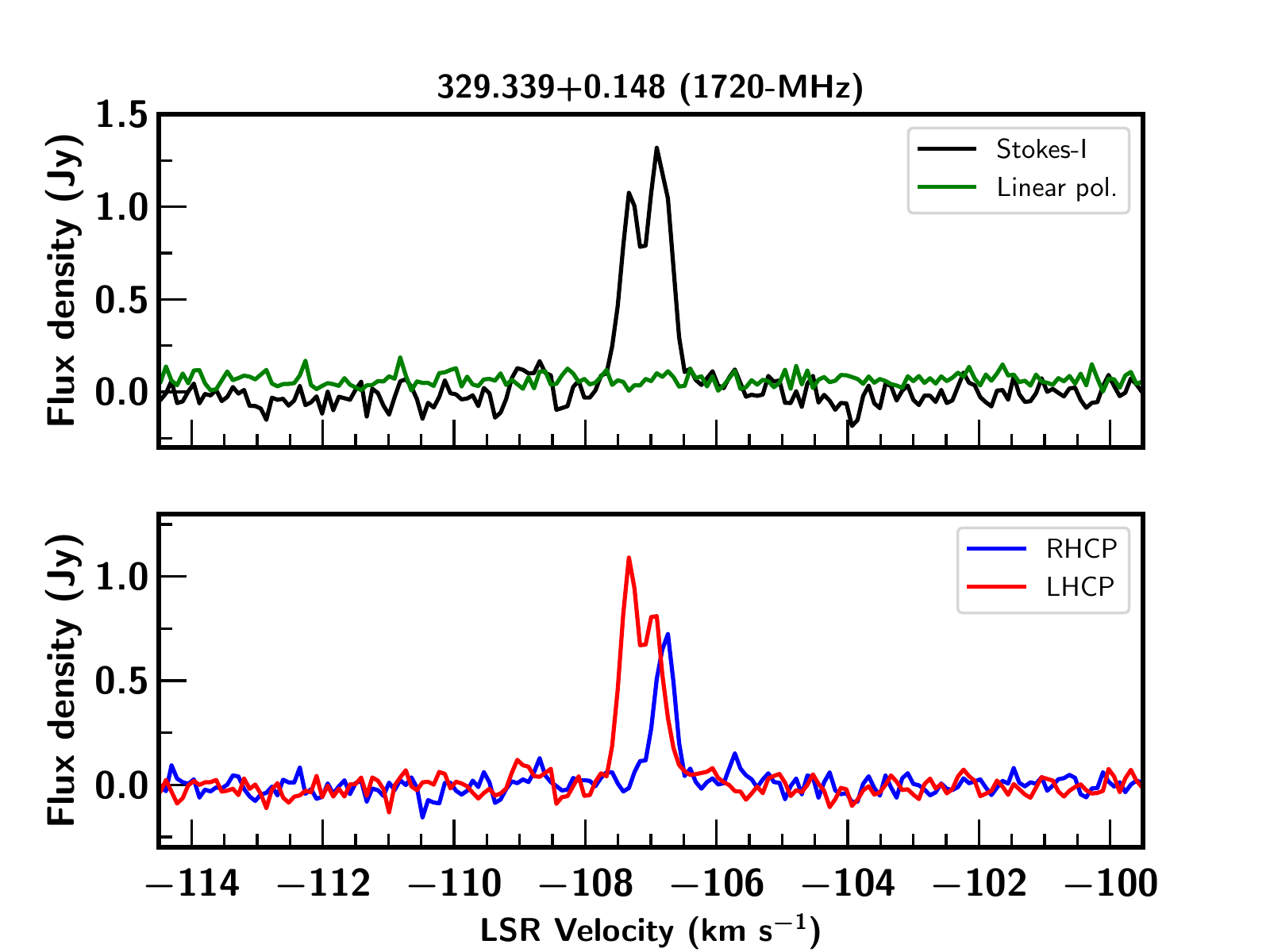}} 
\subfloat{\includegraphics[width = 3.5in]{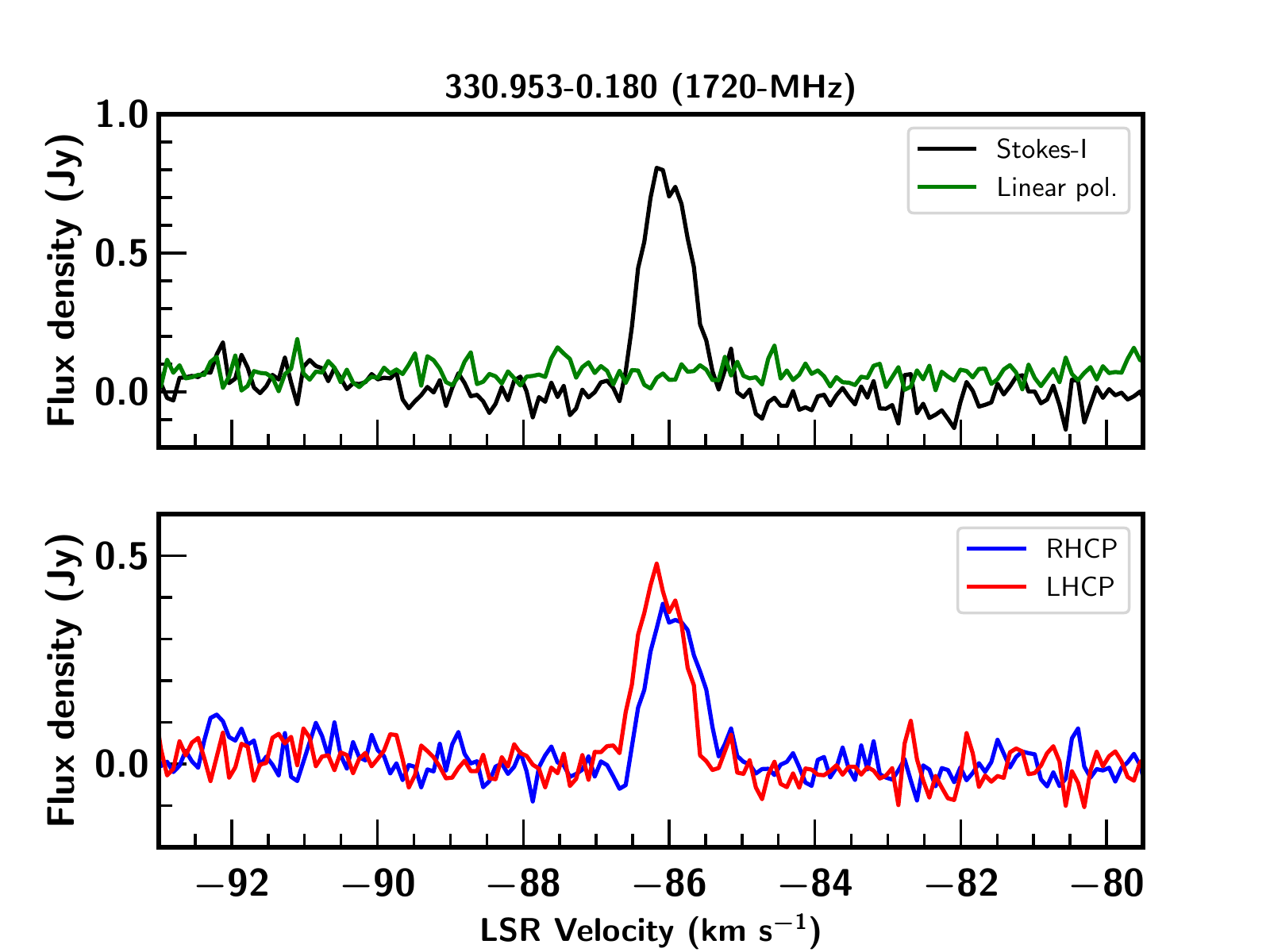}}\\
\subfloat{\includegraphics[width = 3.5in]{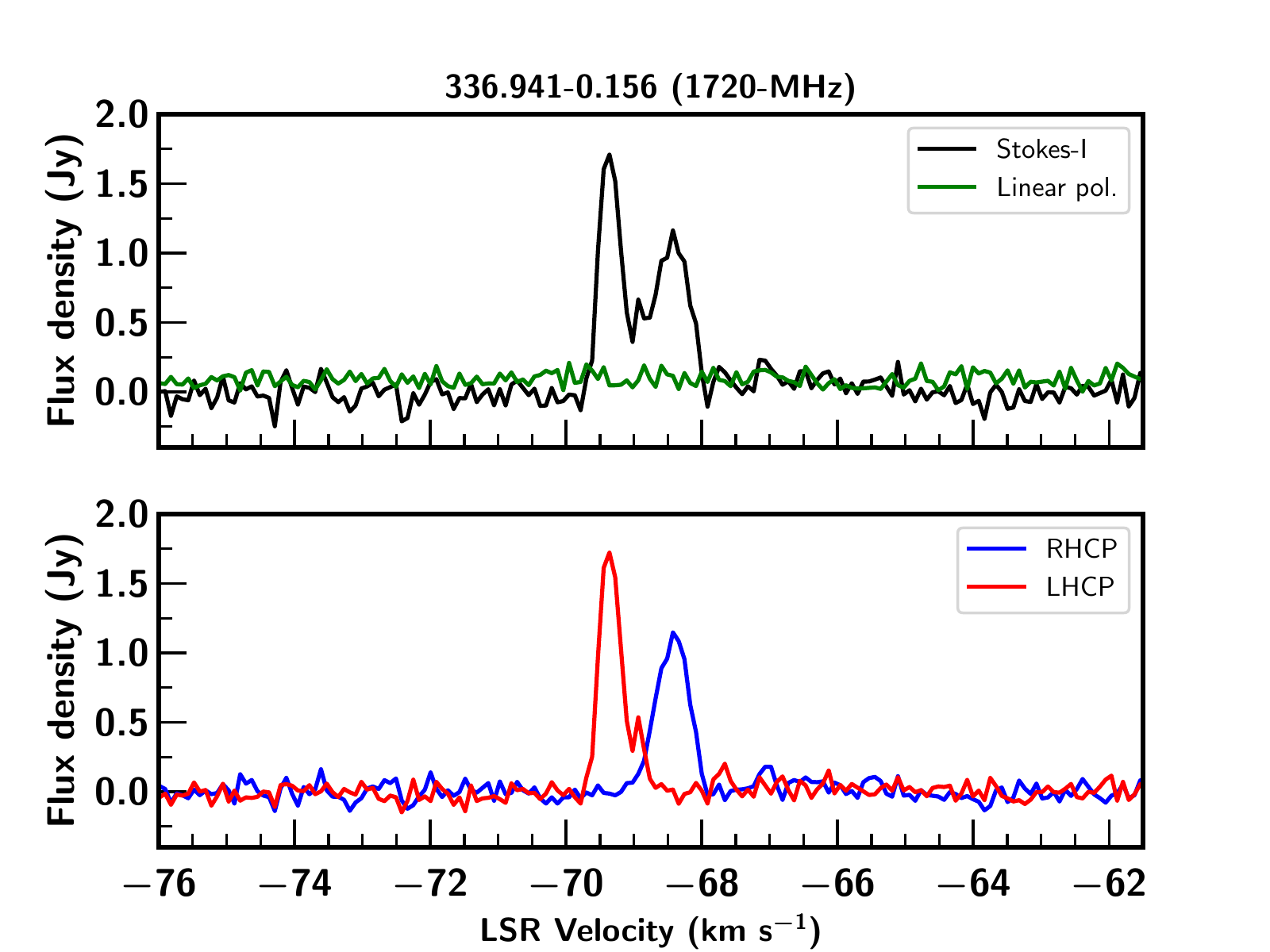}}
\subfloat{\includegraphics[width = 3.5in]{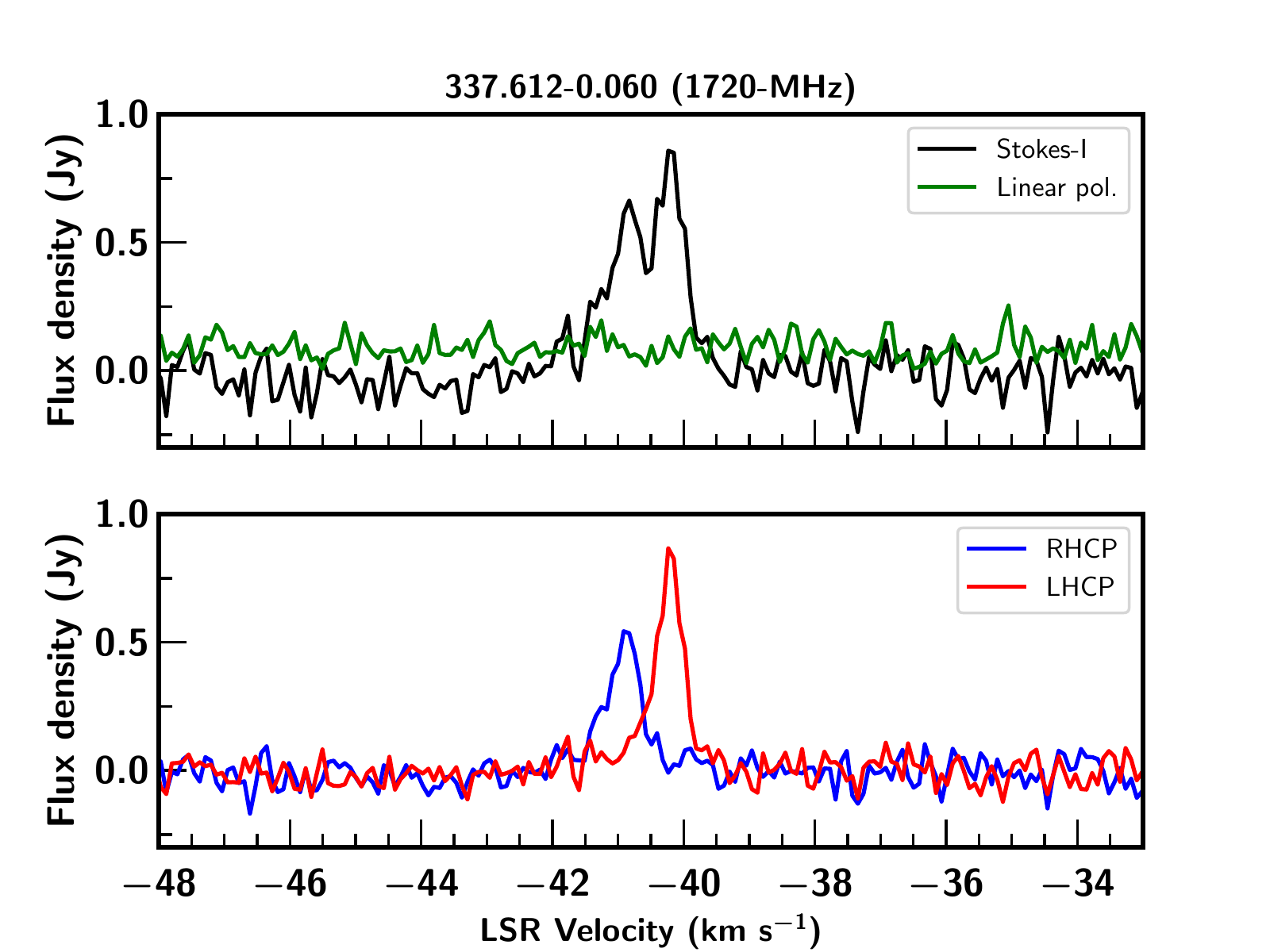}}
\caption{\it{$-$continued}}
\end{figure*}

\clearpage
\newpage


\addtocounter{figure}{-1}
\begin{figure*}

\subfloat{\includegraphics[width = 3.5in]{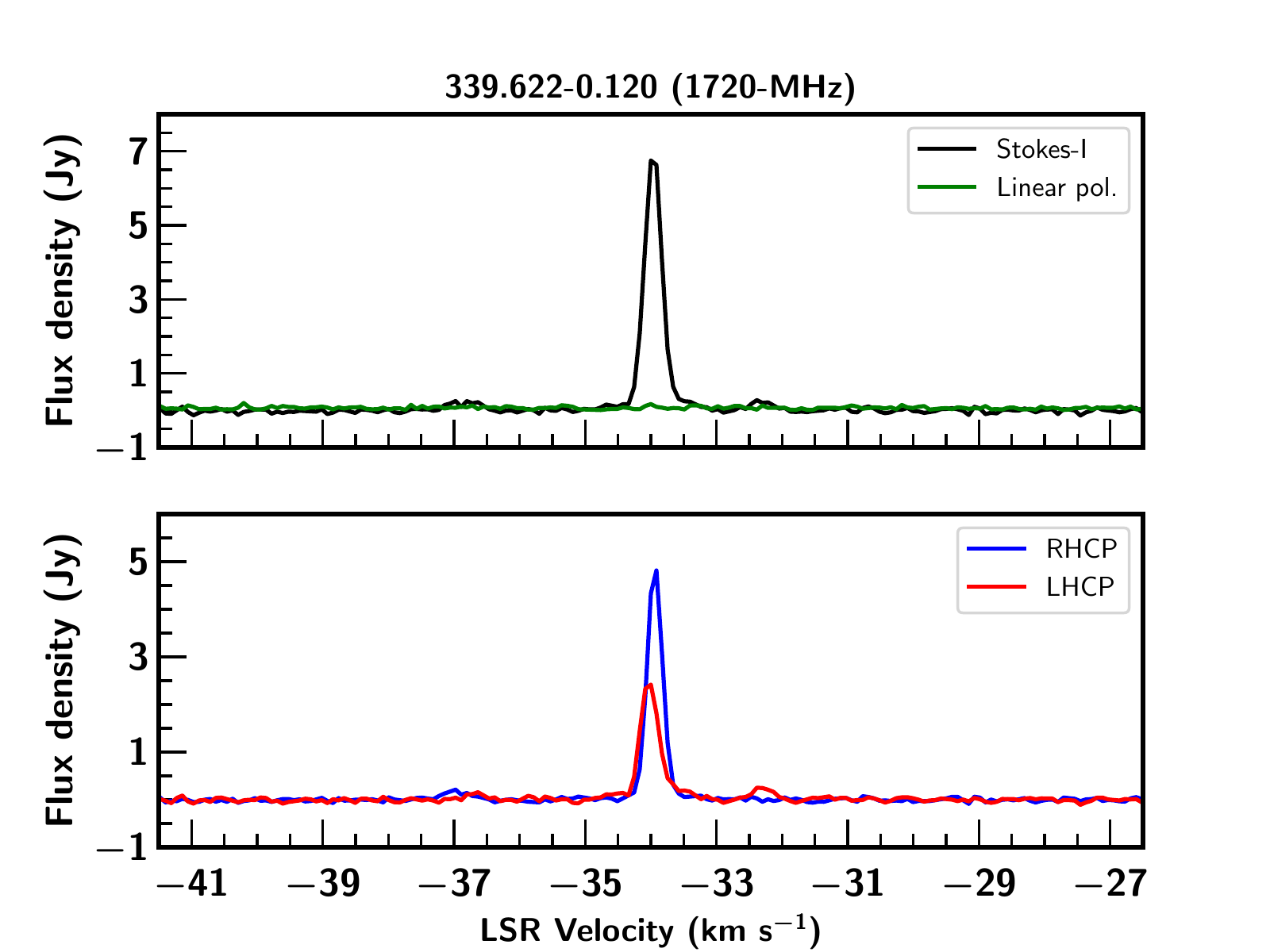}}
\subfloat{\includegraphics[width = 3.5in]{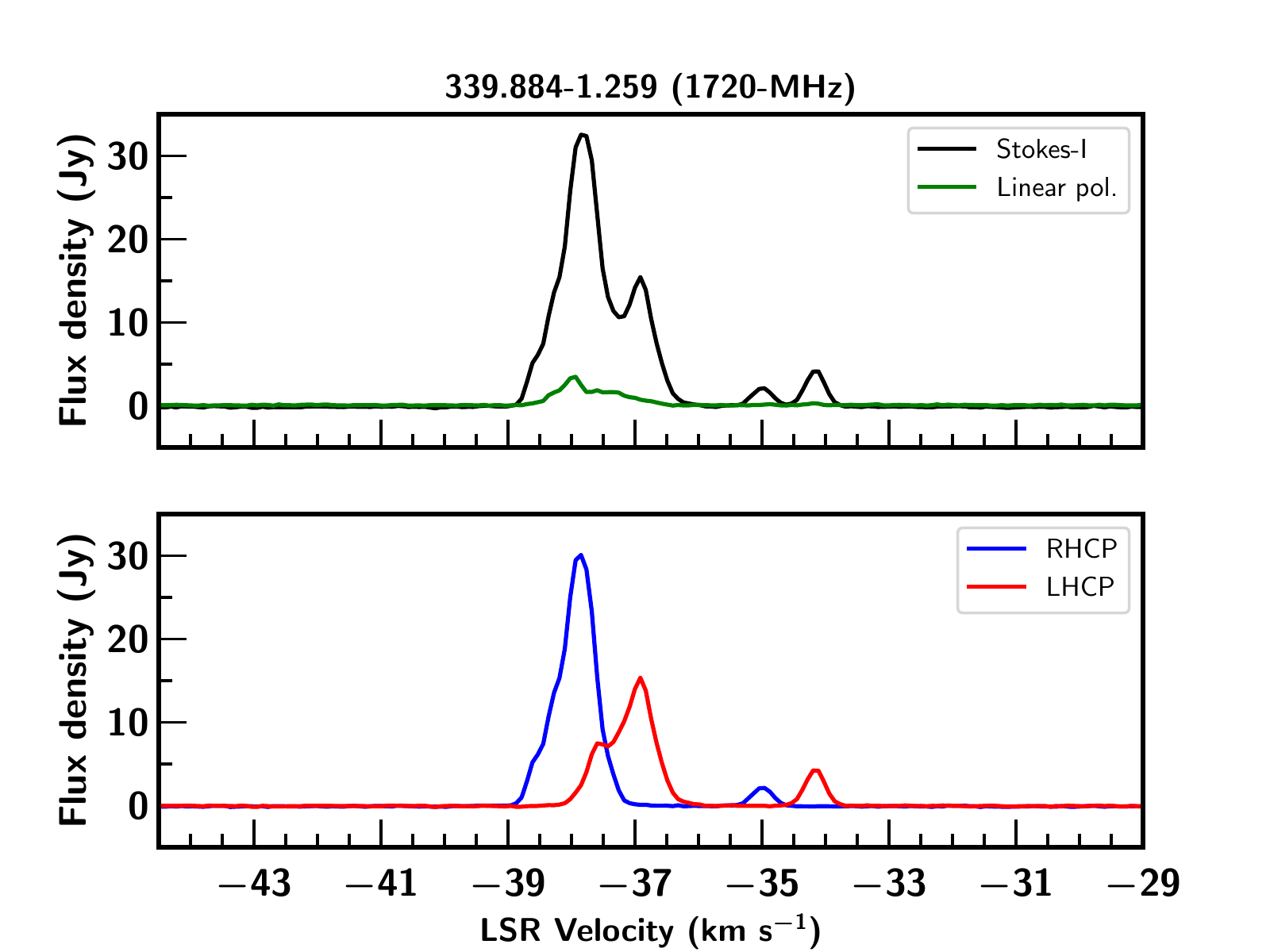}}\\
\subfloat{\includegraphics[width = 3.5in]{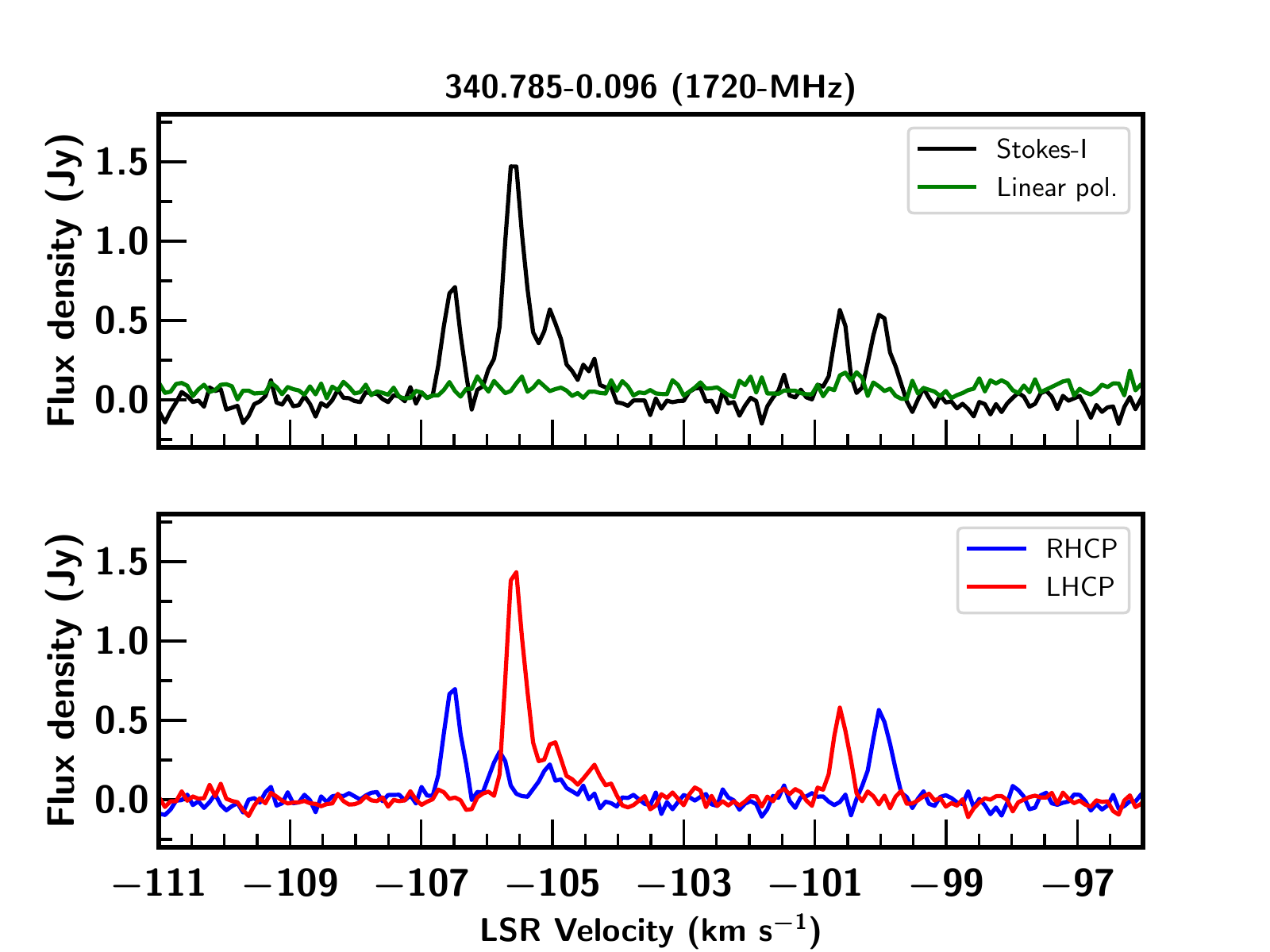}}
\subfloat{\includegraphics[width = 3.5in]{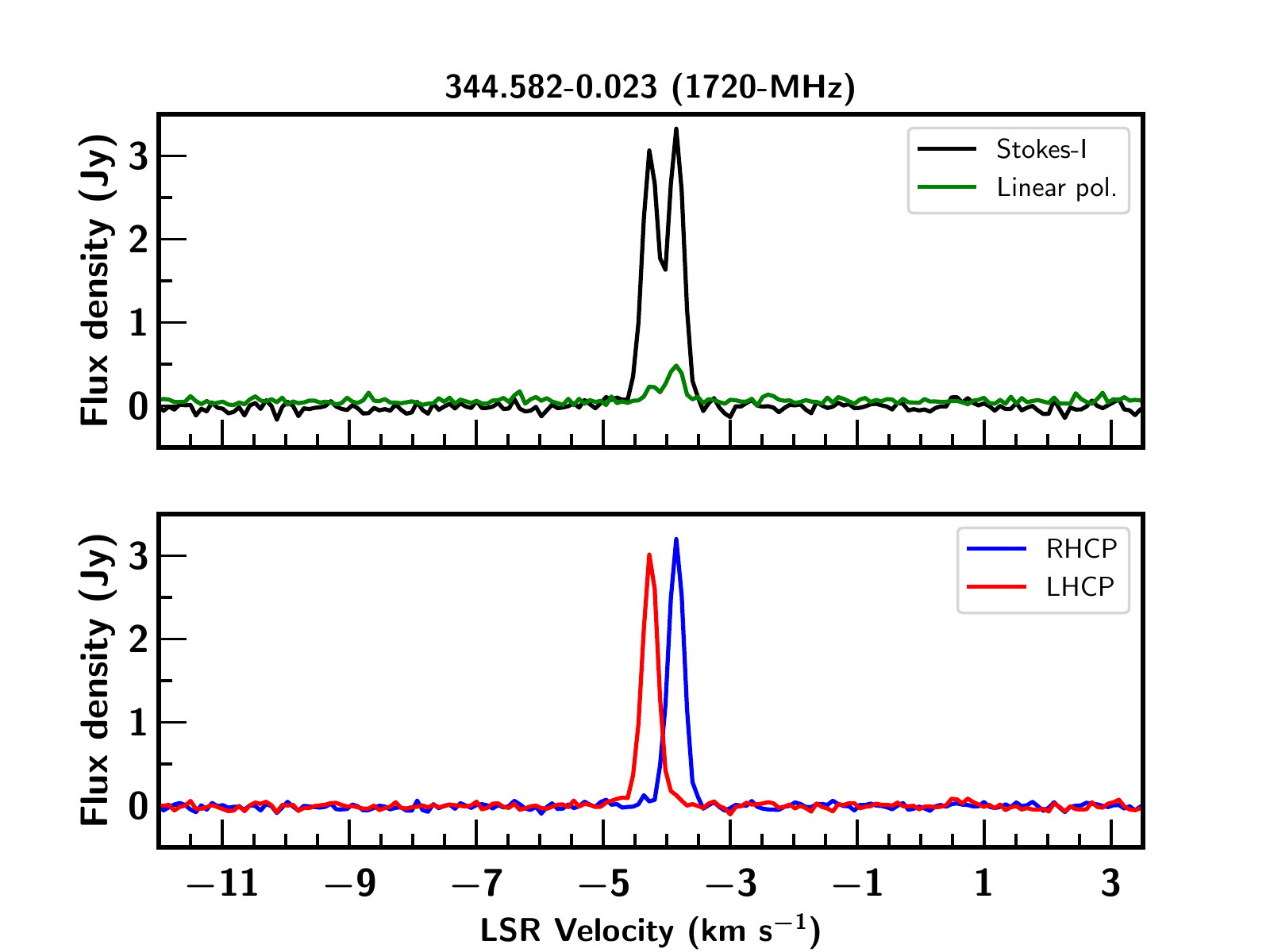}}\\
\subfloat{\includegraphics[width =3.5in]{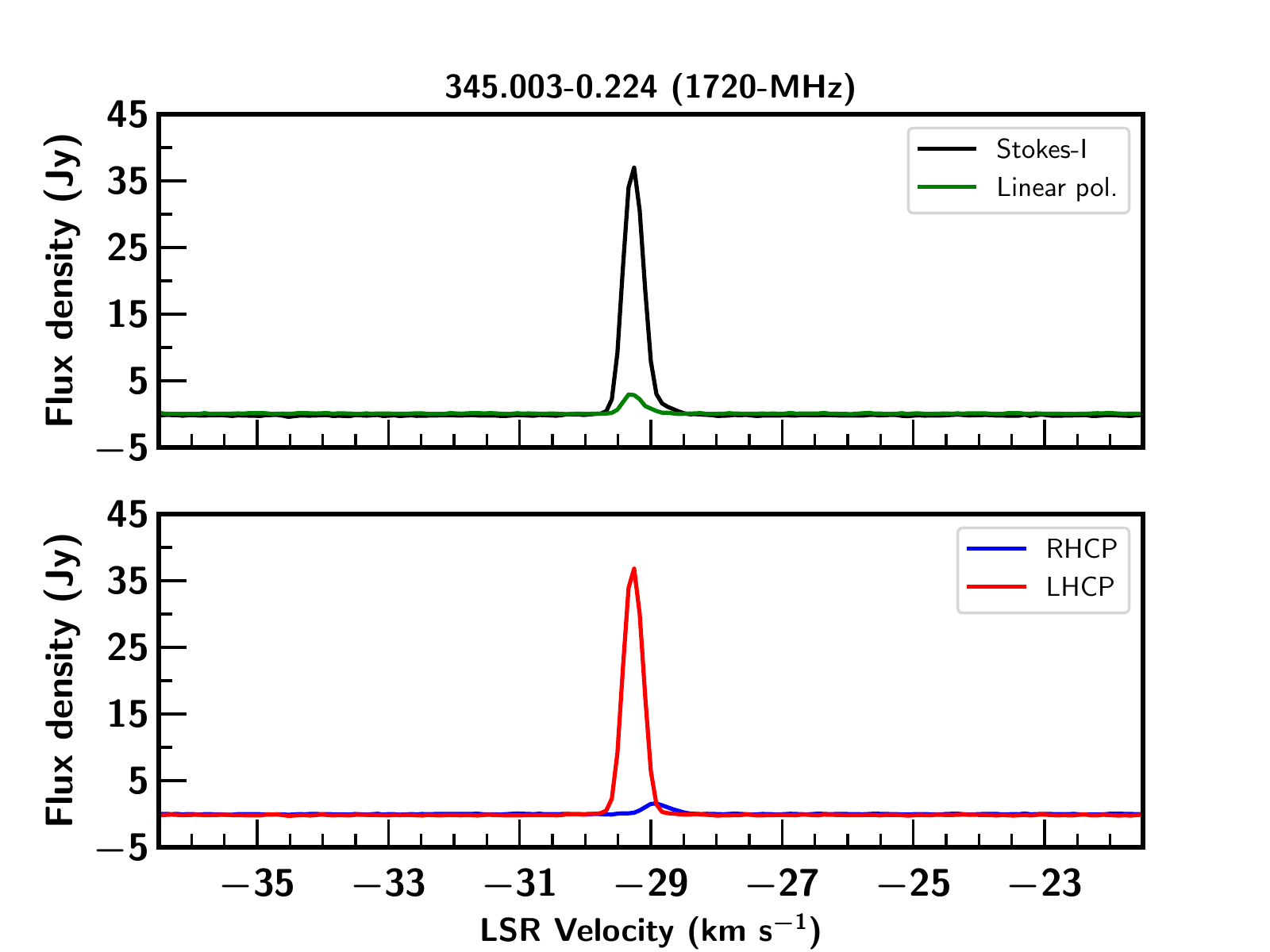}}
\subfloat{\includegraphics[width =3.5in]{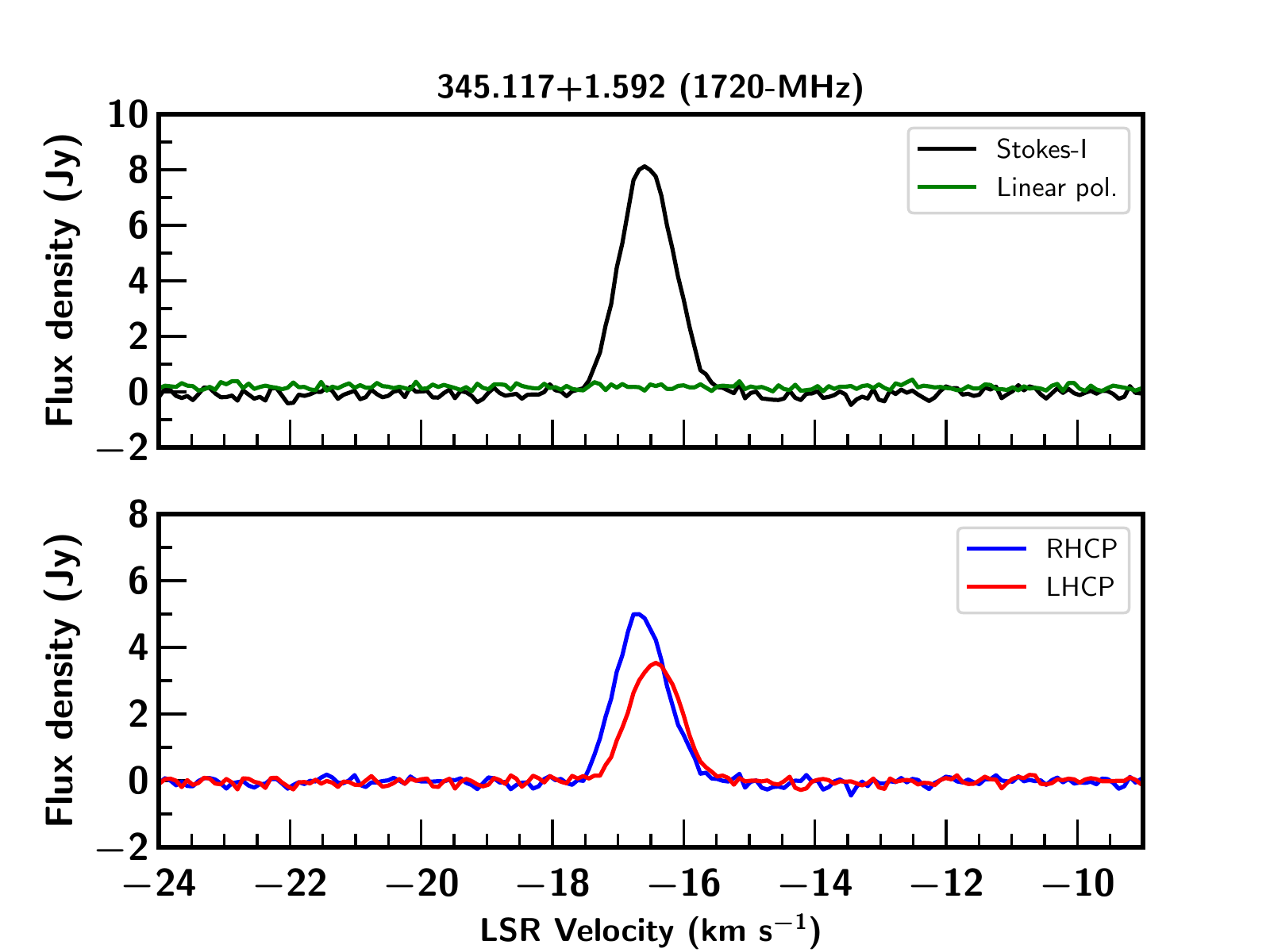}}\\
\caption{\it{$-$continued}}
\end{figure*}

\clearpage
\newpage


\addtocounter{figure}{-1}
\begin{figure*}

\subfloat{\includegraphics[width =3.5in]{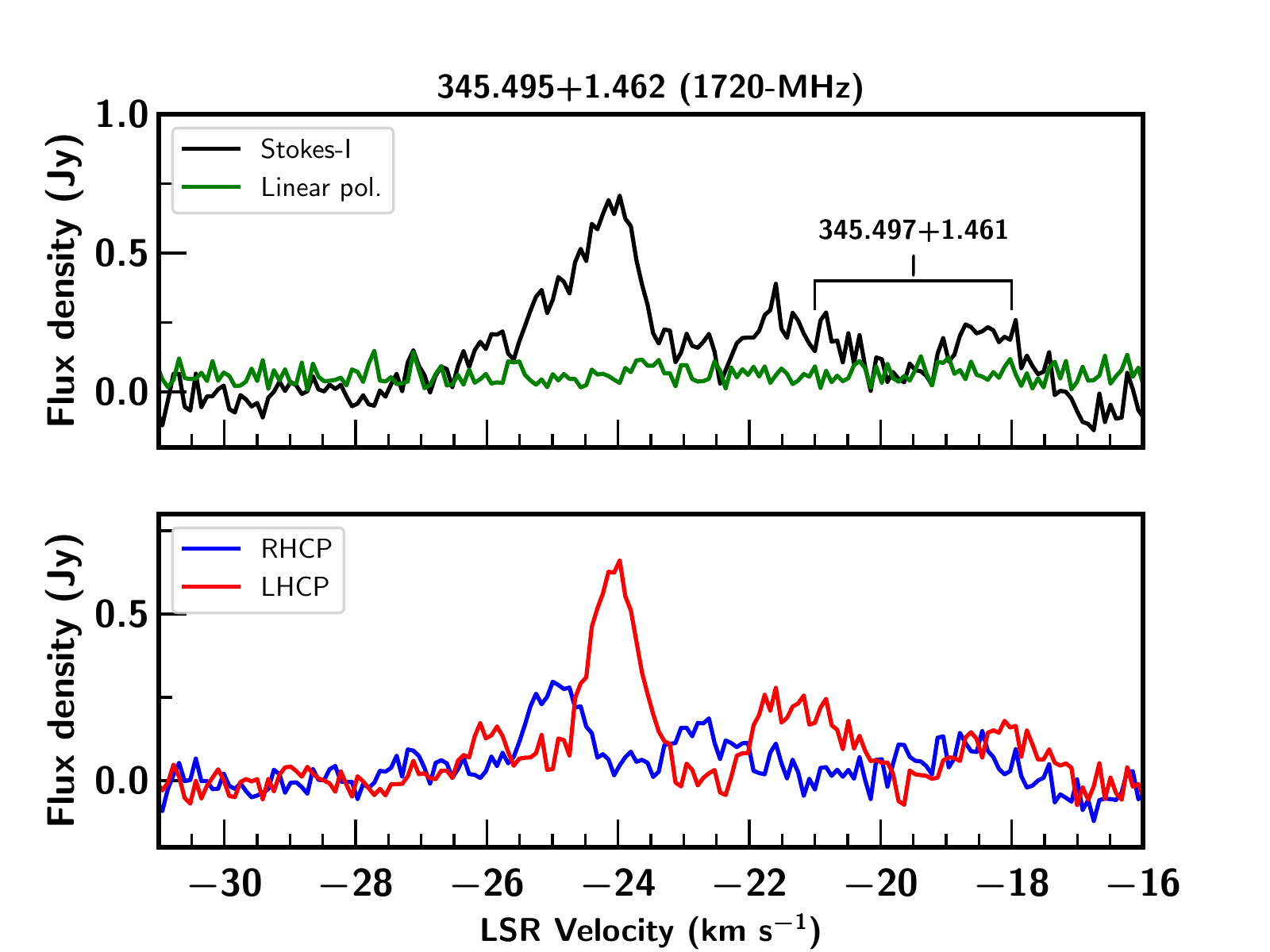}}
\subfloat{\includegraphics[width =3.5in]{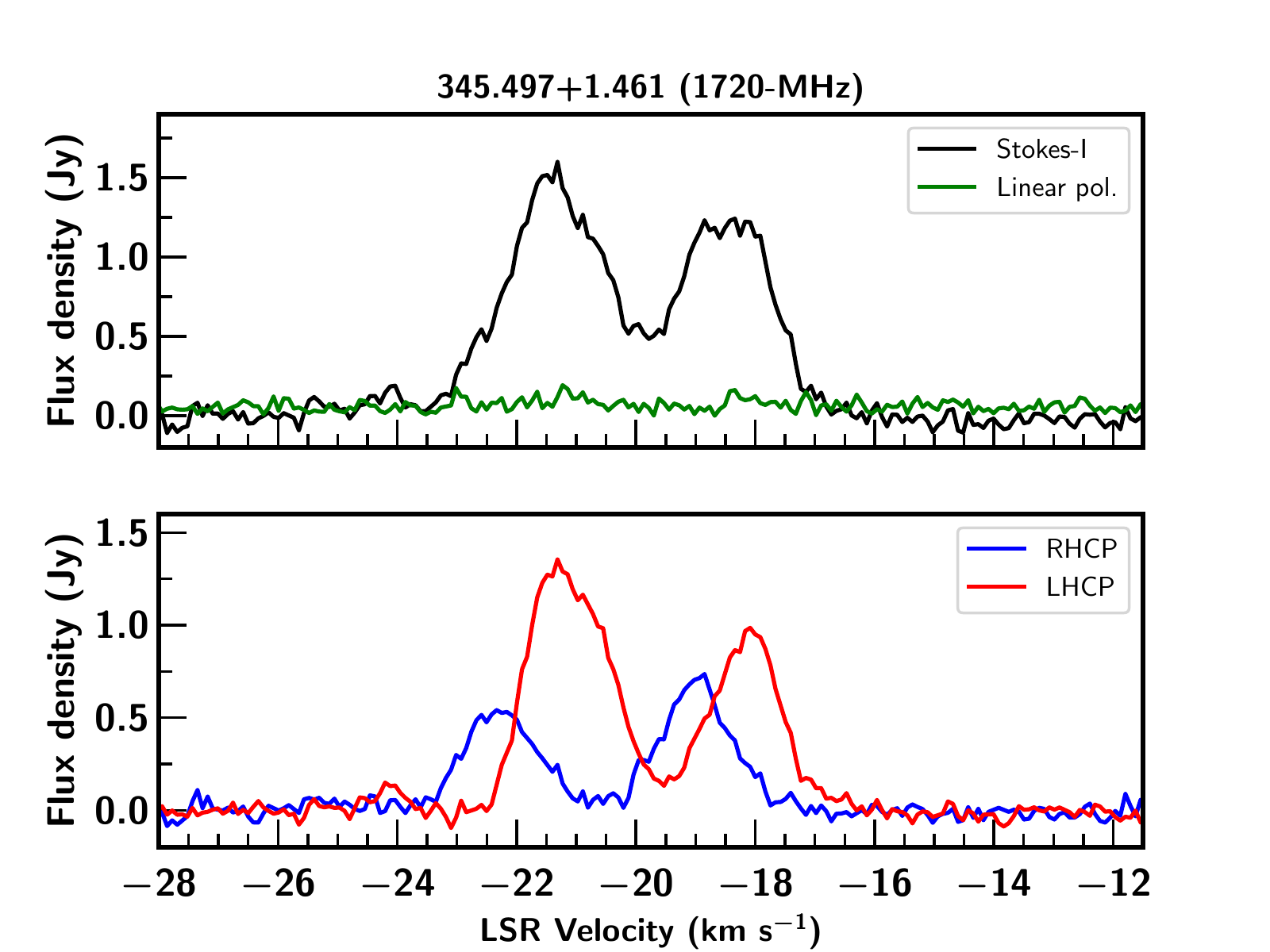}}\\
\subfloat{\includegraphics[width = 3.5in]{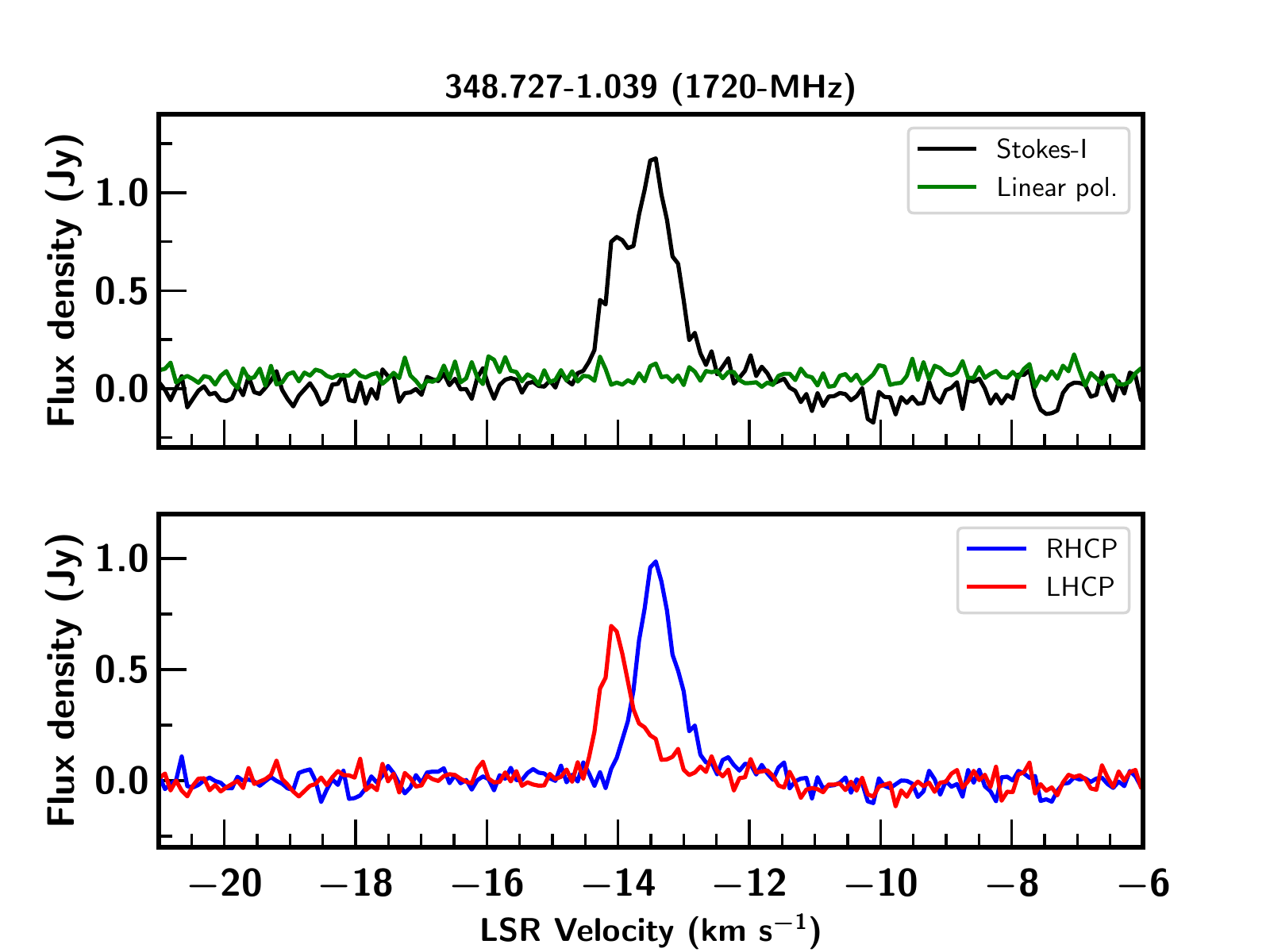}}
\subfloat{\includegraphics[width = 3.5in]{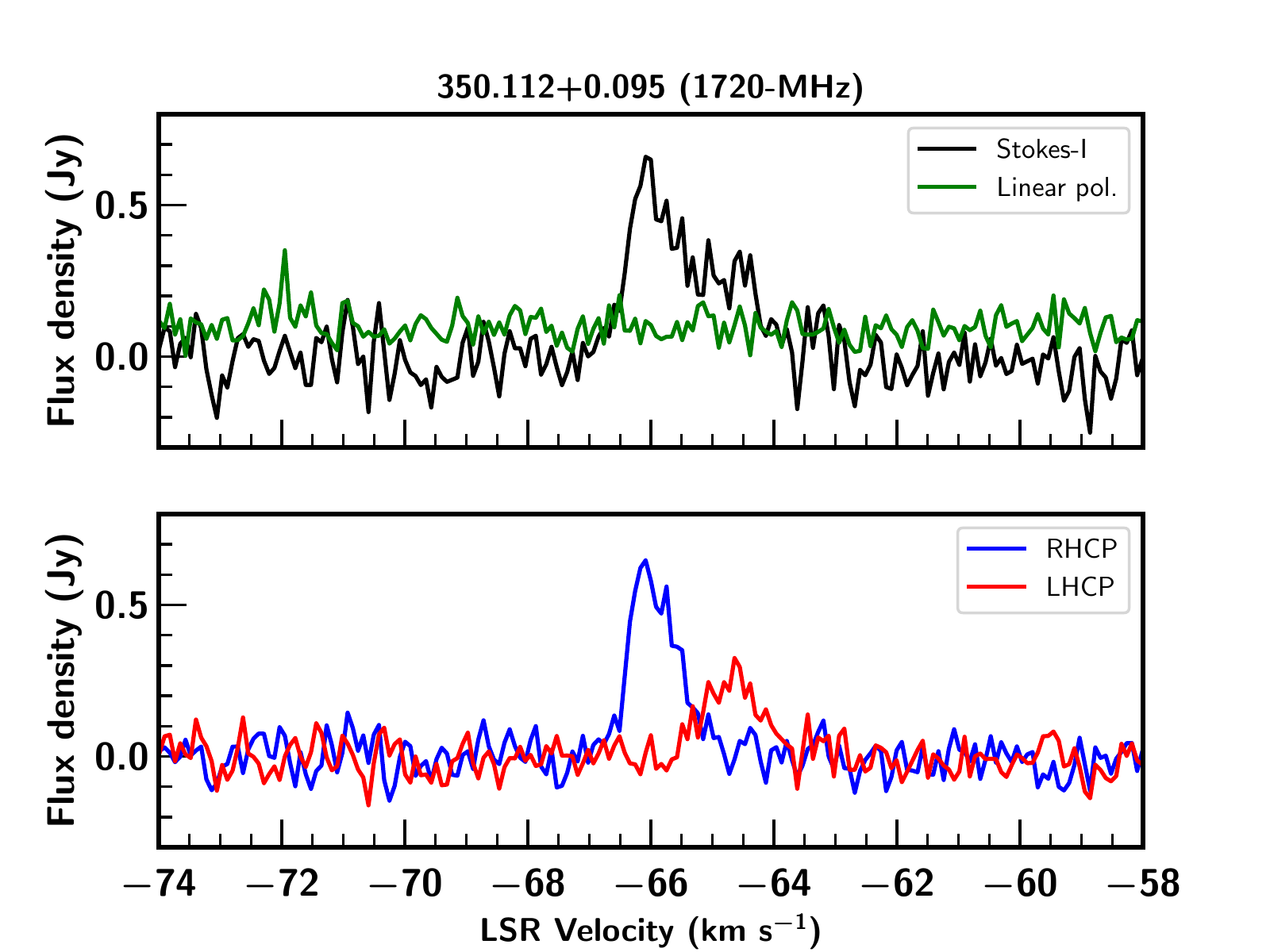}}\\
\subfloat{\includegraphics[width = 3.5in]{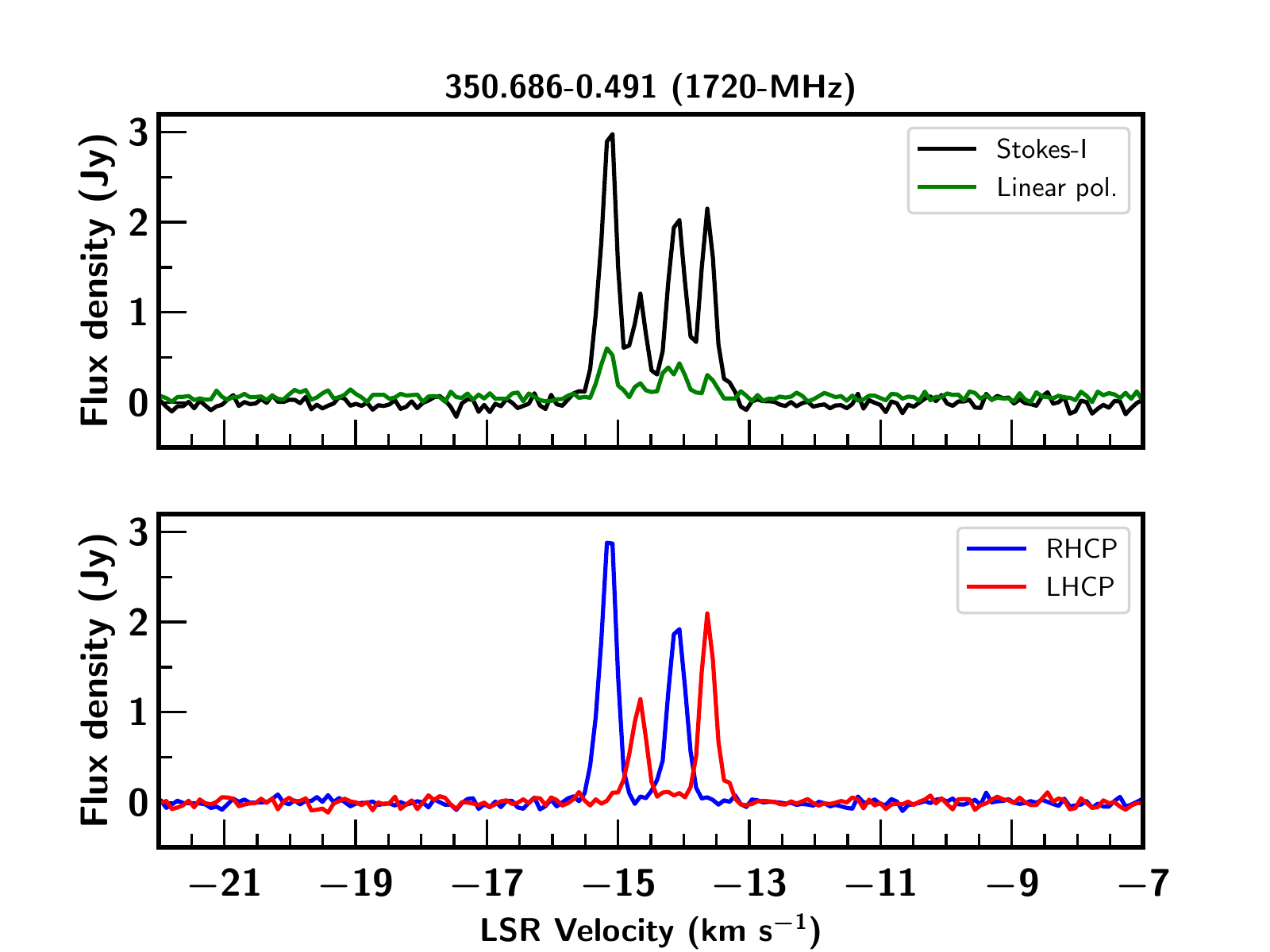}}
\subfloat{\includegraphics[width = 3.5in]{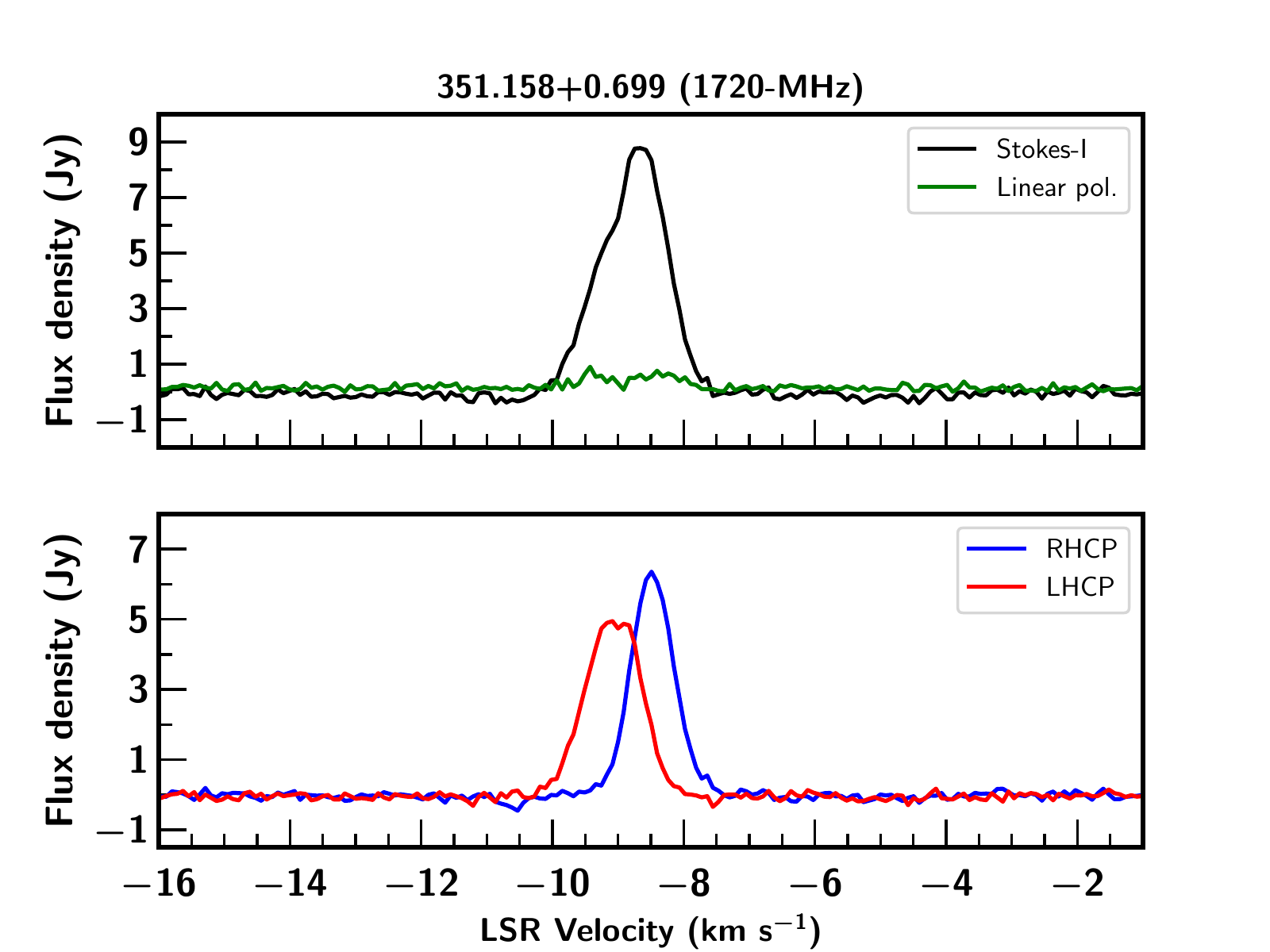}}\\

\caption{\it{$-$continued}}
\end{figure*}

\clearpage
\newpage


\addtocounter{figure}{-1}
\begin{figure*}

\subfloat{\includegraphics[width = 3.5in]{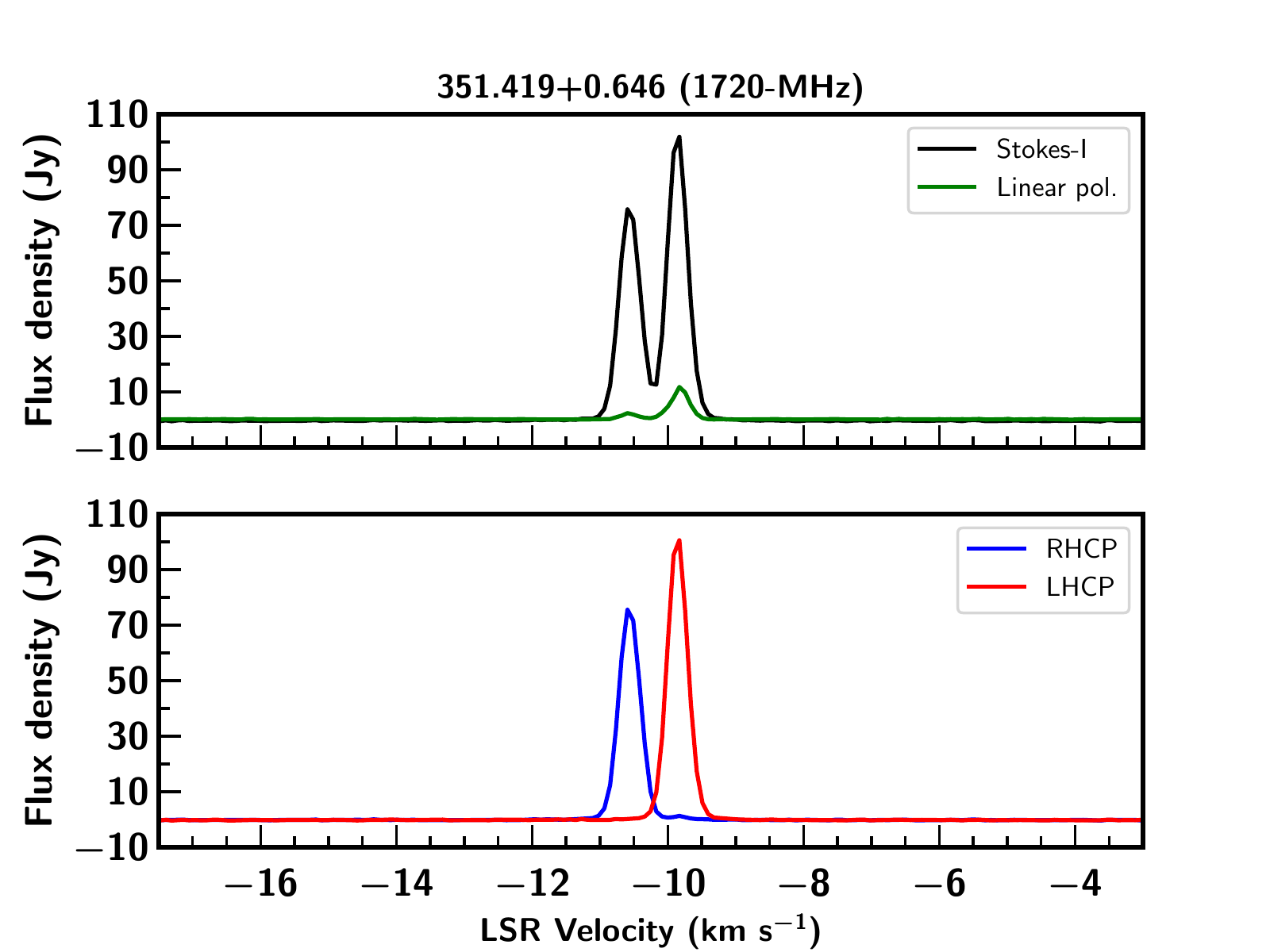}}
\subfloat{\includegraphics[width = 3.5in]{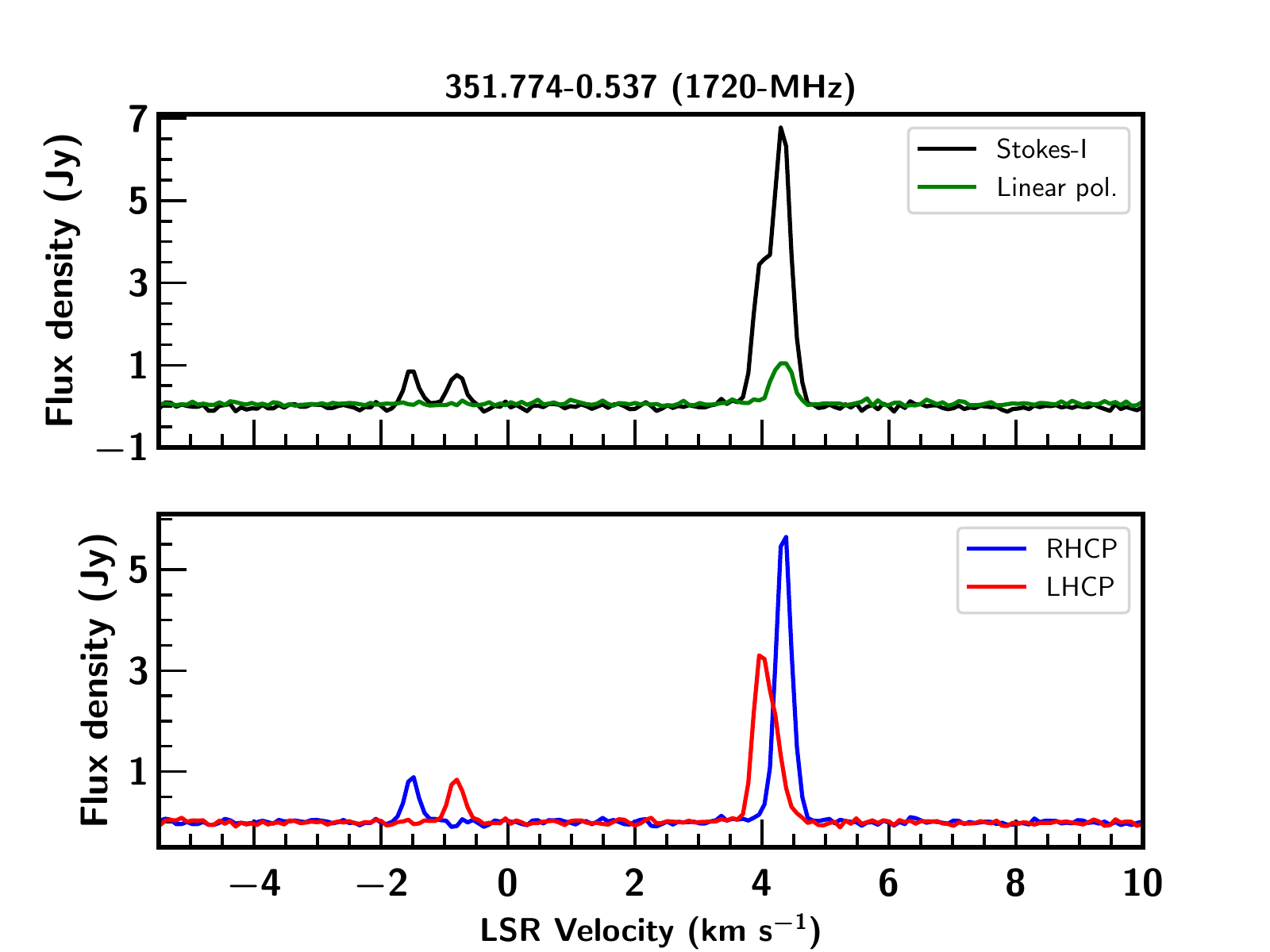}}\\
\subfloat{\includegraphics[width = 3.5in]{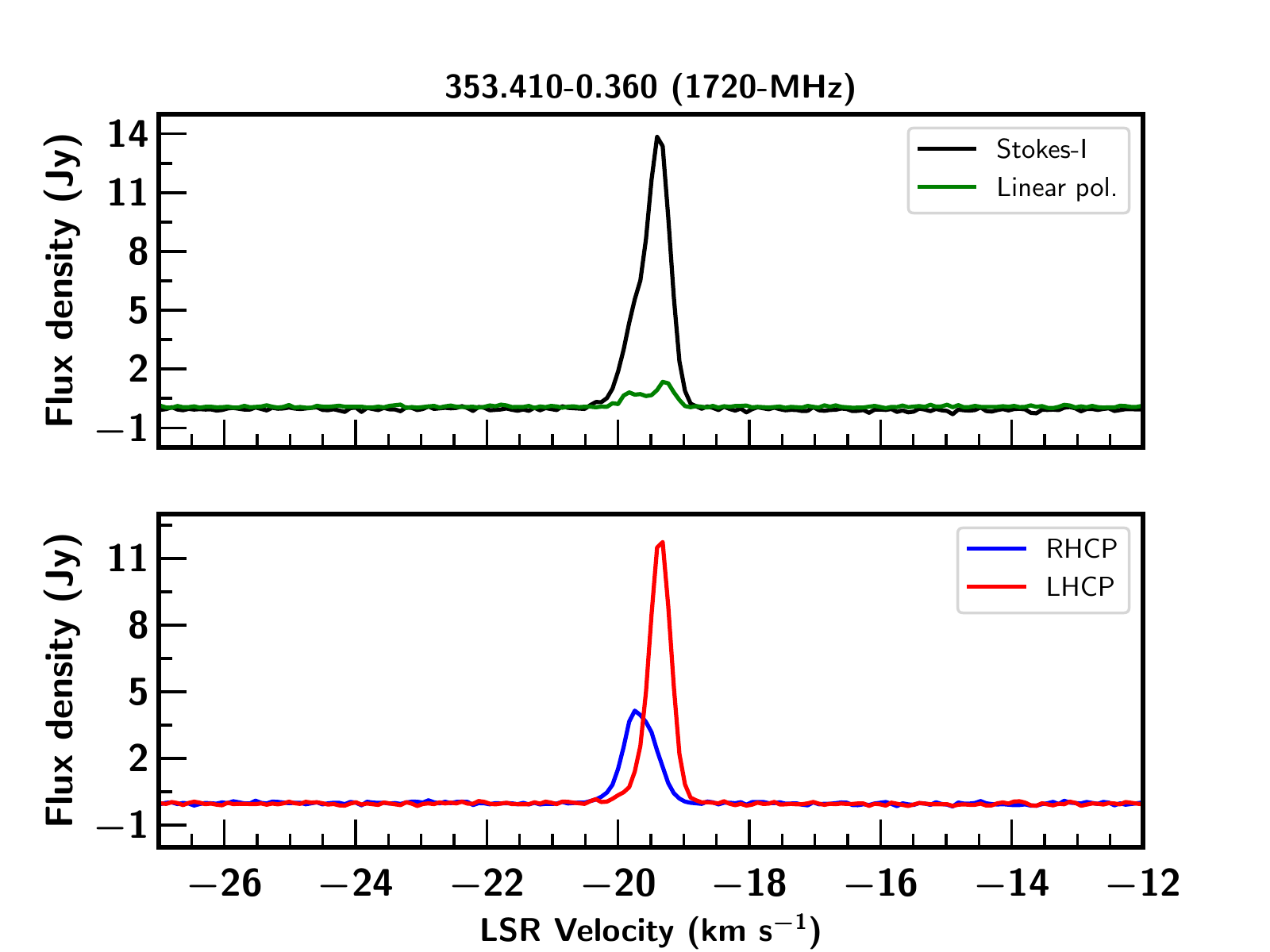}}
\subfloat{\includegraphics[width = 3.5in]{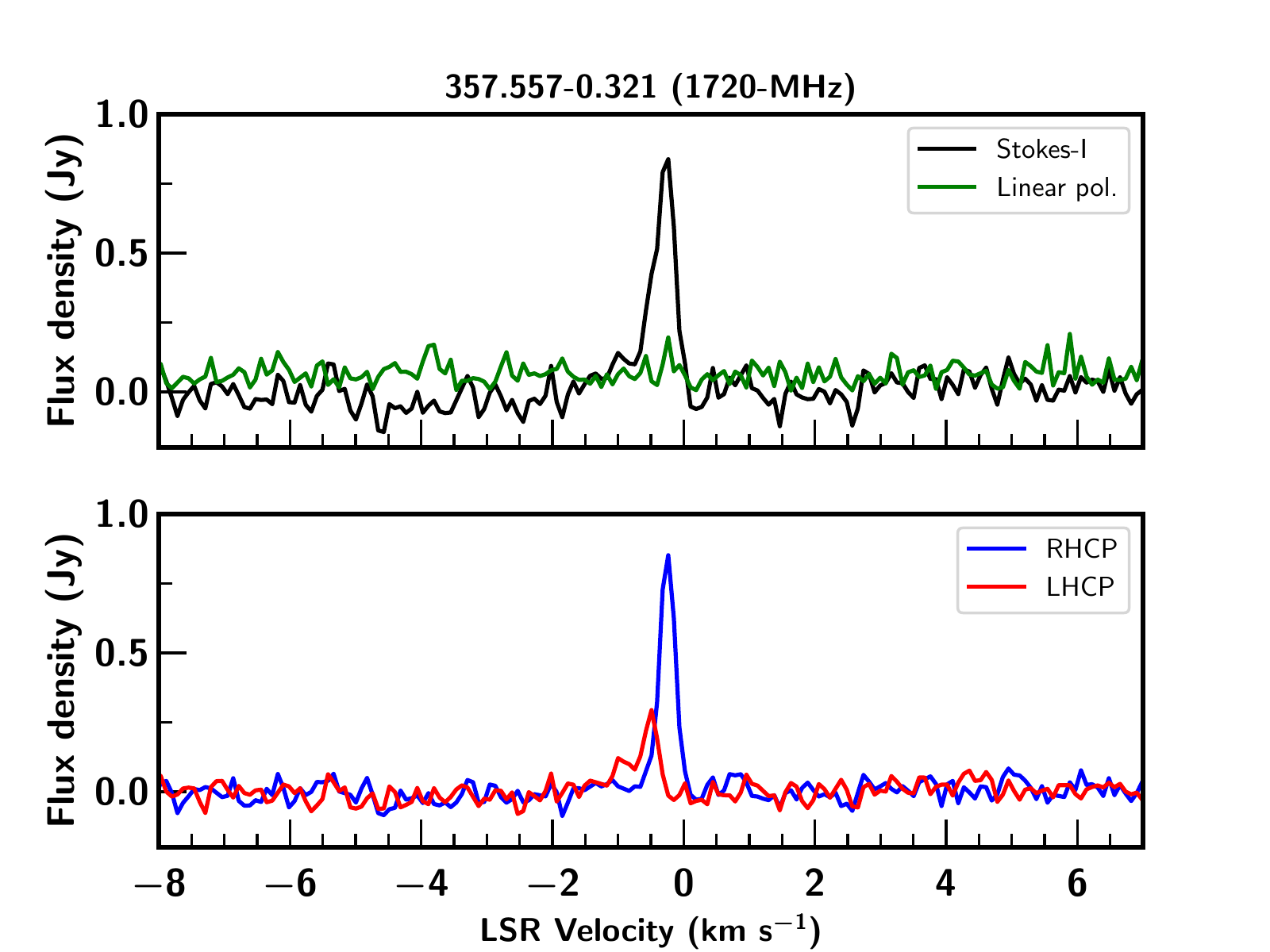}}\\
\subfloat{\includegraphics[width = 3.5in]{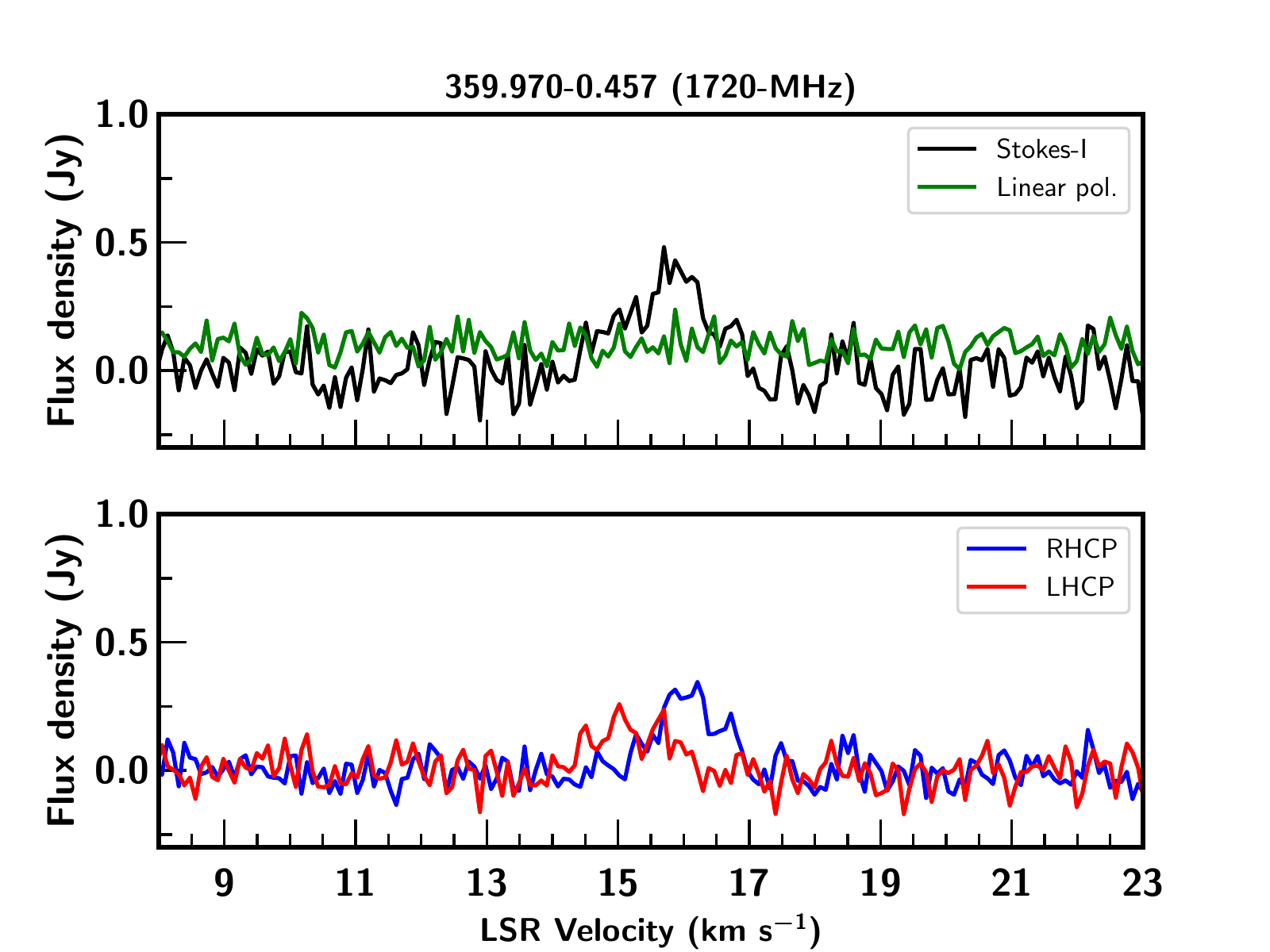}}
\subfloat{\includegraphics[width = 3.5in]{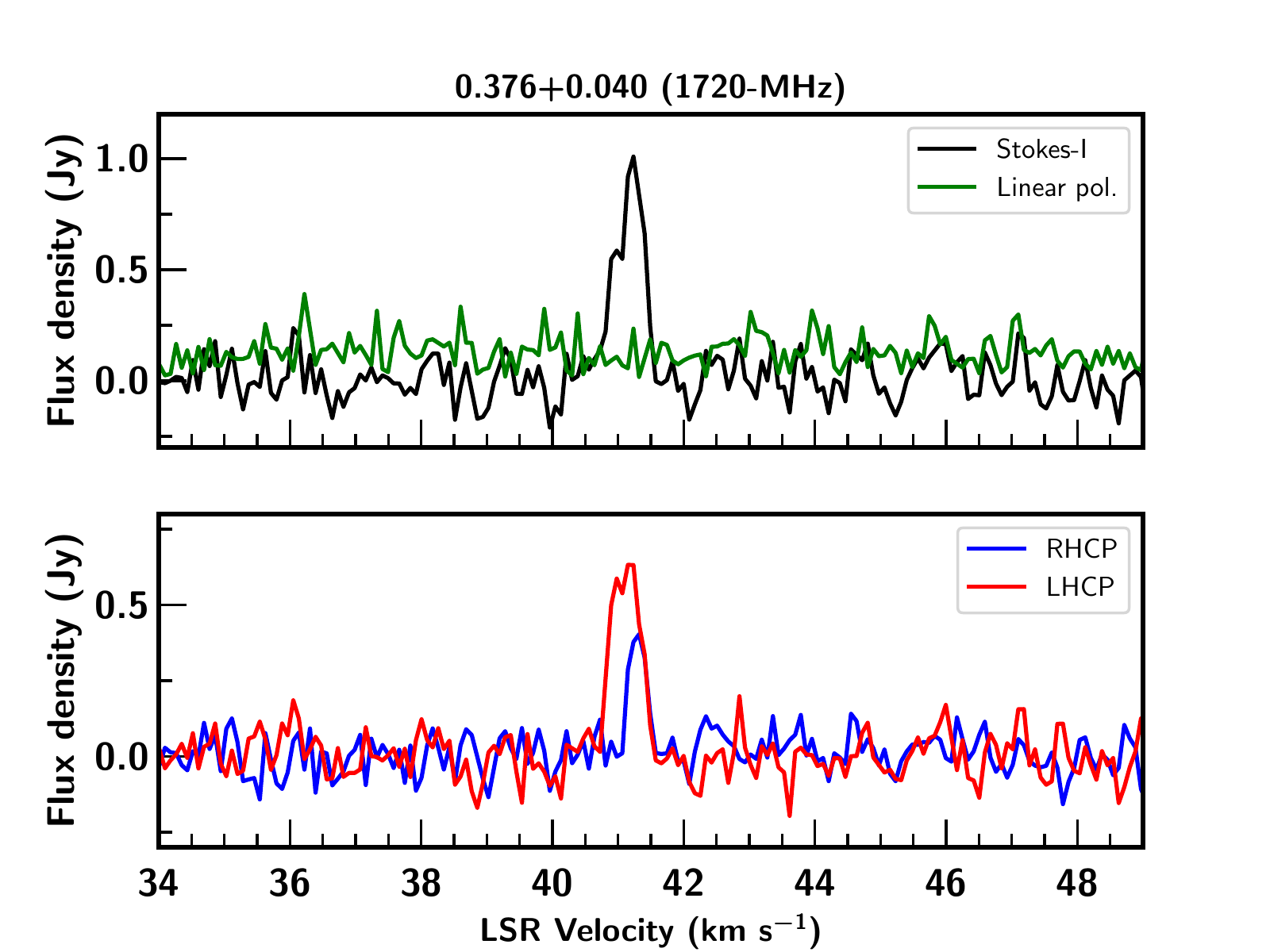}}\\
\caption{\it{$-$continued}}
\end{figure*}

\clearpage
\newpage


\addtocounter{figure}{-1}
\begin{figure*}
\subfloat{\includegraphics[width = 3.5in]{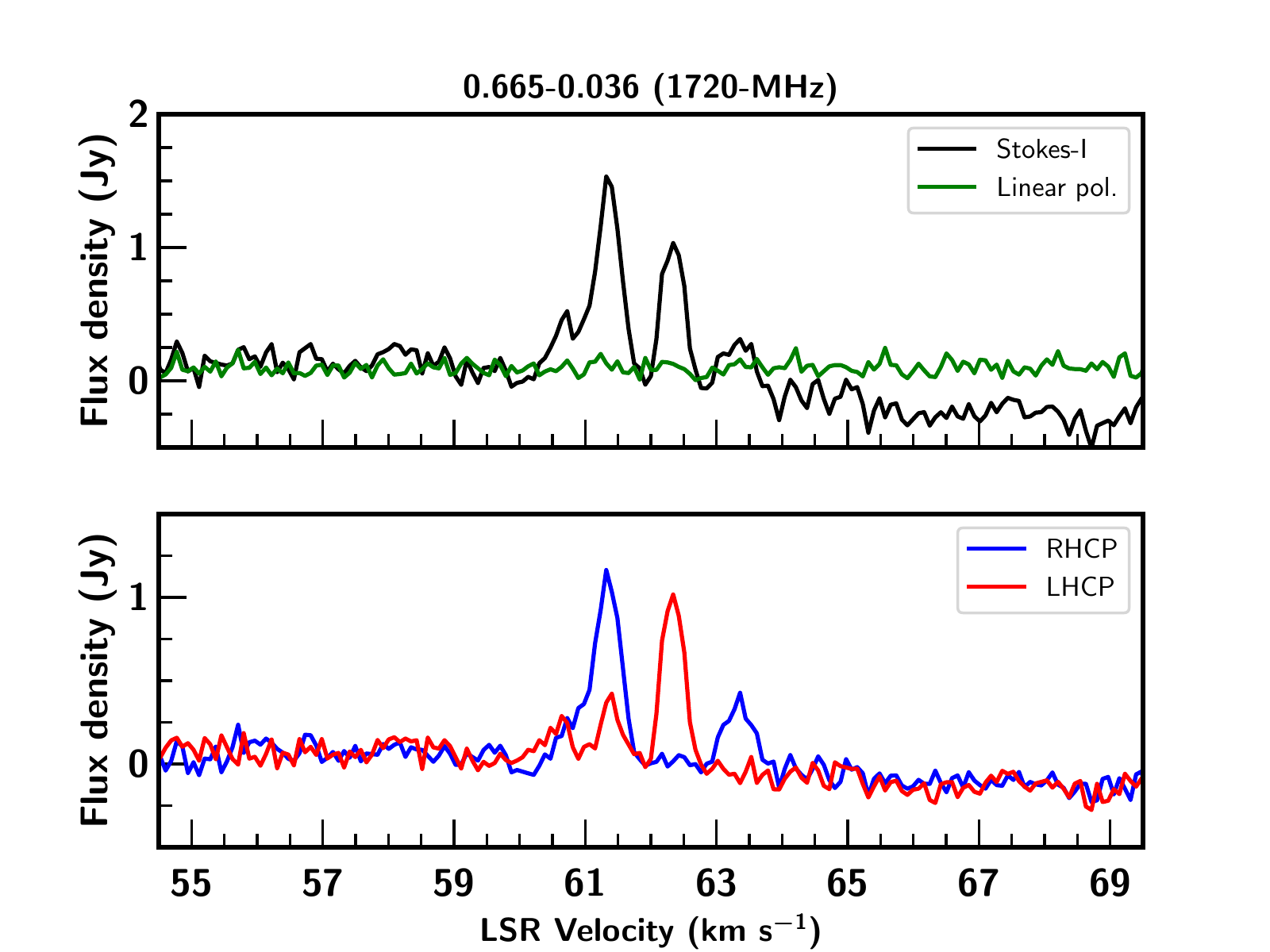}}
\subfloat{\includegraphics[width = 3.5in]{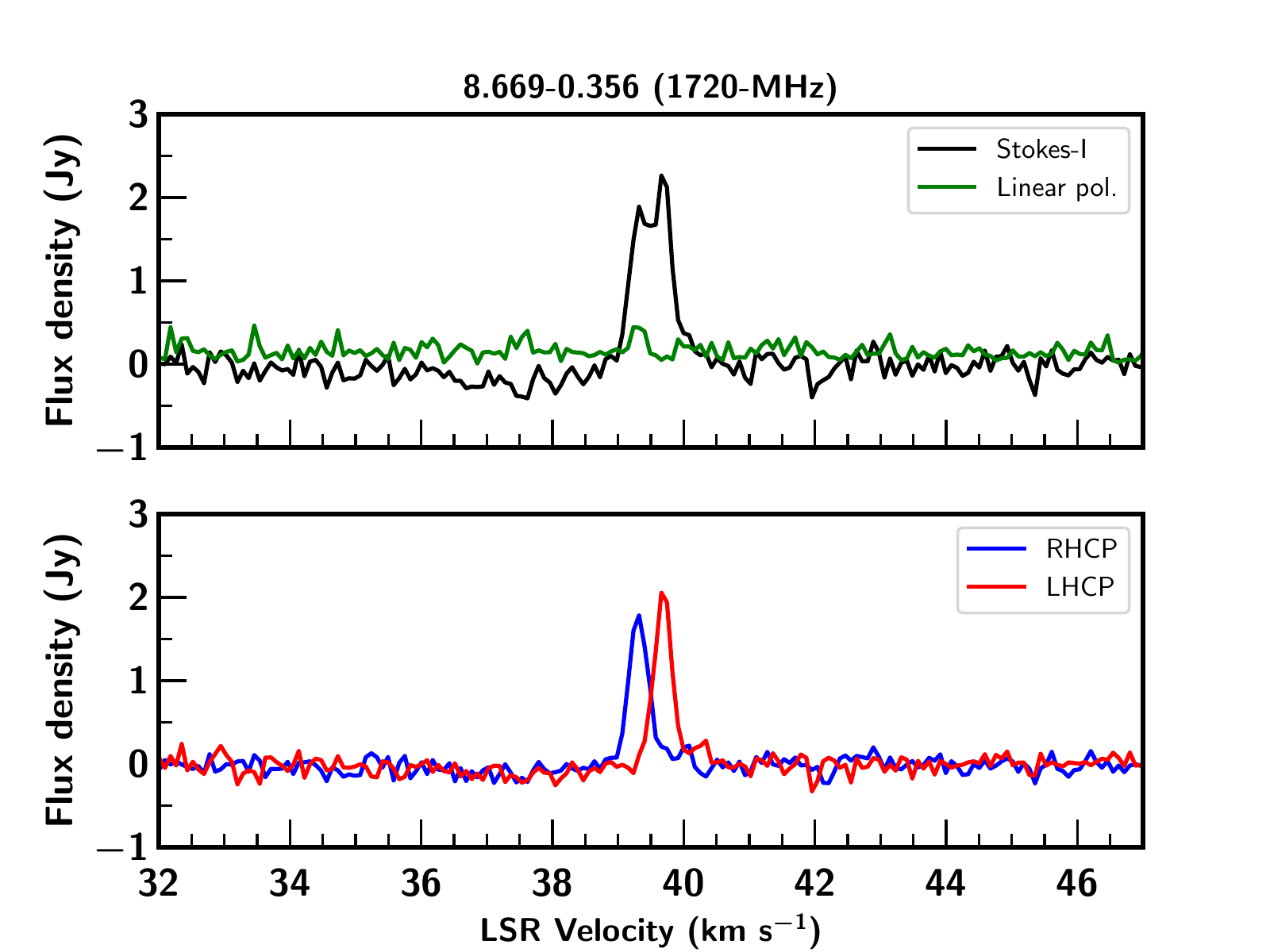}}\\
\subfloat{\includegraphics[width = 3.5in]{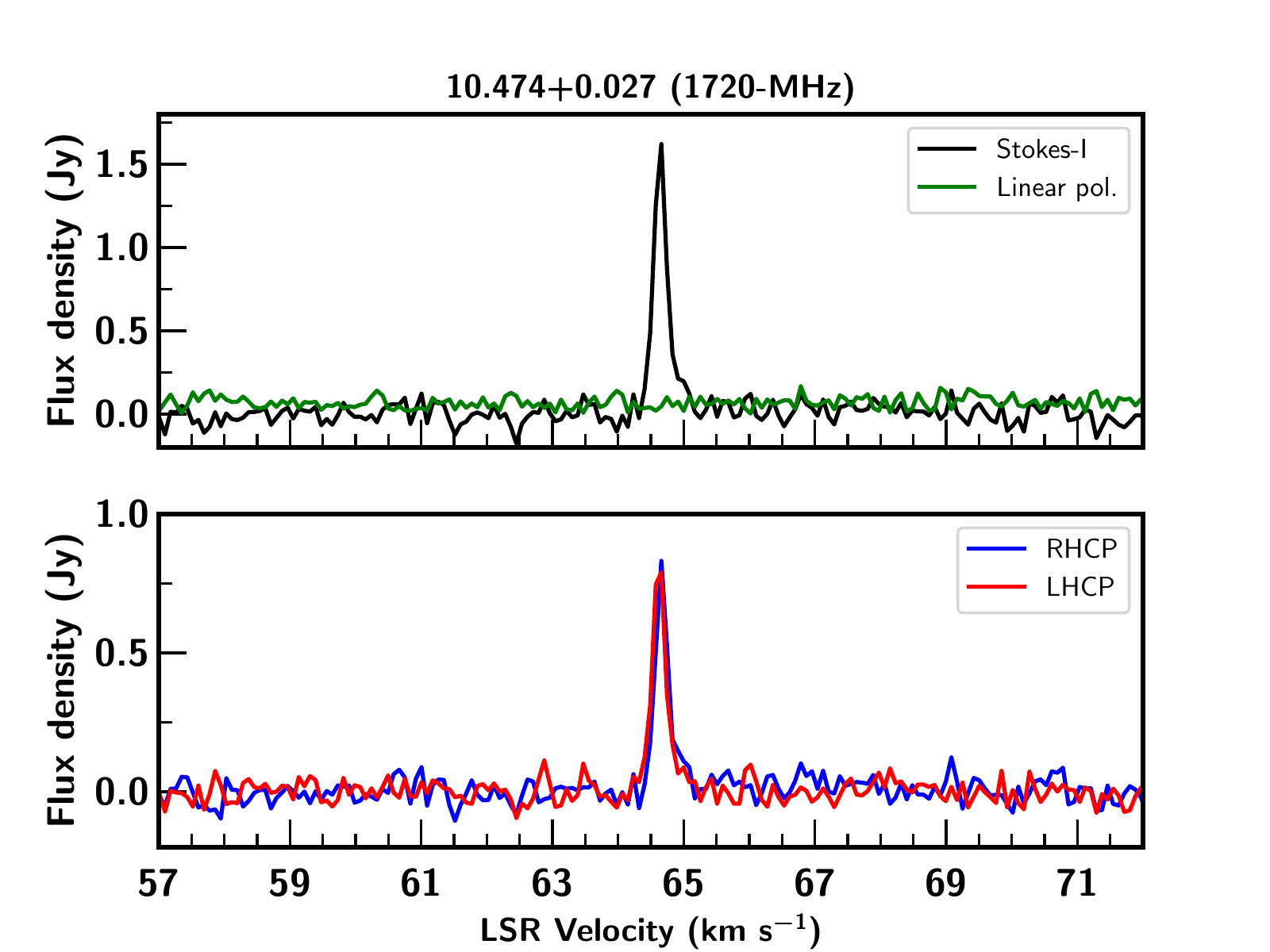}}
\subfloat{\includegraphics[width = 3.5in]{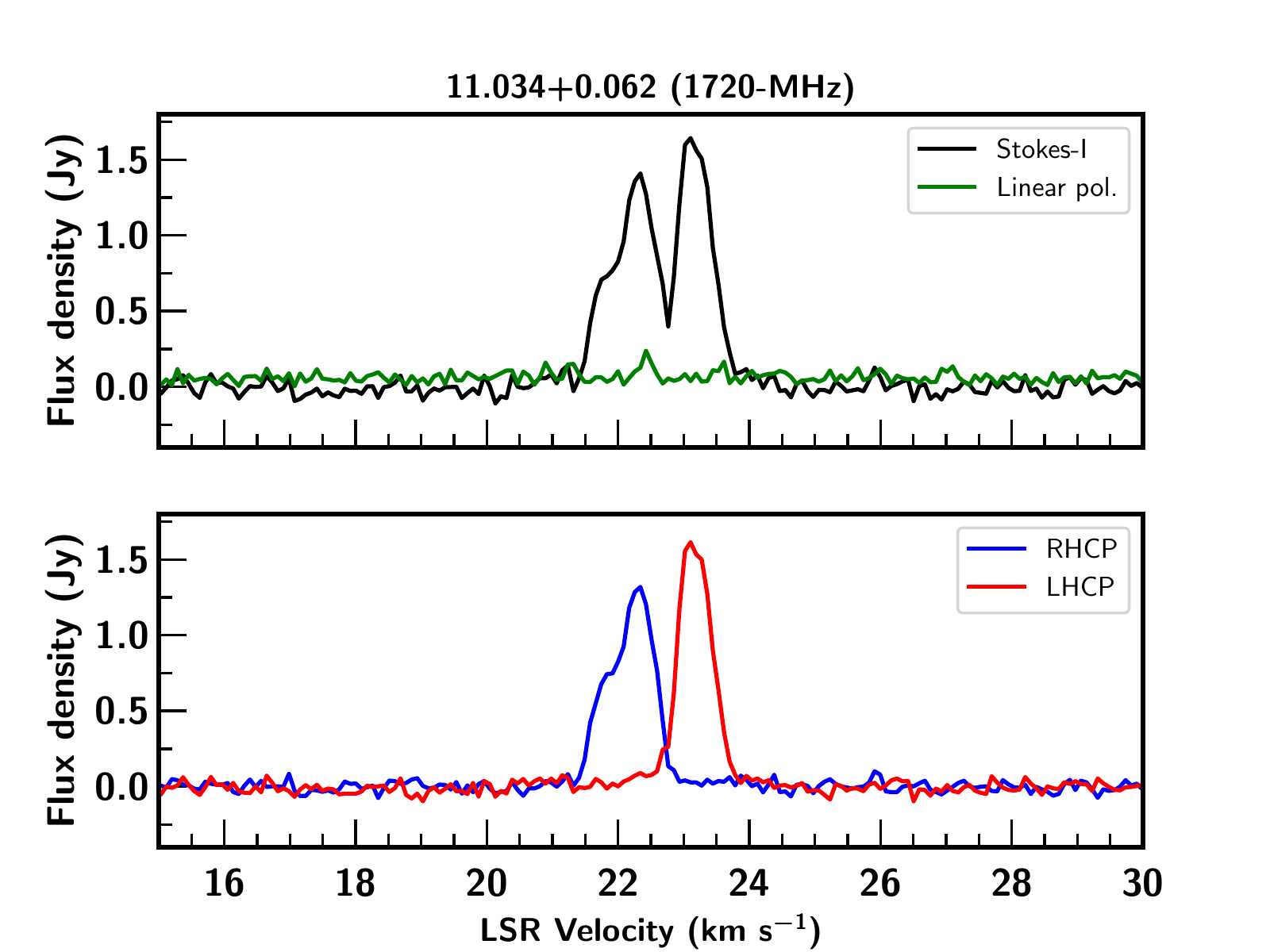}}\\
\subfloat{\includegraphics[width = 3.5in]{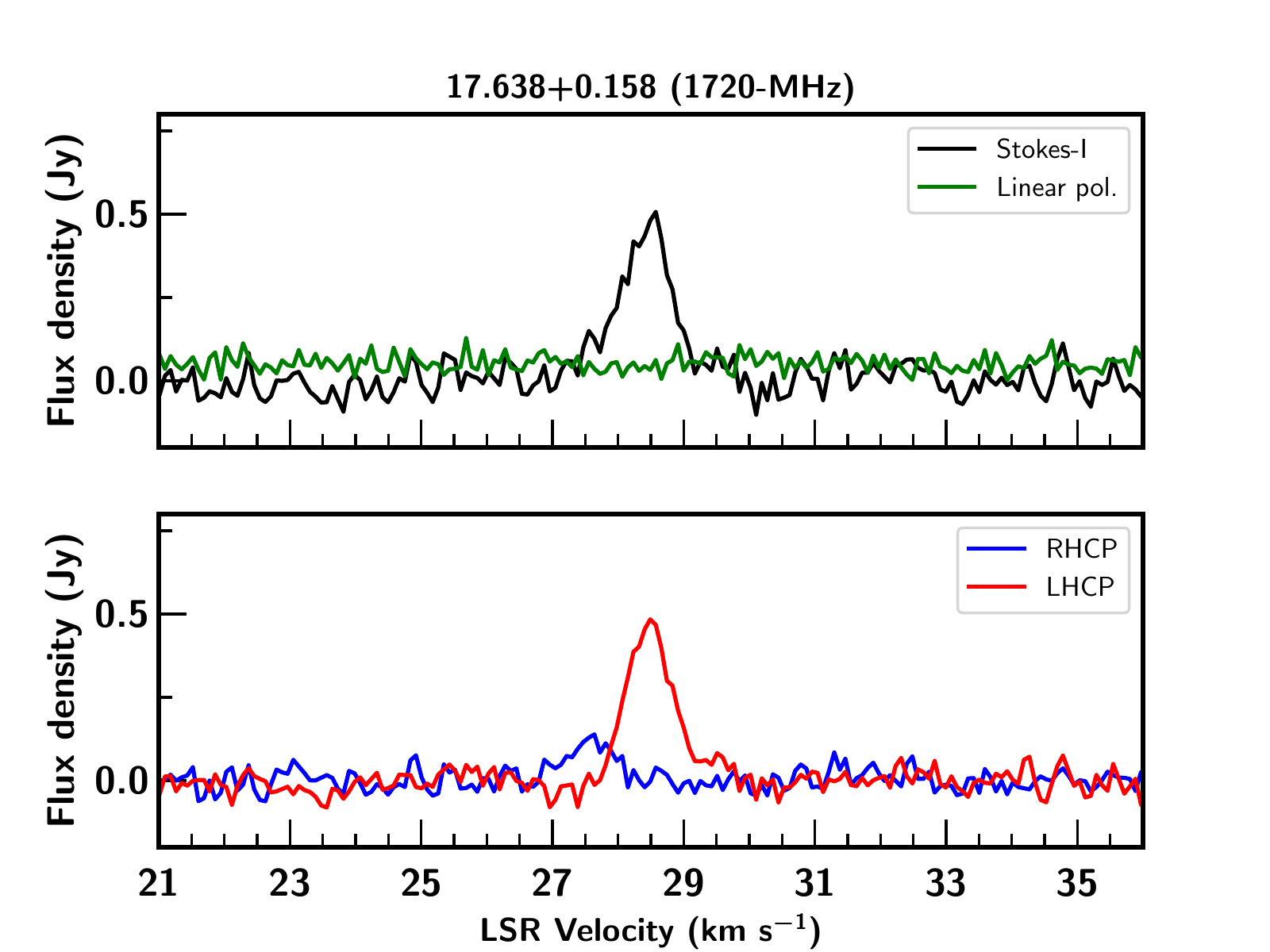}}
\caption{\it{$-$continued}}
\end{figure*}


\begin{figure*}
\subfloat{\includegraphics[width = 3.5in]{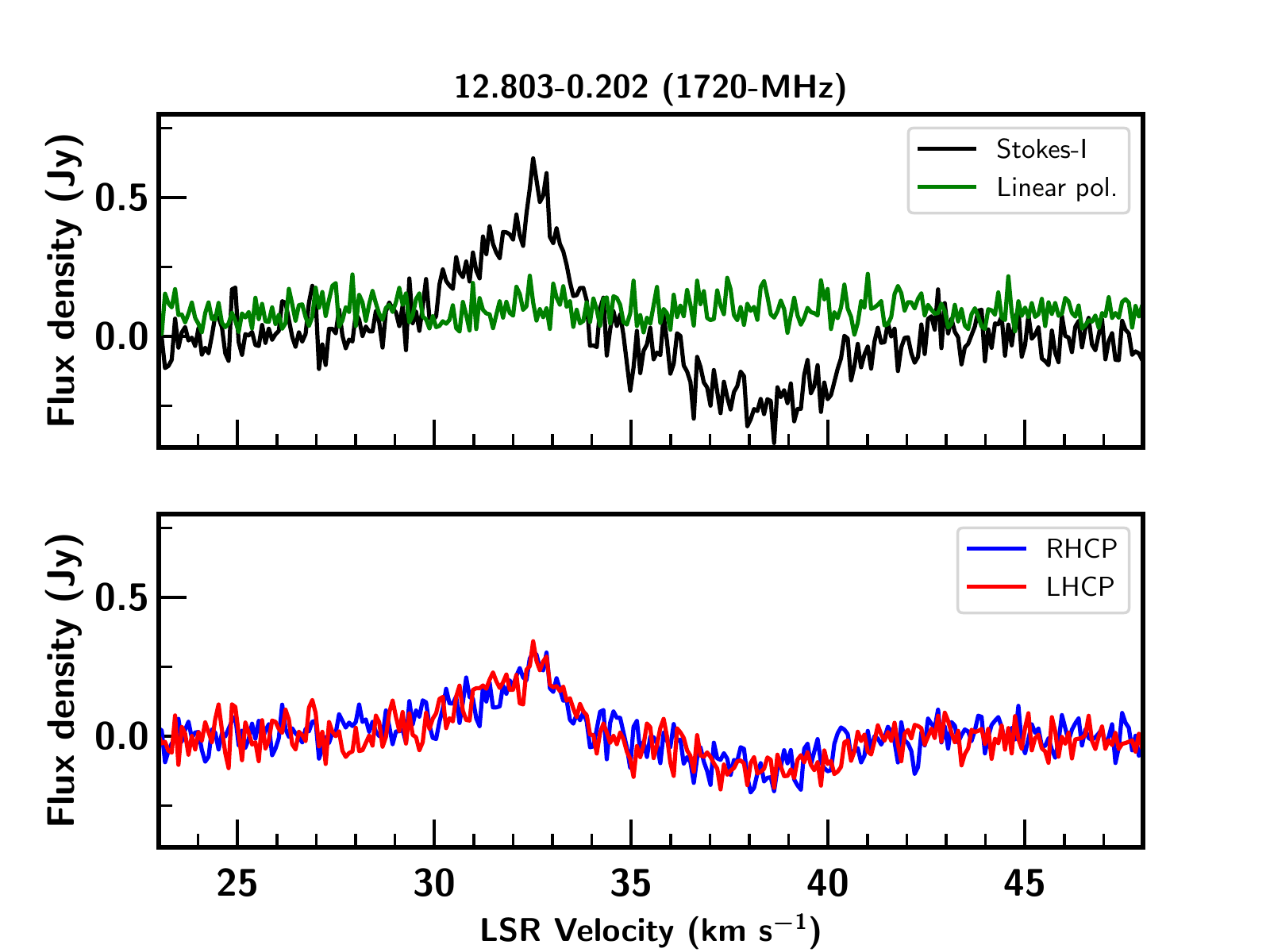}}
\subfloat{\includegraphics[width = 3.5in]{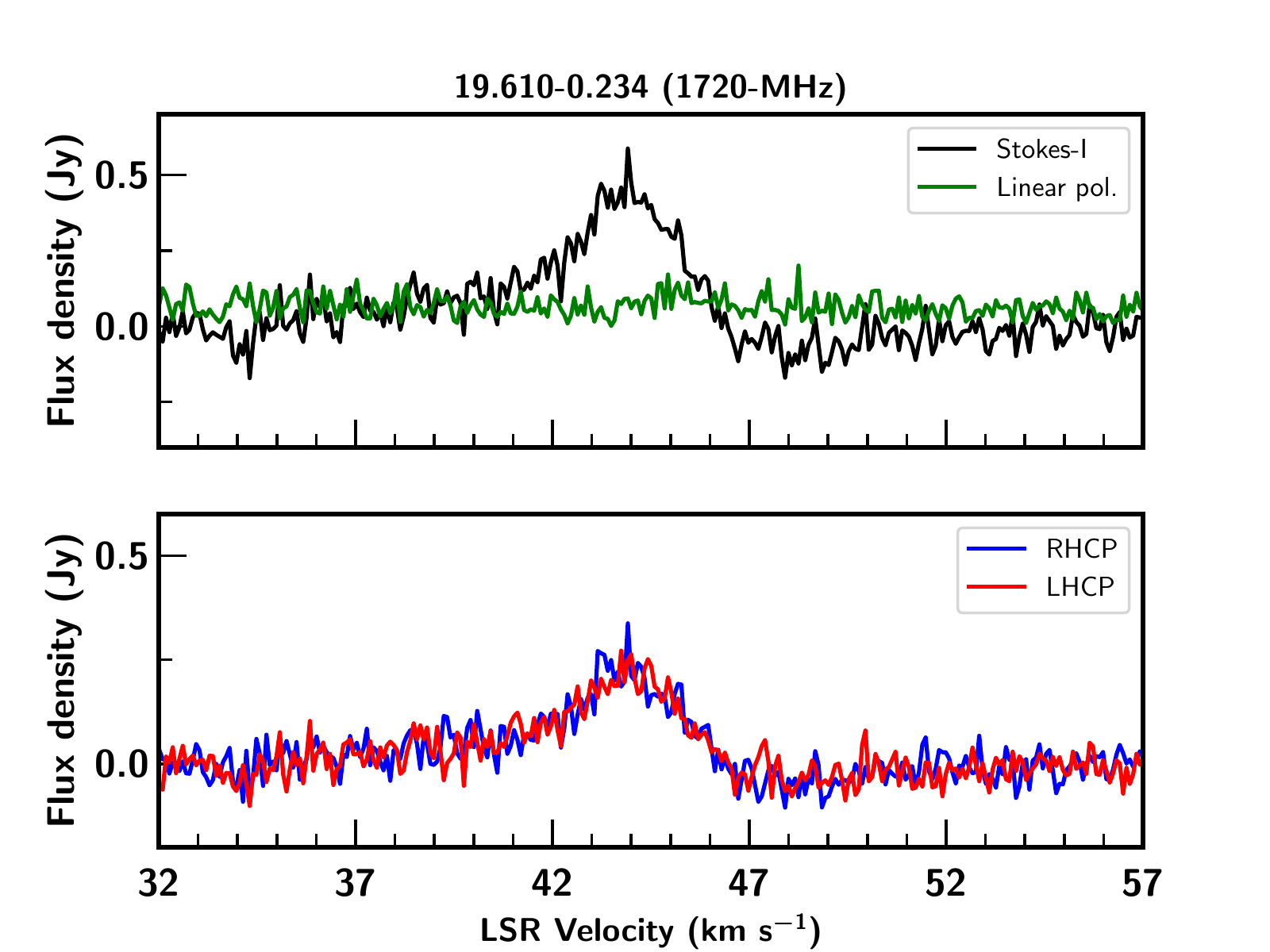}}\\
\caption{Spectra of diffuse OH sources.}
\label{Figure2}
\end{figure*}


\begin{figure*}
\subfloat{\includegraphics[width = 3.5in]{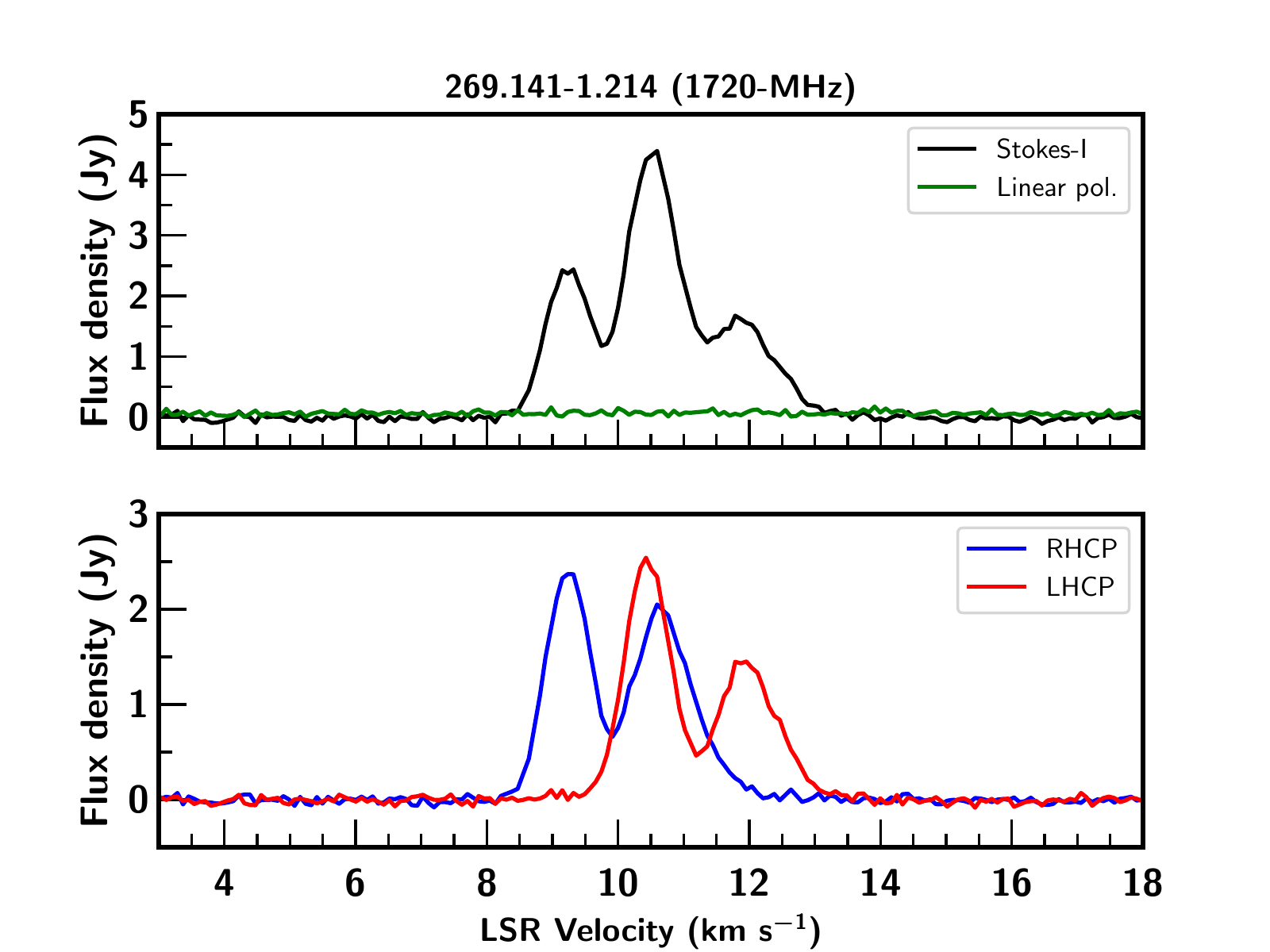}}
\subfloat{\includegraphics[width = 3.5in]{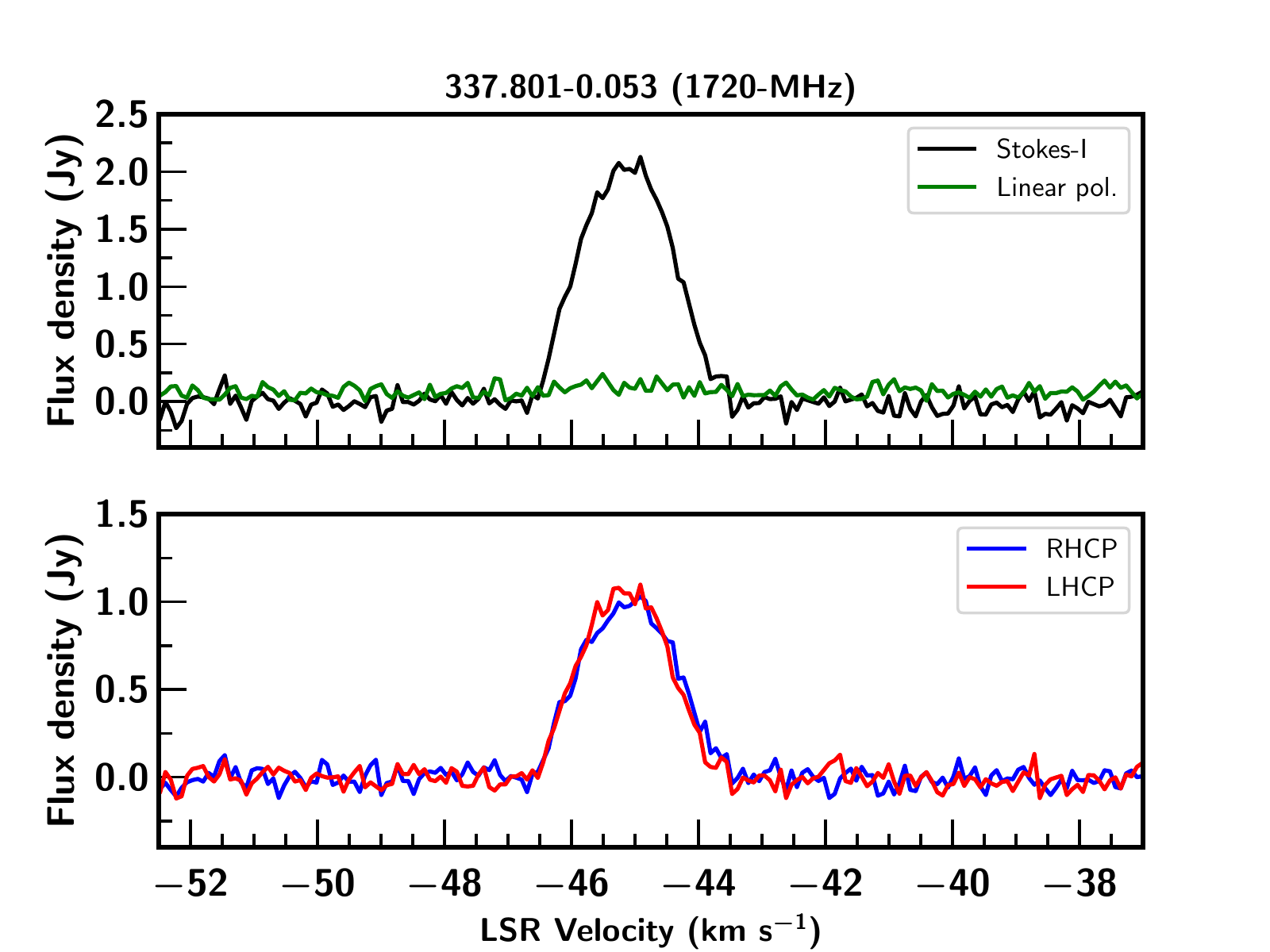}} \\
\subfloat{\includegraphics[width = 3.5in]{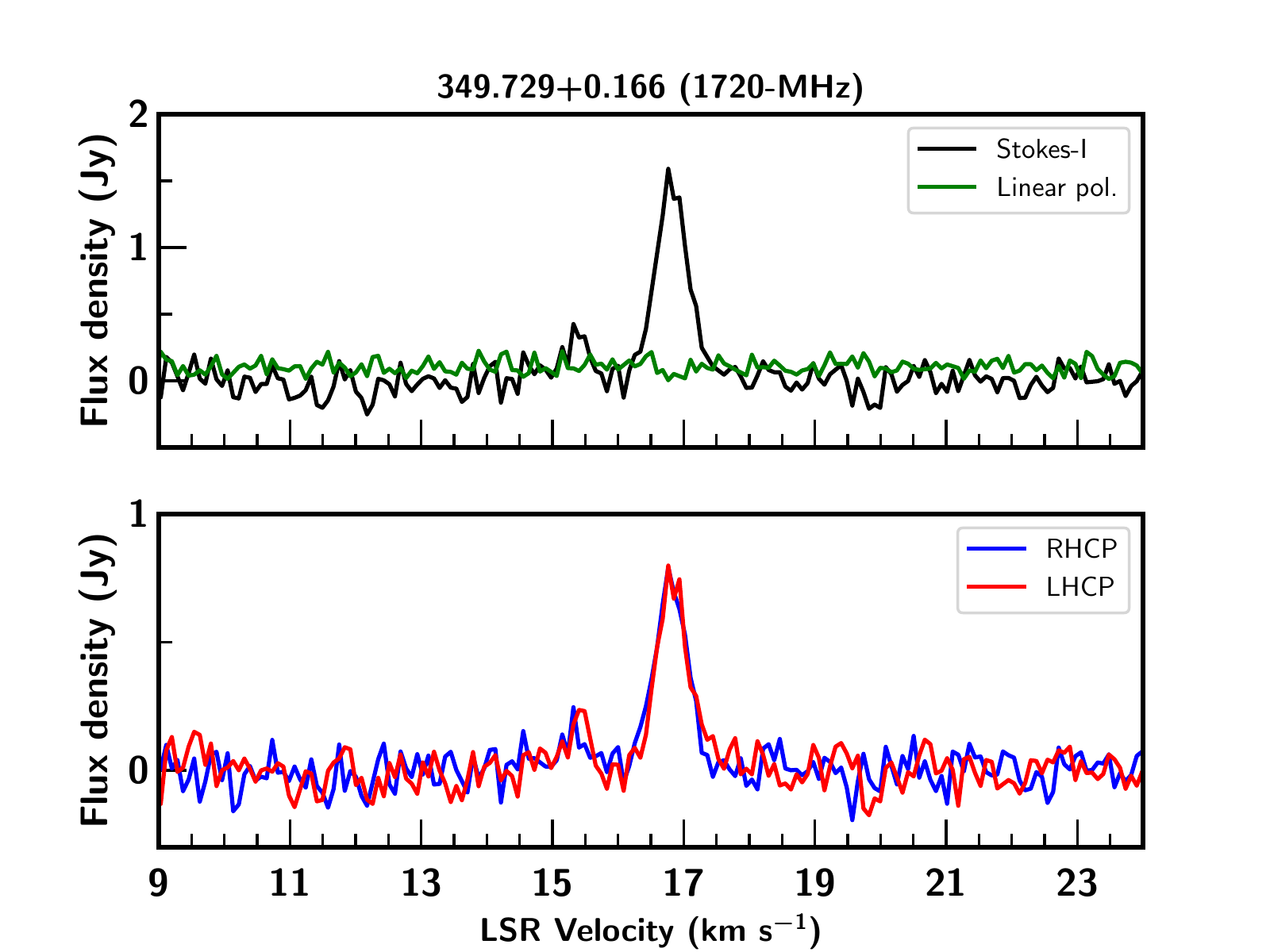}}
\subfloat{\includegraphics[width = 3.5in]{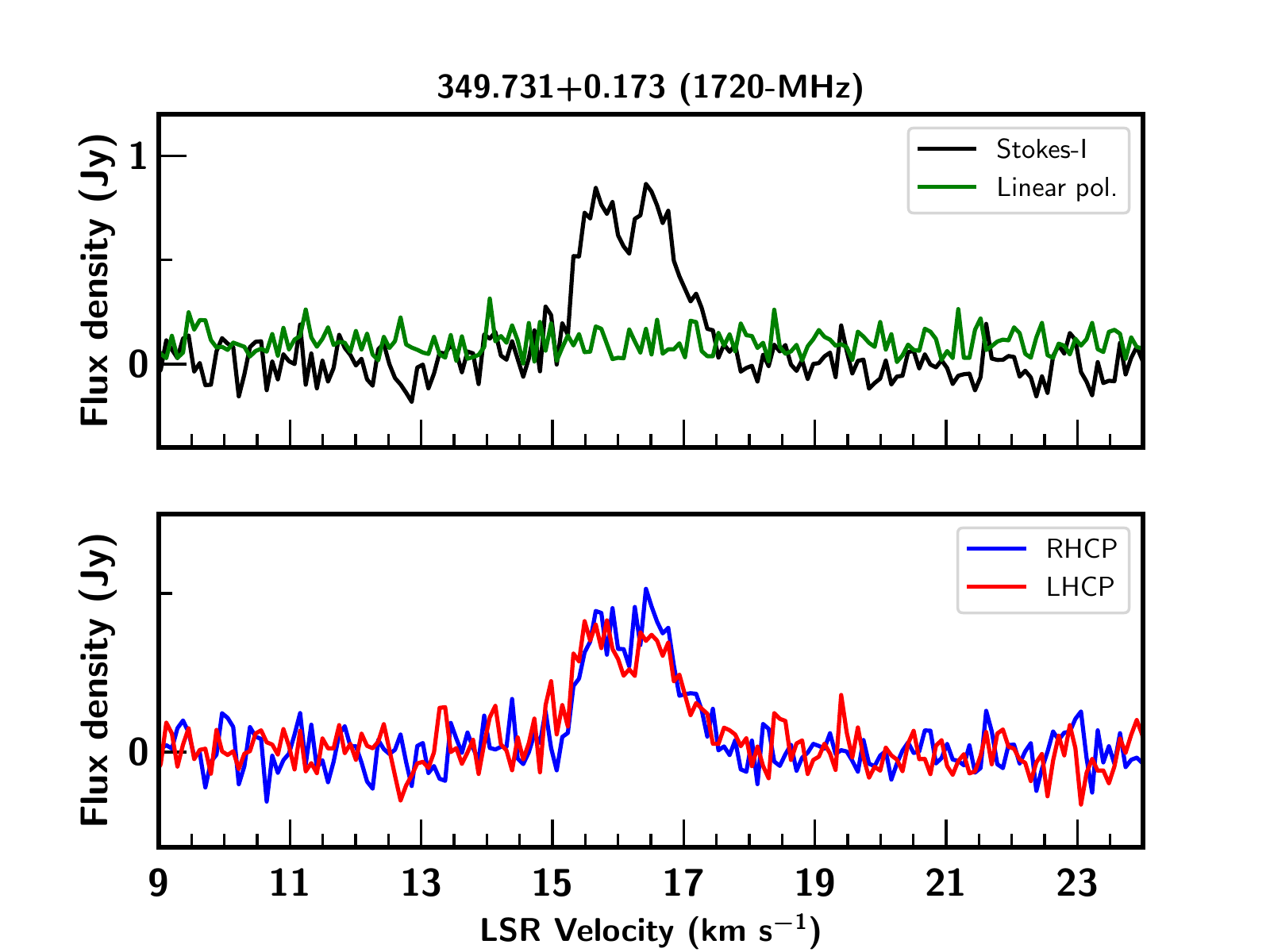}}\\

\caption{Spectra of 1720-MHz OH masers associated with supernova remnants}
\label{Figure3}
\end{figure*}

\addtocounter{figure}{-1}

\begin{figure*}

\subfloat{\includegraphics[width = 3.5in]{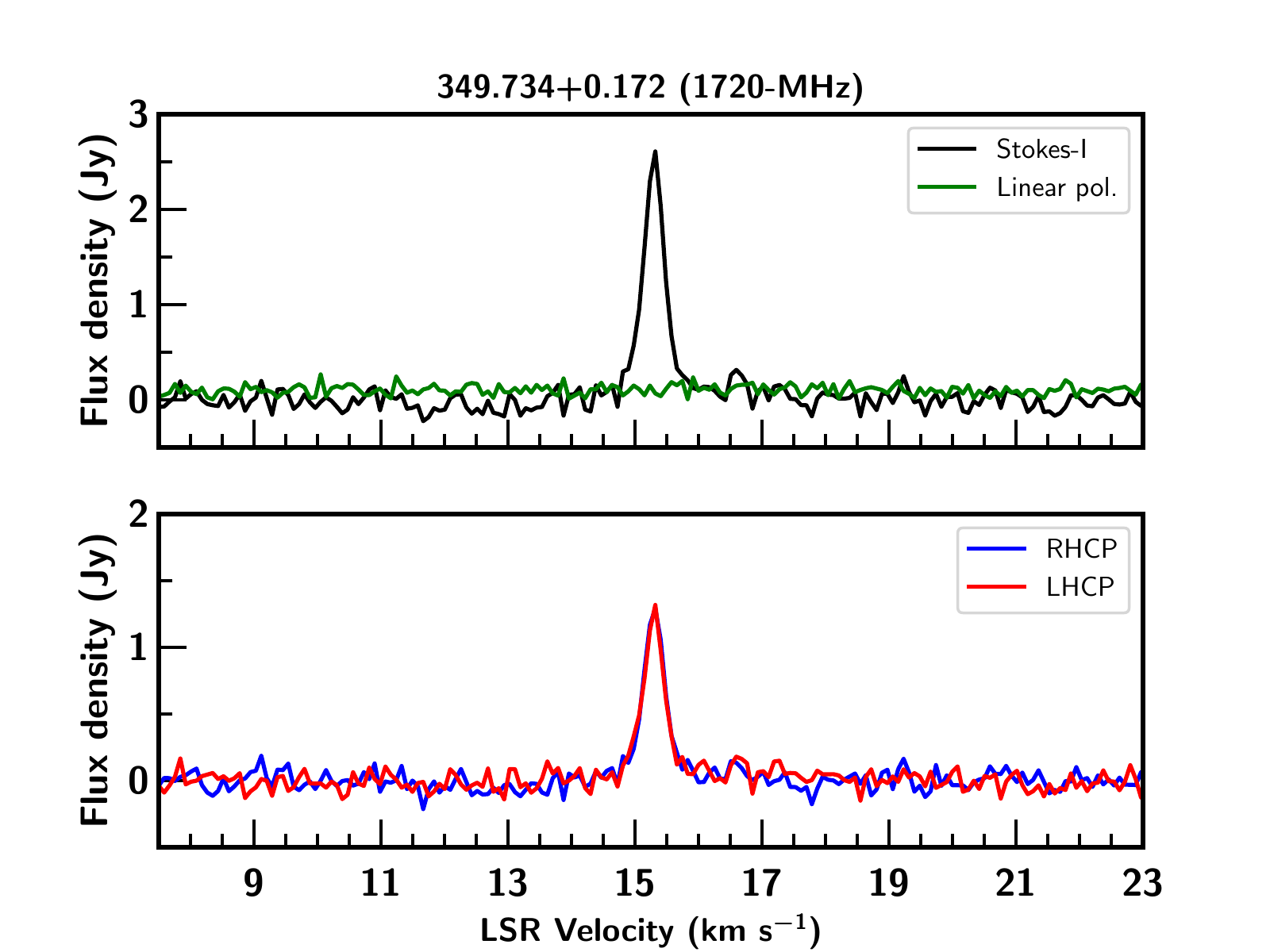}}
\subfloat{\includegraphics[width = 3.5in]{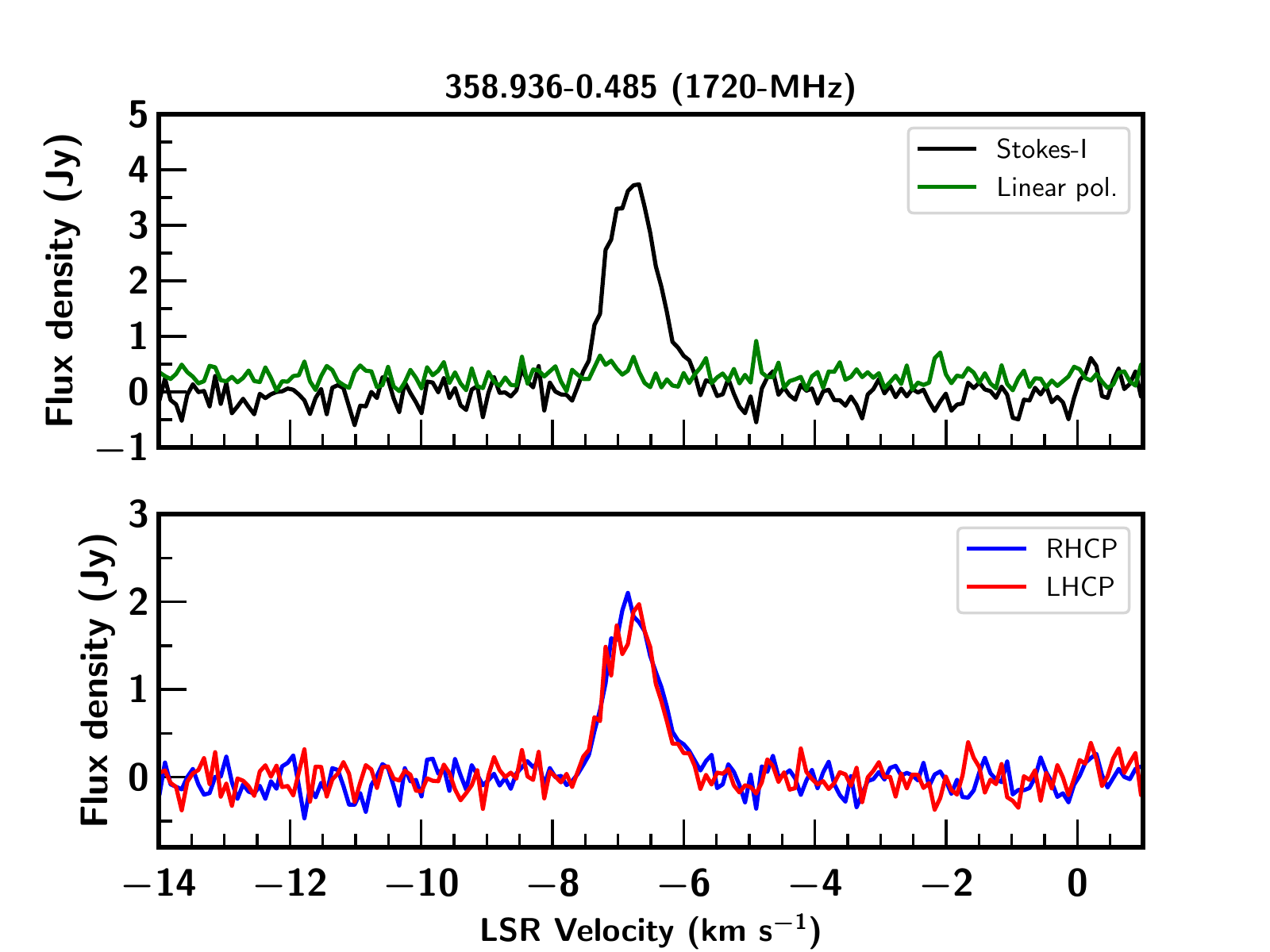}}\\
\subfloat{\includegraphics[width = 3.5in]{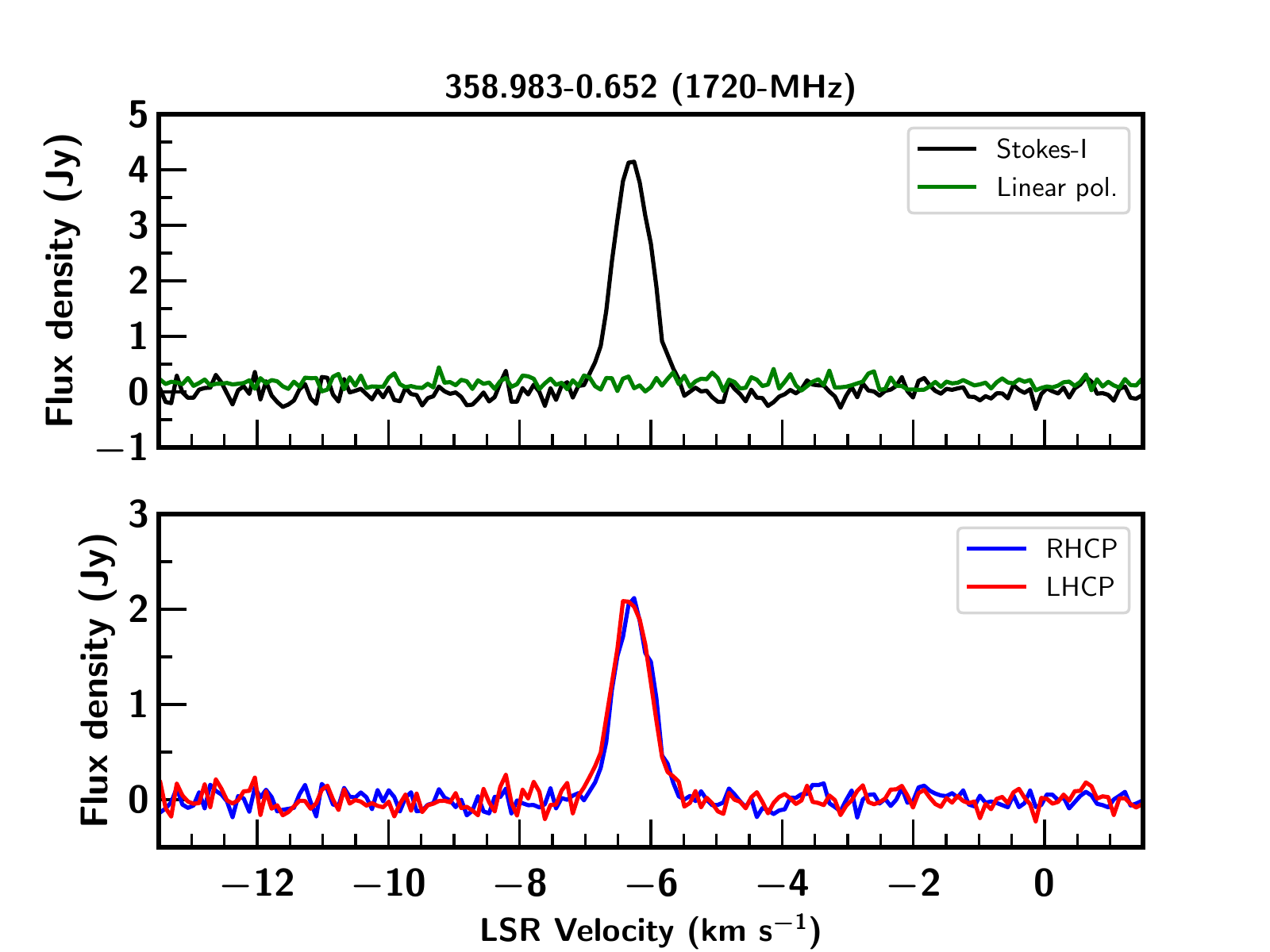}}
\subfloat{\includegraphics[width = 3.5in]{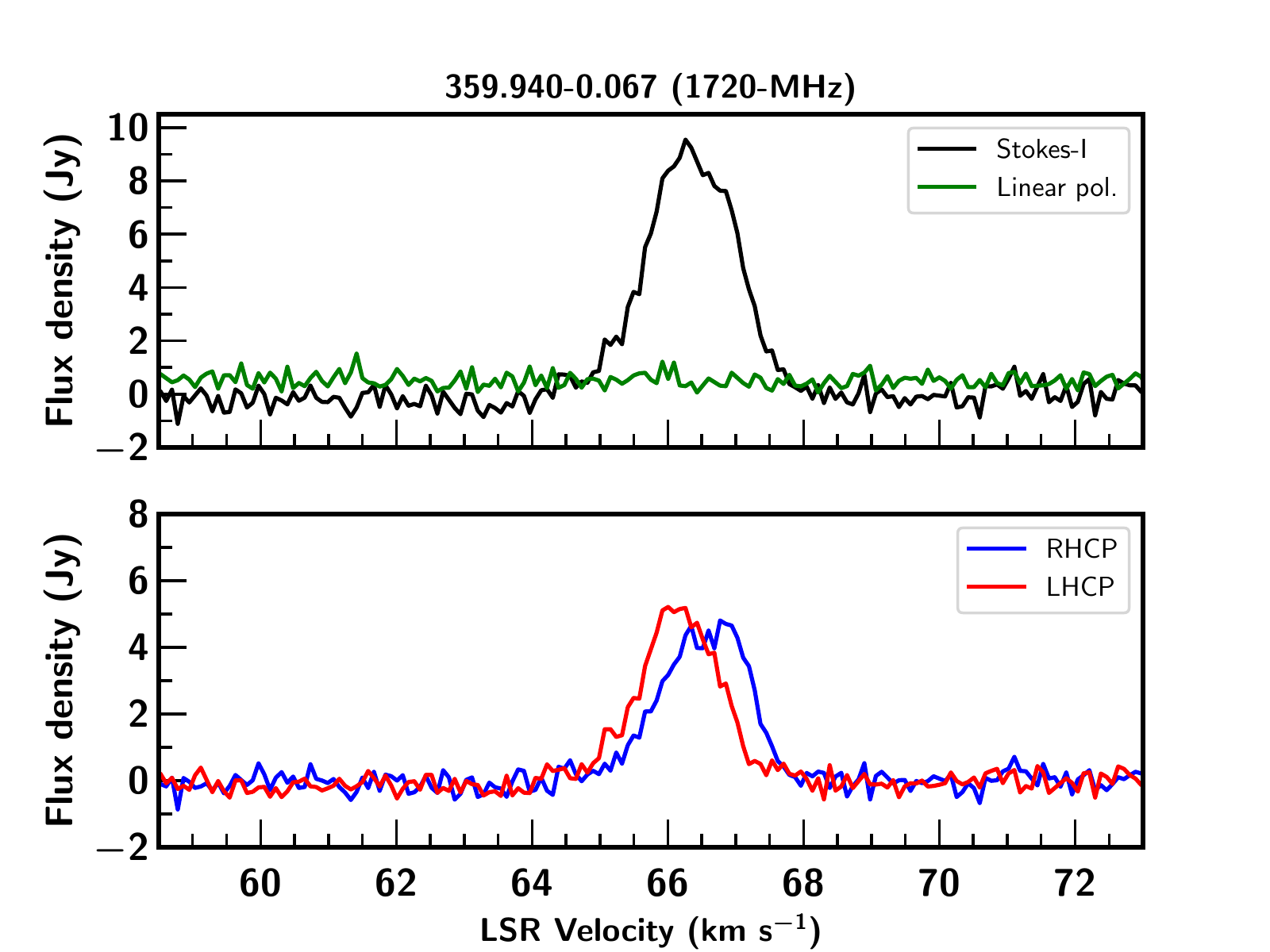}}\\
\subfloat{\includegraphics[width = 3.5in]{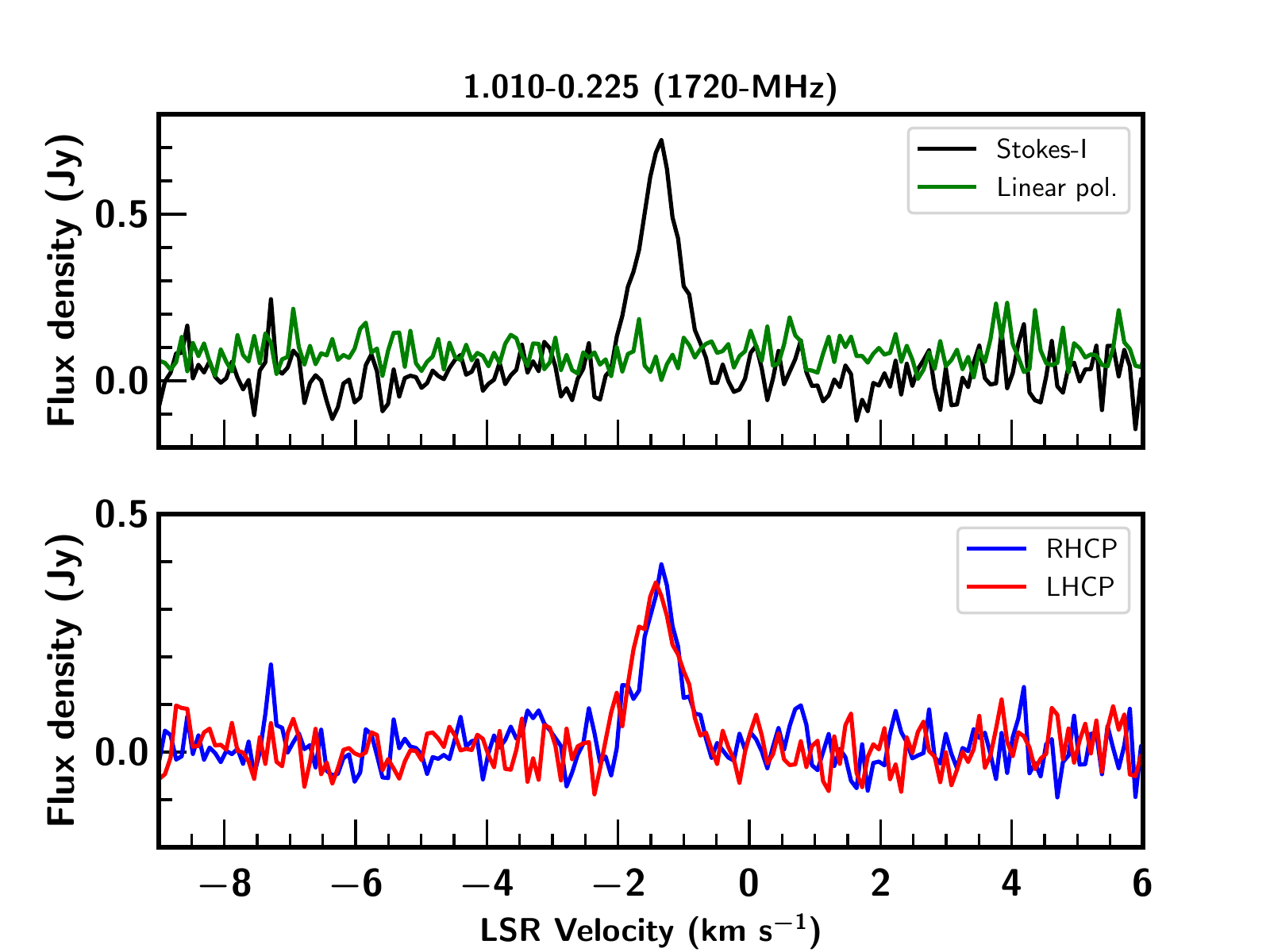}}
\subfloat{\includegraphics[width = 3.5in]{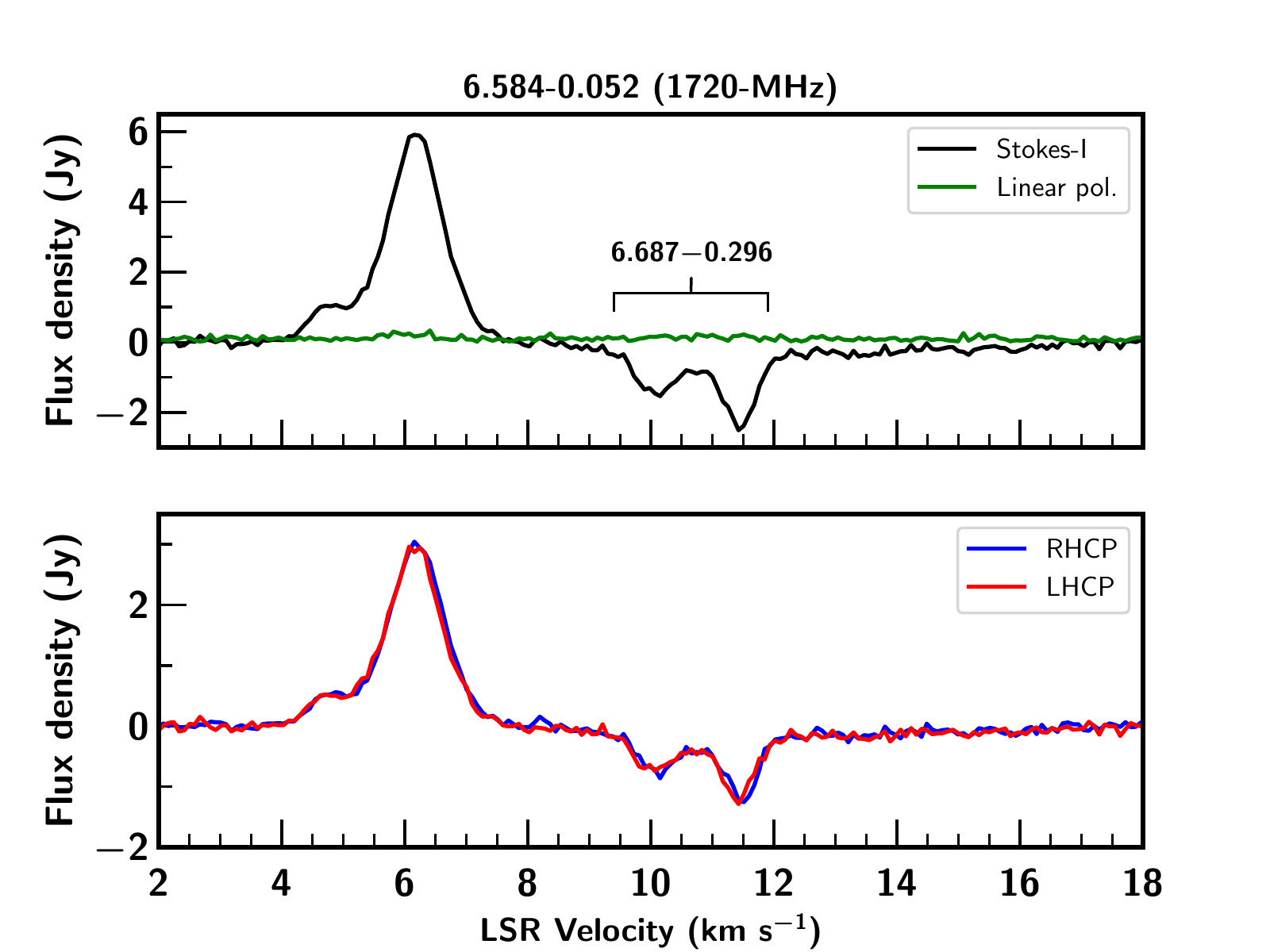}}

\caption{\it{$-$continued}}
\end{figure*}

\addtocounter{figure}{-1}

\begin{figure*}
\subfloat{\includegraphics[width = 3.5in]{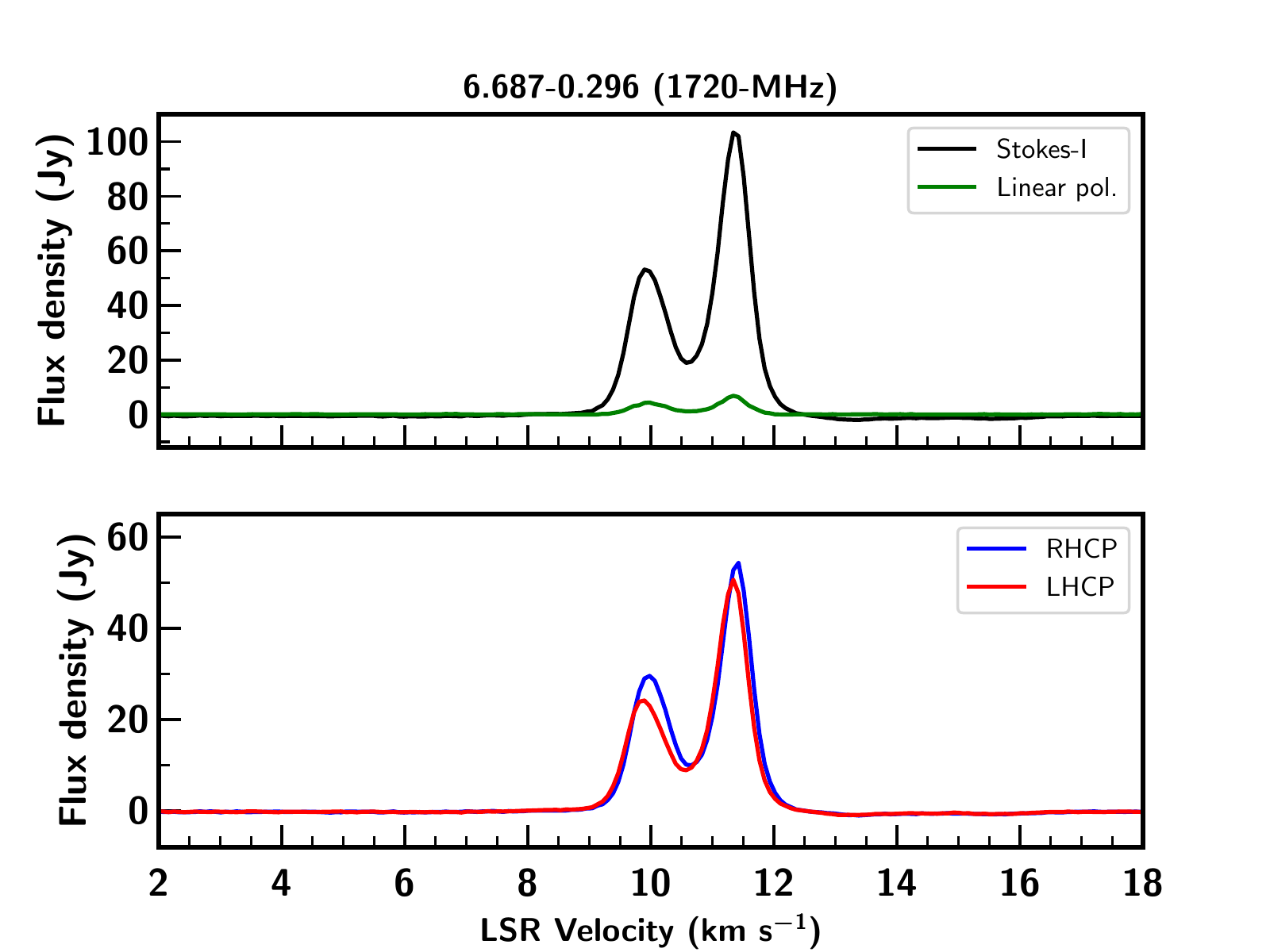}}
\subfloat{\includegraphics[width = 3.5in]{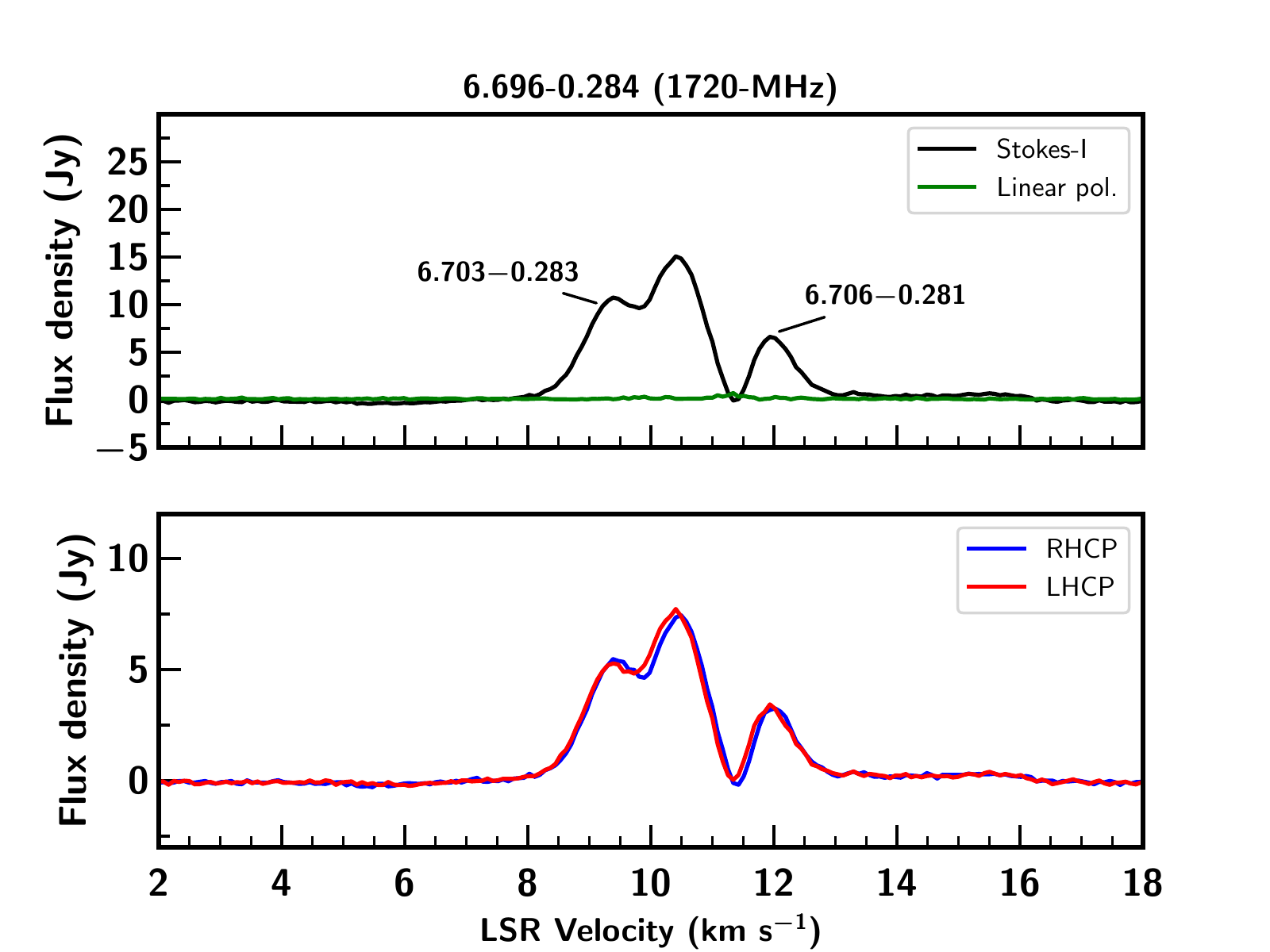}}\\
\subfloat{\includegraphics[width = 3.5in]{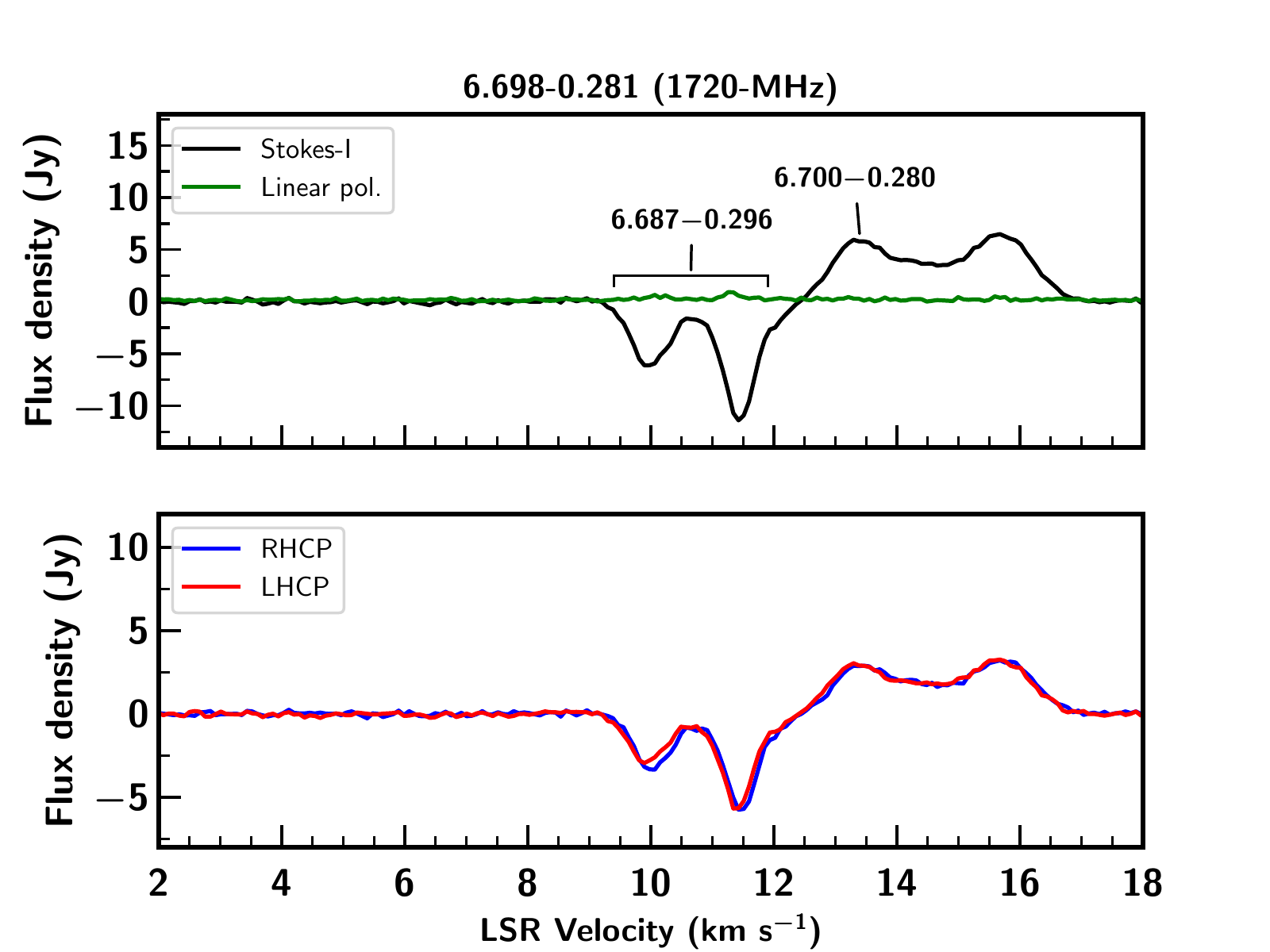}}
\subfloat{\includegraphics[width = 3.5in]{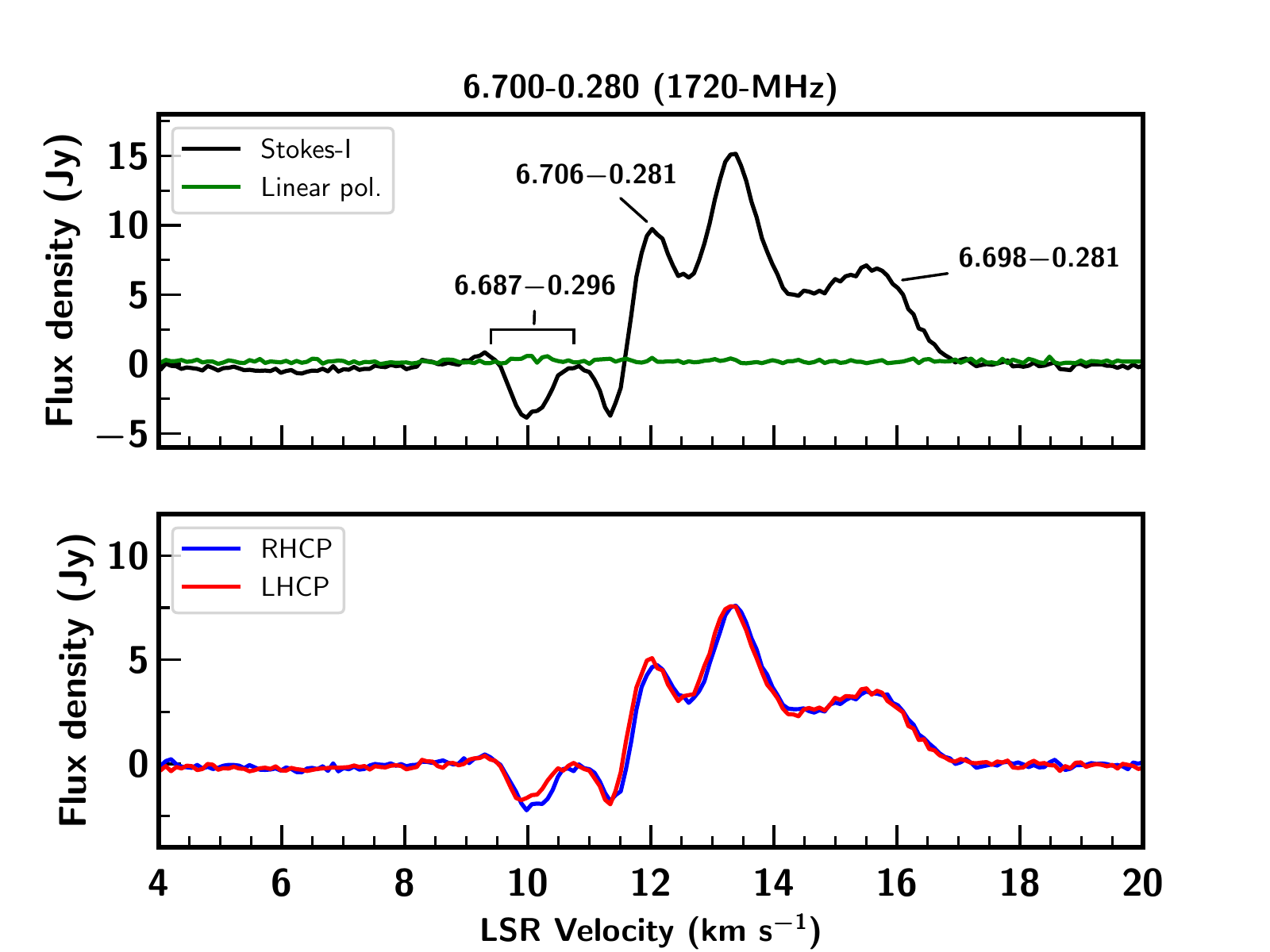}}\\
\subfloat{\includegraphics[width = 3.5in]{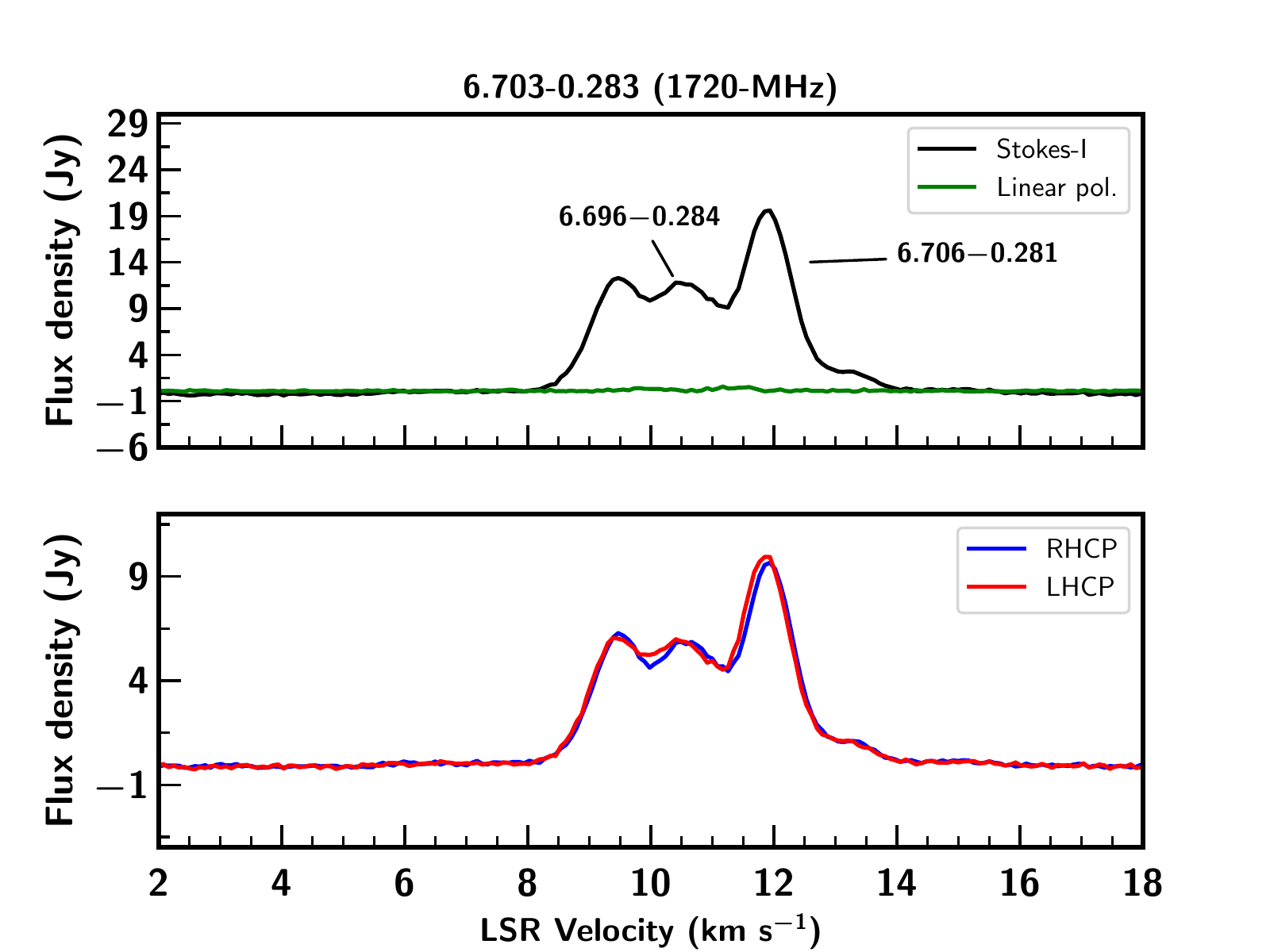}}
\subfloat{\includegraphics[width = 3.5in]{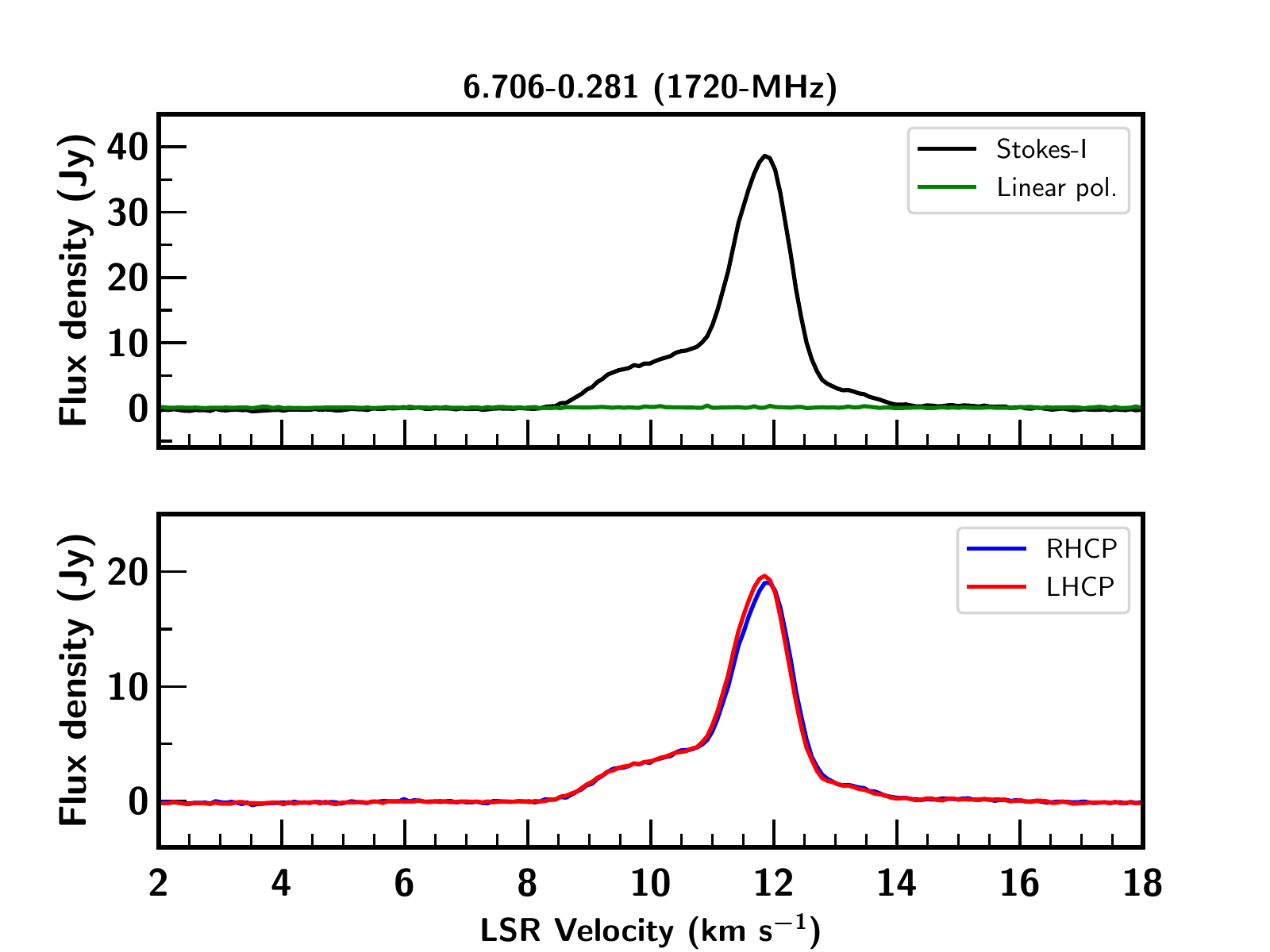}}

\caption{\it{$-$continued}}
\end{figure*}

\clearpage
\newpage




\subsection{Comments on individual sources}\label{sourcenotes}

In this section, we describe the physical and magnetic properties, as well as any other relevant information, of each associated exciting source, and also document associated maser species and transitions. For other maser species and excited-state OH transitions we refer to the literature; for the ground-state OH lines we inspect main line and 1612-MHz data from the current MAGMO dataset and \citet{Caswell98, Caswell99, Caswell2004}. To aid in the determination of associations for star formation masers, we inspect GLIMPSE \citep[Galactic Legacy Infrared Mid-plane Survey Extraordinaire;][]{Benjamin2003,Churchwell2009} 3-colour images, which are presented in Figure~\ref{appendix}. The association of other maser transitions is discussed in detail in Section \ref{association_section}, and associations are tabulated in Table \ref{Table7}.


\subsubsection{Star Formation Masers}\label{sourcenote}

\textbf{189.032+0.809}:
This is a newly detected 1720-MHz OH maser that shows the highest reliable fractional linear polarisation (46$\pm$3\%) in this study. The RHCP and LHCP components are barely split in velocity, and the RCHP component is stronger in flux density. The magnetic field strength is at the 1$\sigma$ level and accordingly, has been categorised as a non-detection. However, the presence of the strong linear component, plus RHCP and LHCP, makes this source a good candidate for an unresolved Zeeman triplet.  This is one of the few 1720-MHz SFR maser sources without an associated methanol maser, nor any detected association with main lines or excited OH masers \citep{Avison2016}. Given its coincidence with a high-mass YSO,  0.27 arcsec offset from the peak of G189.0307$+$00.7821 \citep{Cooper2013}, we suggest that it belongs to the proposed sub-class of 1720-MHz OH SFR-masers distinct for their lack of associated methanol or 1665-MHz OH masers \citep{Caswell2004}. This is discussed in Section \ref{newclass}.\\

\textbf{189.776+0.346:}
This new maser detection exhibits distinct Zeeman splitting despite having an ambiguous identification of matching Zeeman pairs. Selected pairs account for strong absolute field strengths of 7.5 and 10.4 mG. \citet{Green2012MMB} identified an unassociated methanol maser located 8 arcsec from the 1720-MHz OH maser. It is unclear whether the water maser detected towards \textit {IRAS}\,06055+2039 by \citep{Palagi93} is associated with our 1720-MHz source due to the inadequate resolution of water maser the observations. Both of these maser detections are contained within the sub-millimetre dust clump AGAL189.778+00.345 \citep{Urquhart2015}. This host dust clump is also associated with broad winged emission lines in the SiO thermal transitions (2$-$1) and (3$-$2), together with identified HCO$^{+}$ emission \citep{Lopez2011}, indicating outflow activities in the host source. \\

\textbf{306.322$-$0.334:}
The pattern of spectral components is consistent with \citet{Caswell2004}, which also reveals the same triple-peaked RHCP profile we see here, together with only two clear LHCP components.
\citet{Caswell2004} reported variability in the source's peak flux density between 1.8 and 3.1~Jy during the 3 epochs observed between November 1997 to April 2002. We detect a 2.04~Jy source, showing that the source has not undergone extreme variability. Variability has also been observed in the 6.7-GHz methanol line, with flux densities of 0.4, 1.0 and 0.5~Jy reported by \citet{Caswell95}, \citet{Caswell09} and \citet{Green2012MMB}. 
From the MAGMO data we confirm an associated 1665-MHz maser, as also previously reported by \citet{Caswell98}, and also find a coincidence with a compact sub-mm source from the ATLASGAL catalogue, G306.3199$-$0.3328 \citet{Csengeri2014}. \\

\textbf{323.459$-$0.079:}
This 1720-MHz OH maser was first detected with a peak flux density of 0.35~Jy \citep{Caswell2004}, absent from observations conducted four years earlier \citep[detection limit of 0.4~Jy;][]{Caswell99}. We now report an increase in the peak flux density to 1.8~Jy. This source is associated with 6.7-GHz \citep{Green2012MMB} and 12-GHz methanol \citep{Breen2012b} masers, both ground-state OH main lines \citep[MAGMO,][]{Caswell98}, 1612-MHz \citep[MAGMO,][]{Caswell99} and excited-state OH lines at 6030- and 6035-MHz \citep{Avison2016,Caswell1995,Caswell03}. VLBI maps of the region in the 1665- and 1667-MHz ground- as well as the 6035-MHz  excited-state OH transitions were presented by \citet{Caswellvlbi}, who also reported magnetic field strengths ranging between 1.47 and 4.13 mG within the region. The non-detection of $0.4\pm0.4$ mG found in the present work from the 1720-MHz line is consistent with \citet{Caswell2004}, who also report a value of 0.5 mG. A high-mass YSO which likely hosts these masers was identified by \citet{Mottram2007}, which is also coincident with the dust clump AGAL323.4589$-$0.0800 \citep{Contreras2013}. \\

\textbf{328.165+0.586:}
This new 1720-MHz detection is 3 arcsec offset from its methanol maser target 328.164+0.587 \citep{Green2012MMB} which is also the only maser species detected in the immediate region. Examination of the 4.5 \micron\, band in the GLIMPSE image (see Figure \ref{appendix}) indicated that the 1720-MHz and methanol masers are likely excited by the same infrared source.
The 1720-MHz and 6.7-GHz methanol masers are likely associated with G328.16+0.59 - an Extended Green Object (EGO) designated as a ``possible'' outflow candidate by \citet{Cyganowski2008} and likely related to the dust clump object G328.1650+0.5865 \citep[whose peak position is offset by 4 arcsec from the 1720-MHz maser][]{Csengeri2014}. At 13.2 $\pm$ 0.4 mG, this source shows the strongest absolute magnetic field strength in the current study.\\

\textbf{328.808+0.633:}
This 1720~MHz OH maser has been previously detected by \citet{Caswell2004}, and in the ground state OH main lines by \citet{Caswell98} (as well as in the current observations). The single Zeeman pair implies a magnetic field strength of 1.4 mG, and according to \citet{Faundez2004} and \citet{Csengeri2014}, there is a compact dust clump (G328.8087+0.6328) 1.4 arcsec offset from its position. Also associated with the 1720-MHz and main lines are 6030 \citep{Caswell2001} and 6035-MHz excited-state OH \citep{Avison2016,Caswell2001,Caswell1995}, 22-GHz water \citep{Breen2010water}, 6.7-GHz \citep{Green2012MMB}, 12-GHz \citep{Breen2012b}, 36-GHz and 44-GHz methanol \citep{Voronkov2014} masers.\\

\textbf{329.339+0.148:}
1720-MHz emission previously classified as star-formation-associated \citep{Caswell2004} shows two peaks in the LHCP profile, but shows only a single feature in  RHCP.  
1665-MHz OH \citep[MAGMO,][]{Caswell98}, 6030- \citep{Caswell03} and 6035-MHz excited-state OH \citep{Caswell2001,Avison2016,Caswell03}, and 6.7-GHz \citep{Green2012MMB} and 12-GHz \citep{Breen2012b} methanol masers are also associated with the source. The measured magnetic field strength of $4.9\pm0.4$ mG is similar to the value of 4 mG reported by \citep{Caswell2004}. From weak maser detections, the magnetic field derived from the 6035-MHz line was substantially higher \citep[13 mG][]{Caswell03}. The masers are coincident with an UCH{\sc ii} region \citep{Bronfman96} and a dust clump from the ATLASGAL catalogue \citep[G329.3382+0.1471;][]{Csengeri2014}. \\

\textbf{330.953$-$0.180:}
The Stokes $I$ flux density and magnetic field strength of this source (0.8\,Jy, 1.6 mG) are approximately consistent with the last observation in 2001 \citep[0.8\,Jy, 2.0 mG,][]{Caswell2004}. The broad thermal emission and the weaker spectral feature seen between $-$83 and $-$84 \kms\ in the 2001 spectrum observed using the Parkes telescope are not detected in this study; this is presumably because the broad features in \citet{Caswell2004} are extended diffuse OH resolved out by our interferometric observations. The 1720-MHz maser is associated with the same exciting mid-infrared source hosting the 22-GHz water \citep[e.g.][]{Breen11}, 1665- and 1667-MHz ground-state OH \citep{Caswell98}, 6035-MHz excited-state OH \citep{Caswell2001,Caswell03,Avison2016} and 6.7- \citep{Caswell2011}, and 44-GHz methanol \citep{Voronkov2014} masers. For the 6035-MHz OH line, \citet{Caswell03} report a field strength of $-$2.5 mG. The 1720-MHz OH maser appears to be projected against the same dust source \citep[G330.9545$-$0.1828,][]{Csengeri2014} as the other maser species. The higher resolution 4.5 $\micron$ GLIMPSE image and the designation of an associated EGO object, G330.95$-$0.18, as a ``possible'' outflow candidate in \citet{Cyganowski2008}, suggest that the 1720-MHz OH maser is likely associated with the outflow from the central exciting mid-infrared source.\\

\textbf{336.941$-$0.156:}
With a peak Stokes $I$ flux density of 1.7~Jy, this source shows an increase in flux density compared to a previous ATCA measurement of 0.6\,Jy \citep{Caswell2004}. We measure a magnetic field strength of 8.1~mG which is slightly higher than the previous measurement of 6 mG made by \citet{Caswell2004}. This is likely due to a slight change in the spectral profile from the earlier observations, whereby a secondary feature, barely detectable in \citet{Caswell2004}, has risen disproportionately in RHCP, resulting in a slightly larger apparent split. \citet{Csengeri2014} listed a compact dust clump G336.9420$-$0.1556 whose centre is about 6 arcsec away from the maser position. This 1720-MHz source is accompanied by emission in the main lines \citep[MAGMO, 1665-MHz also detected by][]{Caswell98}, both the 6030- and 6035-MHz excited-state OH lines \citep{Avison2016,Caswell2001,Caswell03}, as well as the 6.7-GHz \citep{Caswell2011} and 12-GHz methanol \citep{Breen2012a} masers. \\

\textbf{337.612$-$0.060:}
First identified by \citet{Caswell2004} with reported flux density and magnetic field strength consistent with this study \citep[0.83~Jy and $-$6.2~mG in the current work compared with 0.44~Jy and $-$5.7~mG in][]{Caswell2004}. At 870 $\micron$, \citet{Contreras2013} identified a compact dust continuum source AGAL337.612-00.059, the peak of which is 2.4 arcsec offset from the maser position. This 1720-MHz OH maser is associated with 1665-MHz OH maser emission \citep[MAGMO,][]{Caswell98}, 6035-MHz exited-state OH \citep{Caswell03, Avison2016}, 22-GHz water \citep{Breen2010water}, 6.7- \citep{caswell11} and 12-GHz methanol \citep{Breen2012a} masers.\\

\textbf{339.622$-$0.120:}
This is a known 1720-MHz OH maser that has been detected on a number of occasions \citep[e.g.][]{Macleod97,Caswell99,Caswell2004}. Our spectrum reveals that this source has continued to show significant temporal variability, also evident in the previous observations. Most notably we find a new spectral feature at $-$34.0 kms$^{-1}$, sandwiched between two features that have been previously reported at least one epoch \citep[$-$36.8 and $-$32.3 \kms;][]{Caswell2004,Macleod97,Caswell99}. Both features had flux densities greater than 1\,Jy in the most recent previous observations made by \citet{Caswell2004}, but now have 0.25 and 0.28\,Jy respectively.
The new feature is within the velocity range of the OH absorption feature in the single-dish spectrum of \citet{Caswell2004}, and has weak spectral splitting yielding no significant magnetic field strength. In contrast, the field strength of the feature at $-$36.8 km\,s$^{-1}$, which now shows a clear splitting, is 2.7 mG. Towards the weak detection at $-$32.4 km\,s$^{-1}$, we detect no RHCP emission \citep[similarly to][]{Caswell2004}, resulting in a high circular polarisation fraction. The RMS survey indicates a nearby infrared source offset by $\sim$1 arcsec which is classified as a high-mass YSO \citep{Mottram10}. As shown on the GLIMPSE image (see Figure~\ref{appendix}), there is positional coincidence between the 1720-MHz maser and other maser species -- 1665, 1667-MHz \citep[MAGMO,][]{Caswell98} and 6035-MHz OH \citep{Caswell03,Avison2016,Caswell1995}, 22-GHz water \citep{Breen2010water}, 6.7- \citep{Caswell2011} and 12-GHz methanol \citep{Breen2012a}.\\

\textbf{339.884$-$1.259:}
This source, first reported by \citet{Caswell2001} in both the 1720- and 6035-MHz OH transitions, was also detected in the 1720-MHz transition by \citet{Caswell2004}. It has four prominent Stokes $I$ features at $-$37.8, $-$36.9, $-$35.0 and $-$34.1 km\,s$^{-1}$. The previous 1720-MHz observations show peak emission at $-$34~km~s$^{-1}$, varying significantly in peak flux density \citep[1.9\,Jy in 2000 Nov, 3.3\,Jy in 2001 Mar, 10\,Jy in 2001 Aug, 10\,Jy in 2001 Dec, and 12\,Jy in 2002 Apr,][]{Caswell2004}. The current 1720-MHz peak of $\sim$32~Jy is at a velocity of $-$37.8~km\,s$^{-1}$ and has shown a significant flux density increase from $\sim$1.5~Jy in 2002 Apr \citep{Caswell2004}. The location of the 1720-MHz emission is offset by less than 1 arcsec from an UCH{\sc ii} region \citep{Walsh98} and falls within a dust clump in the ATLASGAL compact source catalog \citep{Contreras2013}. This source interestingly exhibits emission in 1665/1667-MHz \citep[MAGMO,][]{Caswell98}, 6030/6035-MHz \citep[e.g.][]{Avison2016,Caswell03}, 22-GHz water \citep[e.g.][]{Breen2010water}, and four methanol transitions, 6.7-GHz \citep{Caswell2011}, 12-GHz \citep{Breen2010a} and 36/44-GHz \citep{Voronkov2014}. \\

\textbf{340.785$-$0.096:}
Several clearly resolved Zeeman pairs are seen in the spectrum of this strongly circularly polarised source, which was originally reported by \citet{caswell83} and later by \citet{Caswell1995} and \citet{Caswell2004}. A set of three components with peak velocities between $-$104 to $-$107~\kms\ all have the same line-of-sight magnetic field direction ($-$8.4, $-$6.6, $-$6.0~mG), contrary to the field reversal seen in the other feature at $-$100 \kms ($+$5.5~mG). This field reversal was also reported by \citet{Caswell2004} for the 1720-MHz transition and by \citet{Caswell1995} for 6035-MHz excited-state OH. 
\citet{Csengeri2014} reports a dust clump (G340.7848-0.0968), with a peak that is 2.45 arcsec offset from the maser position. The main lines \citep[MAGMO,][]{Caswell98}, 6035-MHz excited-state OH \citep{Caswell1995,Caswell03,Avison2016}, 22-GHz water \citep{Breen2010water} and 6.7- \citep{Caswell2011}, 12- \citep{Breen2012a} and 36-GHz \citep{Ellingsen2018} methanol masers are also associated with this source.\\

\textbf{344.582$-$0.023:}
We report the first detection of this 3.2~Jy (Stokes $I$), highly circularly polarised 1720-MHz OH maser. It is likely associated with an Extended Green Object \citep[located within 2 arcsec][]{Cyganowski2008}, and is also coincident with a dust clump \citep[G344.5816-0.0235][]{Csengeri2014} as well as water \citep{Breen2010water}, OH main line \citep[MAGMO,][]{Caswell98} and 6.7-GHz methanol \citep{Caswell2011} masers. Compared to the 1720-MHz field strength of 3.6 mG measured in this work, the split 1665-MHz OH lines reported by \citet{Fish03b} trace a magnetic field strength of $-$2.8~mG, notably propagating in the opposite direction.  \\

\textbf{345.003$-$0.224:}  
Spectral similarities between the current study and previous studies of this maser source \citep{Gaume87,Caswell2004} show disproportionate flux densities between RHCP and LHCP. The spectrum in Figure \ref{Figure1} shows that the LHCP at 37.79\,Jy is more dominant than the RHCP profile at 1.54\,Jy. Clear Zeeman splitting reveals a magnetic field strength of 3.5\,mG and this source exhibits significant linear and circular polarization relative to the uncertainties. This 1720-MHz source is associated with 1665-MHz \citep[MAGMO,][]{Caswell98}, 6035-MHz excited-state OH line \citep{Caswell98, Caswell2004, Avison2016}, 6.7-GHz \citep{Caswell2011} and 12-GHz \citep{Breen2012a} masers. The maser sits 2 arcsec offset from the peak of dust clump G345.0029-0.2241 \citep{Csengeri2014}. \\

\textbf{345.117+1.592:}
This maser source is significantly offset from the target methanol maser (345.010+1.792) by 818 arcsec. However, given the absence of any nearby SNR and the sighting of enveloping H{\sc ii} regions, \citet{Caswell2004} assigned an SFR-type status to this source. Furthermore, \citet{Caswell2004} identified this source as one of the isolated SFR 1720-MHz sub-class, and as commonly observed in masers in that sub-class, this source (with a FWHM of 0.93 \kms) has a broad linewidth feature and also doesn't have any observed associated maser species either from MAGMO observations (1612-MHz and main lines) or from literature.  \\

\textbf{345.497+1.461 and 345.495+1.462:}
\citet{Caswell2004} first detected these 1720-MHz OH masers that are separated by 5.2 arcsec. Both have seen a decrease in peak flux density from 4.9 to 1.57\,Jy for 345.497+1.461 and 2.3 to 0.69\,Jy for 345.495+1.462. 345.495+1.462 appears to have undergone a reduction in magnetic field strength from $-$12 to $-$7.9 mG. \citet{Guzman2010} identified a jet associated with high-mass star forming region, \textit{IRAS}\,16562$-$3959, which we find has radio bright spots at the locations of the 1720-MHz OH masers. While significantly offset from the 1720-MHz OH masers, the exciting source from which the jet originates is coincident with 1665- and 1667-MHz OH \citep[MAGMO,][]{Caswell98}, 6035-MHz OH \citep{Avison2016} and water maser emission \citep{Breen2010water}. As shown in Figure~\ref{appendix}, the closest methanol maser is associated with a separate nearby infrared source.\\

\textbf{348.727$-$1.039:}
This new detection in the 1720-MHz OH transition shows clear Zeeman splitting that corresponds to a field strength of 5.3~mG. This 1720-MHz OH maser is offset from a prominent green source (see Appendix~\ref{appendix} which is associated with both 6.7- \citep{Caswell2010} and 12-GHz \citep{Breen2012a} methanol maser emission as well as 1665-MHz OH maser emission \citep[MAGMO,][]{Caswell98}. A water maser is present in the vicinity, but offset from both the 1720-MHz and methanol masers \citet{Forster1989,Breen2010water}. This 1720-MHz OH maser is offset 3.6 arcsec from the peak of a compact dust clump AGAL348.726-01.039 \citep{Contreras2013}.\\

\textbf{350.112+0.095:}
The RHCP and LHCP spectral profiles of this new 1720-MHz detection show distinct and significant Zeeman splitting implying an absolute magnetic field strength of 11\,mG, and the source also exhibits strong fractional circular polarization. The nearest 6.7-GHz methanol maser catalogued in the MMB survey \citep{Caswell2011}, is offset by 43 arcsec, however, the GLIMPSE image in  Figure \ref{appendix} shows associated main lines \citep[MAGMO,][]{Caswell98, Argon2000} and 6035-MHz excited \citep{Caswell98, Avison2016} OH lines. The 1720-MHz source and the associated maser species are nearby the ATLASGAL complex dust clump G350.1113+0.0906 \citep{Csengeri2014}. \\

\textbf{350.686$-$0.491:}
First detected in the 1720-MHz transition by \citet{Caswell2004}, the Stokes I spectrum shows four components with relatively high levels of linear and circular polarised emission. They are split into two distinct Zeeman pairs with corresponding magnetic field strengths of 4.0 and 4.2 mG. These masers are most likely hosted by a high-mass double-core system, likely a binary high-mass star forming region \citep{Chen2017}. 1665, 1667-MHz OH \citep[MAGMO][]{Caswell98,caswell2013}, 6035-MHz OH \citep{Avison2016,Caswell2004,Caswell1995}, 22-GHz water \citep{titm16}, and 6.7-/12-GHz methanol \citep{Caswell2010,Breen2012a} masers are all associated with the 1720-MHz source.\\

\textbf{351.158+0.699:} 
This is a new 1720-MHz detection located within the NGC 6334 H{\sc ii} region complex \citep{Willis13,Kwon13}. The GLIMPSE image in Figure \ref{appendix} appears to have a bi-polar outflow feature with the 1720-MHz maser located at one tip, and the other identified maser species - OH main lines \citep[MAGMO,][]{Caswell98}, 22-GHz water \citep{Breen2010water}, 6.7-GHz \citep{Caswell2011}, and 36/44-GHz methanol \citep{Voronkov2014} masers - positioned on or close to a central driving source. We believe that the 1720-MHz maser is likely associated with the outflow of the driving source that apparently hosts the other maser species. We see a similar scenario in 345.497+1.461 and 345.495+1.462. The separation of the Zeeman pairs reveal a field strength with a magnitude of 5.2 mG. \\

\textbf{351.419+0.646:}
NGC6334F is a well known star formation region, with previously detected 1720-MHz OH maser emission \citep[e.g.][]{Lo-1975,Gaume87,Caswell2004}. This star formation region \citep[identified as dust clump AGAL351.416+00.646][]{Contreras2013} is known to host many maser transitions \citep[e.g.][]{Caswell2004}, including the OH ground-state main lines \citep[e.g. MAGMO,][]{Caswell98}, excited-state OH emission at both 6030- and 6035-MHz \citep{Avison2016,Caswell03,Caswell1995}, 6.7-GHz methanol \citep[e.g.][]{Caswell2010}, 12-GHz methanol \citep[e.g.][]{Breen2012a}, 36- and 44-GHz methanol \citep[e.g][]{Voronkov2014} and 22-GHz water masers \citep[e.g.][]{titm16}. 

This strong circularly polarised 1720-MHz OH maser shows a magnetic field strength of $-$6.2~mG, which agrees with the estimated range of $-$6.0 to $-$6.4 mG measured with the Australian Long Baseline Array \citep{Chanapote2019}.
The flux density appears to be variable, with peak measurements in the range $\sim$15 to 70~Jy previously reported \citep{Lo-1975,Gaume87,Caswell2004}. The current observations find a component at $-$10.6 \kms\ with a Stokes I flux density of $\sim$76\,Jy and $\sim$102\,Jy for the component at $-$9.9 \kms\, making it to be the strongest source detected in this study.

More recent observations by \citet{Macleod2018} during a maser flaring event (in methanol, OH and water) show an even higher peak flux density (up to $\sim$130~Jy in LHCP) and the detection of a new feature at $-$7.7~\kms\ which disapeared 247 days later. This transient feature showed a magnetic field strength of $+$1.3~mG based on a Lande splitting factor of 0.236 kms$^{-1}$ mG$^{-1}$ \citep[this corresponds to $+$2.7~mG using the splitting coefficient used in this work][]{Macleod2018}. The maser flaring event was suggested to be as a result of episodic accretion producing powerful outburst of radiation, driving the maser flaring activity \citep[e.g][]{Macleod2018,Hunter-2018}. \\

\textbf{351.774$-$0.537:} 
This 1720-MHz OH maser was first discovered by \citet{Macleod97} and then positioned by \citet{Caswell99}. Like \citet{Caswell99}, we find that the 1720-MHz OH maser emission is somewhat spatially distributed (over 0.7 arcsec in the current work). The RHCP and LHCP components of the source at $\sim$$-$1.5 \kms\ have approximately equal amplitude of  $\sim$0.9\,Jy, in contrast to the spectrum in \citet{Caswell2004}, where the RHCP appears much weaker. Like 340.785$-$0.096, this source shows a field reversal.

\citet{Beuther2017} used the 0.06 arcsec resolution of ALMA to clearly resolve the high-mass dust clump in the region \citep[identified as high-mass proto-cluster by][]{Urquhart2013}, revealing four cores contained within a few arcsec. However, only one of these cores \citep[core-2 in][]{Beuther2017} is associated with 6.7-GHz methanol masers detected by \citet{Walsh98}. The 1720-MHz OH maser position is offset by $>$2 arcsec from any of the four cores so it is difficult to say which of the objects is exciting the 1720-MHz OH emission (if any). A possible scenario might be the 1720-MHz OH maser emission is originating from the shocked region associated with an outflow from one of the cores \citep[CO observations of][show that core 1 might be the most likely candidate]{Beuther2017}.

Based on the proximity of known 22-GHz water \citep{titm16}, OH main lines \citep[e.g. MAGMO,][]{Caswell98}, the 1612-MHz OH satellite line \citep[e.g. MAGMO,][]{Sevenster1997}, 6035-MHz OH \citep{Avison2016}, 6.7- \citep{Caswell2010}, 12-GHz \citep{Breen2012a}, and 36/44-GHz methanol masers \citep{Voronkov2014}, we list them as associated with the 1720-MHz OH masers, but note that it is unclear that they are all associated with the same \citet{Beuther2017} core.\\ 

\textbf{353.410$-$0.360:}
The spectral profile and measured magnetic field towards this previously detected 1720-MHz OH maser \citep[e.g.][]{Gaume87} are consistent with the previous measurement made by \citet{Caswell2004}. The 1720-MHz OH maser is associated with an UCH{\sc ii} region \citep[e.g.][]{Garay-2006,Walsh98} embedded within dust clump AGAL353.409$-$00.361 \citep{Contreras2013}. 

The 1720-MHz OH maser emission is accompanied by 1665-MHz OH emission \citep[MAGMO,][]{Caswell98}, 6035/6030-MHz OH \citep[e.g.][]{Avison2016,Caswell2004}, 6.7-GHz \citep{Caswell2010} and 12-GHz methanol maser transitions \citep{Breen2012a}, but the nearest water masers \citep{Breen2010water} are offset (see GLIMPSE images in Figure~\ref{appendix}).\\

\textbf{357.557$-$0.321}
This is a newly detected 1720-MHz OH maser which is within a similar velocity range, and is spatially consistent with, a methanol maser from the MMB survey \citep{Caswell2010}. We also see association with the main lines also obtained from MAGMO data, and a 22-GHz water maser \citep{titm16}.  \cite{Fontani2005} identified the host WISE object as a high-mass protostar coincident with compact dust source G357.5577$-$0.3217 \citep{Csengeri2014}.\\

\textbf{359.970$-$0.457:}
This new 1720-MHz OH maser detection is coincident with 1665-MHz OH (MAGMO), 6.7-GHz methanol \citep{Caswell2010} and water maser \citep{titm16} emission, all falling within the dust clump, G359.9703-0.4570 \citep{Csengeri2014}. The Stokes I spectrum shows a relatively broad single feature and the RHCP and LHCP profiles show a clear split corresponding to a magnetic field strength of 7.8~mG.\\

\textbf{000.376+0.040:}
This new detection with implied magnetic field strength of +1.5 mG shows a peak flux density at approximately 1\,Jy. This OH maser falls within dust clump, G000.3756+0.0409 \citep{Csengeri2014} and is associated with OH main lines \citep[MAGMO,][]{Caswell98}, water \citep{Walsh2014,titm16}, and 6.7-GHz methanol \citep{Caswell2010} masers.\\

\textbf{000.665$-$0.036:}
This known 1720-MHz SFR maser source \citep{Gaume87, Caswell2004, Qiao18}, is located within the Sgr B2 H{\sc ii} D region \citep{Benson84}. Comparison with \citet{Qiao18} indicates a decreased flux density of the only reported feature (at 62.6 \kms), from 1.03$\pm$0.07 Jy\,(current study observed in 01/2011) to 0.65 Jy\, (observed in 02/2016). Figure \ref{Figure1} shows blended multiple peaks in the Stokes-$I$ spectrum. We also report a new RHCP component at $\sim$60.6 \kms. The estimated absolute magnetic field strength is 8.8 mG for the more separated of the two identified Zeeman pairs. The GLIMPSE image in the Figure \ref{appendix} shows that the 6.7-GHz methanol maser \citep{Caswell2010} is the only associated maser species in the current study.  At 8 arcsec offset from the peak position, this 1720-MHz source is most likely associated with the dust clump G000.6652-0.0335 \citep{Csengeri2014}. \\

\textbf{008.669$-$0.356:}
The current study shows an increase in flux density of the source from 1.0\,Jy reported in \citep{Caswell2004} to 2.2\,Jy, while the magnetic field measurements are consistent. The 1720-MHz source is coincident with the dust clump G008.6702-0.3557 \citep{Csengeri2014}. The position of this 1720-MHz maser is 1.2 arcsec offset from a source identified by  \citet{Bronfman96} as an  Ultra-compact H{\sc ii} region. Maser positions indicate that the 1720-MHz source is associated with 1612-MHz (MAGMO), 1665-MHz \citep[MAGMO,][]{Caswell98} and 6035-MHz \citep{Caswell95, Caswell2001, Avison2016} OH, and 6.7-GHz \citep{Green2010} and 12-GHz \citep{Breen2012a} methanol masers. As seen in the GLIMPSE image in Figure \ref{appendix}, the presence of multiple infrared sources makes the association between the 1720-MHz and 1612-GHz (MAGMO), and 22-GHz water \citep{Breen2010water,titm14} masers ambiguous. \\

\textbf{10.474+0.027:}
First detected by \citet{Caswell99} with a total intensity of 0.4\,Jy, we report an increased flux density of $\sim$1.6\,Jy. With no velocity shift between the RHCP and LHCP components, a magnetic field non-detection is reported in this work. This source is coincident with a complex high-mass star forming region \citep{Pascucci2004}, identified as a 25,000--35,000 M$_{\odot}$ high-mass proto-cluster G010.4722+0.0277 \citep{Ginsburg2012,Urquhart2013} which hosts at least one very young
late O, or early B star that may be actively accreting \cite{Titmarsh2013}. 

The GLIMPSE image shows two separate mid infrared sources with coincident maser emission, one with just the 1720-MHz OH maser and the other with the OH main lines at 1665- and 1667-MHz \citep[MAGMO,][]{Caswell98}, 6.7-GHz methanol \citep{Green2010} and a water maser \citep[that holds the record for highest velocity emission of any water maser associated with star formation][]{Titmarsh2013}. \\

\textbf{11.034+0.062:} 
At 1.63\,Jy, this strongly circularly polarised 1720-MHz OH maser shows an increase in peak flux density from 0.45\,Jy \citep{Caswell2004}. The weaker second feature present at 28 \kms\ is sitting on a broad absorption dip in the Parkes spectrum of \citet{Caswell2004}, and is not seen in our spectrum. \citet{Forster2000} identify the pumping source as a high-mass star forming region. The 1720-MHz source is associated with 22-GHz water \citep{titm14}, 1665-MHz  \citep[MAGMO,][]{Caswell98} and 6035-MHz OH \citep{Caswell03,Avison2016} and 6.7-GHz methanol maser \citep{Green2010} transitions. SiO (2--1) thermal emission has also been detected in the high-mass SFR \citep{Csengeri2016}. \\

\textbf{017.638+0.158:}
The Parkes single-dish spectrum of \citet{Caswell2004} contains three spectral features at 20.5, 28.5 and 35 \kms, but in this study, consistent with \citet{Beuther_Thor}, we only detect the feature at 28.5 \kms. The position and velocity of the 1720-MHz feature at 20.5 \kms\ from \citep{Caswell2004} indicates coincidence with the peak of the central exciting source \citep{Maud2018}, 6.7-GHz methanol \citep{Caswell95_meth,Green2010}, 1665-MHz OH \citep[MAGMO,][]{Argon2000,Beuther_Thor,caswell2013}, 1667-MHz OH (MAGMO), and water \citep{Breen2010water,titm14} masers.  On the other hand, the 1720-MHz feature at 28.5 \kms\ is offset by 4 arcsec from both the peak of the exciting source and the other maser species. Figure 2 in \citet{Maud2018} indicates that the position and velocity of the 28.5 \kms\ feature coincides with the ALMA $^{13}$CO emission (red contour moment maps in the referenced figure), tracing the red-shifted (velocity range of 24.7 -- 35.0 \kms) outflow of the exciting source. \citet{Caswell2004} reported varying flux densities for the feature at 28.5 \kms\ across three epochs: <0.15\,Jy (2001), 0.3\,Jy (2002) and 0.8\,Jy (2003). The current study (observed in 2012) reports a flux density of 0.48\,Jy.\\ Interestingly, we note that although the spectral profile at 28.5 \kms\ in \citet{Caswell2004}, and the one reported in this study, appear comparable in total intensity, and are within the same velocity range, they differ in polarisation handedness. The spectral feature (28.5 \kms) in Figure 1 of \citet{Caswell2004} shows prominent RHCP while our study reveals prominent LHCP. As a consequence of this apparent reversal in polarisation handedness, the measured magnetic field vector we report is $-$8.1$\pm$0.5\,mG in contrast to +14\,mG measured by \citet{Caswell2004}. The field orientation of the feature at 20.5 \kms\ ($-$2\,mG) measured by \citet{Caswell2004} is the same as the field orientation of other maser species at the same velocity range, 1665-MHz ($-$1\,mG) \citep{Argon2000,Caswell2004,caswell2013} and 1667-MHz ($-$1\,mG) \citep{caswell2013}. The consistency of these features  suggests that the polarisation handedness in the spectrum in \citet{Caswell2004} is correct, and affirms the observed opposite handedness for the feature at 28.5 \kms. By comparing our measurements with other sources observed the same day, and also with literature, we also confirmed the handedness of our observations. As such there appears to be a genuine field reversal between the time of  \citet{Caswell2004}'s measurements and ours. The observed magnetic field reversal together with the demonstrated flux density variability of the 28.5 \kms\ feature, suggests short time-frame variation in the localised masing environment  associated with the outflow (and potentially a complex magnetic field structure).\\

\subsubsection{Supernova Remnant Masers}

\textbf{269.141$-$1.213:}
This is a new 1720-MHz OH SNR maser detection. The combined evidence of the proximity of this source to the Vela SNR \citep{Duncan1996} and the absence of OH main-line transitions at the MAGMO detection limit, make a strong case for designating it a SNR source. Additionally, the velocity of the maser source, 10.6 \kms\, roughly falls within the LSR velocity range of 0--10 \kms\ covered by $^{12}$CO(J=1--0) emission in the region \citep{yama99}. Although not necessarily indicative of the absence of high-mass star formation, the significant spatial difference between this maser and the closest MMB methanol maser \citep[309 arcsec, ][]{Green2012MMB} is also notable. By using a multi-wavelength approach (radio, optical and x-ray), \citet{Yamaguchi1999} showed that some parts of the molecular clouds traced by $^{12}$CO are interacting with the SNR. Typical kinematic behaviour and the physical conditions of SNR-molecular clouds environments of 1720-MHz SNR-type masers are discussed in details in \citet{Brogan2013} and references therein.  However, further verification is advisable, as the 1720 MHz maser shows very strong circular polarisation and magnetic field strengths (10.7 and 11.4 mG) not typical of SNR-masers (as can be seen in Table~\ref{Table3}). \\

\textbf{337.801$-$0.053:}
This 1720-MHz OH maser was first detected by \citet{Green-1997} towards SNR G\,337.8$-$0.1 and later observed (and positioned) by \citet{Caswell2004}. \\

\textbf{349.734+0.172:}
The spectrum shows a single spectral feature of narrow linewidth and has no indication of Zeeman splitting. This maser is one of five distinct 1720-MHz OH maser detections made by \citet{Frail96} towards SNR G\,349.7+0.2. Comparison between the location of the 1720-MHz OH maser and radio continuum emission by \citet{Frail96} showed that the masers were all located within the interior of the SNR, and that the brightest masers were found towards the emission ridge. Its relatively far distance ($\sim$20.4~kpc) makes it one of the highest luminosity 1720-MHz OH masers detected towards a SNR.\\

\textbf{358.936$-$0.485:}
This 1720-MHz source is associated with the SNR G359.1-0.5 \citep{zadeh95,Wardle2002,Qiao18}. The RHCP and LHCP profiles have the same peak velocity leading to no observed Zeeman splitting. 
\\

\textbf{358.983$-$0.652:}
This is a new 1720-MHz detection and has been categorized as an SNR source in this study. As seen in Figure 22 of \citet{Stupar2011}, the position of this 1720-MHz source is at the edge of the shell (traced by the contours of H$\alpha$ emission) surrounding the SNR G359.1-0.05. The unsplit spectral line also exhibits the larger linewidth typical of SNR masers.\\

\textbf{359.940$-$0.067:}
This 1720-MHz OH maser is the brightest of five detected at the boundary of Sgr A East supernova remnant \citep[see e.g.][]{maeda2001} by \citet{zadeh96}, who named it Sgr A OH1720:A (66). The 1720-MHz OH masers have no main line counterparts (either in this work or in \citealt{zadeh2001}), and are interpreted as arising from the shock interaction between Sgr A East and the M$-$0.02$-$0.07 molecular cloud \citep{zadeh96,zadeh2001}. 
Our measured magnetic field strength (3.5 mG) is consistent with the value of \citet[][3.7 mG]{zadeh96}, and is one of the highest SNR-maser B-field values in this work.\\

\textbf{1.010$-$0.225:}
This 1720-MHz OH maser \citep[named Sgr D OH1720 A1 in][]{zadeh99} is located close to the Galactic centre in the Sgr D region, positionally coincident with a known SNR \citep[see][and references within]{zadeh99}.\\

\textbf{6.584$-$0.052, 6.687$-$0.296, 6.696$-$0.284, 6.698$-$0.281, 6.700$-$0.280, 6.703$-$0.283, and 6.706$-$0.281:}
These 1720-OH masers are all associated with the W28 supernova remnant, towards which \citet{Claussen97} previously detected six separate regions of 1720-MHz OH maser emission. Although we do not report reliable magnetic field measurements from these sources (due to their weak Zeeman splitting being insufficient to measure a clear velocity separation), extensive VLBI polarization studies of these 1720-MHz masers by \citet{Hoffman2005} reported a magnetic field strength of |B| $\approx$ 0.75 mG. Previous measurements of the magnetic field have been quoted within the range of 0.09 -- 0.45 mG \citep{Claussen97} and (in a subset) 2.0 mG \citep{Claussen99}.

\subsubsection{Diffuse OH Sources}

\textbf{12.803$-$0.202:}
This broad and weak (FWHM $\sim3$\kms, $\sim0.6$\,Jy) detection is associated with a H{\sc ii} region within the W33 complex. 
The 1720-MHz OH line is detected across the full spatial extent of a compact ($\sim$1.5$\times$0.5 arcmin) 18\,cm continuum source \citep[catalogued at 20\,cm as G\,12.805$-$0.201 by][]{white05}, and shows neighbouring emission and absorption peaks with conjugate detections in the 1612-MHz transition (also from the MAGMO dataset).  
This pattern of weak emission in one of the satellite lines and conjugate absorption in the other is typical of diffuse OH, and is seen throughout much of the Inner Galaxy, where the background continuum emission at 18\,cm is non negligible \citep[e.g.][]{turner79,dawson14}. While the presence of an emission line against continuum background may signify maser amplification of the incident continuum radiation, this interpretation is not certain without measurements of the line excitation temperature, which is generally non-LTE and may be high enough to produce emission without maser action \citep[e.g.][]{guibert78}. For this reason, and because this kind of weak masing is usually considered distinct from the classical, high-gain, compact and narrow-line masers featured in this work, we classify this detection as a ``diffuse OH'' source. As expected for the extended molecular ISM, the source shows no measurable Zeeman splitting and no significant polarisation fractions.
\\

\textbf{19.610$-$0.234:}
Similarly to 12.803$-$0.202, this diffuse OH detection is broad (FWHM greater than 5 \kms), weak ($\sim0.6$\,Jy) and shows neighbouring emission and absorption features with conjugate detections confirmed in the 1612-MHz transition (also from the MAGMO observations). It also is detected across the full extent of an associated radio continuum source, and shows no detectable Zeeman splitting and no significant measured polarisation fractions. The background H{\sc ii} region in this case is G\,19.6$-$0.2 \citep[see e.g.][]{garay98}, with an angular size of $\sim$1.0$\times$0.5 arcmin in our observations (derived from a preliminary continuum reduction of the 1720-MHz zoom band data), and which forms part of a highly-active high-mass star forming region with main line OH \citep{Ho81,Forster1999}, water \citep{titm14}, methanol \citep{Green2010} and rare SiO \citep{Cho2016}, 9.9-GHz methanol \citep{Voronkov-2010} as well as ammonia (11, 9) and (8, 6) maser emission \citep{Walsh2011}.


\begin{figure}

	\includegraphics[width=\columnwidth]{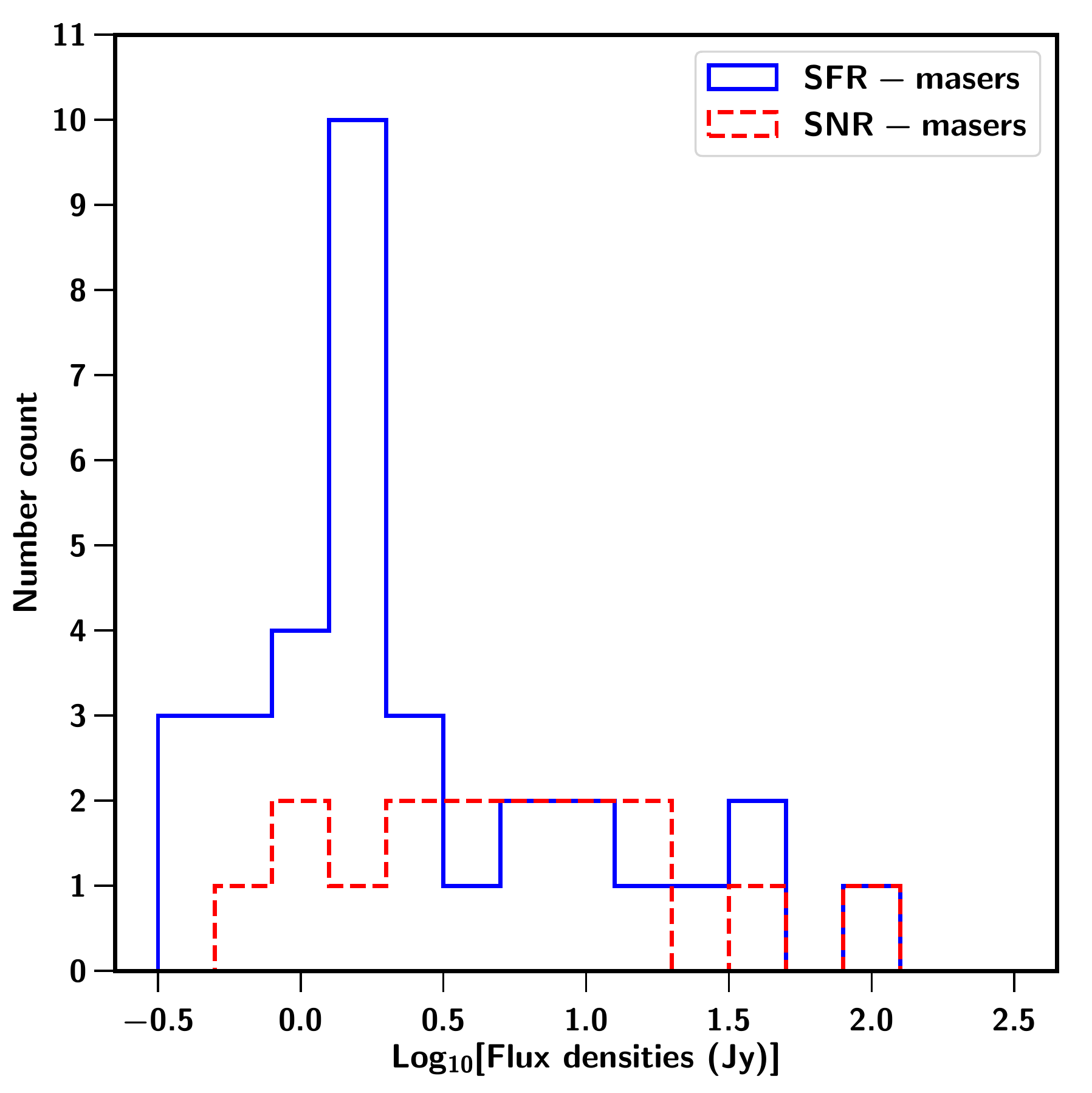}
    \caption{Peak flux density distribution of all SFR (blue) and SNR (red) masers detected in this study.}
    \label{Figure4}
\end{figure}

\begin{figure}
	\includegraphics[width=\columnwidth]{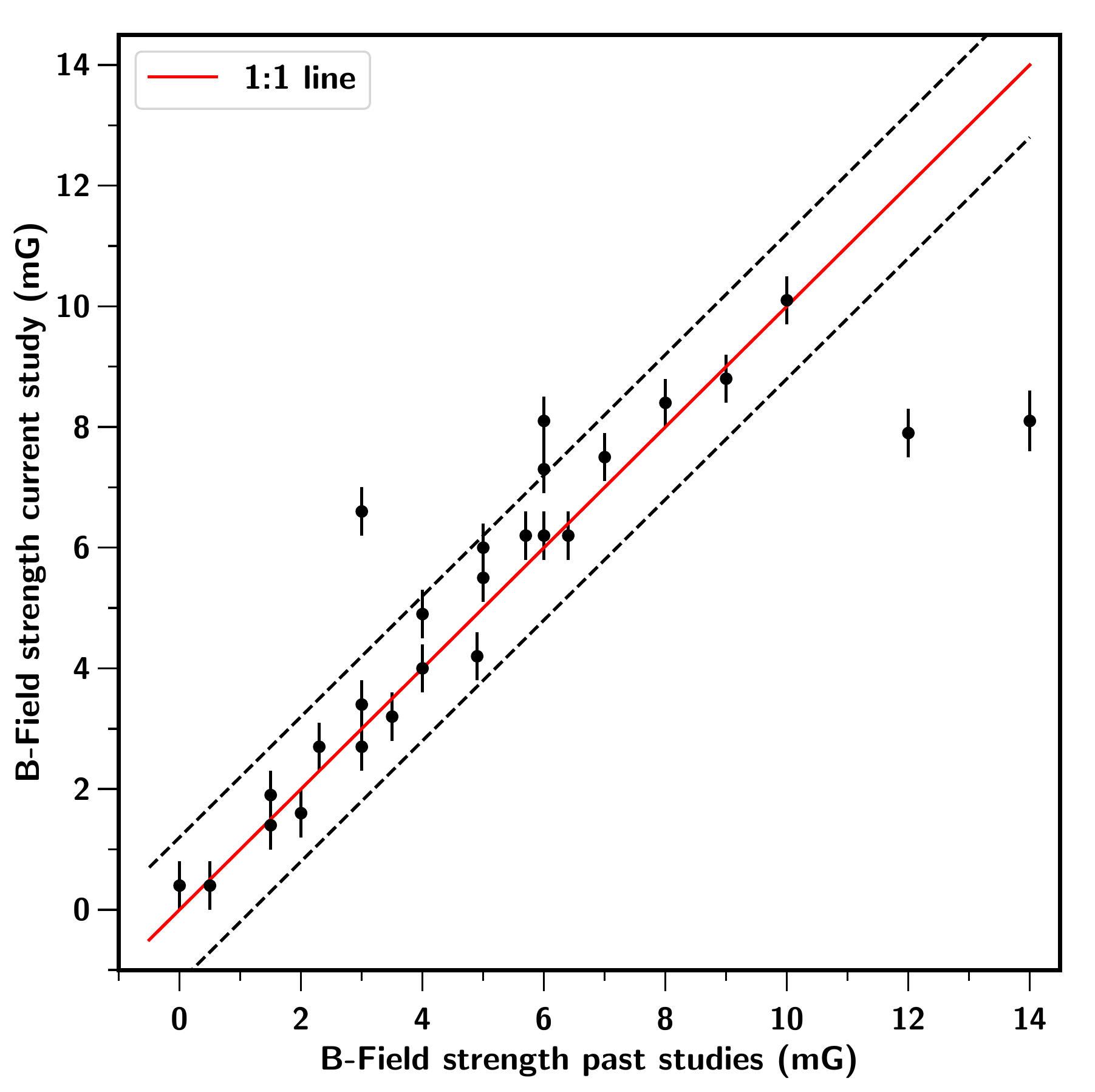}
    \caption{A comparison of magnetic field strengths of common sources obtained in this work and those measured by \citet{Caswell2004}. The error bars are the $1\sigma$ measurement errors on our measured magnetic field strengths. The unbroken red line shows a one-to-one relationship, while the dashed lines represent $\pm3\sigma$ offsets from the red line.}
    \label{Figure5}
\end{figure}


\section{Discussion}\label{section4}

\subsection{Association with other maser species} \label{association_section}

Here  we  discuss  the  association  of  the  1720-MHz  OH  star formation masers with main line ground-state OH, excited-state OH, class  I/II  methanol  and  water  masers. We have searched for OH main line emission from the MAGMO dataset only in the same fields as the 1720-MHz sources solely for ascertaining associations, and leave detailed discussion of these sources for a future publication. Associations between masers are determined first by proximity, using initial thresholds dependent on typical relative distances of the different transitions from their exciting sources; and secondly by utilising the visual aid of maser positions overlaid on background infrared images of their associated star formation sites. For the first step: Given that the pumping mechanism of both the 6.7-GHz and 12-GHz methanol masers is dependent on a nearby radiating source \citep{Sobolev1997,Cragg2005}, we expect to find them close to the peak of their exciting infrared source. We note that while precise positions for the 6.7-GHz methanol masers are available over much of the southern Galaxy \citep[e.g.][]{Caswell2010,Green2010,Caswell2011,Green2012MMB,Breen2015}, the 12-GHz follow-up work makes the reasonable assumption that they will arise from the same location as the 6.7-GHz sources \citep[e.g.][]{Breen2012a,Breen2012b,Breen2014}. Likewise, since the OH main lines \citep{Litvak1969,Elitzur1978} and 6030-/6035-MHz excited OH masers \citep{Cesaroni1991,Gray92} are also radiatively pumped, we also expect to find them close to their exciting infrared source. On the other hand, water \citep{Elitzur89}, class I methanol \citep{Cragg1992,Voronkov2005}, and 1720-MHz (while occasionally radiatively pumped) have collisonal pumping mechanisms, and are often found in outflows from their exciting sources. Taking these factors into account, we adopted initial association thresholds for the different transitions, which are discussed in detail in the following subsections. For the second step: We compiled RGB-colour GLIMPSE \citep[Galactic Legacy Infrared Mid-plane Survey Extraordinaire;][]{Benjamin2003,Churchwell2009} images for each of the SFR-maser regions, to visually aid our check for association between maser species, and also with their hosts. Each composite RGB colour image was created by combining 8.0 \micron, 4.5 \micron, and 3.6 \micron\ images, which were then overlaid with markers representing the maser positions obtained in this work and from the literature. See Figure \ref{appendix} for the images. Shown in Table \ref{Table7} is the list of associations of 1720-MHz with other transitions and maser species.

\subsubsection{Association with ground-state OH masers}

With the same procedure and detection limit used ($\sim$40 mJy at 1$\sigma$) for the 1720-MHz transition (described in Section \ref{section2}), we reduced the MAGMO data and searched for 1612/1665/1667-MHz emission in the same fields as the 1720-MHz sources, solely for ascertaining associations (see Table \ref{Table7}). From the current study, we only detect 1612-MHz emission near 351.774$-$0.537, 323.459$-$0.079 and 008.669$-$0.356 -- and in no other field hosting 1720-MHz masers. Utilising 1 arcsec offset as the initial association threshold, we find that 16 out of the 33 1720-MHz SFR masers were less than 1 arcsec offset from the nearest 1665- or 1667-MHz OH maser. With extra information provided by the GLIMPSE images, we are also able to identify instances where the 1720-MHz sources are either clearly associated with outflows or appear to be associated but slightly offset from the exciting source. Including these sources, we find that a total of 25 of the 1720-MHz SFR population are associated with the 1665-MHz maser transition, of which 17 of those are cases where the 1665- and 1667-MHz are both present. None of the 1720-MHz sources showed only a 1667-MHz detection. Detailed analysis and discussion of the detected main line sources is left for a future publication.

\subsubsection{Association with Methanol Masers} \label{1720-MMB}

\textbf{Class II:} As we have searched for 1720-MHz SFR masers towards 702 6.7-GHz class II methanol maser sources, detected in the MMB maser survey \citep{Caswell2010,Green2010,Caswell2011,Green2012MMB,Breen2015}, we can examine their association statistics. The positions of the 6.7-GHz methanol masers are generally known to within 0.4 arcsec (like the 1720-MHz OH masers presented here) so we use an initial association threshold of 1 arcsec and further consider cases on an individual basis outside this range. 
Based on the 1 arcsec threshold, out of the 33 1720-MHz SFR maser sources identified in this work, 19 have angular offsets less than or equal to one arcsec from the associated methanol maser. With the aid of the GLIMPSE images we also associate 017.638$+$0.158, 328.165+0.587, 330.953$-$0.180, 351.158$+$0.69, 351.774$-$0.537 and 359.970$-$0.457 with methanol masers due to their apparent association with the same exciting object. This leaves eight sources with 1720-MHz OH masers (189.032+0.809, 189.776+0.346, 345.117$+$1.592, 345.495$+$1.462, 345.497$+$1.461, 348.727-1.039, 350.112$+$0.09, 010.474$+$0.027) but devoid of methanol maser counterparts. A total of 25 of 702 6.7-GHz methanol maser targets therefore have associated 1720-MHz emission, resulting in an overall detection statistic of $\sim$4\%. 

For the 12.2-GHz class II methanol maser associations, we utilise a targeted search conducted with the Parkes radio telescope towards the same 6.7-GHz methanol masers detected by the MMB in the Galactic longitude 186$^{\circ}$ through the Galactic centre to 20$^{\circ}$ as in this work. The 12.2-GHz search detected 315 masers within this range of Galactic longitudes \citep{Breen2012a,Breen2012b,Breen2014}. As stated in Section~\ref{association_section}, this work does not derive precise positions, rather making the justifiable assumption that the 12.2-GHz methanol maser emission arises from the same location at the target 6.7-GHz methanol masers \citep[e.g.][and references therein]{Breen2012a,Breen2012b,Breen2014}. We therefore adopt the same methodology used for the 6.7-GHz methanol masers to determine associations and find that 15 out the 33 1720-MHz SFR masers have associated 12.2-GHz methanol maser counterparts.\\

\textbf{Class I:}
\citet{Voronkov2014} made targeted observations of class I methanol masers towards 6.7 GHz methanol masers, and reported that more than 70\% of the class I sources detected in that sample are within 30 arcsec of 6.7-GHz masers. Given this large association radius, and since both class I and 1720 masers may be offset from a central exciting source, we examine each instance as the case presents, seeking to classify as ``associated'' masers that share the same exciting source. High spatial resolution observations of class I are generally limited to targeted searches, and although \citet{Jordan2015} was an unbiased survey for 44-GHz methanol masers, it has a modest detection limit (1$\sigma$ of 0.9~Jy) and only covered longitudes between 330$^\circ$ and 335$^\circ$ and latitudes of $\pm$0.5$^{\circ}$. Although 36- and 44-GHz observations are not available for our full sample of 1720-MHz OH masers, we find eight associations between our 1720-MHz OH masers and class I methanol masers \citep[at 36- and 44-GHz;][]{Voronkov2014,Jordan2015,Jordan2017,Ellingsen2018}. We note that one additional previously-detected 1720-MHz source 338.925+0.557, which was not detected in this work (see Table \ref{Table1}) was previously found to share the same exciting source as both 36- and 44-GHz class I methanol maser emission \citet{Voronkov2014}.

The overall lack of class I methanol maser survey coverage restricts further attempts to quantify the association between 1720-MHz and class I methanol masers, but we note a tendency for these masers to be associated with 1720-MHz OH emission. This is consistent with their shared underlying pumping mechanism, which depends of the presence of shocks.

\subsubsection{Association with excited-state OH masers}\label{4.1.2}

Alongside the 6.7-GHz methanol maser observations in the MMB survey, a concurrent complete survey for 6035-MHz excited-state masers was conducted \citep[1-$\sigma$ sensitivity of 0.17~Jy; e.g.][]{Green2009}. Follow-up observations of the 6035-MHz detections (both with Parkes and the ATCA) often also revealed 6030-MHz detections, both of which are presented in \citet{Avison2016}. Since the 6035-MHz observations were simultaneously observed with the 6.7-GHz, the observations have the same Galactic coverage and can be directly compared with our 1720-MHz OH maser detections. We supplement this large complete survey for 6035-MHz OH masers with targeted searches for 6030- and 6035-MHz OH maser emission available in the literature \citep{Caswell1995,Caswell2001,Caswell03}.

As with the ground state main line transitions, the positional uncertainty of the 6035- and 6030-MHz excited-stated OH masers is 0.4 arcsec \citep[See][for position references]{Avison2016}. Using the same methodology as for the main lines, we find that 19 out of the 99 6035-MHz excited-state OH masers in the longitude range of our present survey have an associated 1720-MHz OH maser (11 of which have an angular separation of less than 1 arcsec). In addition, five of the 19 sources also have associated 6030-MHz emission. Previous targeted observations of 1720- and 6035-MHz OH masers have shown a moderate association rate between the two lines -- \citet{Macleod97} found 18 6035-MHz masers towards 29 1720-MHz targets (62\%), where as \citet{Caswell2004} found 24 1720-MHz OH maser detections towards 72 6035-MHz OH masers (33\%). In the current observations we find that $\sim$58\% of the 1720-MHz detections have associated 6035-MHz OH maser emission.

\subsubsection{Association with water masers}\label{water}

There are 323 MMB 6.7-GHz methanol masers from Galactic longitude 341$^{\circ}$ through the Galactic centre to 20$^{\circ}$ that were targeted both in the current 1720-MHz study and the 22-GHz maser survey of \citet{titm14,titm16}. With an association criteria of $\leq$ 3-arcsec angular separation between 6.7-GHz methanol and water masers, \citet{titm14,titm16} found that 156 out of 323 6.7-GHz methanol masers had associated water maser emission. We utilize the 3 arcsec as an initial threshold for association between 1720-MHz and water masers, and then further inspected the GLIMPSE images for cases beyond 3 arcsec radius. We find that 15 out of the 20 1720-MHz OH masers within this longitude range have associated water maser emission. 
We have supplemented the \citet{titm14} and \citet{titm16} targeted observations with other water maser data reported in the literature \citep[][]{Palagi93,Walsh2014,Breen2010water} and find that overall 21 of our 1720-MHz maser detections are associated with known water maser sources. Towards 189.776+0.346, \citet{Palagi93} detected a further water maser in single dish observations, but with their 1.9 arcmin beam we are unable to confirm an association.

\subsubsection{1720-MHz OH masers devoid of 6.7-GHz methanol maser counterparts} \label{newclass}

From their search, \citet{Caswell2004} reported eight 1720-MHz OH star-formation masers that were either devoid of, or found at positions offset from, 1665-MHz OH and 6.7-GHz methanol maser targets. These 1720-MHz OH masers had broad linewidths ($\sim$1 km\,s$^{-1}$), approximately twice the FWHM of most of the 1720-MHz SFR masers in the \citet{Caswell2004} sample. \citet{Caswell2004} investigated the counterparts of these masers and proposed that they represented a distinct sub-class of SFR 1720-MHz OH masers. Adopting the \citet{Caswell2004} definition, we find that 12 of our 1720-MHz SFR masers (see Table \ref{Table6}) would be classified as belonging to this sub-class. Four of these masers were previously identified as belonging to the sub-class (345.117$+$1.592, 345.495$+$1.461, 345.497$+$1.461, 17.638$+$0.158) and the remaining sources that we find were either not observed or detected (six sources; 189.032$+$0.809, 189.776$+$0.346, 328.165$+$0.586, 348.727$-$1.039, 350.112$+$0.095, 351.158$+$0.699) or detected  but not considered to be part of the subclass (two sources; 330.953$-$0.180, 10.474+0.027), because precise positions of the respective masers were unknown at the time. We note that three of the initial eight masers that \citet{Caswell2004} classified as belonging to this subclass (290.375$+$1.666, 310.146$+$0.760 and 329.426$-$0.158) are undetected in the present work.

Using GLIMPSE data and the literature, we have been able to investigate the local environments of the 12 sub-class members we identified. Many of these are offset in position from the centre of an exciting source (and from other radiatively pumped maser species) making it possible that they are located within the interaction zone of shocks with the surrounding environment. We find that four masers show strong evidence of association with outflow activity: 345.495$+$1.462, 345.497$+$1.461 and 351.158$+$0.699 show clear morphological signatures of outflows in the GLIMPSE images, with the 1720-MHz masers associated with the edges of the outflow emission; 17.638+0.158 is similarly associated with the edge of a bipolar outflow mapped in $^{13}$CO with ALMA \citep{Maud2018}. An additional six sources (189.776$+$0.346, 328.165$+$0.586, 330.953$-$0.180, 345.117$+$1.592, 348.727$-$1.039 and 350.112$+$0.095) are associated with prominent (and often extended) 4.5\,$\micron$ emission, which commonly indicates the presence of outflows \citep[e.g.][]{Cyganowski2008,Simpson2012}. Of these, three (328.165$+$0.586, 330.953$-$0.180 and 348.727$-$1.039) are associated with 4.5\,\micron\ sources designated as ``possible'' outflow candidates by \citet{Cyganowski2008}. 
The remaining two sources (189.032$+$0.809, and 10.474$+$0.027) both appear to be truly isolated 1720-MHz masers, positioned close to a central exciting source and with no other associated masers. 

With the benefit of the GLIMPSE data and subsequent high-resolution maser surveys we have therefore gleaned that the majority of the members of the \citet{Caswell2004} 1720-MHz sub-class are offset from the SFR that is exciting the maser emission, explaining why they are devoid of masers that are radiatively excited (class II methanol or main line OH). Our associations listed in Table~\ref{Table7} capture the fact that often radiatively excited masers are still associated with the same exciting source as a 1720-MHz OH maser, and so while their specific location is devoid of the accompanying emission, the source is not. We therefore suggest that the bulk of sources that satisfy the subclass requirements defined by \citet{Caswell2004} are associated with the interaction zone of outflows (or other shocks) emanating from a young high-mass star. These masers tend to have linewidths similar to SNR-associated 1720-MHz masers -- the ten 1720-MHz masers we find to be offset from their exciting sources have FWHM between 0.73--1.65\,\kms\ (with a median of 0.87\,\kms).

Conversely, the two masers that appear closely associated with their exciting objects (189.032$+$0.809 and 10.474$+$0.027) have FWHM of 0.23 and 0.26\,\kms. These are truely devoid of class II methanol and main-line OH maser emission and possibly indicate a specific evolutionary phase, as \citet{Caswell2004} initially hypothesised.

\begin{table}
\fontsize{7}{7}\selectfont
	\centering
	\caption{1720-MHz SFR-masers that are not co-located with either 1665-MHz main line OH or 6.7-GHz methanol maser emission \citet[proposed as a sub-class of 1720-MHz OH masers by][]{Caswell2004} found in \citet{Caswell2004} (left column) and this work (right column).	Sources listed in italics in the left-hand column were detected by \citet{Caswell2004}, but not initially recognised as a member of this sub-class. The source with $\ast$ was included in this category by \citet{Caswell2004}, but has been removed since the MMB detected an associated 6.7-GHz methanol maser. The superscripts refer to the epochs at which the sources were observed: $^{1}$January 1985, $^{2}$November 1997, $^{3}$November 2000, $^{4}$August 2001, $^{5}$December 2001, $^{6}$April 2002, $^{7}$September 2002, $^{8}$March 2003. See Table \ref{Table3} for dates of observation of sources listed in the right column.}
	\label{Table6}
	\begin{tabular}{cc} 
	\hline
		\multicolumn{1}{c}{Source} & \multicolumn{1}{c}{Source} \\
		 \citep{Caswell2004} &  \multicolumn{1}{c}{This study} \\
		\hline
        -       & 189.032+0.809 (1.3\,Jy)  \\
        -       & 189.776+0.346 (1.5\,Jy)   \\
        -       & 328.165+0.586 (0.45\,Jy)   \\
      290.375+1.666 (0.6$^{5}$\,Jy) &       -         \\
	  310.146+0.760 (1.2$^{6}$\,Jy) &       -         \\
	  329.426$-$0.158 (2.5$^{3}$, 3.5$^{4}$\,Jy) &     -        \\
	  \textit{330.953-0.182} (0.8$^{4}$\,Jy) & 330.953$-$0.180 (0.7\,Jy) \\
	  $\ast$345.003$-$0.224 (20$^{1}$\,Jy) &    -  \\
	  345.118+1.592 (6.2$^{6}$\,Jy) &   345.117+1.592 (8.1\,Jy)      \\
	  345.495+1.462 (2.5$^{4}$\,Jy) &   345.495+1.462 (0.7\,Jy)  \\
      345.497+1.462 (4.9$^{4}$\,Jy) &   345.497+1.461 (1.6\,Jy)  \\ 
      -                         & 348.727$-$1.039 (1.1\,Jy) \\
      -                         & 350.112+0.095 (0.7\,Jy) \\
      -                         & 351.158+0.699 (8.8\,Jy) \\
      \textit{010.473+0.027} (0.4$^{2}$\,Jy) &  010.474+0.027 (1.6\,Jy) \\
      17.639+0.158 (<0.15$^{4}$, 0.3$^{7}$, 0.8$^{8}$ \,Jy) &17.638+0.158 (0.5\,Jy) \\

		\hline
	\end{tabular}
\end{table}

\begin{table*}
\fontsize{9}{7}\selectfont
\centering
\caption{Table of association between 1720-MHz OH and other maser transitions. Detections of 1612-, 1665- and 1667-MHz OH maser transitions listed in columns 2, 3 and 4 are obtained from the full MAGMO dataset and \citet{Caswell98,Caswell99}. The 6035/6030-MHz transitions from \citet{Avison2016,Caswell03,Caswell95} are given in columns 5 and 6. Water maser entries (including the non-detections) in column 7 are from \citet{Breen2010water} except for those with superscripts $\it{p}$, $\it{T}$, $\it{t}$, and $\it{w}$ which are from \citet{Palagi93}, \citet{Titmarsh2013}, \citet{titm14,titm16} and \citet{Walsh2014}. The 6.7-GHz methanol masers in column 8 are all from the MMB survey, while the 12-GHz methanol listed in column 9 are taken from \citet{Breen2012a,Breen2012b,Breen2014}. The last column has the 36/44-GHz methanol maser transitions and are taken from \citet{Voronkov2014} except for the ones with superscripts $\it{j}$ and $\it{e}$ which are taken from \citet{Jordan2017} and \citet{Ellingsen2018}. Non-detections are denoted by their detection limits in Janskys. The \textit{italicised} sources are those that we do not consider associated with the nearest identified 1720-MHz OH maser, but which are found within the region captured in the GLIMPSE images Figure~\ref{appendix}.}  
\label{Table7}
\renewcommand{\arraystretch}{1.5}
\begin{adjustbox}{width=\textwidth}
\begin{tabular}{ lcccccccccc  }

\hline
\hline

\multicolumn{6}{c}{OH} &
\multicolumn{1}{c}{Water}  & 
\multicolumn{3}{c}{Methanol} \\

\cline{1-6}
\cline{8-10}

\multicolumn{1}{c}{} & 
\multicolumn{1}{c}{} &
\multicolumn{1}{c}{} & 
\multicolumn{1}{c}{} &  
\multicolumn{1}{c}{} &
\multicolumn{1}{c}{} &
\multicolumn{1}{c}{} &
\multicolumn{1}{c}{} &
\multicolumn{1}{c}{} & 
\multicolumn{1}{c}{} \\

\multicolumn{1}{l}{1720-MHz} & 
\multicolumn{1}{l}{1612-MHz} &
\multicolumn{1}{l}{1665-MHz} &
\multicolumn{1}{l}{1667-MHz} &
\multicolumn{1}{c}{6035-MHz} & 
\multicolumn{1}{c}{6030-MHz} &
\multicolumn{1}{c}{22-GHz} &
\multicolumn{1}{c}{6.7-GHz} &
\multicolumn{1}{c}{12-GHz} & 
\multicolumn{1}{c}{36/44-GHz}\\

\multicolumn{1}{c}{} &
\multicolumn{1}{c}{} &
\multicolumn{1}{c}{} &
\multicolumn{1}{c}{} & 
\multicolumn{1}{c}{3$\sigma$ (Jy)} & 
\multicolumn{1}{c}{3$\sigma$ (Jy)} &
\multicolumn{1}{c}{3$\sigma$ (Jy)} &
\multicolumn{1}{c}{3$\sigma$ (Jy)} &
\multicolumn{1}{c}{5$\sigma$ (Jy)} & 
\multicolumn{1}{c}{3$\sigma$ (Jy)} \\

\hline

189.032+0.809 & <0.2 & <0.2 & <0.2 &  <0.5 & <0.5 & - & <0.5 & <0.5 &  - \\

189.776+0.346 & <0.2 & <0.2 & <0.2 & <0.5 & <0.5 & \it{189.778+0.347} $^{p}$ & \it{189.778+0.345} & <0.5 & - \\

306.322$-$0.334 & <0.2 &  306.322$-$0.334 & <0.2 & <0.5 & <0.5 & <0.15  & 306.322$-$0.334 & <0.8 & - \\

323.459$-$0.079 & 323.459$-$0.079 &  323.459$-$0.079 & 323.459$-$0.079 & 323.459$-$0.079 & 323.459$-$0.079 & <0.15 & 323.459$-$0.079 & 323.459$-$0.079 & - \\

328.165+0.586 & <0.2 &  <0.2  & <0.2 & <0.5 & <0.5 & <0.15 & 328.164+0.587 & <0.5 &  \\

328.808+0.633 & <0.2 &  328.808+0.633 & 328.808+0.633 & 328.808+0.633 & <0.5 & 328.808+0.633 & 328.808+0.633 & 328.808+0.633 & 328.81+0.63 \\

329.339+0.148 & <0.2 &  329.339+0.148 & <0.2 & 329.339+0.148 & <0.5 & <0.15 & 329.339+0.148 & 329.339+0.148 & - \\

330.953$-$0.180 & <0.2 &  330.953$-$0.182 & 330.953$-$0.182 & 330.953$-$0.182 & <0.5 & 330.954$-$0.182 & 330.953$-$0.182 & <0.5 & 330.95$-$0.18 $^{j}$ \\

336.941$-$0.156 & <0.2 &  336.941$-$0.156 & 336.941$-$0.156 & 336.941$-$0.156 & 336.941$-$0.156 & <0.15 & 336.941$-$0.156 & 336.941$-$0.156 & - \\

337.612$-$0.060 & <0.2 &  337.612$-$0.060 & <0.2 & 337.606$-$0.052 & <0.5 & 337.612$-$0.060 & 337.613$-$0.060 & 337.613$-$0.060 & - \\

339.622$-$0.120 & <0.2 &  339.622$-$0.120 & 339.622$-$0.120 & 339.622$-$0.120 & <0.5 & 339.622$-$0.120 & 339.622$-$0.120 & 339.622$-$0.120 & - \\

339.884$-$1.259 & <0.2 &  339.884$-$1.259 & 339.884$-$1.259 & 339.884$-$1.259 & 339.884$-$1.259 & 339.884$-$1.259 & 339.884$-$1.259 & 339.884$-$1.259 & 339.88$-$1.26 \\

340.785$-$0.096 & <0.2 &  340.785$-$0.096 & 340.785$-$0.096 & 340.785$-$0.096 & <0.5 & 340.785$-$0.096 & 340.785$-$0.096 & 340.785$-$0.096 & 340.785$-$0.096$^{e}$ \\

344.582$-$0.023 & <0.2 &  344.581$-$0.023 & 344.581$-$0.023 & <0.5 & <0.5 & 344.582$-$0.024 & 344.581$-$0.024 & <0.7 & - \\

345.003$-$0.224 & <0.2 & 345.003$-$0.224  & <0.2 & 345.003$-$0.224 & <0.5 & 345.004$-$0.224 & 345.003$-$0.224 &  345.003$-$0.224 & 345.00$-$0.22 \\

345.117+1.592 & <0.2 & <0.2  & <0.2 & <0.5 & <0.5 & <0.15 & <0.5 & <0.7 & - \\

345.495+1.462 & <0.2 &  345.494+1.469 & 345.498+1.467 & 345.495+1.469 & <0.5 & 345.495+1.473 & <0.5 & <0.7 & - \\ 
 
345.497+1.461 & <0.2 &  345.494+1.469 & 345.498+1.467 & 345.495+1.469 & <0.5 & 345.495+1.473 & <0.5 & <0.7 & - \\

348.727$-$1.039 & <0.2 &  \it{348.727$-$1.037} & <0.2 & <0.5 & <0.5 & \it{348.726$-$1.038} & \it{348.727$-$1.037} & \it{348.727$-$1.037} & - \\

350.112+0.095 & <0.2 & 350.113+0.095 & 350.113+0.095 & 350.113+0.095 & <0.5 & 350.113+0.095 & <0.5 & <0.7 & - \\

350.686$-$0.491 & <0.2 &  350.686$-$0.491 & 350.686$-$0.491 & 350.686$-$0.491 & <0.5 & 350.686$-$0.491 & 350.686$-$0.491 & 350.686$-$0.491 & - \\

351.158+0.699 & <0.2 & 351.160+0.697  & 351.160+0.697 & <0.5 & <0.5 & 351.160+0.696 & 351.161+0.697 & <0.7 & 351.16+0.70 \\

351.419+0.646 & <0.2 &  351.417+0.645 & 351.417+0.645 & 351.417+0.645 & 351.417+0.645 & 351.417+0.646 & 351.417+0.646 & 351.417+0.645 & 351.42+0.65 \\

351.774$-$0.537 & 351.774$-$0.536 &  351.774$-$0.536 & 351.774$-$0.536 & 351.775$-$0.536 & <0.5 & 351.775$-$0.536 & 351.775$-$0.536 & 351.775$-$0.536 & 351.77$-$0.54 \\

353.410$-$0.360 & <0.2 &  353.410$-$0.360 & <0.2 & 353.410$-$0.360 & 353.410$-$0.360 & \textit{353.411$-$0.362} & 353.410$-$0.360 & 353.410$-$0.360 & - \\  

357.557$-$0.321 & <0.2 &  357.557$-$0.321 & 357.558$-$0.321 & <0.5 & <0.5 & 357.558$-$0.321 $^{t,w}$  & 357.558$-$0.321 & <0.7 & - \\

359.970$-$0.457 & <0.2 &  359.970$-$0.457 & <0.2 & <0.5 & <0.5 & 359.969$-$0.457  & 359.970$-$0.457 & <0.7 & - \\

000.376+0.040 & <0.2 &  000.376+0.040 & 000.376+0.040 & <0.5 & <0.5 & 000.376+0.040 & 000.376+0.040 & <0.7 & - \\

000.665$-$0.036 & <0.2 & \textit{000.666$-$0.036}  & \textit{000.666$-$0.035} & \textit{000.666$-$0.035} & <0.5 & \textit{000.668$-$0.035} & 0.665$-$0.036 & \textit{0.667$-$0.034} & - \\

008.669$-$0.356 & 008.669$-$0.355 & 008.669$-$0.356  & <0.2 & 008.669$-$0.356 & <0.5 & 008.670$-$0.356 & 008.669$-$0.356 & 008.669$-$0.356 & - \\

010.474+0.027 & <0.2 &  \it{010.472+0.027}& \it{010.473+0.027} & <0.5 & <0.5 & \it{010.473+0.027   $^{T}$ } & \it{010.472+0.027} & \it{010.472+0.027} & - \\

011.034+0.062 & <0.2 &  011.034+0.062 & <0.2 & 011.034+0.062 & <0.5 & 011.034+0.062  & 011.034+0.062 & <0.7 & - \\

017.638+0.158 & <0.2 & 017.638+0.157 & 017.638+0.157 & <0.5 & <0.5 & 017.638+0.156 & 017.638+0.157 & <0.7 & - \\

\hline
\hline

\end{tabular}
\end{adjustbox}
\end{table*}

\subsection{Polarimetry of 1720-MHz star formation masers}

\begin{figure}

	\includegraphics[width=\columnwidth]{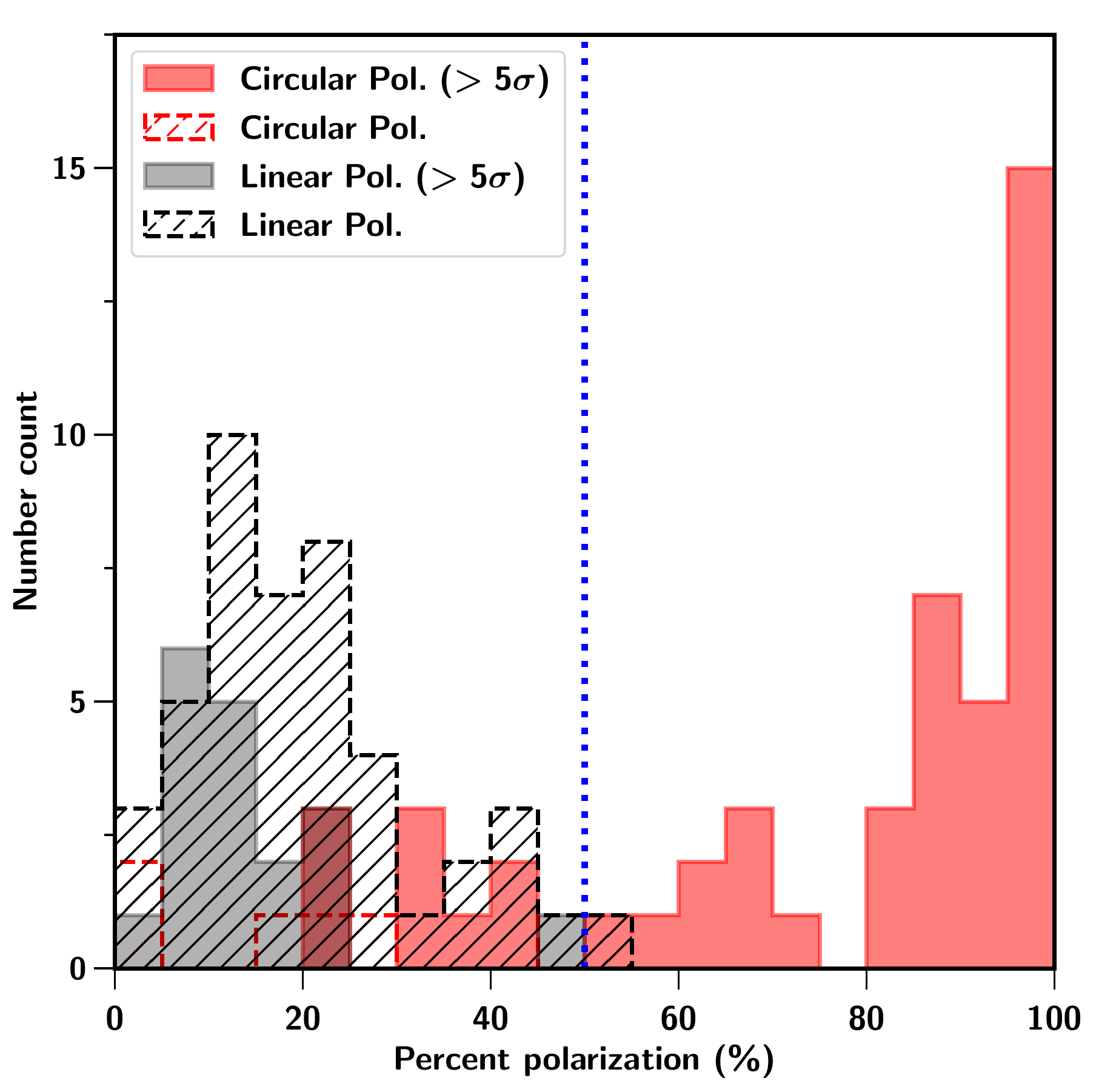}
    \caption{A histogram showing the distribution of percent fractional circular (red) and linear (black) polarisations of SFR 1720-MHz OH maser components, divided by those that are statistically significant (solid colour) and those that are not (hatched colour). The blue dashed line represents the 50\% polarisation point. Note that the majority of the linear measurements are less than 5$\sigma$, but are included here to demonstrate the tendency of the 1720-MHz masers to show linear polarisation percentages less than 30\% (and for those that are statistically significant to mostly fall below 20\%).}
    \label{Figure6}
\end{figure}

\subsubsection{Polarisation properties}
Based on our sample of star formation 1720-MHz maser components (Table \ref{Table4}),  Figure \ref{Figure6} shows a histogram of the degree of both circular and linear polarisation for the 1720-MHz OH masers, divided by statistical significance. For the linear polarisation, the plot clearly indicates the majority of star formation masers have linear polarisation fractions less than 20\%.
For the reliably measured circular polarisation fraction (i.e. greater than 5$\sigma$), f84\% have circular polarisation fractions greater than 50\%; with a number of components consistent with 100\% circular polarisation (in sources 306.322$-$0.344, 329.339+0.148, 336.941$-$0.156, 339.622$-$0.120, 339.884$-$1.259, 340.785$-$0.096, 351.774$-$0.537, and 357.557$-$0.321). 

The predominance of strongly circularly polarised features, and comparatively fewer (and weaker) linearly polarised features, exhibited by these 1720-MHz SFR masers demonstrate consistency with maser propagation theories, e.g. \citet{Goldreich1973,Gray92,Gray1994,Gray1995,Elitzur1996,Elitzur98,Watson2001}. According to these maser theories, of the three components produced in Zeeman splitting, the strong circular polarisation components are most clearly evident when the maser propagation is parallel to the magnetic field, and conversely the strong linear polarisation component is evident when the field is orthogonal. The fact that we see mainly circular components, with only occasional linears (which is similar to higher frequency OH transitions of \citealt{green2015ex}), indicate that, according to \citet{Gray1994}, the masers are preferentially beamed when aligned with the magnetic field vector.

\subsubsection{Magnetic field properties}
Our reliably measured (greater than 3$\sigma$) magnetic field strengths in star formation masers range between 1.4$-$13.2 mG (listed in Table \ref{Table3}), and have mean and median values of 6.2 and 6.2 mG. Excluding sources with magnetic field strength measurements below 3$\sigma$,  
we find (using the IEEE convention) that 45\% of the SFR maser Zeeman pairs have their field direction pointing away from us, while the other 55\% have their field direction oriented towards us. Excluding sources with unreliable magnetic field measurements, there are eight sources in our sample with more than a Zeeman pair. From these we find that the absolute field strength dispersion within any one source ranges from 0.2$-$3.5 mG. We further note that the two highest dispersions are from the two sources (340.785-0.096, 351.774-0.537) which exhibit field reversals (as already stated in the source notes).

\begin{figure}

	\includegraphics[width=\columnwidth]{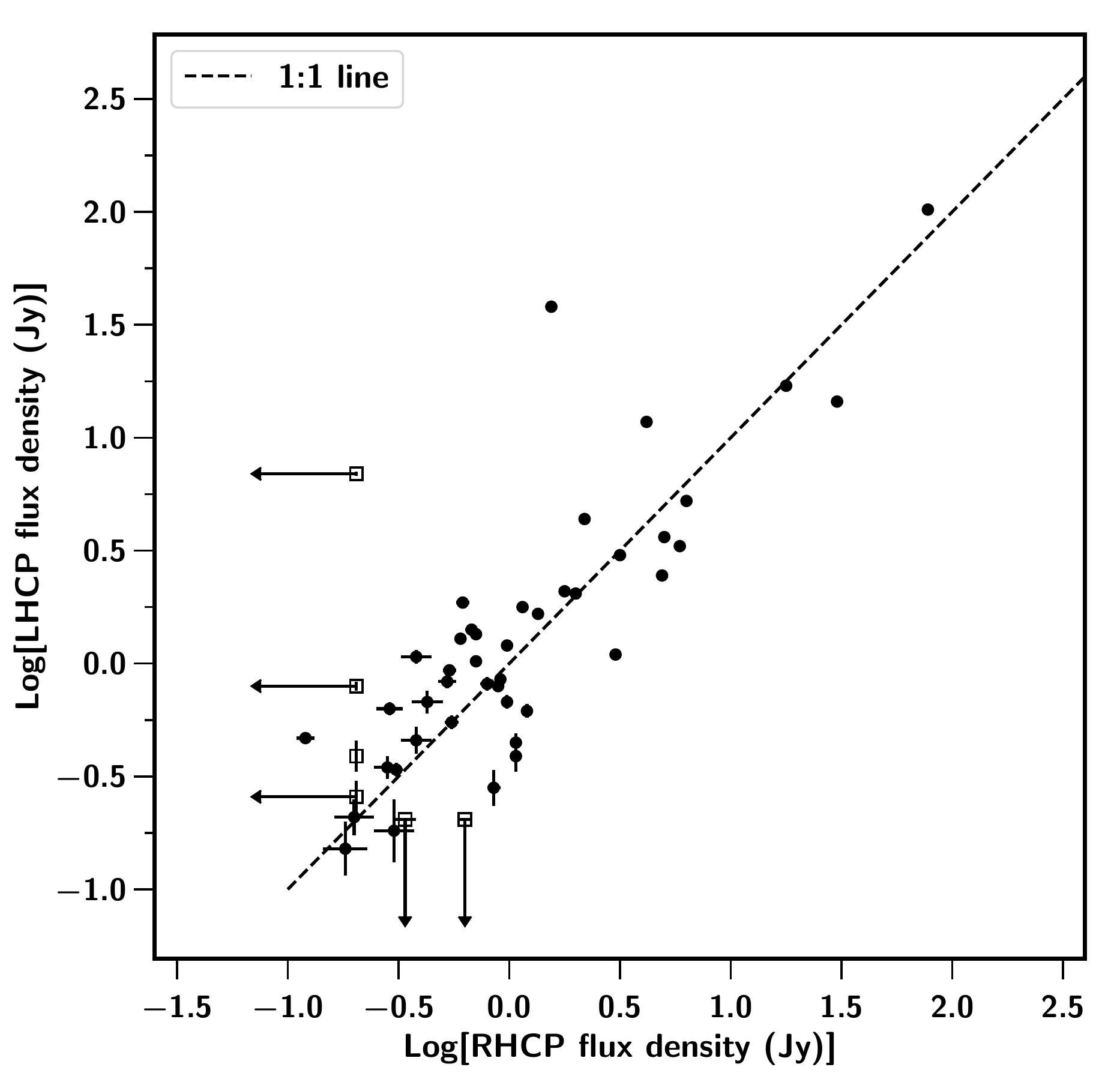}
    \caption{Comparison of RHCP and LHCP flux densities for the 1720-MHz SFR OH maser Zeeman pairs represented by the filled circles. The squares represent components with only a RHCP or LHCP profile, and in such cases detection upper limit have been used for either the RHCP or LHCP when undetected. The two hands of polarisation, representing the $\sigma ^{+}$ \&  $\sigma ^{-}$ components of Zeeman splitting, are expected to have similar fluxes for strongly circularly polarised masers.}
    \label{Figure7}
\end{figure}

We further compared the RHCP and LHCP flux densities of each Zeeman pair. This is represented in the scatter plot of Figure \ref{Figure7}.
The model of \citet{Gray1995} predicts that strong circularly polarised Zeeman pairs of 1720-MHz OH masers should on the whole behave similarly, implying generally equal flux densities for RHCP and LHCP line profiles. The overall linearity seen in Figure \ref{Figure7} broadly supports the expected similar flux density behavior between components of 1720-MHz OH masers, noting that we see 13 of the 44 Zeeman pairs have components differing by a factor greater than two.

\subsubsection{Excited-state OH Magnetic field comparison}
We collated the associated 6035-MHz OH maser Zeeman pairs from \citet{Caswell1995}, and compared the field strength measurements. We find 15 sources in \citet{Caswell1995} which are associated with 1720-MHz OH masers from the current study, but only 11 of which had listed magnetic field strengths. Of these 11, nine have the same field orientation as the 1720-MHz OH masers. The two that do not are 330.953--0.180 and 351.774--0.537, with the former showing --2.5 mG at 6035-MHz and +1.6 mG at 1720-MHz, and the latter showing --3.3 mG at 6035-MHz and both +2.7 mG and --6.2 mG at the 1720-MHz.
The nine that show the same orientation have field strengths that vary by up to 2 mG.

\subsection{Distribution of SFR 1720-MHz OH masers and magnetic fields}

\begin{figure*}
    \includegraphics[width=\textwidth]{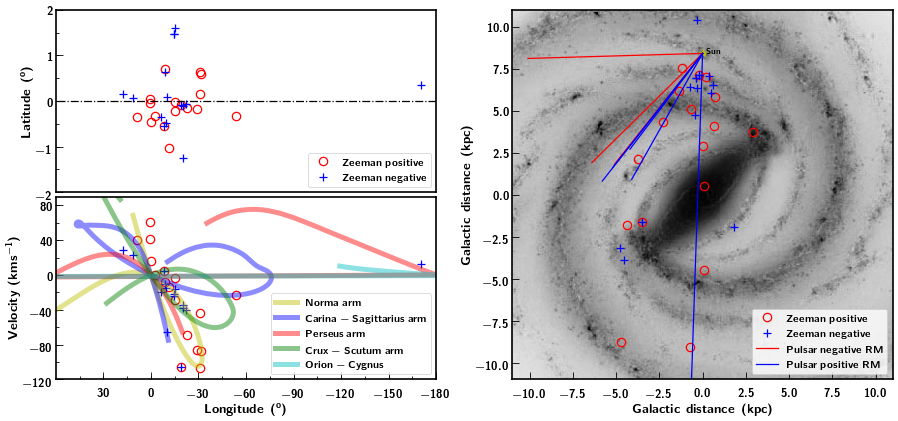}
    \caption{\textbf{Left top panel:} Distribution of SFR 1720-MHz OH masers from the current study in Galactic longitude and latitude.  As in all panels, red circles denote measured Zeeman B-fields  directed away from us and blue crosses denote measured Zeeman B-fields directed towards us. The broken black line runs through the Galactic plane at latitude 0$^{\circ}$. \textbf{Left bottom panel:} Plot of longitude-velocity (LSR) distribution of the masers overlaid with coloured loci of the Galactic spiral arms, using the rotation curve model of \citet{Brand93} and the spiral arm model of \citet{Taylor1993} (a modification of \citealt{Georgelin1976}). \textbf{Right panel:} 
    Positions indicating field direction from Zeeman measurements of the masers overlaid on an artist's impression of the Milky Way (R. Hurt: NASA/JPL-Caltech/SSC.). Red lines represent the magnetic field direction indicated by pulsar Rotation Measure ($-$RM) observations directed away from the sun, while the blue lines represent pulsar $+$RM fields directed towards the sun from \citet{Nota2010}. The length of the line shows  the distance to the pulsar from the sun, thus representing the entire line of sight which contributes to the RM.}
    \label{Figure8}
\end{figure*}

\subsubsection{Galactic distribution}   
The top left panel of Figure \ref{Figure8} shows the spatial distribution ($l$-$b$) of SFR 1720-MHz OH maser detections from this work, while the bottom left panel shows the longitude-velocity ($l$-$v$) distribution of the same OH masers. As expected they are clustered about the Galactic mid-plane, within the $|b| < 2^{\circ}$ latitudinal range of the survey limits. By Galactic quadrant, we find: 6 sources in the first quadrant ( 3.6\%; from 168 targeted 6.7-GHz methanol masers); two sources in the third quadrant (9.5\%; from 21 targeted 6.7-GHz methanol masers); and as expected the largest population of 25 sources fall in the 4th quadrant (4.9\%; from 513 targeted 6.7-GHz methanol maser sources).
Considering the magnetic field direction inferred from individual Zeeman pairs by Galactic quadrants shows: quadrant 1 has three field measurement directed away from us and three directed towards us; quadrant 3 has 2 Zeeman pairs with one magnetic field directed away and the other undetermined as it shows no Zeeman splitting; and in quadrant 4 there are 15 Zeeman pairs with the field directed away from us and 19 towards. The $l$-$b$ and $l$-$v$ plots together reflect Galactic structure, for example we see that the Norma arm tangent indicates a field towards us, and potentially that the Crux-Scutum arm has a field away from us (but in this case more sources and clearer association will be required).

\subsubsection{Field direction comparison with large-scale fields determined by Faraday rotation measures.}

Past studies have indicated the coherence of the magnetic field orientation across maser sources/molecular clouds of the order of a few kpcs in scale \citep{Davies1974,reidsilver90,Fish05,Bartkiewicz05}. This was seen in the MAGMO pilot sample between Galactic longitudes 280$^{\circ}$ and 295$^{\circ}$ (the Carina-Sagittarius spiral arm tangent; \citealt{Green2012pilot}). Past studies have also shown that individual regions of star formation may exhibit apparent field reversals within them \citep[e.g.][]{Fish03b}, which require VLBI resolution to interpret, for example the studies of 351.417+0.645 by \citet{Chanapote2019}.

It was suggested by \citet{Fish03b} that three major conditions were required for the Zeeman splitting of OH masers to be reliably used as a tracer of the Galactic-scale magnetic field: firstly, such a galaxy-scale magnetic field must exist; secondly, the direction of the magnetic field in the diffuse medium before cloud collapse must be retained by the OH clumps after collapse; and thirdly, the local field reversals within star forming sites must be able to be mitigated statistically. For the first condition, large-scale Galactic magnetic field measurements exist \citep[e.g.][]{Han2006,brown2007,Mao2010,Nota2010,Vaneck2011,schnitzeler2019}, and suggest that a large-scale coherent field does exist within the Milky Way disk.
Exploring the second and third conditions are part of the motivation of the MAGMO project. Any individual star forming region, a few thousand AU in scale, does not sufficiently represent the large-scale field orientation of a spiral arm, which is believed to be $\sim$1 kpc in width (see for example underlying image of Figure \ref{Figure8}), but collating many such regions across the spiral arms allows this to be probed (noting that we see two out of the 33 with indications of field reversals) and the 1720-MHz OH masers represent a small portion of the overall MAGMO catalogue (which will include the far more populous main line transitions).

A comparison of the large-scale field orientation revealed by rotation measures (RMs) and the field orientation probed by masers, within the tangent of Carina-Sagittarius spiral arm, indicated opposing field orientations between the methods \citep{Green2012pilot}. Given the small sample size of SFR 1720-MHz OH masers and their broad distribution across Galactic longitude (see Figure \ref{Figure8}), examining coherence within any specific region or Galactic arm will have limited statistical significance. However, we can briefly compare our in-situ small-scale magnetic field measurements from SFR 1720-MHz OH maser polarimetry with the large-scale Galactic magnetic field obtained from the RMs of Galactic pulsars. For consistency with the MAGMO pilot paper \citep{Green2012pilot}, we adopt the pulsar RM results of \citet{Nota2010}. We only selected the pulsar RMs with sight-lines passing through H{\sc ii} regions. This was on the basis of finding lines that were likely to be influenced by the same environment as the maser host: pulsar rotation measurements are distance-dependent and represent an integration across the entire line of sight, hence they are heavily influenced by any intervening H{\sc ii} region along the path, where electron densities will be high. Studies \citep[e.g.][]{Mitra2003, Harvey-Smith2011} have demonstrated coherent large-scale line of sight magnetic fields within H{\sc ii} regions, and thus these results enable comparison with the in-situ small-scale magnetic field measurements from Zeeman splitting.
We note that the pulsar distances reported in \citet{Nota2010} were mostly stellar and kinematic distances quoted without uncertainties, but we are aware that the ambiguities and inherently significant uncertainties associated with these distance measurements will tend to affect the interpretation of the pulsar RMs. 
In Figure \ref{Figure8} we overlay positions and field directions from our Zeeman measurements of SFR 1720-MHz OH maser components on an artist's impression of the Milky Way (R. Hurt: NASA/JPL-Caltech/SSC.), together with the pulsar RM-derived field directions. The RM fields are represented on the overlay as lines so as to indicate the integrated line-of-sight nature of the magnetic fields, with the length corresponding to the distance to the pulsar. Three out of the eight integrated sight-lines to the pulsars have fields oriented away from us (red lines in Figure \ref{Figure8}), and five have magnetic fields oriented towards us (blue lines in Figure \ref{Figure8}). This figure shows both agreement and apparent contradiction between the Zeeman measurements and the RMs, although it is clear that the cases of contradiction may be due to the difference in the maser location and the majority of electrons contributing to the RM line-of-sight measurement. For example, as per \citet{Green2012pilot}, we see both positive and negative Zeeman measurements apparently located within the Carina-Sagittarius arm, whilst two red and all five blue lines from RM measurements pass through the same arm. This clearly needs a more comprehensive (both clustered and widely distributed) sample of Zeeman measurements and RM field directions to draw statistically significant conclusions across the Galaxy, which we hope to achieve with the the main line OH transitions in a follow-up publication.



\section{Summary}

We have performed a targeted search for 1720-MHz OH masers towards Southern hemisphere star forming regions, as traced by 6.7-GHz methanol masers.
We report a total of 51 1720-MHz OH maser detections in the vicinity of the 702 target sources.
In addition to 10 new detections and 23 re-detections towards star forming regions, we also report two new and 14 previously-known masers towards supernova remnants, and two detections of diffuse OH towards compact continuum sources. 12 of the 33 star formation masers are devoid of either 6.7-GHz methanol or 1665-MHz OH masers, and we suggest that ten of these sources may be associated with outflow activity from a central exciting source. We also presented association statistics for 1720-MHz masers with other maser species including excited and other ground-state OH maser transitions in the vicinity of their common star formation hosts.

By utilizing the polarimetric information revealed by the Zeeman splitting of the maser emission, we determined the strengths and line-of-sight directions of the in-situ magnetic field of the star formation and SNR masers. For each source we calculated both linear and circular polarisation fractions, and found (as expected) the latter to be higher; of those SFR masers with reliably measured Zeeman splitting, 84\% exhibited $>$50\% fractional circular polarisation. 
Comparable with previous studies of main line ground-state and excited-state OH, we found $\sim$28\% of the SFR maser sample exhibited detectable linearly polarised features, suggestive of the presence of $\pi$ components. Consistent with theoretical predictions, we confirm no large systematic differences between the right handed and left handed circularly polarised flux densities. 

We compared the positions and in-situ magnetic field directions of our masers with previously measured integrated line-of-sight magnetic fields from pulsar Faraday rotation measures. 
We found suggestions of coherence between the two methods, but conclusions are restricted by the small sample size and distance errors, and so large-scale magnetic field structure will be explored further with the larger main line MAGMO catalogue.

\section*{Acknowledgements}
This work is presented as part of C.S.O's doctoral program funded by the International Macquarie Research Excellence Scholarship program (iMQRES). C.S.O. is also a recipient of the CSIRO Astronomy and Space Science Student Program grant for which C.S.O. is grateful for. J.R.D. acknowledges the support of an Australian Research Council (ARC) DECRA Fellowship (project number DE170101086). The ATCA is part of the Australia Telescope National Facility and funded by CSIRO. This research made use of APLpy, an open-source plotting package for Python \citep{aplpy2012}.




\bibliographystyle{mnras}
\bibliography{mnras_template} 





\appendix

\twocolumn[\section{GLIMPSE images of 1720-MHz OH star formation masers}\label{appendix}]

\begin{figure*}
\caption{Overlaid positions of maser species and transitions on RGB-composite images of star forming regions obtained from GLIMPSE survey.  Please see Table 6 for references.}
\subfloat{\includegraphics[width = 3.1in]{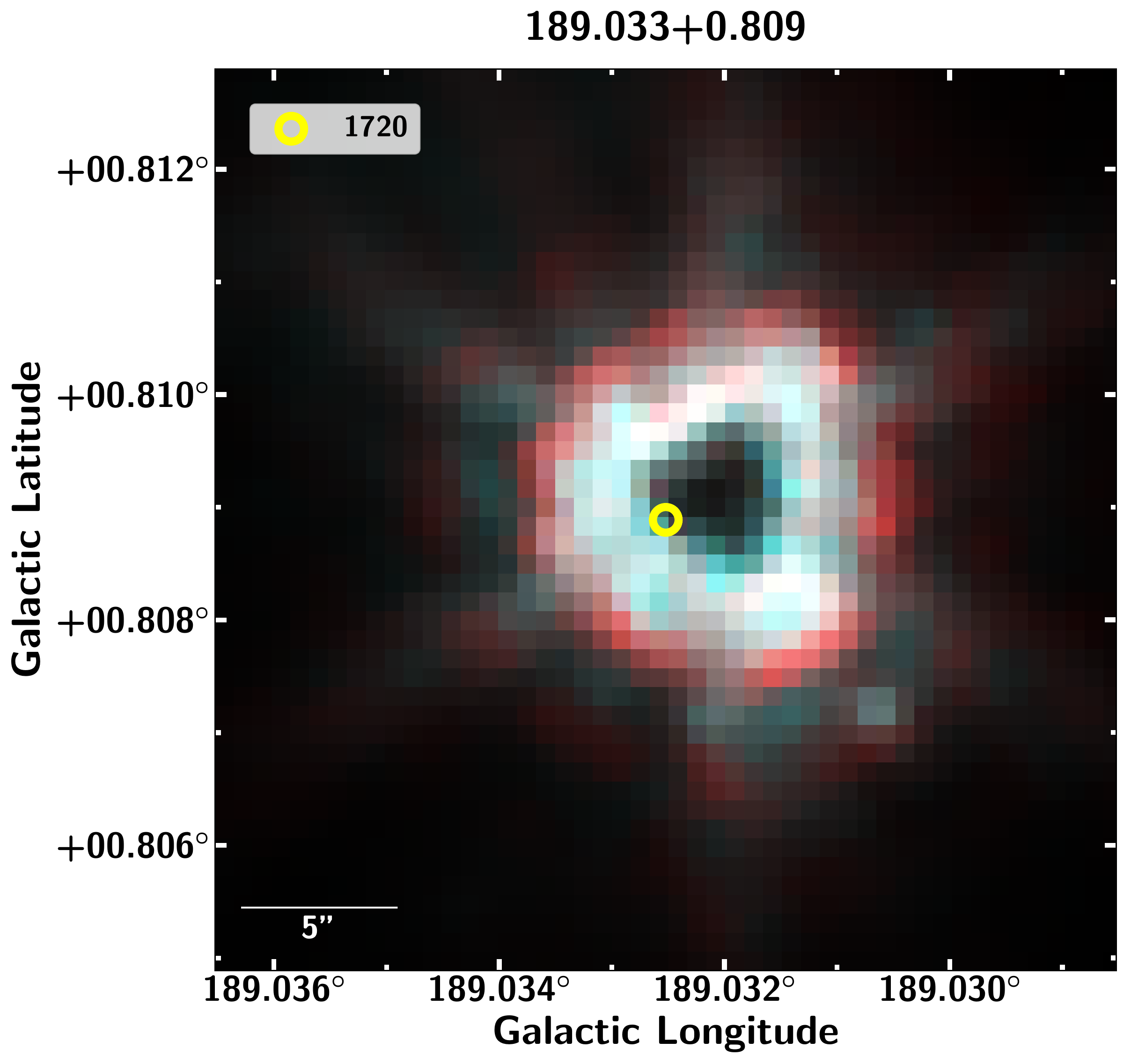}}
\subfloat{\includegraphics[width = 3.1in]{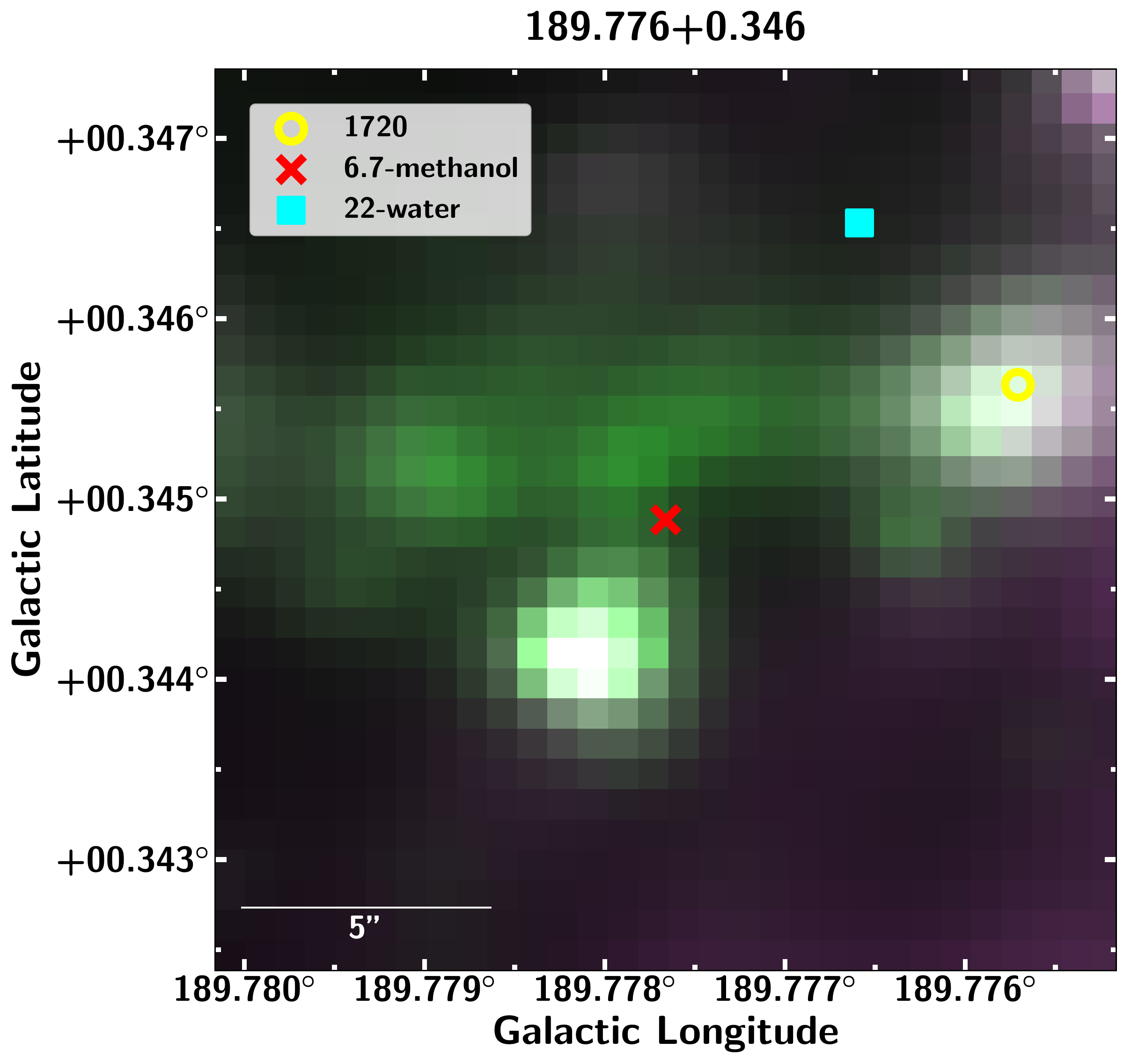}}\\
\subfloat{\includegraphics[width = 3.1in]{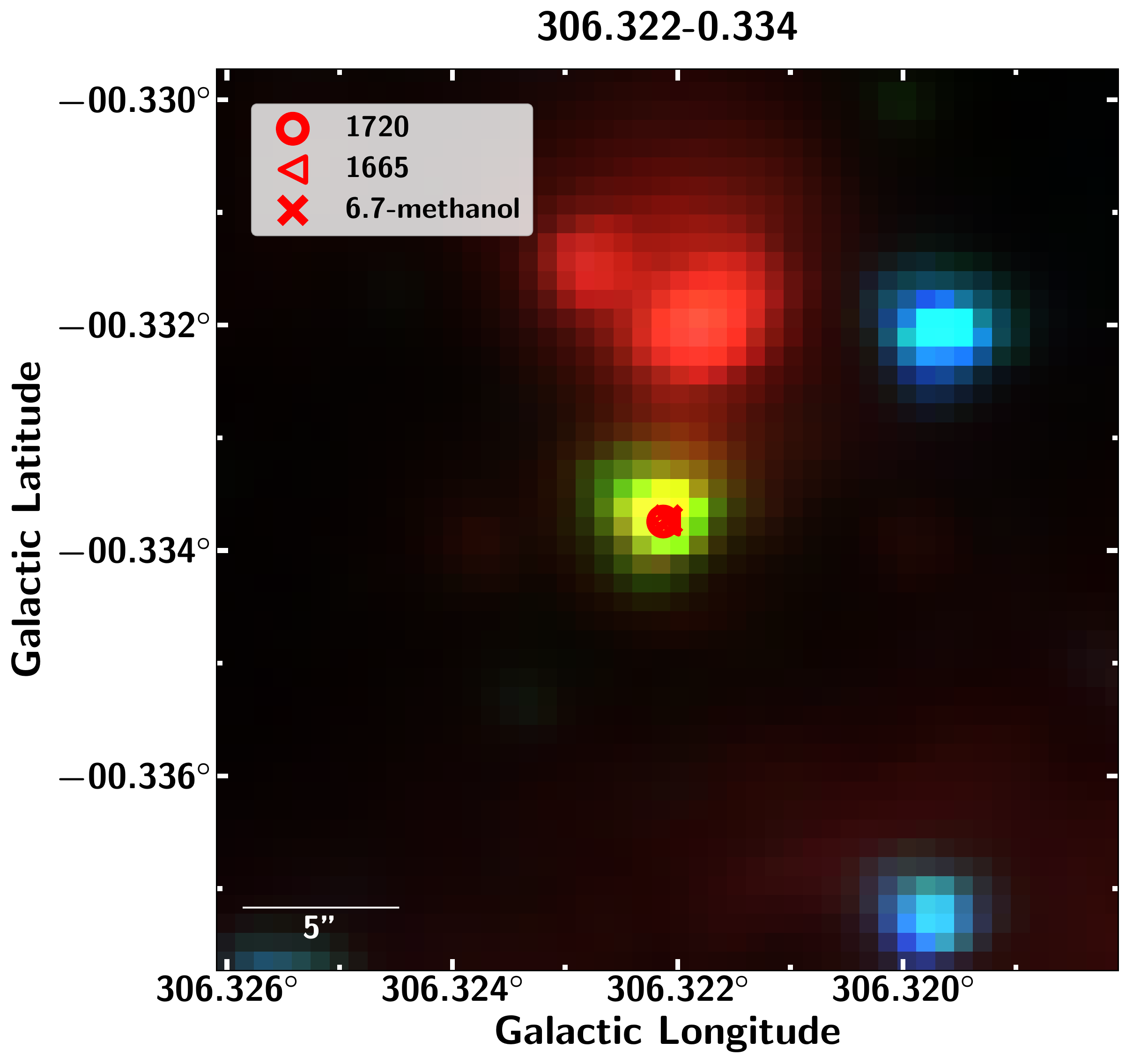}}
\subfloat{\includegraphics[width = 3.1in]{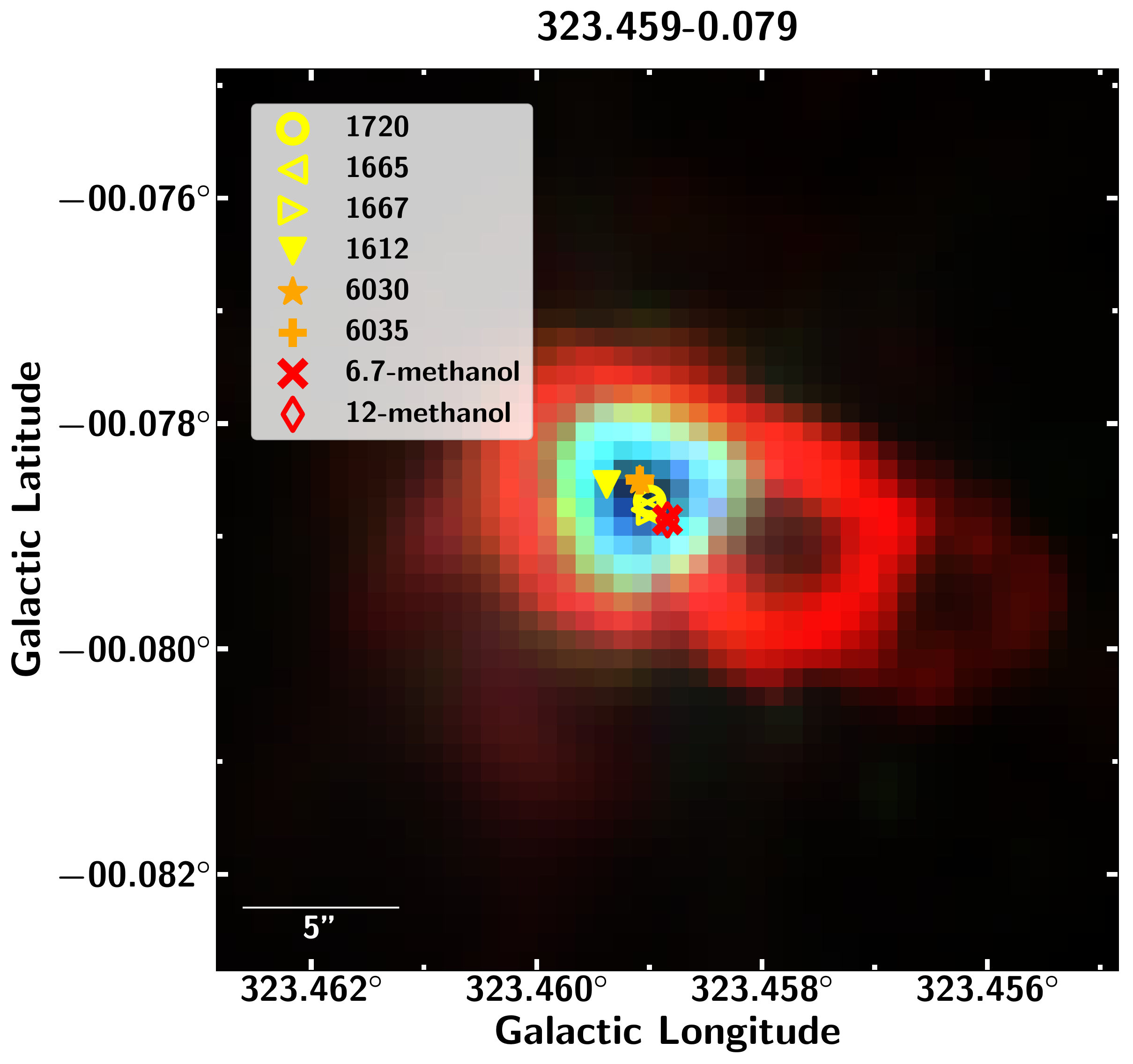}}\\
\subfloat{\includegraphics[width = 3.1in]{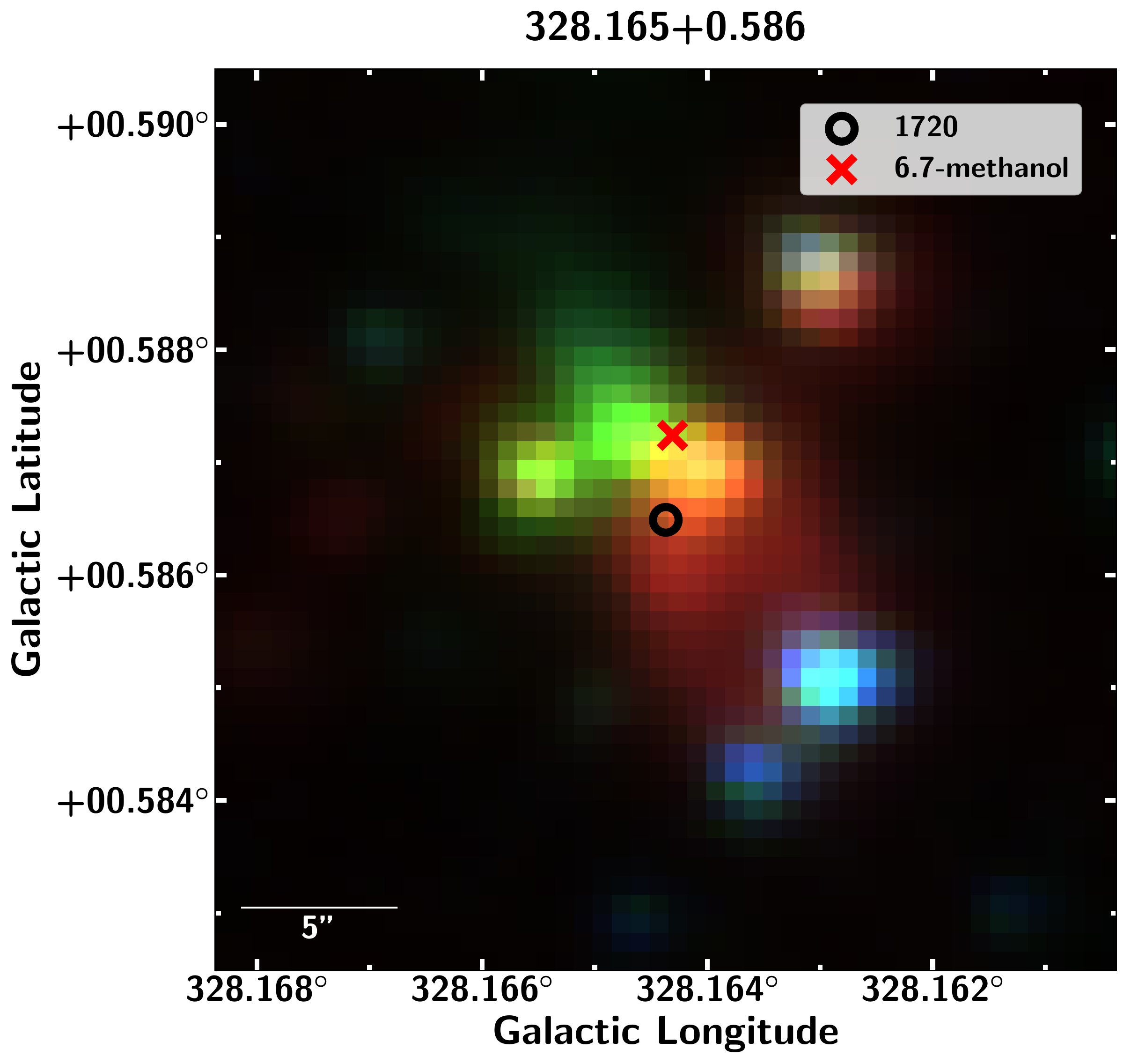}}
\subfloat{\includegraphics[width = 3.1in]{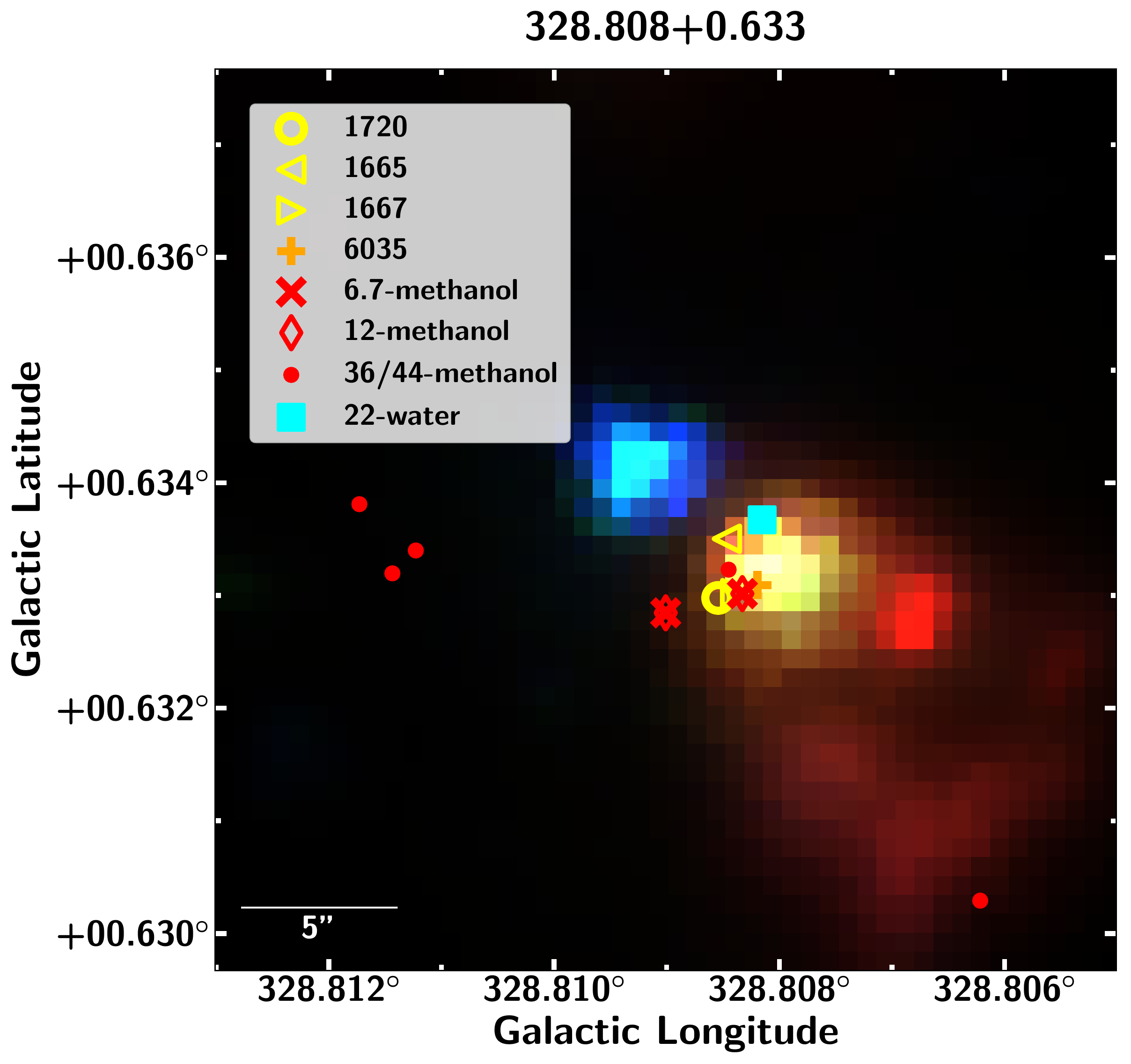}}\\
\end{figure*}

\begin{figure*}

\subfloat{\includegraphics[width = 3.1in]{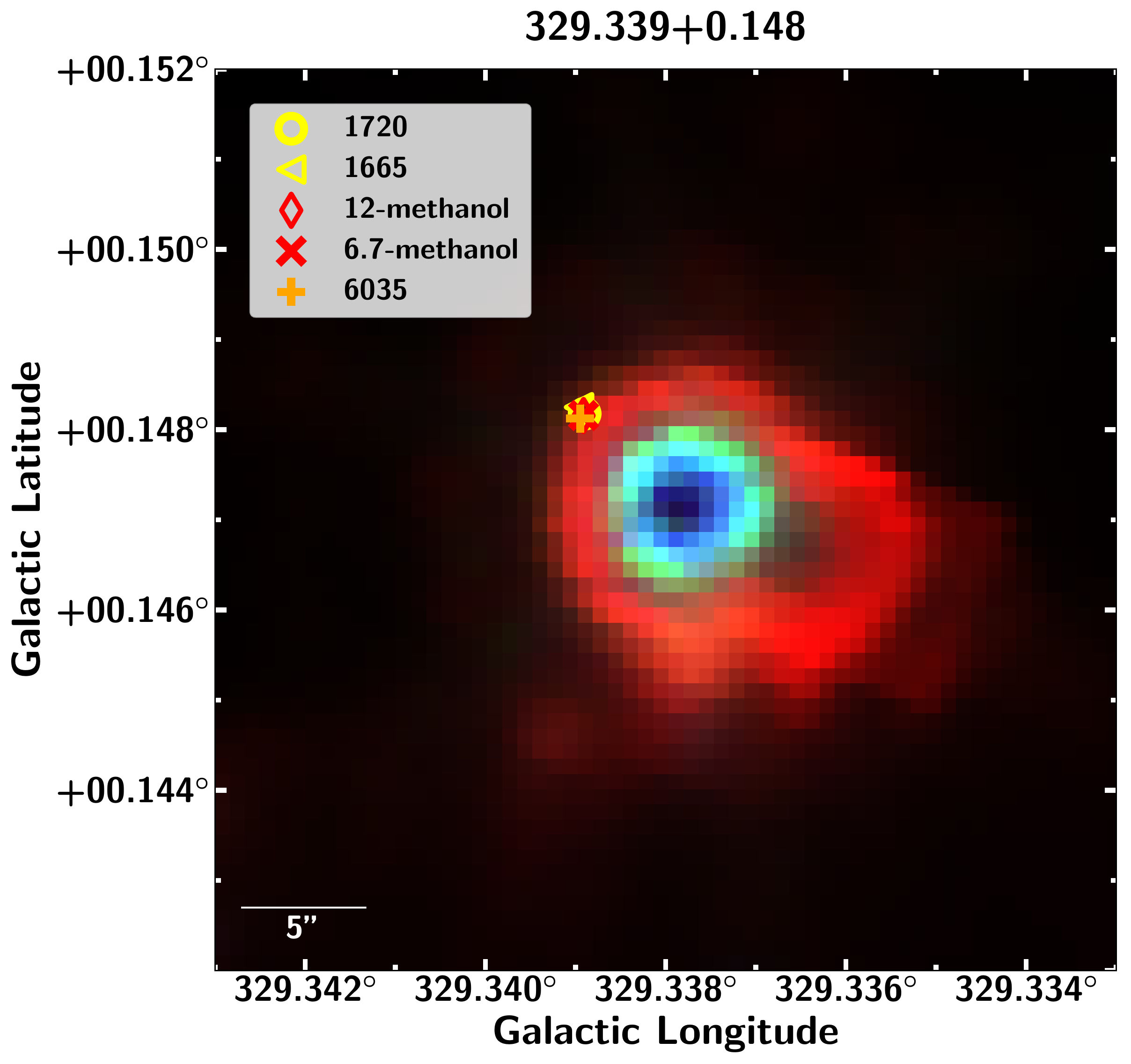}}
\subfloat{\includegraphics[width = 3.1in]{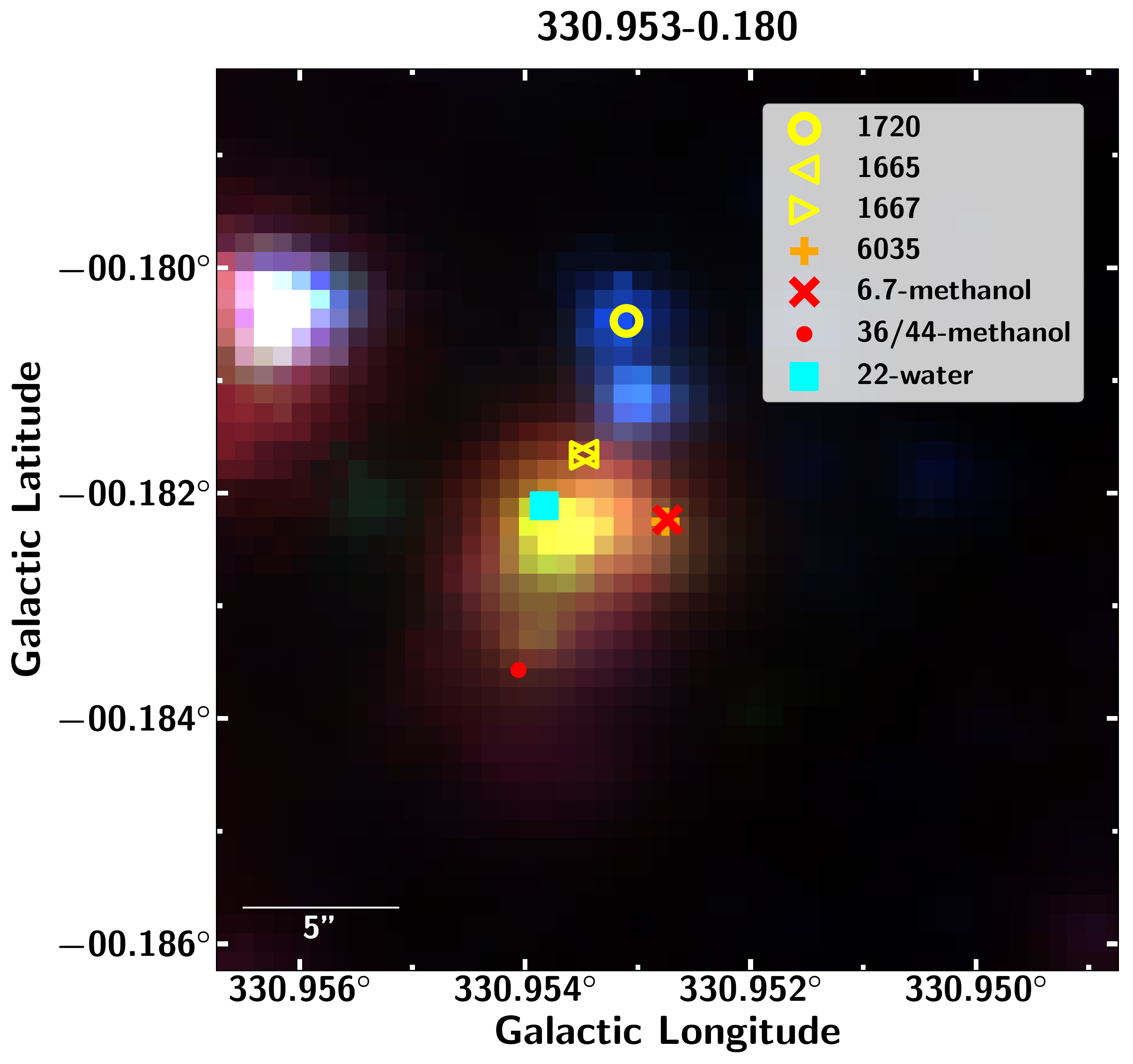}}\\
\subfloat{\includegraphics[width = 3.1in]{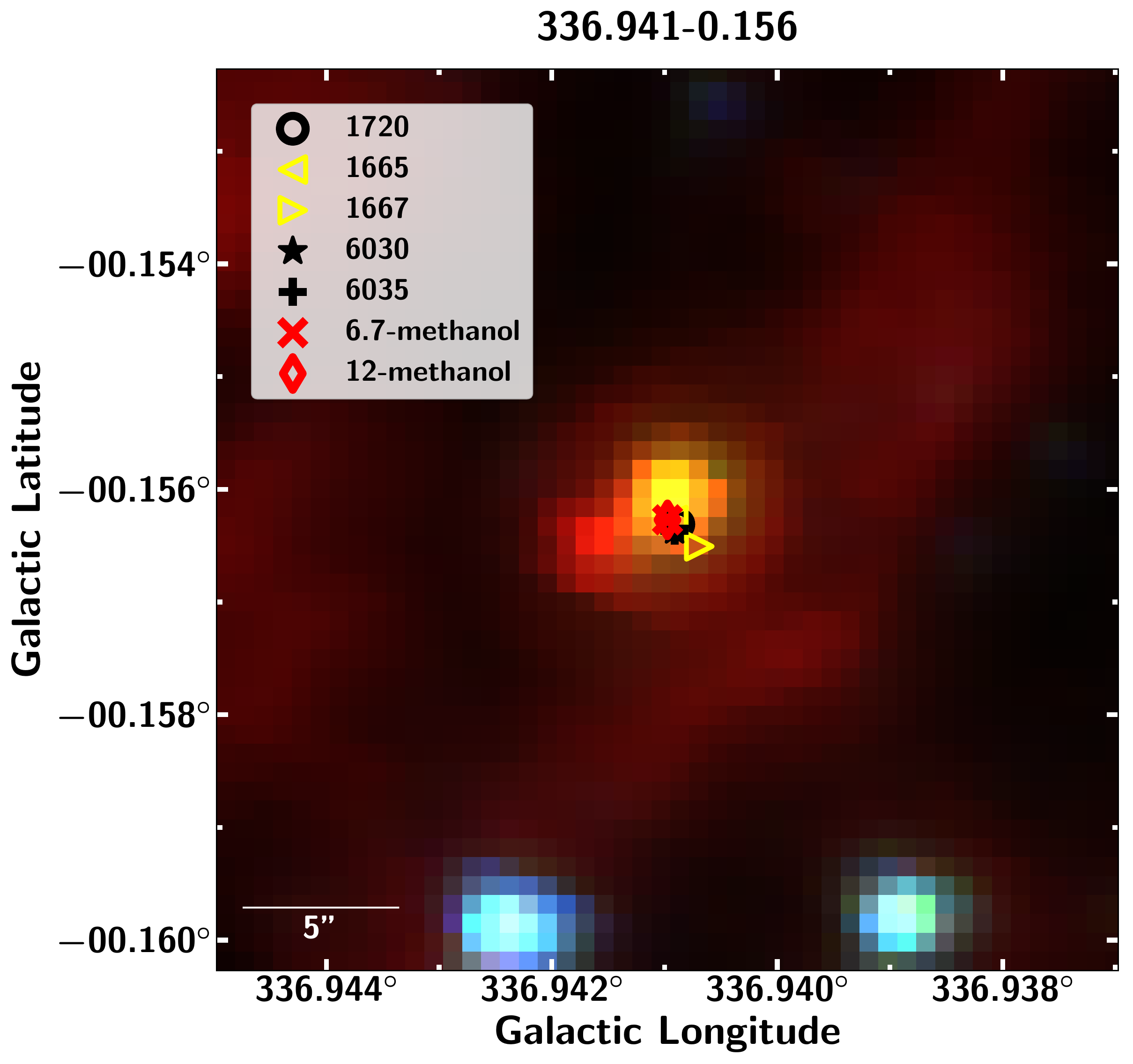}}
\subfloat{\includegraphics[width = 3.1in]{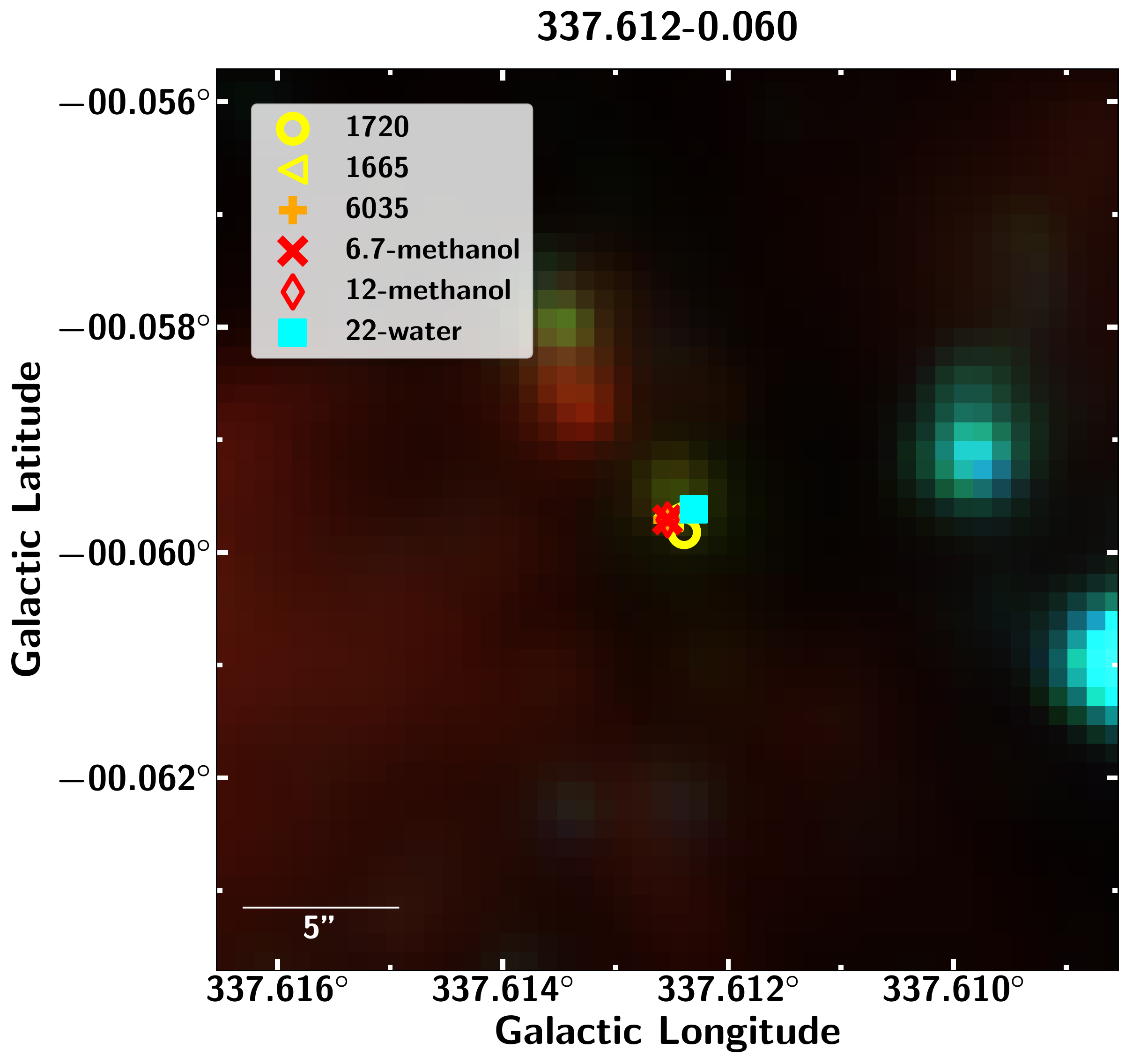}}\\
\subfloat{\includegraphics[width = 3.1in]{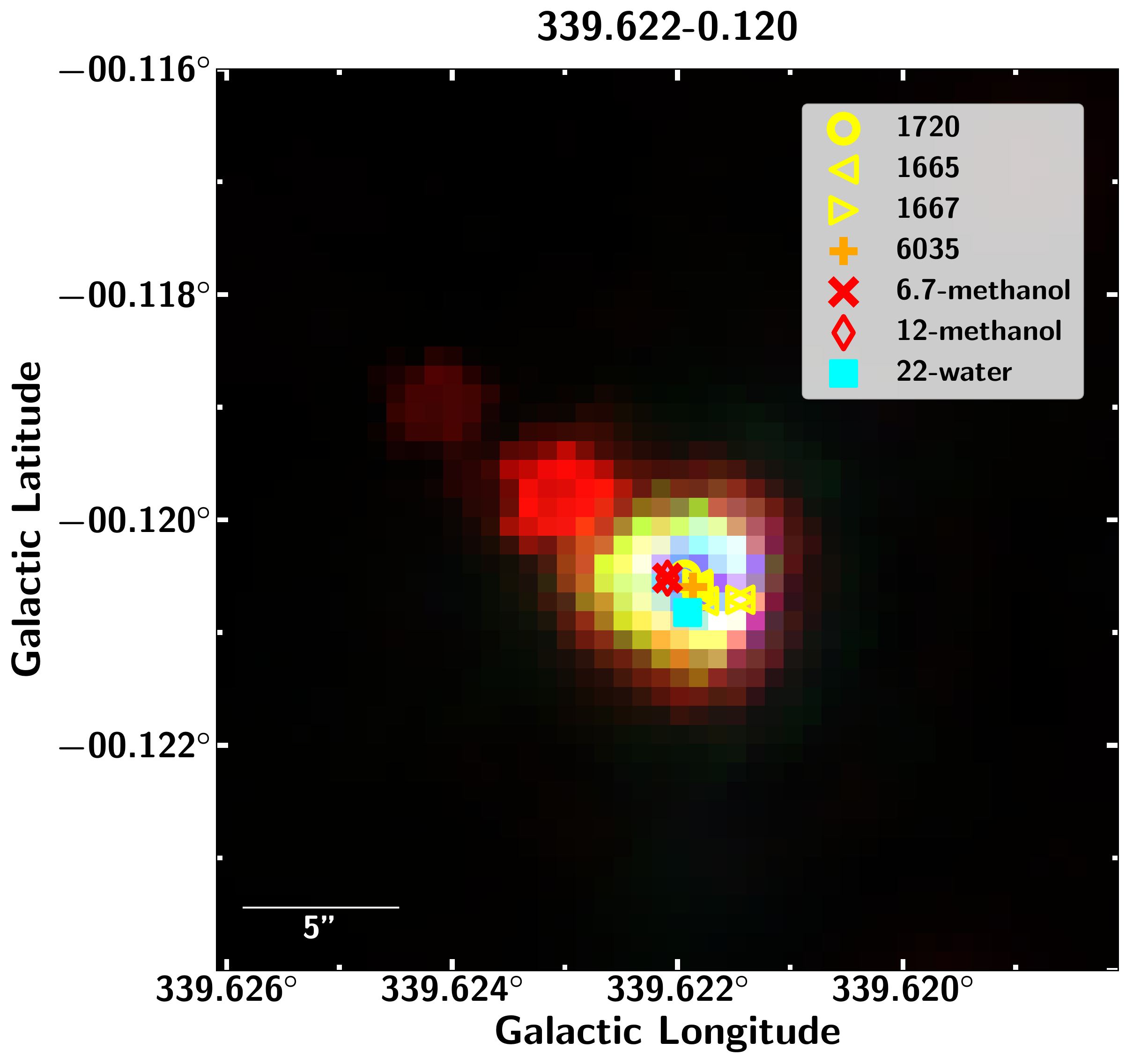}}
\subfloat{\includegraphics[width = 3.1in]{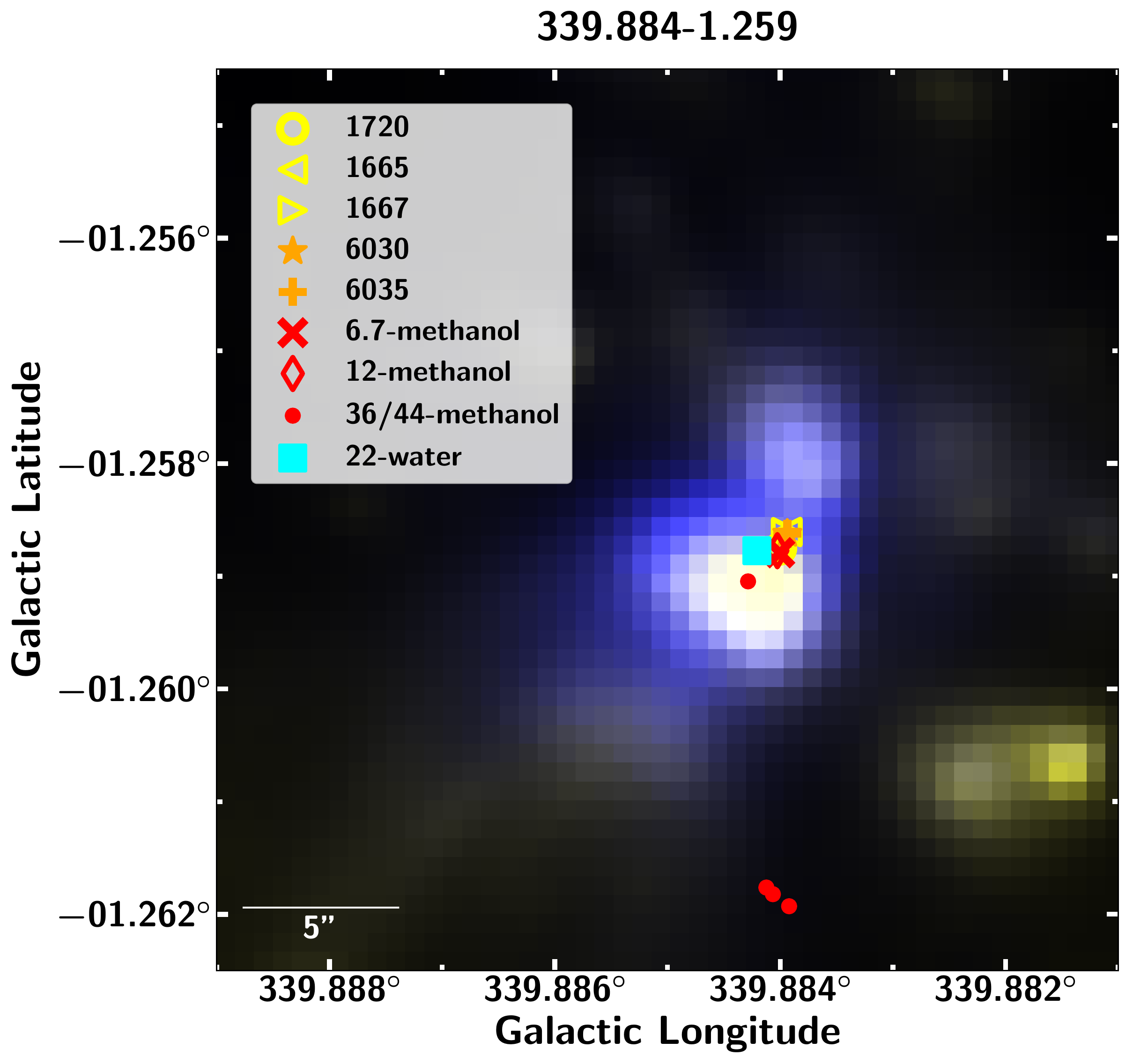}}\\

\end{figure*}

\begin{figure*}

\subfloat{\includegraphics[width = 3.1in]{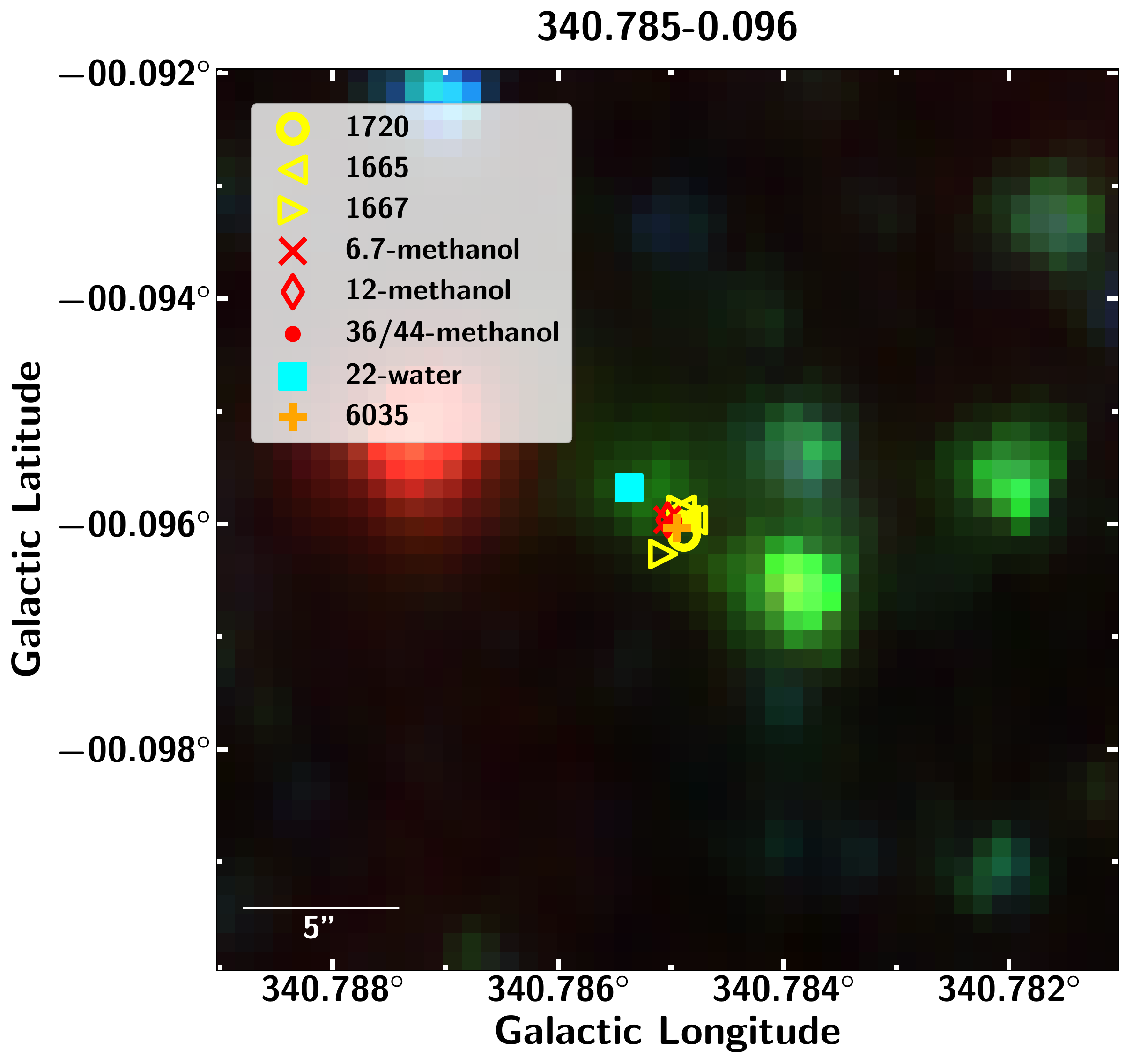}}
\subfloat{\includegraphics[width = 3.1in]{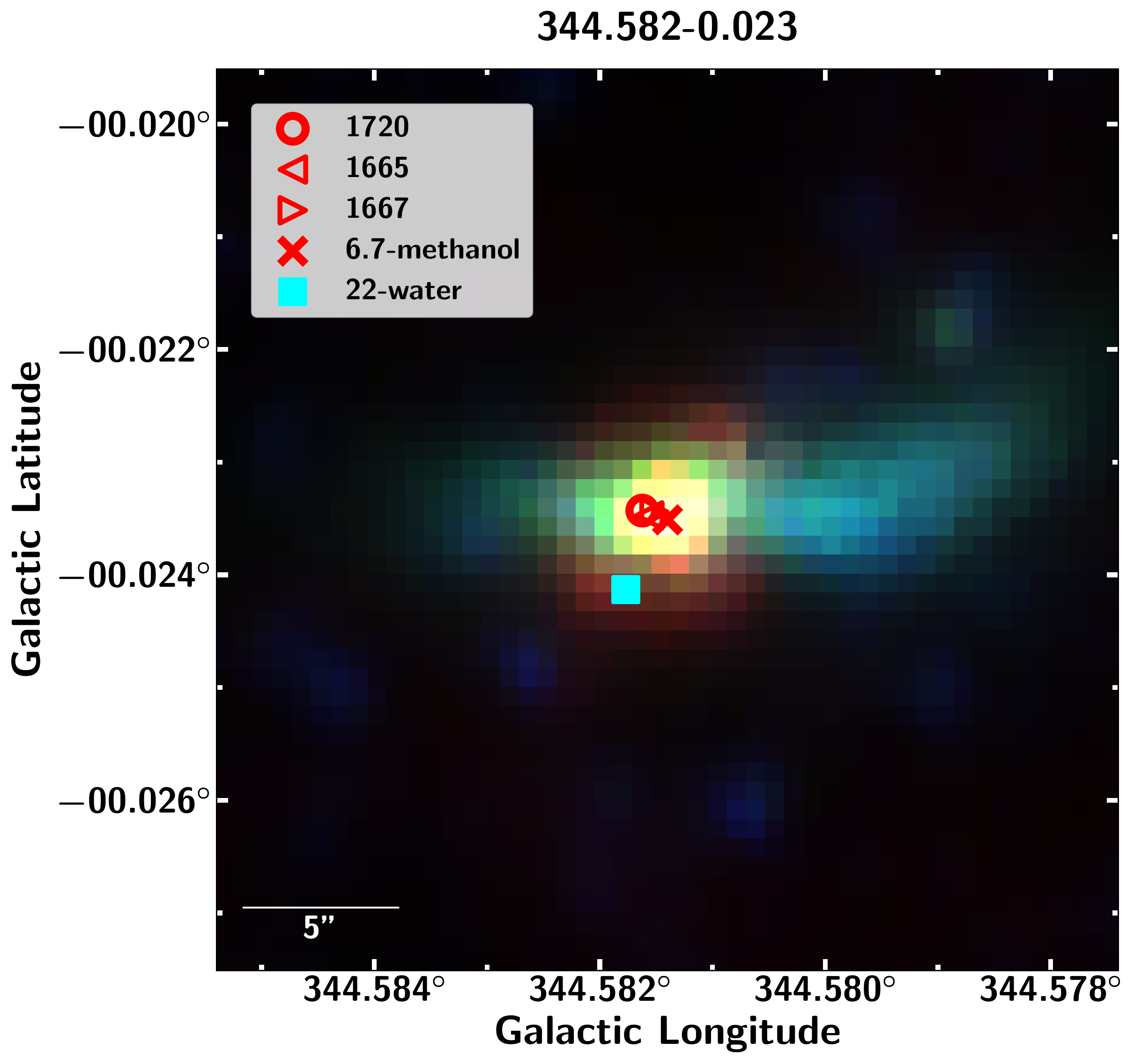}}\\
\subfloat{\includegraphics[width = 3.1in]{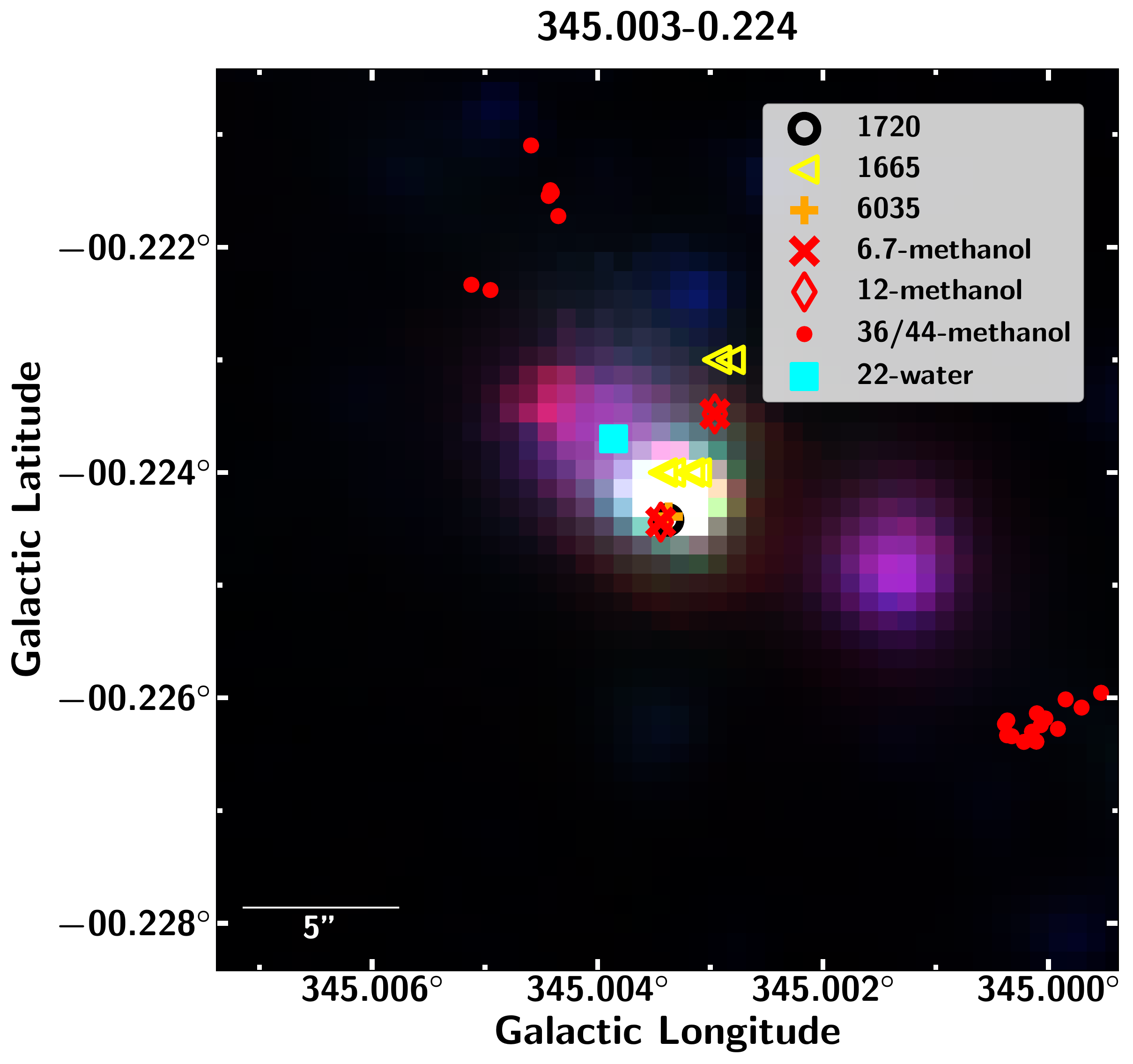}}
\subfloat{\includegraphics[width = 3.1in]{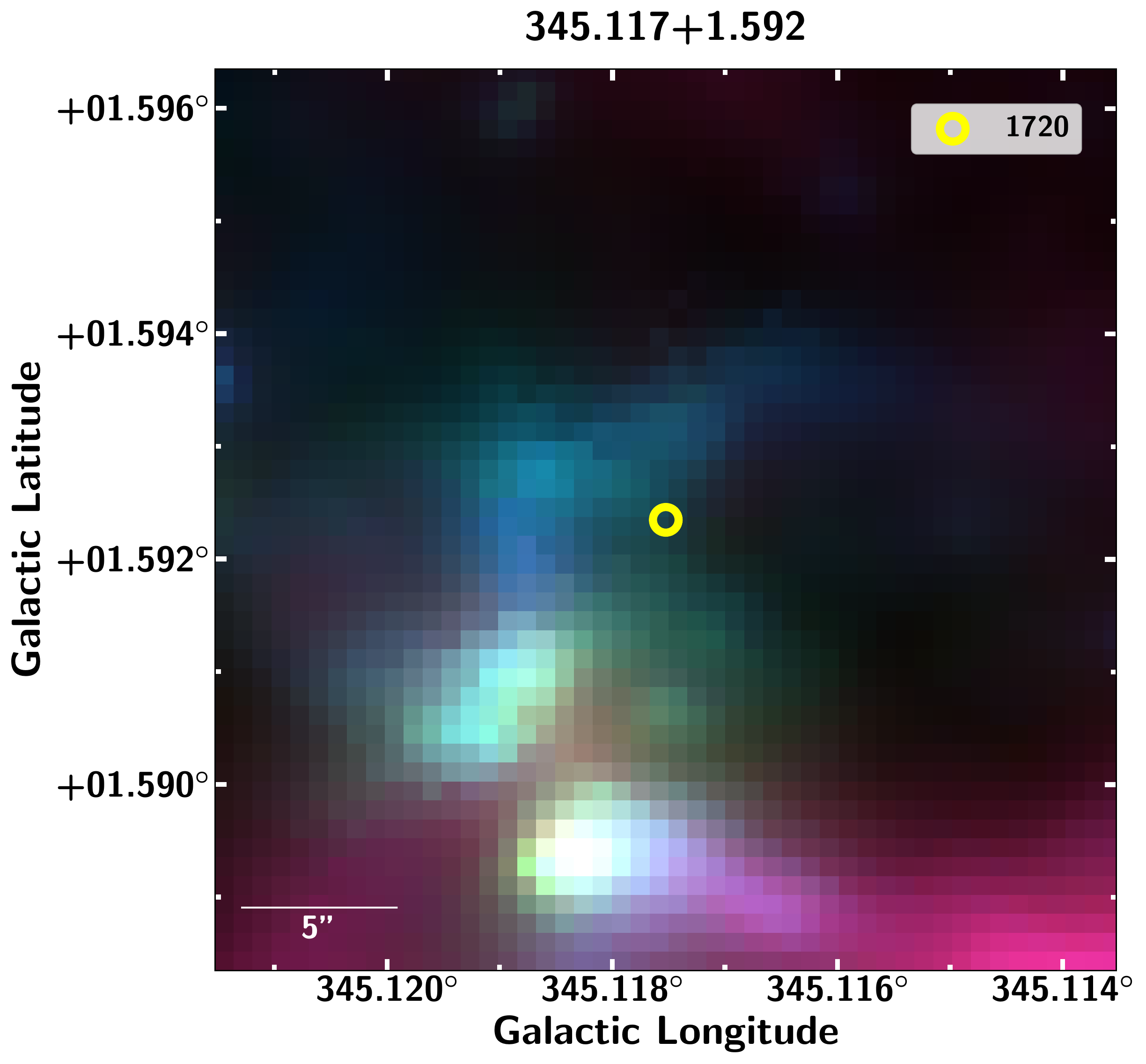}}\\
\subfloat{\includegraphics[width = 3.1in]{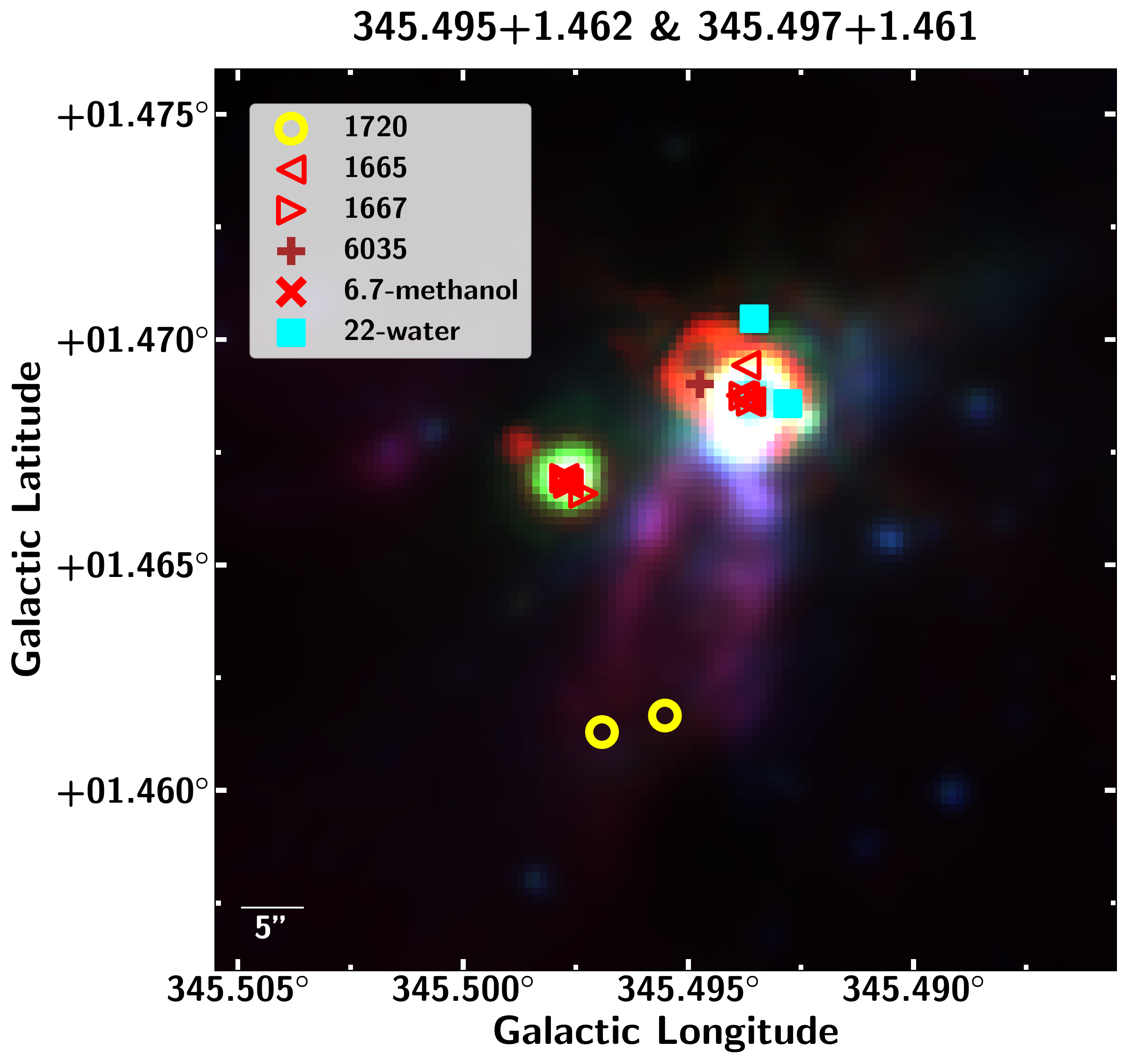}}
\subfloat{\includegraphics[width = 3.1in]{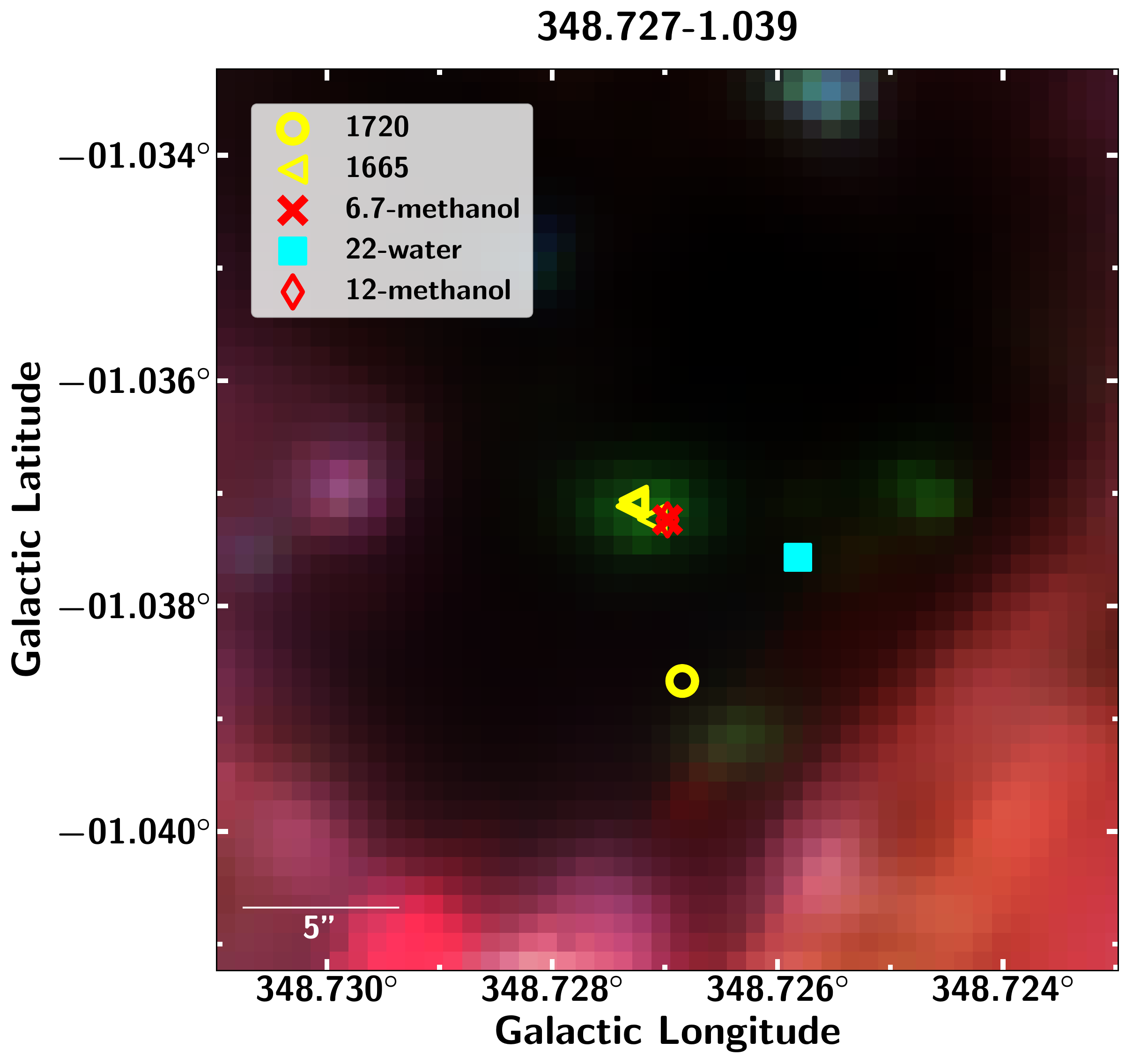}}\\
\end{figure*}

\begin{figure*}

\subfloat{\includegraphics[width = 3.1in]{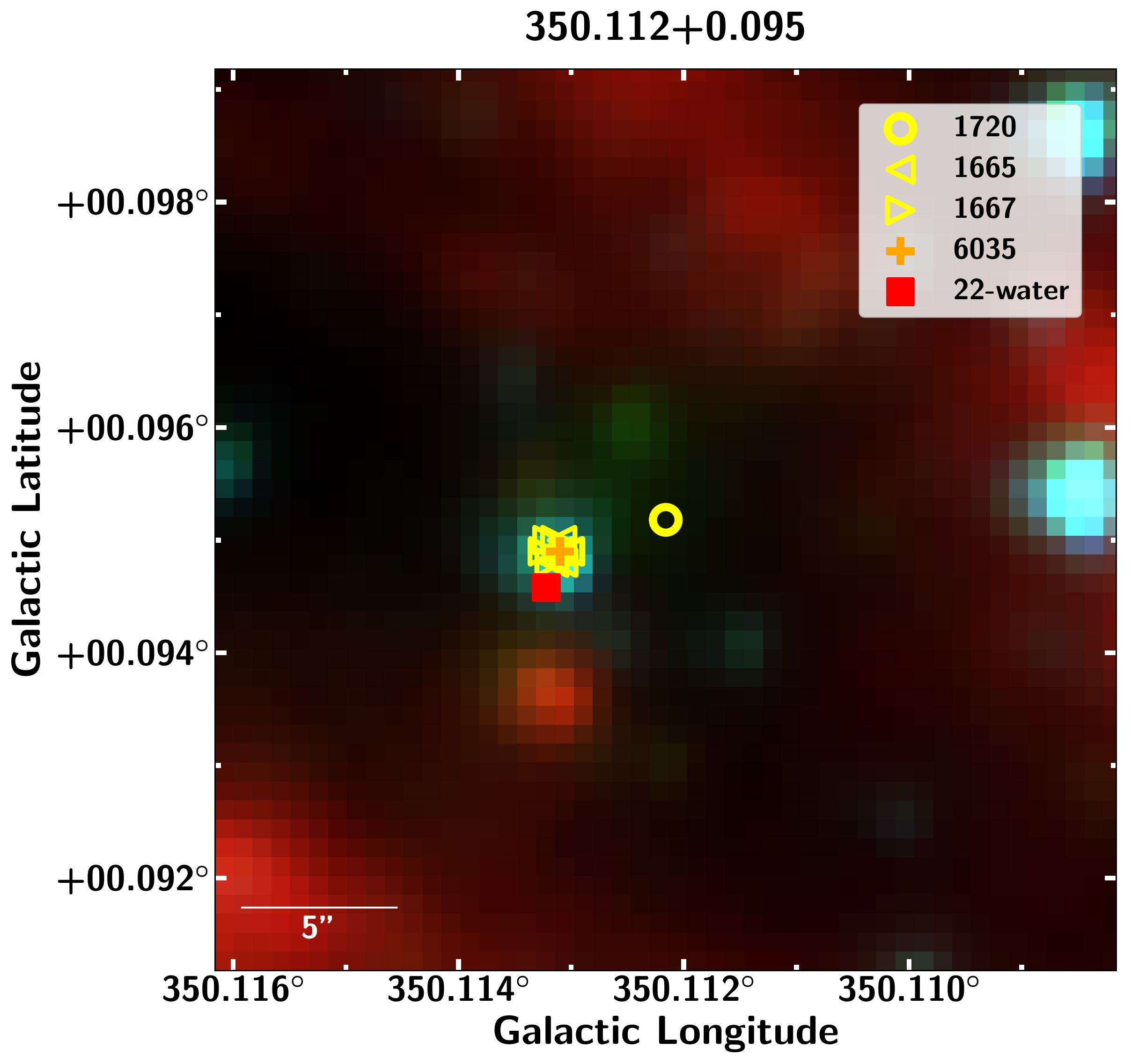}}
\subfloat{\includegraphics[width = 3.1in]{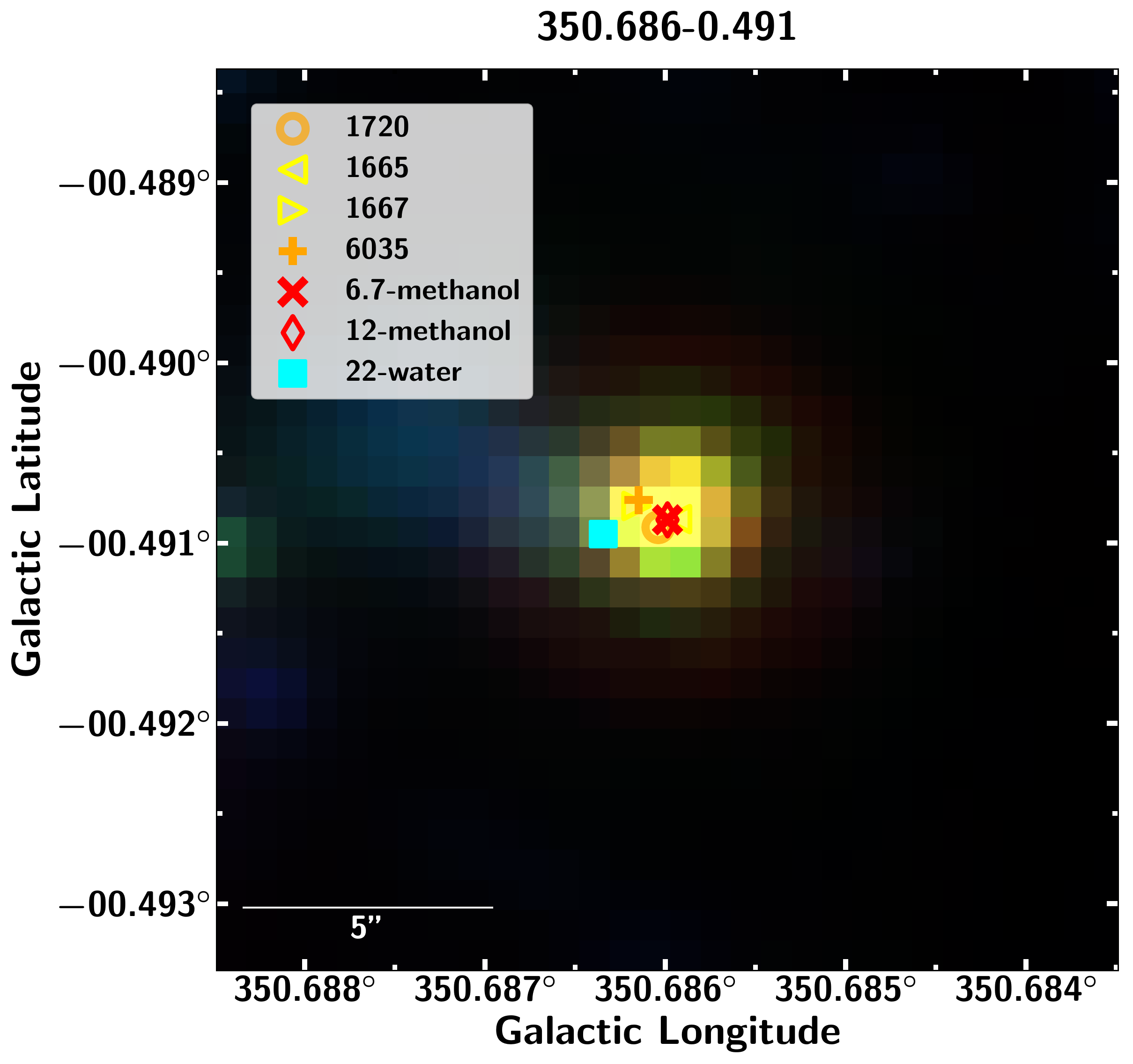}}\\
\subfloat{\includegraphics[width = 3.1in]{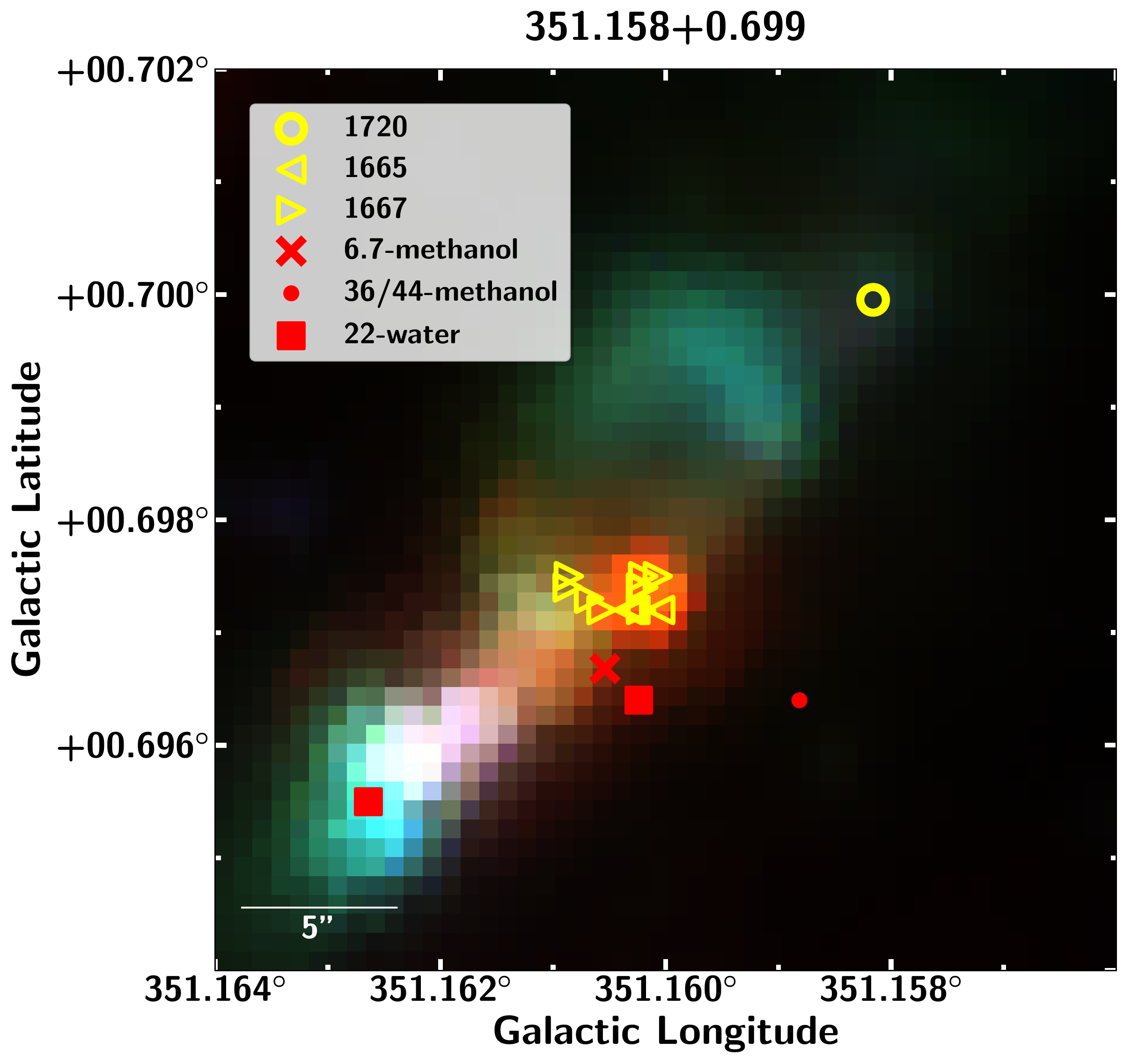}}
\subfloat{\includegraphics[width = 3.1in]{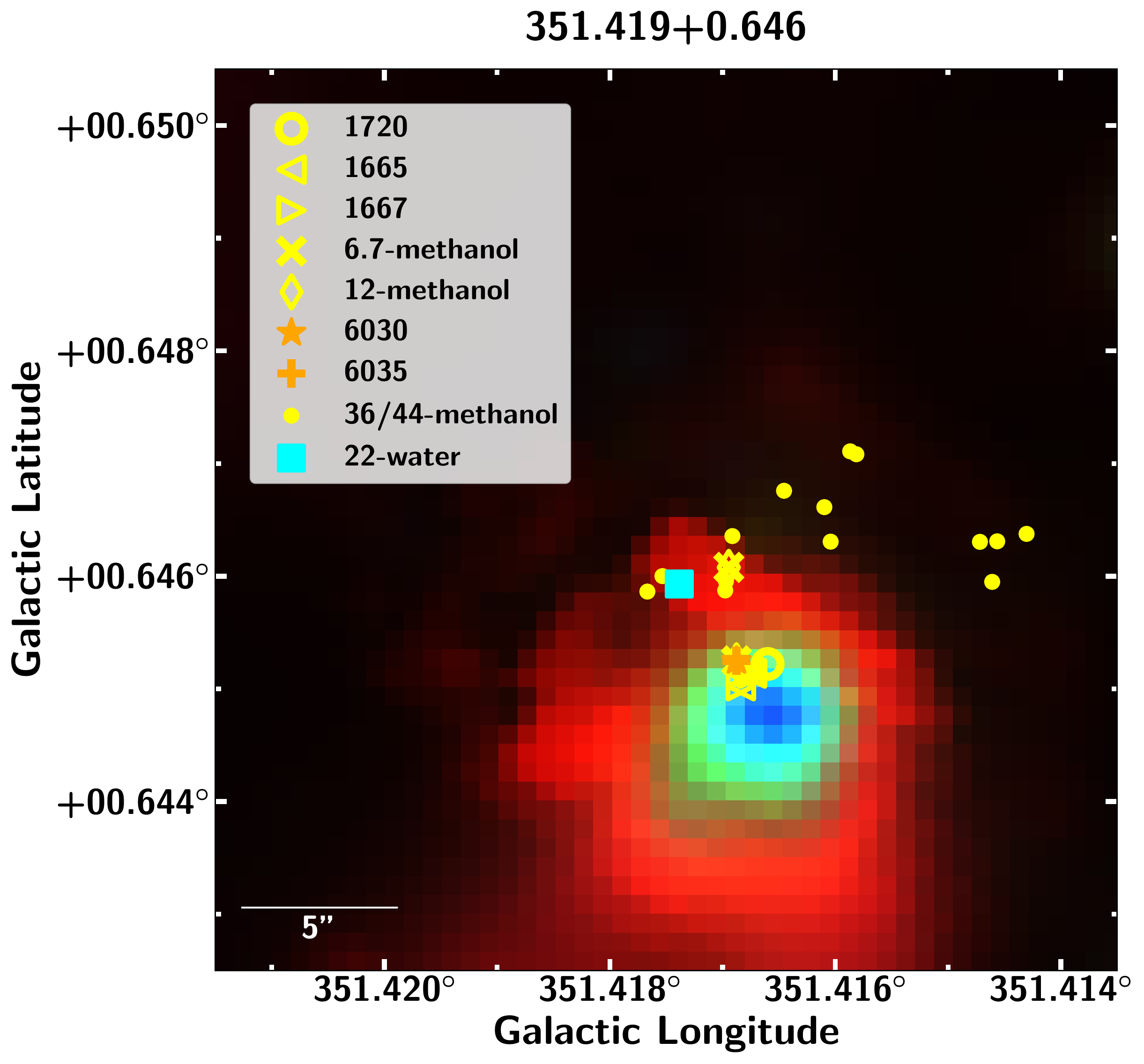}}\\
\subfloat{\includegraphics[width = 3.1in]{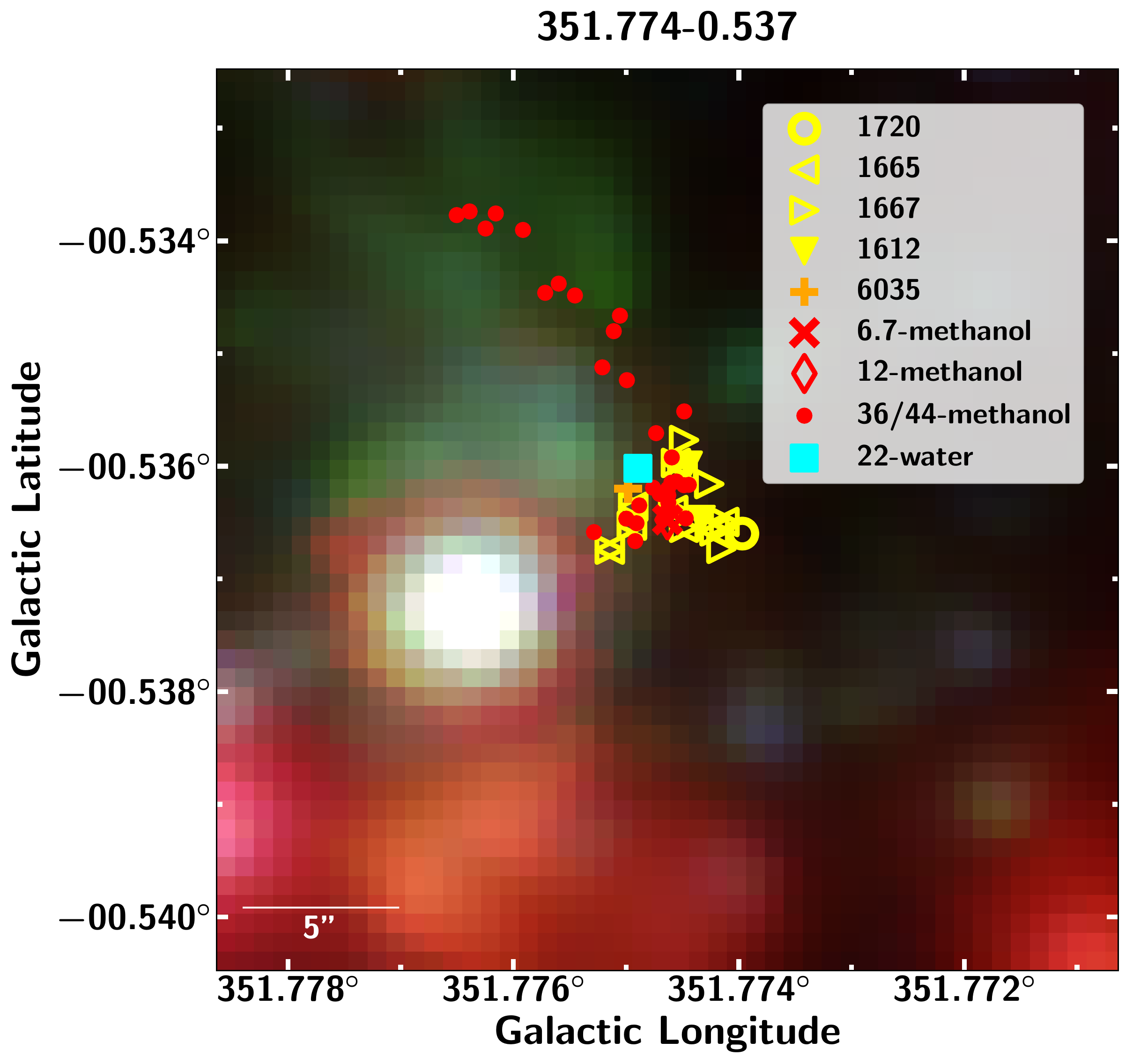}}
\subfloat{\includegraphics[width = 3.1in]{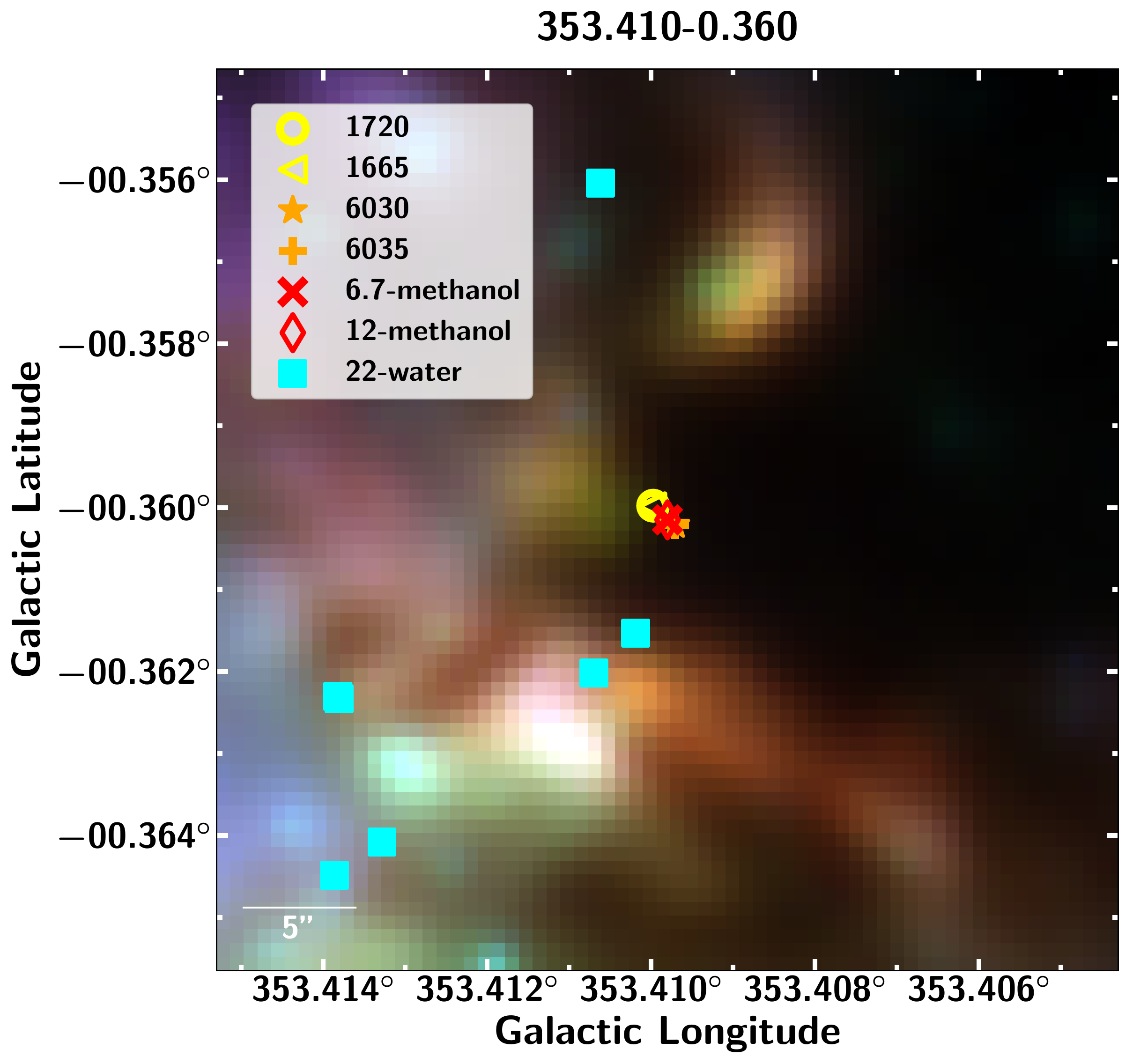}}\\
\end{figure*}

\begin{figure*}

\subfloat{\includegraphics[width = 3.1in]{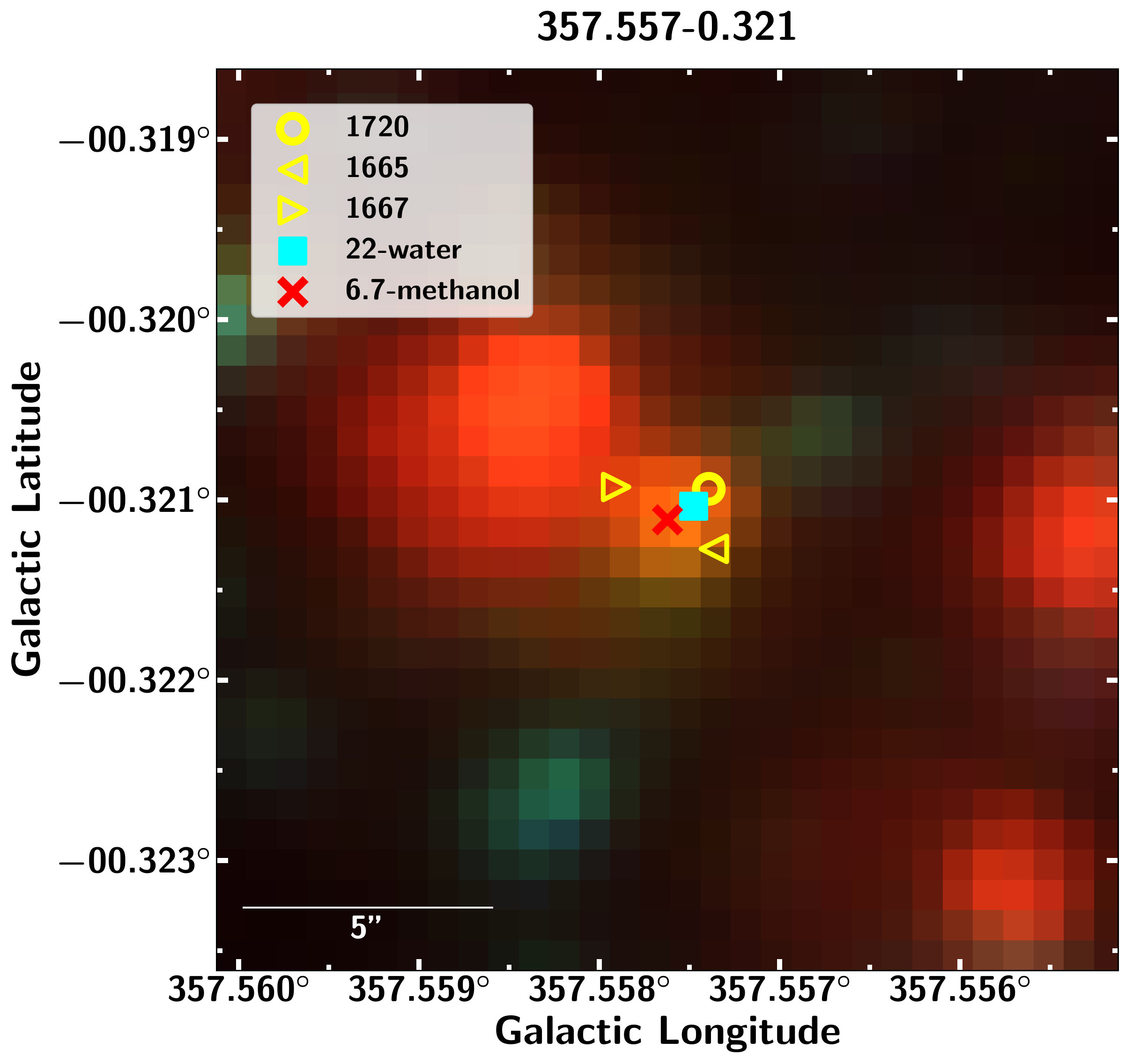}}
\subfloat{\includegraphics[width = 3.1in]{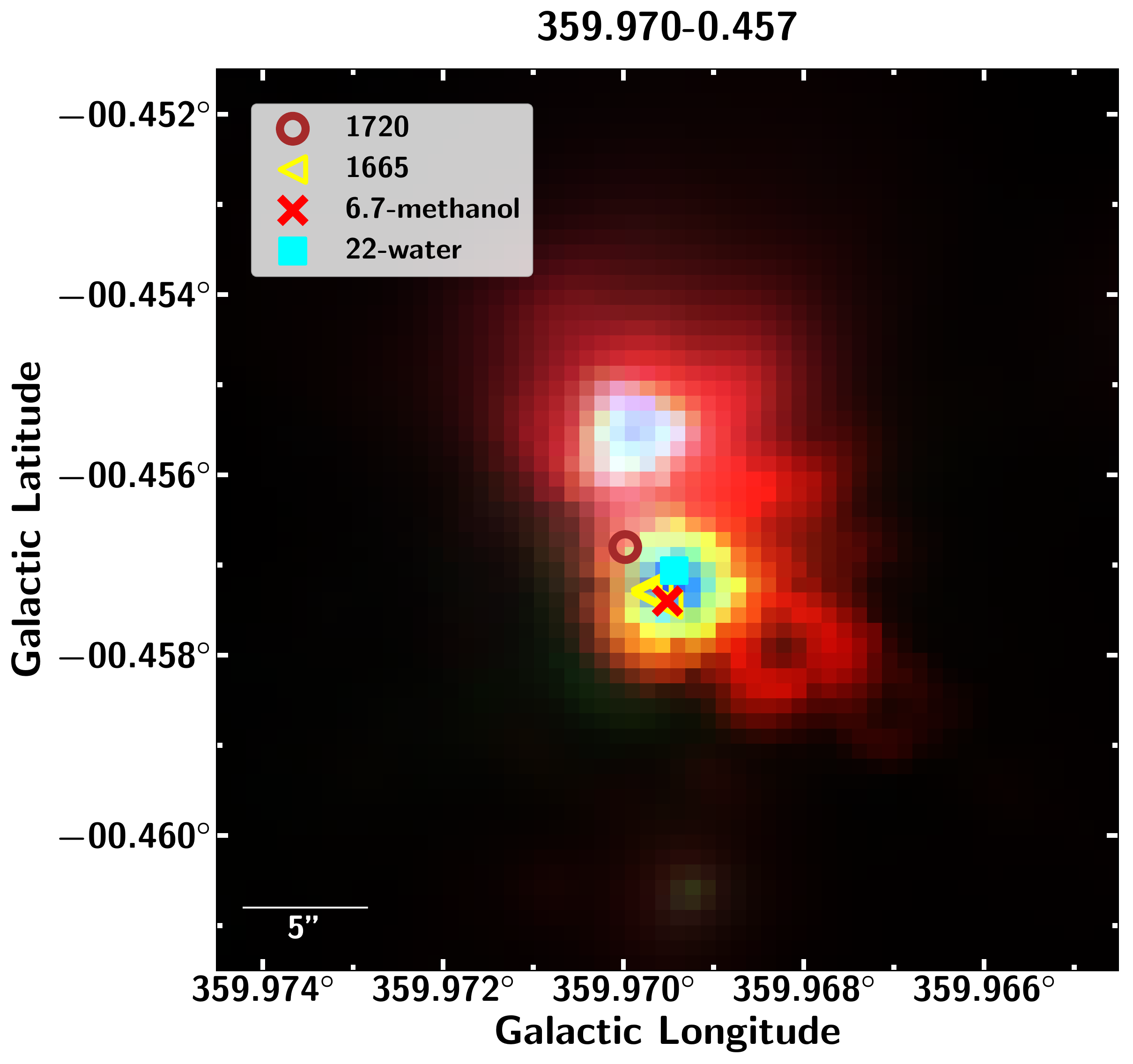}}\\
\subfloat{\includegraphics[width = 3.1in]{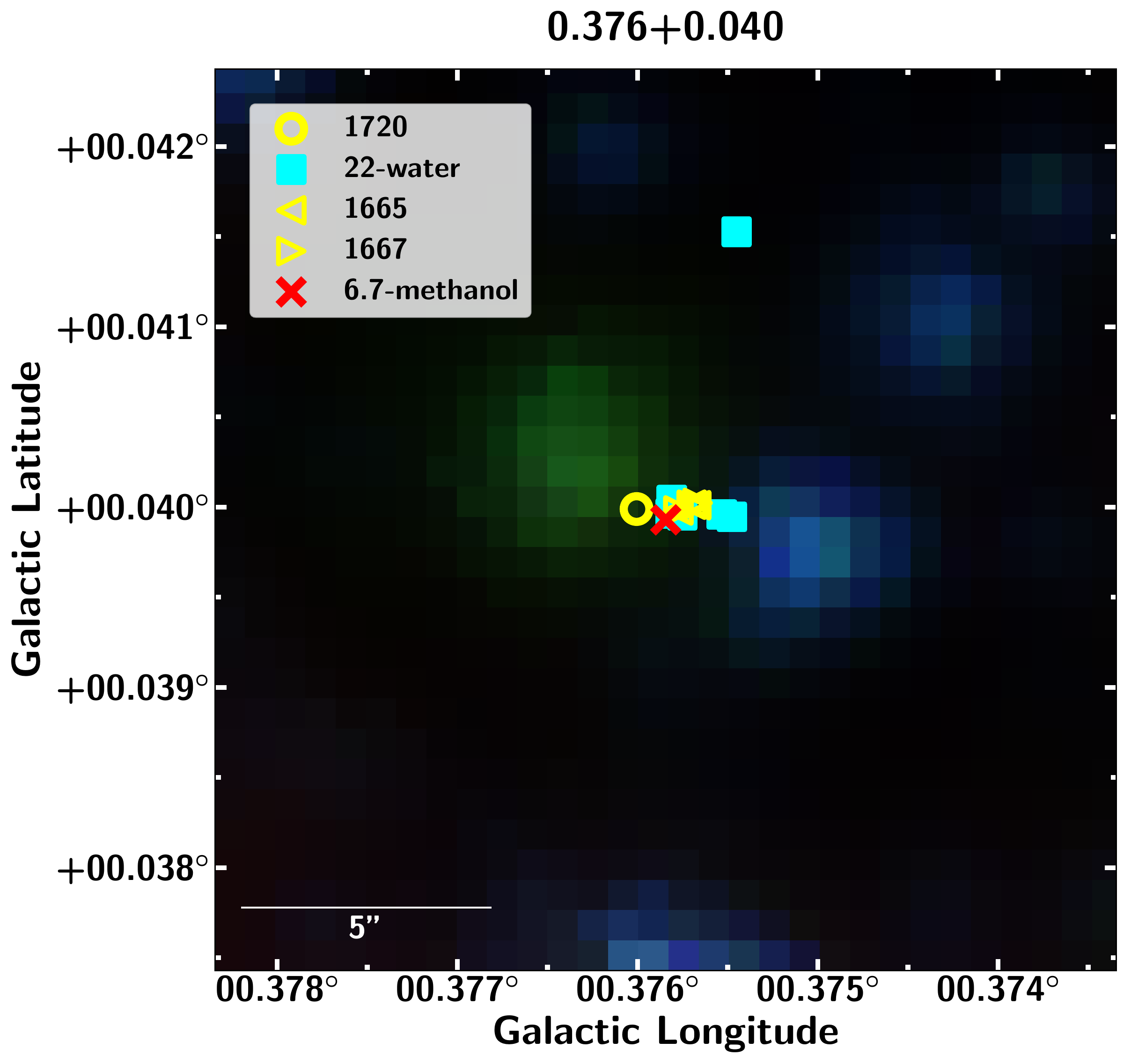}}
\subfloat{\includegraphics[width = 3.1in]{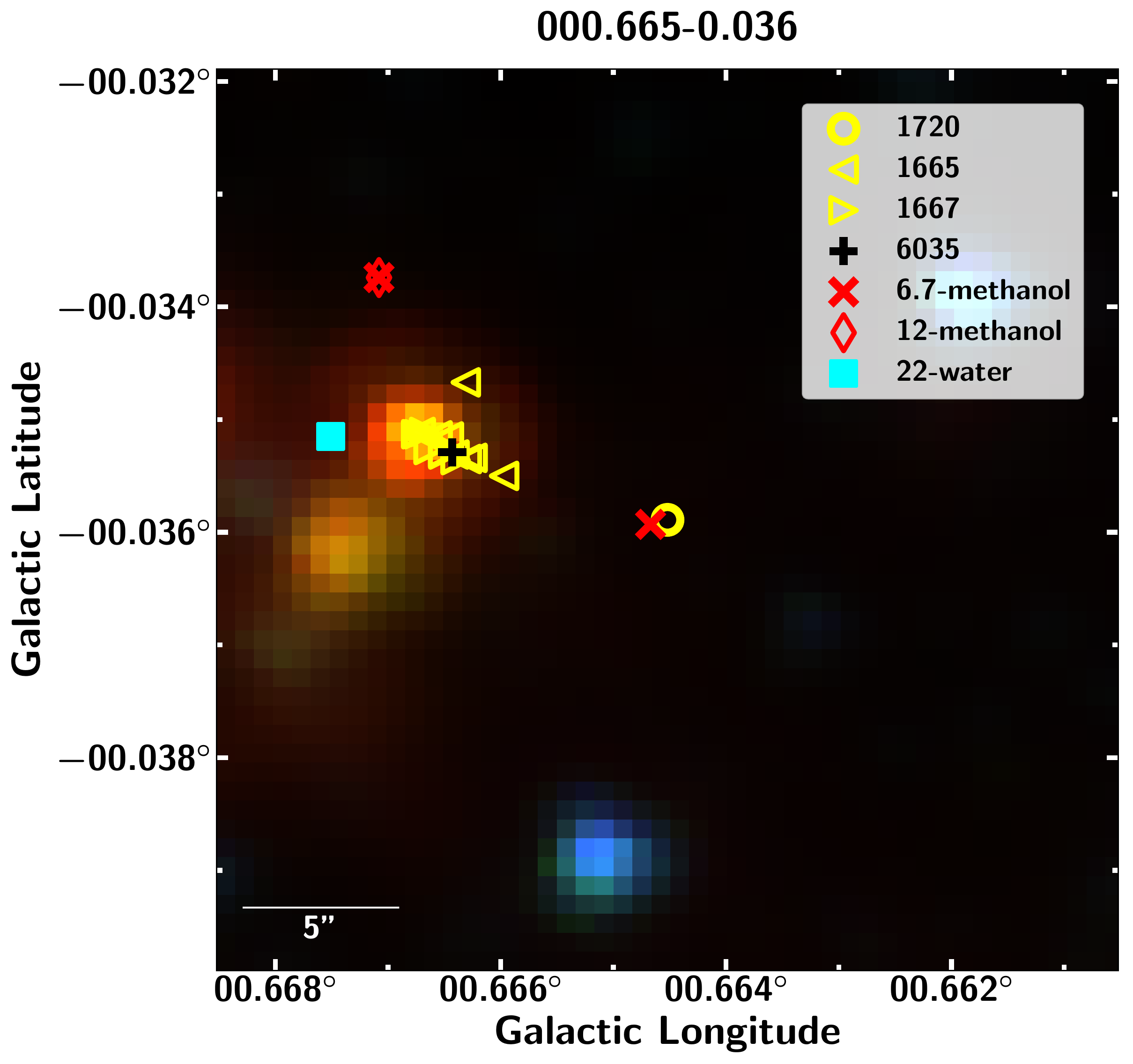}}\\
\subfloat{\includegraphics[width = 3.1in]{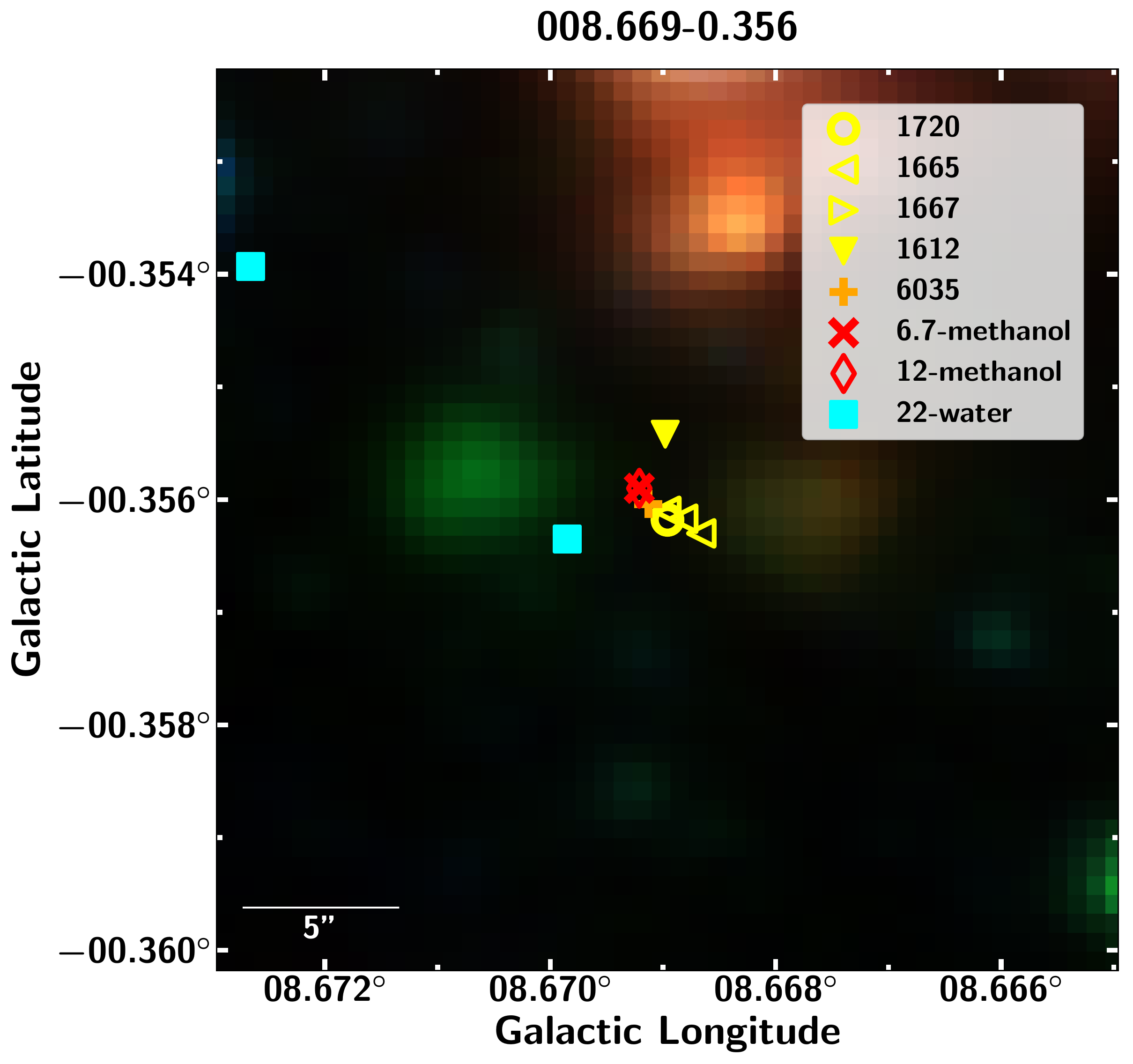}}
\subfloat{\includegraphics[width = 3.1in]{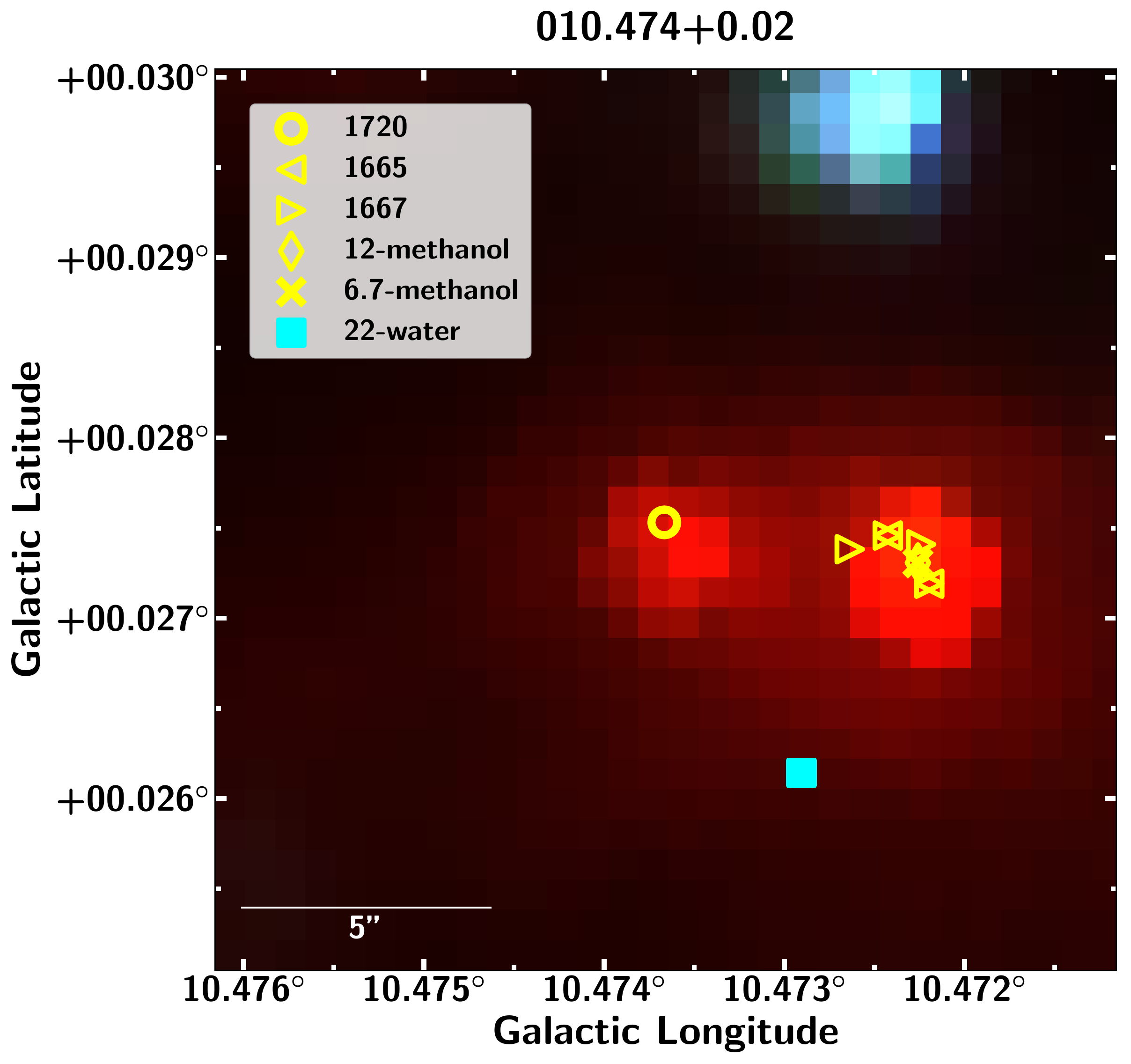}}\\
\end{figure*}

\begin{figure*}

\subfloat{\includegraphics[width = 3.1in]{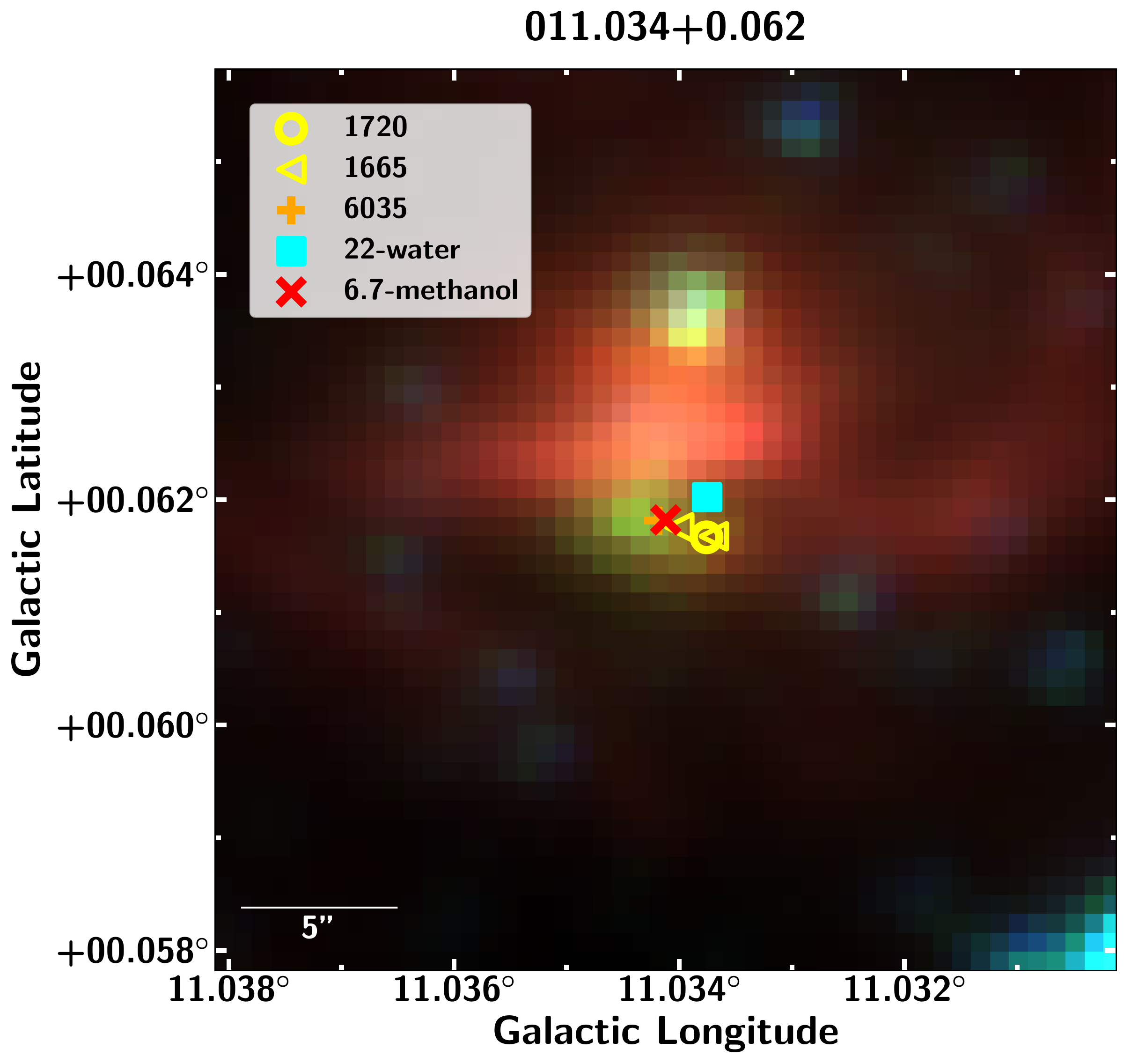}}
\subfloat{\includegraphics[width = 3.1in]{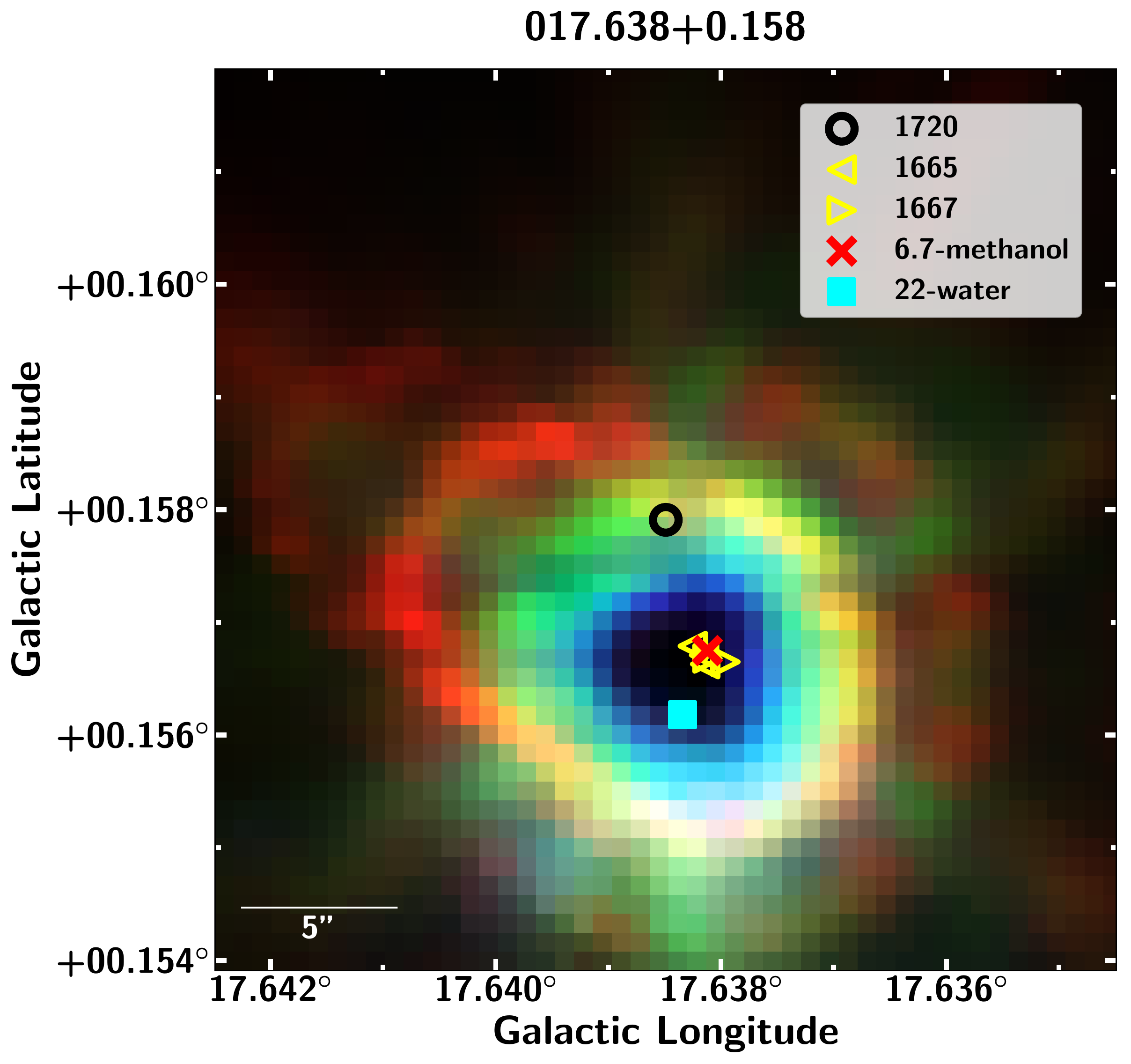}}\\

\end{figure*}





\bsp	
\label{lastpage}
\end{document}